\documentclass[letterpaper,11pt]{report}

\usepackage{fullpage}
\usepackage[parfill]{parskip}    
\usepackage{graphicx}
\usepackage{amssymb}
\usepackage{amsmath}
\usepackage{amsthm}
\usepackage{url} 
\usepackage{lineno} 
\usepackage{endnotes} 
\usepackage{longtable} 
\usepackage{booktabs}
\usepackage[font=small,labelfont=bf]{caption} 
\usepackage{graphicx}
\usepackage[all]{xy} 
\usepackage{color} 
\usepackage{datetime} 
\usepackage{boxedminipage}

\usepackage[T1]{fontenc}
\usepackage{currvita}

\usepackage{appendix}

\newboolean{figurelist}
\setboolean{figurelist}{false}

\definecolor{colora}{rgb}{.9,0,0}
\definecolor{colorb}{rgb}{0,0,0.9}
\definecolor{colorc}{rgb}{0,0.9,0}

\newenvironment
	{changea}%
	{%
	\color{black}%
	}%
	{%
	\color{black}%
	}%

	{%
	\color{black}%
	}%
	{%
	\color{black}%
	}%

	{%
	\color{black}%
	}%
	{%
	\color{black}%
	}%

\newcommand
	{\textbox}
	[2]
	{
	\setlength{\fboxsep}{0.5cm}
	\setlength{\fboxrule}{1pt}
	\begin{figure}
		\begin{boxedminipage}{1.0 \textwidth}
			\begin{center}
				\section*{#1}
			\end{center}
			\small
			#2
		\end{boxedminipage}
	\end{figure}
	}

\newcommand{\boxbreak}{\\*[3pt]}

\newcommand	
	{\incfig}	
	[3]	
	{
	\ifthenelse{\boolean{figurelist}}
		{
		\immediate\write\outstream{cp fig-#1.pdf arXiv}
		}
		{}
	\begin{figure}
		\begin{center}
			\includegraphics
				[width=#2\textwidth]	
				{fig-#1}
			\caption
				{#3}
			\label
				{fig:#1}	
		\end{center}
	\end{figure}
	}

\newcommand 
	{\incxy}
	[4] 
	[@C=8pt] 
	{
	\begin{figure}
	\begin{displaymath}
   		 \xymatrix#1{#4}
	\end{displaymath}
		\caption{#3}
		\label{fig:#2}
	\end{figure}
	}

\newcommand 
	{\tib}[1]
	{\small{\textbf{#1}}} 
\newcommand 
	{\tibs}[1]
	{\footnotesize{\textbf{#1}}} 

\newcommand	
	{\comment}	
	[1]	
	{
	\begin{samepage}
	\begin{center}
	\framebox{
		\parbox{13cm}{
		\textbf{Comment:}
			#1
		}
		}
	\end{center}
	\end{samepage}
	}
	
\newcommand 
	{\doctable}
	[5] 
	{
	\begin{table}[t]
	\centering
	#1
	\caption{#3}
	\label{tbl:#2}
	\medskip
	\begin{tabular}{#4}
	#5
	\end{tabular}
	\end{table}
	}

\newcommand 
	{\citeref}
	[1] 
	{\cite{RefWorks:#1}}

\newcommand 
	{\citerefWloc}
	[2] 
	{\cite[#2]{RefWorks:#1}}
	
\newcommand 
	{\citereftwo}
	[2] 
	{\cite{RefWorks:#1, RefWorks:#2}}

\newcommand 
	{\citerefthree}
	[3] 
	{\cite{RefWorks:#1, RefWorks:#2, RefWorks:#3}}
	
\newcommand 
	{\citereffour}
	[4] 
	{\cite{RefWorks:#1, RefWorks:#2, RefWorks:#3, RefWorks:#4}}

\newcommand 
	{\citereffive}
	[5] 
	{\cite{
	RefWorks:#1, RefWorks:#2, RefWorks:#3, RefWorks:#4,RefWorks:#5}
	}
	
\newcommand 
	{\citerefsix}
	[6] 
	{\cite{
	RefWorks:#1, RefWorks:#2, RefWorks:#3,RefWorks:#4,RefWorks:#5,RefWorks:#6}
	}

\newenvironment{makefigurelist}
	{
	\ifthenelse{\boolean{figurelist}}
		{
		\newwrite\outstream
		\immediate\openout\outstream=figure_list
		\immediate\write\outstream{!/bin/bash}
		\immediate\write\outstream{}
		}
		{}
	}
	{
	\ifthenelse{\boolean{figurelist}}
		{
		\immediate\closeout\outstream
		}
		{}
	}
	
\newcommand{\convolve}{\otimes}						
\newcommand{\cc}[1]{{#1}^{\ast}}                            			

\newcommand{\prob}[1]{ \text{Pr} \left\{ #1 \right\} }			

\newcommand{\snr}{\text{SNR}}						
\newcommand{\energy}{\mathcal E}						

\newcommand{\power}{\mathcal P}						
\newcommand{\powerpeak}{\power_{\text{peak}}}

\usepackage{mathpazo}
\usepackage[scaled=0.87]{berasans}
\usepackage[scaled=0.87]{beramono}
\normalfont
\usepackage[T1]{fontenc}
\usepackage{textcomp}

\newcommand{\expec}[1]{\mathbb E \left[ #1 \right]}
\newcommand{\mgf}[2]{\mathbb M_{#1} \left[ #2 \right]}
\newcommand{\var}[1]{\text{Var}\left[#1\right]}
\newcommand{\dm}{\text{DM}}
\newcommand{\deltadm}{\dm_{\Delta}}
\newcommand{\dc}{\mathcal{D}}

\newcommand{\pe}{P_{\text{\,E}}}
\newcommand{\psymbol}{P_{\text{S}}}

\newcommand{\snrin}{\snr_{\text{in}}}
\newcommand{\rate}{\mathcal R}
\newcommand{\ratenorm}{\tilde{\mathcal R}}
\newcommand{\capacity}{\mathcal C}

\newcommand{\pfa}{ P_{\text{FA}} }
\newcommand{\pd}{ P_{\text{D}}}
\newcommand{\pb}{ P_{\text{B}}}
\newcommand{\po}{P_{\text{O}}}

\newcommand{\vs}{\sigma_s^2}
\newcommand{\rf}{r_{\text{F}}}
\newcommand{\rdiff}{r_{\text{diff}}}
\newcommand{\rref}{r_{\text{ref}}}
\newcommand{\powere}{\mathfrak P}
\newcommand{\powerenoise}{\mathfrak P_{\text{max}}}

\newcommand{\energybit}{\mathfrak E_{\text{b}}}
\newcommand{\spectrale}{\mathfrak S}
\newcommand{\energyeff}{\energy_{\text{eff}}}
\newcommand{\oot}{\lambda_{\text{O}}}

\newcommand{\ecrs}{\xi_{\text{s}}}
\newcommand{\ecrszero}{\xi_{\text{s},0}}
\newcommand{\ecrsone}{\xi_{\text{s},1}}
\newcommand{\ecrb}{\xi_{\text{b}}}
\newcommand{\ecro}{\xi_{\text{b,o}}}
\newcommand{\ecreff}{\xi_{\text{eff}}}
\newcommand{\recenergy}{\energy_{h}}
\newcommand{\flux}{\mathcal{F}}
\newcommand{\pfaa}{P_{\text{FA,a}}}
\newcommand{\pda}{P_{\text{D},a}}
\newcommand{\Btotal}{B_{\text{total}}}

\title{\huge End-to-end interstellar communication system design for power efficiency}
\author{David G. Messerschmitt\\
Department of Electrical Engineering and Computer Sciences\\
University of California at Berkeley}

\date{\textbf{
	Version:} ArXiv Version 2 \\ 
	\textbf{Date}: \today \\ 
	\textbf{Time}: \xxivtime \\
	\copyright \   2013 David Messerschmitt. This work is licensed under a Creative Commons Attribution-NonCommercial-ShareAlike 3.0 Unported License.
	\url{http://creativecommons.org/licenses/by-nc-sa/3.0/}
	}

\begin{document}

\maketitle

\newpage
\tableofcontents

\newpage

\begin{makefigurelist}


\chapter*{Abstract}
\addcontentsline{toc}{chapter}{Abstract}

\begin{changea}
Radio communication over interstellar distances is studied, accounting for noise, dispersion, scattering and motion. Large transmitted powers suggest maximizing power efficiency (ratio of information rate to average signal power) as opposed to restricting bandwidth. The fundamental limit to reliable communication is determined, and is not affected by carrier frequency,  dispersion, scattering, or motion. The available efficiency is limited by noise alone, and the available information rate is limited by noise and available average power. A set of five design principles (well within our own technological capability) can asymptotically approach the fundamental limit; no other civilization can achieve greater efficiency. Bandwidth can be expanded in a way that avoids invoking impairment by dispersion or scattering. The resulting power-efficient signals have characteristics very different from current SETI targets, with wide bandwidth relative to the information rate and a sparse distribution of energy in both time and frequency. Information-free beacons achieving the lowest average power consistent with a given receiver observation time are studied. They need not have wide bandwidth, but do distribute energy more sparsely in time as average power is reduced. The discovery of both beacons and information-bearing signals is analyzed, and most closely resembles approaches that have been employed in optical SETI. No processing is needed to account for impairments other than noise. A direct statistical tradeoff between a larger number of observations and a lower average power (including due to lower information rate) is established. The ''false alarms'' in current searches are characteristic signatures of these signals. Joint searches for beacons and information-bearing signals require straightforward modifications to current SETI pattern recognition approaches.
 \end{changea}

\chapter*{Preface}
\addcontentsline{toc}{chapter}{Preface}

This is a report on my original research into interstellar communication at radio frequencies
over the past few years.
The ground covered by this report is extensive, including modeling of
the interstellar channel, power-efficient communications, and
power-optimized discovery of both information-bearing signals and beacons.
These topics have a lot of overlap and are all influenced by
a common set of considerations, which is why I group them in this report form
rather than separate them into individual papers.

To make this extensive subject matter more accessible, I have
started with three increasingly detailed summaries of the overall results,
starting with an executive summary, continuing with an introductory chapter
(Chapter \ref{sec:introduction}), 
and following that with a more detailed and technical summary chapter
(Chapter \ref{sec:powerefficient}).
The remaining chapters cover each constituent topic in greater depth.
A challenge with these remaining chapters is the
''missing the forest for the trees'' phenomenon.
Therefore, I recommend rereading Chapter \ref{sec:introduction} after
finishing Chapter \ref{sec:powerefficient}, and a second time after absorbing
the remaining chapters.
This should contribute to your overall understanding and
appreciation of the relationship among the various more detailed topics.

Another challenge in writing this report is two distinct target audiences,
physicists and astronomers on the one hand and communication engineers on the other.
I think both audiences will benefit from this report, and I want both audiences to
understand all the material.
To address this challenge, I have started from first principles, included more explanation
than usual, and specifically attempted to bolster intuition.
Fortunately the use of a power-efficiency methodology, which I argue is
appropriate for interstellar communications, simplifies the communications design methodology
significantly
(at least relative to terrestrial communications, which consistently employs a
spectrally-efficient design methodology).
Just as fortunately, power-efficient communication design also requires us to avoid
many interstellar impairments through transmit signal design,
which allows less detailed modeling of interstellar impairments
than is typical of the astrophysics literature.
If in spite of my efforts to make this material accessible to all,
should you have difficulty with any of my explanations,
please let me know so that I can improve future versions of this report.

To the communication engineers in our audience,
it isn't often that we get to play with
an entirely new application, one with distinct and interesting challenges.
Interstellar communication is one such opportunity.
I hope that you find this interesting and find yourself motivated
to perform additional research on the interstellar channel.
Given the single-minded emphasis of terrestrial communication on
spectral efficiency, you may find the power efficiency design methodology
here unfamiliar.
If you are like me, it will take you some time and effort to
detach your thinking from our discipline's obsession with
spectral efficiency and redirect it to power efficiency.
In my case this did not occur as a sudden epiphany, 
but rather developed gradually over the past several years of pondering
interstellar communications.
Sensitized to this challenge and opportunity, it should come more easily to you.

\section*{Change log}

As an aid to those who may have read an earlier version, the following
is a list of substantive changes for each version of this report.

\subsection*{Version 2}

\begin{enumerate}
\item
The overall implications of these results have been 
emphasized more strongly in Chapter \ref{sec:introduction}.
Interstellar impairments (other than thermal noise)
have no impact on power efficiency, 
leaving no dependence on carrier frequency.
During discovery, with proper design of the transmit signal 
no processing is required in the transmitter or receiver
to account for interstellar impairments.
During communication, all interstellar impairments can be avoided
except for noise and scintillation.
\item 
While Version 1 emphasized the possibility of reducing
the average power and energy consumption, it is equally possible
to use increased power efficiency as a way to increase the information
rate for a fixed average power.
This version thus emphasizes
(in Chapters \ref{sec:introduction} and \ref{sec:powerefficient}) 
that increased power efficiency is
valuable to both transmitter and receiver designers regardless of whether the energy
resources available to the transmitter designer are restricted or generous. 
\item
A concise summary of all the interstellar impairments that affect
energy bundles was added to Chapter \ref{sec:channelmodel},
in Section \ref{sec:impairmentreview}, as an aid for readers who
do not wish to tackle the detail of Chapter \ref{sec:incoherence}.
\item
A comparison to and some insights arising from optical SETI have
been added in Chapters \ref{sec:introduction}, \ref{sec:powerefficient} and \ref{sec:discovery}.
\item
In Version 1 the possibility of frequency-selective scintillation
was not taken into account in the outage strategy section.
Addressing this shortcoming required several additions:
\begin{itemize}
\item
An explicit modeling of the relationship between energy bundles
that do not overlap one another in time or frequency was added
to Chapter \ref{sec:channelmodel}, in Section \ref{sec:twobundles}, and
to Chapter \ref{sec:incoherence}, in Section \ref{sec:relationship}.
\item
An erasure strategy, appropriate when frequency-selective scintillation is present, was added to
Chapter \ref{sec:infosignals}, in Section \ref{sec:outageerasure}.
\end{itemize}
\item
A new section in which information is embedded in a beacon
without affecting its discoverability was added to
Chapter \ref{sec:discovery}, in Section \ref{sec:beaconWinfo}.
This highlights a design strategy in which discoverability with minimum receiver resources
is placed at a higher priority than power efficiency in conveying information.
\item
The Executive Summary and Chapters \ref{sec:introduction} and
\ref{sec:powerefficient} were updated to reflect these additions.
\end{enumerate}

\section*{Acknowledgements}

This research was supported in part by a grant from the
National Aeronautics and Space Administration to the SETI Institute.
The author is indebted to the following individuals for
generously sharing their time for discussion and comments:
James Benford of Microwave Sciences always made himself available as a sounding board. 
Gerry Harp and Jill Tarter of the SETI Institute and
Andrew Siemion and Dan Werthimer of the Space Sciences Laboratory at Berkeley
have all contributed greatly to my education on the 
characteristics of the interstellar medium.
Samantha Blair of the Joint ALMA Observatory was a collaborator on the issues
surrounding interstellar scattering, and in this she was assisted by William Coles of the
University of California at San Diego.
David Tse of the University of California at Berkeley
was helpful in identifying literature relevant to power-efficient design.
Ian Morrison of the Australian Centre for Astrobiology has been particularly supportive in maintaining an ongoing
dialog on issues surrounding the communications aspects of interstellar communications as well as reading and commenting on drafts of this work.

\chapter*{Executive Summary}
\addcontentsline{toc}{chapter}{Executive Summary}

\begin{changea}
The end-to-end design of an interstellar information-conveying communication system accounting for all known impairments due to radio propagation in the interstellar medium (attenuation, noise, dispersion, and scattering) and source/observer motion at radio wavelengths is undertaken. The emphasis is on modeling the impact of these impairments on communication, without detailed consideration of technological constraints. Numerous authors have noted the motivation to minimize average transmit power in interstellar communication due to large propagation loss, but the full implications have never been explored. Our system design thus revolves around minimizing the average signal power for a given information rate, or equivalently maximizing information rate for a given average power, which we call \emph{power-efficient design}. This design strategy equivalently seeks to minimize the energy expenditure required to convey a message of a given size. In existing SETI observation programs, it is typically assumed that the received signal has narrow bandwidth and is continuously received. We find that power-efficient information-bearing signals will be very different from this, trading a wider bandwidth for lower average power or higher information rate. These signals are also transmitted intermittently in time to conserve energy through either an improvement in power efficiency and/or a reduction in information rate. We also explore information-free beacons that minimize average power for a given receiver observation time, which we call \emph{power-optimized beacon design}. Power-optimized beacons need not have wide bandwidth, but they trade increasing intermittency of transmission for lower average power. In terms of ease of discovery, power-optimized beacons offer no advantage over power-efficient information-bearing signals, and a search for both types of signals can be conducted using a common strategy.

The fundamental limit to communication through the interstellar medium is determined, in the form of the maximum achievable power efficiency (information rate divided by average power) consistent with reliable recovery of information by the receiver.  Maximizing power efficiency is equivalent to minimizing the average power required to support a fixed information rate, maximizing the information rate achievable with a fixed average power, or minimizing the energy required to convey a message of a given size. This fundamental limit constrains any civilization, no matter how technologically advanced, and methodologies for approaching the fundamental limit provide compatible guidance to the transmitter and receiver designers. Power-efficient design is easily within the capabilities of our own current signal processing technology, although we do not consider technology issues surrounding high-power transmission of power-efficient signals.

A notable conclusion is that the fundamental limit on power efficiency is determined only by the thermal noise power, and the maximum achievable information rate is determined only by the average signal power, both measured at the point of baseband signal processing. Other interstellar impairments do not affect the achievable efficiency or rate, including dispersion, scattering, and motion. This we believe is a significant new insight that should have considerable influence on the choice of wavelength (including radio vs. optical) in interstellar communication. Arguments in favor of shorter wavelengths to minimize or avoid numerous impairments do not take into account our ability to design transmit signals that  achieve the same power efficiency as if these impairments were absent.

The fundamental limit demonstrates an unavoidable coupling between higher power efficiency and greater bandwidth. A narrowband signal has an efficiency penalty relative to the fundamental limit on the order of four to five orders of magnitude.  We propose a set of five straightforward power-efficient design principles, and show that they collectively allow the  fundamental limit to be approached asymptotically as the bandwidth is expanded. The remainder of the report focuses on finding practical measures based on these design principles that can substantially increase the power efficiency through the design of the transmit signal with compatible detection strategies at the receiver. Increases in power efficiency on the order of three to four orders of magnitude can be obtained at the expense of relatively wide signal bandwidth and a sparse distribution of signal energy in both time and frequency. Once the power efficiency has been increased in this fashion, the average power can be controlled by adjusting the information rate, with a lower information rate associated with lower average power and with greater sparsity of energy in the time domain.

According to our power-efficient design principles, information is communicated by on-off patterns of bundles of energy. The energy in each bundle is conveyed by a time-limited and bandwidth-limited waveform that must avoid most of the impairments due to interstellar propagation and relative transmitter-receiver motion. This is accomplished by conforming to the time- and frequency-coherence properties of the propagation and motion, so that the waveform arrives at the receiver with insignificant distortion. We call this the ''interstellar coherence hole''. These coherence properties are estimated for all known interstellar impairments, and it is shown that the resulting time-bandwidth product is sufficient to support power-efficient design at all carrier frequencies of interest without compromising the fundamental limit. Power-efficient design requires no processing related to most impairments in either transmitter or receiver; rather, they must be avoided by conscious design of the transmit signal. The two impairments that cannot be avoided in this fashion are noise (due to the non-zero temperature of the cosmos and receiver circuitry) and scintillation (a time variation in the signal flux at the receive antenna due to scattering in the interstellar medium interacting with receiver motion). Scintillation does not numerically affect the fundamental limit, but does profoundly influence the details of power-efficient design.

A power-efficient signal consists of energy bundles sparsely located in frequency, with multiple bits of information conveyed by a single bundle through its the frequency location. This is energy conserving and allows the signal to simultaneously have wide bandwidth (which is fundamentally required for high power efficiency) and avoid impairment due to the frequency incoherence of the interstellar propagation (because each energy bundle has a relatively narrow bandwidth). Each bundle must have sufficient energy to be detected reliably, both for extracting information and for discovery, and must have a time duration that conforms to the time coherence properties imposed by relative transmitter/receiver motion. For any average power and information rate corresponding to a power efficiency less than the fundamental limit, it is shown that the error probability approaches zero as the number of frequency locations (and hence bandwidth) increases and the sparsity of energy bundles in time increases. In this sense, the fundamental limit is approached asymptotically.

Two approaches to countering scintillation are studied. The first approach is time diversity, which can achieve consistent reliability in spite of scintillation and allow the fundamental limit to be approached as just described. Energy bundles are repeated redundantly at time intervals larger than the coherence time, and their respective energy estimates are averaged at the receiver. The second approach  is a simpler outage and erasure strategy, which deliberately sacrifices information arriving during periods of low signal power due to scintillation. This lost information has to be interpolated through redundant transmission of information, which is consistent with a single message repetitively transmitted.

The challenge of discovering information-bearing signals and beacons is also addressed. We propose the approach of searching for individual energy bundles, which does not require knowledge of the higher-level structure of the signal, and in particular is blind to whether the signal is information-bearing or an information-free beacon. This is a search strategy similar to what has been employed in optical SETI, with heterodyne demodulation and energy estimation employing a matched filter in place of incoherent photon-counting.

Because a beacon conveys no information, it may be narrowband. We find that the best way to reduce the average power of a beacon with minimum impact on the receiver is to maintain fixed energy per bundle (so that they can be reliably detected) and transmit bundles less often. An information-bearing signal and beacon with the same average power can be discovered with essentially the same difficulty using the methodology of detecting individual energy bundles and extending observation time in locations where such energy bundles are detected. In both cases, to achieve high discovery reliability at lower average powers multiple observations are required at the receiver to account for the intermittency due to sparsity in time and also due to scintillation. For a fixed number of observations at the receiver, both types of signals can be power-optimized, and in the case of information-bearing signals the information rate is implicitly determined by this optimization. For example, assuming a thousand independent receiver observations the average power of a beacon can be reduced by about three orders of magnitude relative to a continuously transmitted signal of the type that is currently the target of SETI searches. Reducing that number of observations requires signals to have higher average power to be reliably detected, and vice versa. The lowest average power of signals that can be detected in this fashion is limited only by the receiver observation time.

A SETI observation program can reasonably start with relatively few observations at each location (which will detect higher average power signals) and then extend the number of observations as resources permit (enabling the detection of lower average power signals).
Regardless of the average power, the transmitter must insure sufficient energy in each bundle arriving at the receiver (based on an assumption of minimum receive antenna collection area) to insure reliable detection of individual bundles. The transmitter controls the average power by adjusting the average rate at which energy bundles are transmitted, rather than the energy of each bundle. The transmitter can embed information in the signal without an impact on the discovery reliability, albeit with a limit on the power efficiency that can be achieved.

Power-efficient and power-optimized design have profound implications for both METI and SETI. They significantly alter both transmitted signals and receiver discovery techniques. First, multiple observations are required in a discovery search to detect lower average power signals, even as any processing related to interstellar signal propagation and motion is avoided. Second, post-detection pattern recognition must account for sparsity and randomness in frequency as well as time. Although existing and past SETI observation programs will fail to discover power-efficient and power-optimized signals, they can be fairly easily modified to seek such signals. Observation at even a single location (line of sight and frequency) should be considered a long-term project, and a worldwide shared database capturing individual energy bundle detections (today dismissed as ''false alarms'') can guide future observations to the most promising locations, no matter where and by whom they are performed. Compatible conclusions have been reached by other authors coming from different perspectives, including SETI at optical wavelengths and SETI at radio wavelengths taking into account cost considerations that arise in high-power transmitter design.
\end{changea}


\chapter{Introduction}
\label{sec:introduction}

A fundamental characteristic of the Milky Way is its capability to support the sharing of information using electromagnetic radiation among intelligent civilizations separated by interstellar distances. This capability is subject to various intrinsic physical impairments such as large propagation loss, natural sources of noise, dispersion and scattering of electromagnetic radiation in the interstellar medium (ISM) and time-varying effects due to relative source/observer motion. 
Although there is no possibility of direct experimentation, the ISM is fortunately well
characterized due to the modeling efforts of astrophysicists based in part on
observations by pulsar astronomers.
This report explores this information-carrying capability, both in terms of its fundamental limits and in practical terms. This is the first study that comprehensively takes into account impairments in the ISM
and due to motion in an end-to-end design of an interstellar communication system.
\begin{changea}
The impact of transmission impairments on the ability of interstellar space to support information transfer
at radio wavelengths is emphasized.
The technological implications, especially with respect to high-power transmitter design, are
not addressed in this report.
\end{changea}

\section{Transmitter-receiver coordination}

Prominent among the obstacles to establishing communication is the impossibility of prior coordination between the designers of a transmitter and a receiver \citeref{583}. 
The transmitter chooses a specific signal structure and parameterization, and to achieve high sensitivity a receiver discovering that signal (and subsequently extracting information) must also assume a specific signal structure and search over a large parameter space including spatial, frequency, and temporal variables. 
Yet prior to the establishment of communication there is no feasible way to directly coordinate these choices.
Fortunately there are several possibilities for achieving this coordination implicitly.

\subsection{Guidance from impairments and fundamental limits}

One common lament is the unlikely compatibility of technologies and
techniques in use across civilizations widely separated in
their degree of scientific and technological development.
Recent examples include Fridman \citeref{602}, who asserted that
''In each epoch we can only use our current technical knowledge to predict what kind of signals an ETI could send''.
A variation on the same argument is Horvat, Nak\'c, and Oto\v{c}an,
who asserted that there is a requirement for ''technological synchronicity'' 
between two civilizations \citeref{643}.
Coming from the special perspective of communication engineering, we
can disagree with these laments because there has been notable success
in identifying mathematically-provable and technology-independent limits that constrain
any civilization, no matter how advanced their technology may be.
These limits depend on the nature of the impairments encountered
in propagation of a signal from one civilization to the next,
and fortuitously these impairments are observable by both transmitter
and receiver designers and remain  unchanged over
time periods dramatically longer than the joint epoch of a transmitting and receiving
civilization (which is governed by the interstellar propagation delay).
In this report we hope to demonstrate the implicit coordination
attainable through these two invariants.
Furthermore, it will be demonstrated that our own civilization at its current
state of technological development is capable of approaching the
fundamental limit, particularly on the receiver side, with an appropriate set of design principles in place.
\begin{changea}
The technology implications to high-power transmitter design are not considered here,
but worthy of further study.
\end{changea}

\subsection{Guidance from resource constraints}

A less fundamental but nevertheless promising basis for coordination is 
reasonable presumptions about costs and objectives.
For example, both transmitter and receiver will recognize that a serious practical obstacle to communication is the ''needle in a haystack'' challenge in discovering a signal originating from an extraterrestrial source. At minimum the receiver has to search for a signal over starting time and carrier frequency. 
Expanding the dimensions of search requires additional processing and the energy consumption
associated with that processing,
and increased probability of false alarm (reducing sensitivity by necessitating a more strict criterion for detection).
Another example is the large average transmitted power which is necessary to
overcome the large propagation loss at interstellar distances,
suggesting to both transmitter and receiver designers a
motivation to minimize the average power of the transmitted
signal and the resultant energy consumption in the
transmitter.

Even though they impact both transmitter and receiver,
both the energy consumption in the transmitter
and the processing requirements in the receiver are largely dictated by the
transmitted signal design,
and this is a reason that the end-to-end perspective taken here can be valuable.
If both transmitter and receiver designers undertake parallel end-to-end designs,
and in particular appreciate the impact of their decisions on the other,
the interest of implicit coordination between their designs is served.
\begin{changea}
For example, in this report we presume that three coupled issues are of
interest to the transmitter and receiver designers:
\begin{itemize}
\item
There is a desire to maximize the information rate achieved with
reliable recovery of that information by the receiver and consistent with
an assumed expenditure of resources.
\item
There is a desire to minimize the energy consumption in the
transmitter required for the target information rate, or equivalently
a desire to maximize the information rate for a target energy consumption.
The long-term energy consumption is determined by the average transmit power.
\item
There is a desire to minimize the processing resources 
(and related energy consumption) required by the
receiver to reliably discover the signal.
\end{itemize}
\end{changea}
We  also consider the available tradeoffs between energy consumption in the transmitter
and in the receiver, and how these tradeoffs should be managed in light of economic and other considerations.
Whatever resource assumptions are made,
minimizing those resources places substantial constraints on the design,
and this is fortunate because it serves the best interest of implicit coordination.

\subsection{Guidance from average power and bandwidth}

For a given set of impairments in the interstellar medium and due to
relative motion, there is a fundamental limit on the information rate
(usually measured in bits per second)
at which reliable communication can occur.
In communication engineering,
calculation of the fundamental limit usually assumes that two
resources are constrained: the average power of the signal and the bandwidth of the signal.
Subject to these constraints it seeks the highest information rate that can be achieved 
consistent with reliable recovery of the information at the receiver.
The feasible information rate must increase when the constraint on average power
is relaxed, and also when the constraint on bandwidth is relaxed, since the
transmitter always has the option to utilize less power or less bandwidth than permitted
by the constraint.
This implies that increasing average power is always beneficial, and increasing bandwidth is always beneficial
as well, in the sense of increasing the achievable information rate.

The existence of a fundamental limit is of great import to interstellar communication,
for a couple of reasons.
It gives an absolute reference against which to compare any concrete design,
and reveals precisely how much additional performance can be wrung out 
by changes to the design.
More significantly, circumventing the various impairments in interstellar propagation and
due to the relative motion of transmitter and receiver,
with the goal of moving toward the fundamental limit drives the design
in certain directions.
In fact,
the closer a design is able to achieving a 
performance that approaches the fundamental limit, the more
constrained that design becomes.
Thus the fundamental limit serves as source of implicit coordination,
because if the transmitter and receiver designers examine the
end-to-end communication problem independently,
they will arrive at compatible conclusions about design approaches for both transmitter and receiver.
The closer the transmitter seeks to approach the fundamental limit, the greater the receiver's confidence
 in the characteristics of the transmitted signal.

There is one remaining element of uncertainty or dissidence that could
interfere with the strategy of implicit coordination by fundamental limit, and that
is the relative importance to attach to average power and to bandwidth.
These two design parameters play a similar role in communications to
position and momentum in the Heisenberg uncertainty principle of physics.
Just as more accurate knowledge of position of a particle fundamentally implies less
accurate knowledge of its momentum,
constraining the bandwidth of a communication signal
more tightly implies fundamentally that the signal must have
a greater average power, and vice versa.
Depending on whether we place our priority on bandwidth or on average power,
the design is driven in very different directions.
In terrestrial radio communications, priority is virtually always placed on bandwidth
because of the high economic value attached to the radio spectrum.
In interstellar communication, in contrast, we argue that the priority should
be placed on average signal power due to the large energy requirements
placed on radiated power at the transmitter due to the large propagation losses,
and due to the large microwave window bandwidth available for transmission.
This assumption leads to the approach of power efficient design, which is described next.
This is
a very different design approach from that prevalent in terrestrial communication.

\section{Power-efficient design}

The idea behind power efficient design is to assume that average power is the
only limiting resources, and that bandwidth does not need to be constrained.
By eliminating any bandwidth constraint, a global minimum in average
signal power is achieved.
In addition, by eliminating any bandwidth constraint
the communication design problem is greatly simplified,
which is of great benefit in the context of interstellar communication where
explicit transmitter-receiver design coordination is not feasible.

\subsection{Why average power is important}

Authors addressing SETI have noted the importance of energy and power
arising from the challenges of overcoming an extraordinarily large propagation loss
at interstellar distances, as well as the relative unimportance of
signal bandwidth due to the wide microwave window that is available.
For example, H. Jones argued that bandwidth is a relatively free resource in interstellar
communication due to the broad microwave window \citeref{196}.
P. Fridman recently asserted that power is surely a bigger issue than bandwidth \citeref{602}.
\begin{changea}
The Benford's consider the design of microwave high-power transmitters, and find that
peak and average power are a major capital and operational cost factor for our current technology \citereftwo{568}{569}.
All these authors follow up by proposing design approaches to minimizing average power.
This report follows a similar line of reasoning, adding an accounting of the full range of
impairments in the interstellar propagation and due to transmitter/receiver motion
to the mix of design considerations.
\end{changea}

There is an important distinction between average and peak transmitted power.
Average transmit power is directly related to the transmitter's energy consumption,
which is a major part of transmitter operational costs.
As a concrete example, at a modest information rate of one bit per second and a distance of
one thousand light years, the average transmitter radiated power 
for the type of bandwidth-constrained
signals targeted by present SETI searches and assuming
Arecibo-sized transmit and receive antennas is about twice
that of the Large Hadron Collider at CERN.
Average power also determines the thermal management challenges for high-power transmit electronics,
and thus is more significant than peak power in the construction cost of a transmitter \citeref{646}.
If the transmitter is performing a targeted search, average transmit power may have
a large multiplier due to concurrent transmission on multiple lines of sight,
or alternatively a lower gain (more omnidirectional) transmit antenna may be used,
either of which necessitates a commensurately higher average power.

\begin{changea}
Fridman argues that large transmit powers will likely result in low information
rates achievable at interstellar distances \citeref{602}.
We agree.
On the other hand, whatever average power and energy consumption is allocated by
the transmitter, there is a mutual interest on the part of both transmitter and receiver in 
maximizing the information rate.
This is the objective of power-efficient design.%
\end{changea}

\subsection{Why bandwidth is less important}

As long as we deal with some frequency-dispersive effects successfully,
bandwidth is hardly constrained by the long-distance propagation
and impairments in the interstellar medium.
Of course the available bandwidth is not unlimited, because there is a finite
microwave window where the interstellar medium is highly transparent. 
But in this report we assume that the bandwidth is unconstrained.
In practice this means that the bandwidth required will be relatively
large relative to the information rate, but not infinite.
All concrete designs proposed here assume a total bandwidth considerably less
than what is available in the microwave window.

\begin{changea}
The interstellar propagation introduces frequency-dependent effects which will
definitely affect wide bandwidth signals.
A fundamental constraint on the design is to make sure that these effects do not
result in a reduction in power efficiency,
and a practical constraint is to make sure they do not impose large processing
requirements on either the transmitter or the receiver.
We will find that both of these objectives are easily met.
\end{changea}

If we assume that the information
rates will be low relative to what is typical in terrestrial communication,
this implies that the reception time required for a given
size of message is longer, but this also seems perfectly acceptable
for interstellar communication where extreme patience is required
on the part of both transmitter and receiver even in the best of circumstances.
When the information rate is low, 
it follows that bandwidth will be commensurately low
even without going to the trouble to explicitly constrain the bandwidth.
Our own electronics and processing technologies are 
capable of supporting large bandwidths
with relative ease, and this is likely true of other civilizations at least as
technologically advanced as our own.

In the remainder of this discussion, assume that the bandwidth is
unconstrained.
We will simply let the bandwidth fall where it may, since our priority is on achieving
the lowest feasible average power.
Seeking to reduce the bandwidth by constraining it can only increase the
average power that is required, and that is not the direction we wish to go.
Of course if the required bandwidth turns out to be impractically large, 
this strategy can be modified, albeit with a considerable complication of the design.

\subsection{Information rate vs. average power tradeoff}

The average signal power depends directly on the information rate.
This is illustrated in Figure \ref{fig:powerefficiencyconcept}, which illustrates the special case
where the information rate and average power are proportional to one another.
That this dependence is typically linear is not surprising.
A typical scheme translates a group of information bits into some waveform that
is transmitted, and
adjusts the information rate by adjusting how often that waveform 
(with different information bits) is transmitted.
With this basic scheme,
adjusting the rate of transmission (and hence the information rate) translates directly to a commensurate adjustment to the average power.

\incfig
	{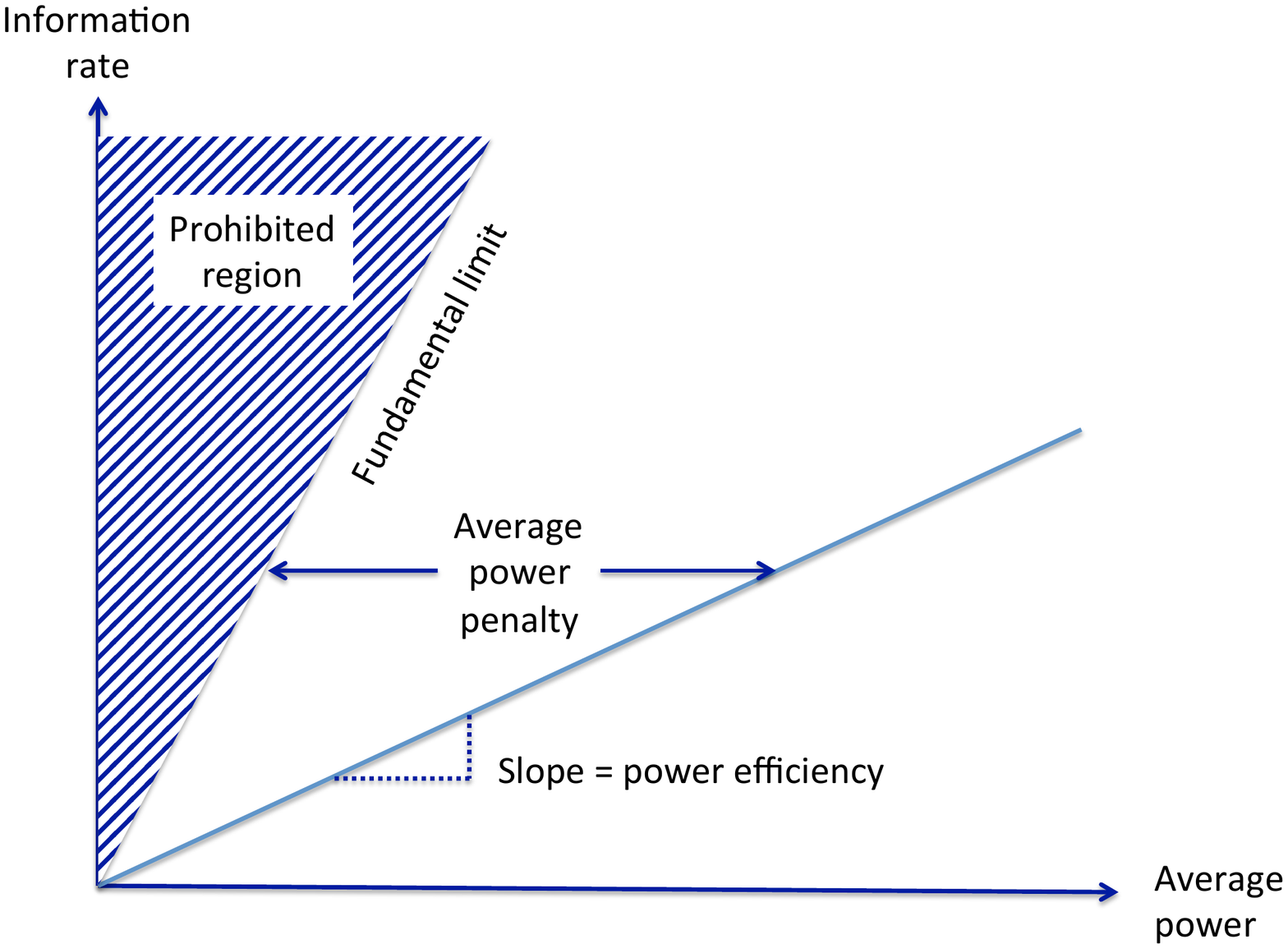}
	{1}
	{An illustration of the relationship between information rate and
	average power.
	There is a prohibited region where communication is not possible
	because the average power is insufficient to support the information rate.
	The lower boundary of that prohibited region is called the fundamental limit.}

Also shown is a prohibited region where reliable communication is not feasible because
the average power is not large enough
to overcome the noise and other impairments and still achieve reliable extraction of
the information at this rate.
The lower boundary of this region is called the \emph{fundamental limit}..
This fundamental limit applies to any civilization, no matter how technologically advanced.
The existence of such a prohibited region was established by the seminal work
of Claude Shannon in the late 1940's \citeref{164}.
A mathematical model for the end-to-end propagation of the signal from transmitter to receiver
is called a communication \emph{channel}.
Shannon's fundamental limit is therefore called the \emph{channel capacity}, and its details
depends on the specific combination of impairments encountered
in the communication channel, so every case has to be considered separately.
The fundamental limit is determined in this report for the specific impairments encountered on the interstellar channel with unconstrained bandwidth,
and the not-too-surprising conclusion is that this fundamental limit on information rate
is also linear in average power, just
like the rate-vs-power dependence of a typical concrete communication system design.

Since the variation in information rate is linear in average power,
for both the fundamental limit and for typical concrete designs,
it is convenient to characterize the performance in terms of a single parameter, the slope of that line.
That slope is called the \emph{power efficiency}, and it is defined as
\begin{equation*}
\text{Power efficiency} = \frac{\text{Information rate}}{\text{Average power}} \,.
\end{equation*}
The power efficiency, which is a constant that does not depend on either average power or
information rate (their ratio remains fixed even as they both vary), is a measure of how efficiently additional average
power is converted into an increase in information rate.
The higher power efficiency associated with the fundamental limit is a
desirable objective because it permits a higher information rate for a given expenditure of
average power.
At a given information rate there is a penalty in power efficiency for any
concrete design relative to the fundamental limit, and hence a larger-than-necessary
average power.

Figure \ref{fig:powerefficiencyconcept} is intended to convey the concept of power efficiency,
but is misleading as to scale.
Over the range of signal designs considered in this report, the power efficiency can range over
four to five orders of magnitude.
Thus, the slope of the rate-vs-power curve may be dramatically smaller than the illustration
for a given concrete design.
If power efficiency is considered an important metric of performance,
the chosen specifics of a concrete design can make a major difference.

The main competitor to power efficiency is spectral efficiency,
which is defined as
\begin{equation*}
\text{Spectral efficiency} = \frac{\text{Information rate}}{\text{Signal bandwidth}} \,.
\end{equation*}
In communications the fundamental tradeoff between
average power and bandwidth is easily reformulated as a tradeoff between
power efficiency and spectral efficiency.
Design techniques aimed at increasing one inevitably reduce the other.
In this report power efficiency is emphasized, which implies that the
spectral efficiency will be relatively poor.

\subsection{Analog signals}

Our description thus far assumes the information being conveyed from transmitter to receiver is
digitally encoded.
In contrast, there has been experimentation with
the analog interstellar transmission of a signal (such as music) \citeref{586}.
Shannon dealt with this case, showing that analog signals can be digitally encoded, decoded, and
recovered, and there is a fundamental limit to the fidelity with which this process occurs and a tradeoff
between this fidelity and the digital information rate required.
Thus, the fundamental limit on power efficiency dealt with here applies
equally to analog signals,
with an additional dependence between the fidelity with which the signal is recovered
and the information rate and average signal power required for its transmission.

A major reason why digital representations of analog signals are preferred
in practice is that the average power necessary to transmit such a signal
digitally is dramatically lower than would be possible with
the direct transmission of the analog representation.
This difference is primarily due to compression (redundancy removal) possible 
during the
conversion from analog to digital as well as the availability of techniques
for increasing the power efficiency for the digital transmission
like those discussed in this report.
\begin{changea}
Thus digital representations of analog signals (like music)
combined with digital transmission are considered to be overwhelmingly
advantageous in terrestrial communication, and are employed in
all recent communication system designs.
It is likely that those advantages transfer over to interstellar communication,
although the lack of explicit coordination is an additional factor that
must be taken into account.
This lack of coordination argues against techniques that are complicated
and hence difficult to ''reverse engineer'' by the receiver.%
\end{changea}

\subsection{Channel coding: changing the power-rate tradeoff}

\begin{changea}
For a concrete design methodology,
in the absence of a bandwidth objective or constraint
there is typically a linear tradeoff between the average power of the
received signal and the information rate at which reliable extraction
of that information can be achieved by the receiver.
This is illustrated by the dashed lines in Figure \ref{fig:channelcodingconcept}.
In view of this linear tradeoff,
\emph{any} average power can be achieved by
adjusting the information rate,
or \emph{any} information rate can be 
achieved by adjusting the average power.
The slope of this linear tradeoff is equal to the
power efficiency characteristic of that particular design methodology.
Our goal in power-efficient design is to increase that slope,
so that it is closer to the slope associated with the fundamental limit.
In that case the tradeoff becomes more efficient, meaning
a higher information rate can be achieved for a fixed average power,
or a lower average power can be achieved for a fixed information rate.
No matter what energy resource is available to
the transmitter designer, whether it be generous or limited,
higher power efficiency is advantageous
to both the transmitter and receiver as it results in a higher information rate.
\end{changea}

\incfig
	{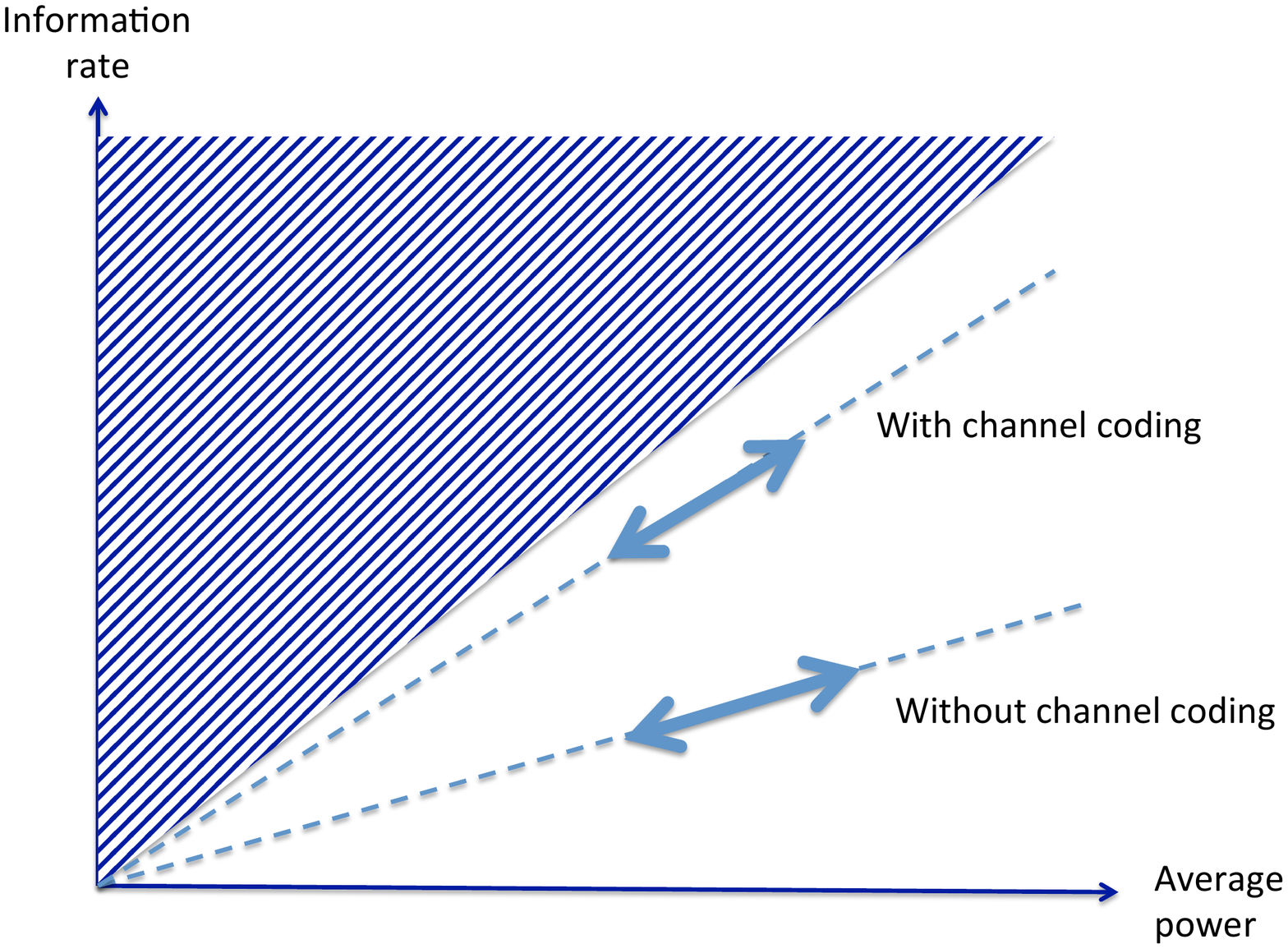}
	{1}
	{
The motivation behind channel coding is to
increase the power efficiency,
which practically speaking means increasing the 
slope of the linear
tradeoff between average power and information rate.
	}

Shannon proved that it is feasible to approach the fundamental limit on power efficiency by using
\emph{channel coding}.
What is channel coding?
The simplest communication schemes we might conceive might map
information bits into a waveform transmitted on the channel, and do so one bit at a time.
For example, an approach called on-off keying (OOK) transmits a signal
waveform to convey a ''one'' bit, and the absence of a signal waveform to convey a ''zero'' bit.
Another example would be phase-shift keying (PSK), which transmits a sinusoid
with zero phase or a sinusoid shifted in phase by 180 degrees.
The idea behind channel coding is a straightforward but powerful generalization of
this simple approach.
While one bit represents two alternatives (like ''on'' or ''off''),
a group of $M$ bits represents $2^M$ alternatives.
So after dividing the information into groups of $M$ bits,
each group of $M$ bits can be represented by transmitting one of $2^M$ waveforms,
which are called \emph{codewords}.
The receiver reverses this process, deciding which of the $2^M$
waveforms is the most likely to have been transmitted, and mapping the
result into $M$ bits.
The essential characteristic of channel coding is that the receiver cannot discern individual
bits by looking at the waveforms in a simpler way, but is forced to recover or extract
all $M$ bits together.
A very simple example would be a variant on OOK, in which the transmitter
starts with two bits, and transmits one of four OOK sequences,
either $\{ \text{ON}, \text{OFF}, \text{OFF}, \text{OFF} \}$
or $\{ \text{OFF}, \text{ON}, \text{OFF}, \text{OFF} \}$
or $\{ \text{OFF}, \text{OFF}, \text{ON}, \text{OFF} \}$
or $\{ \text{OFF}, \text{OFF}, \text{OFF}, \text{ON} \}$.

\begin{changea}
What does it mean to approach the fundamental limit?
Specifically, Shannon proved that for any information rate
and average power corresponding to a power efficiency 
outside the prohibited region in Figure \ref{fig:powerefficiencyconcept}, 
and for any (arbitrarily small) error rate, there exists a channel code
which can achieve that error rate.
A caveat is that the fundamental limit plays a role analogous to the
 speed of light in special relativity, in that
approaching closer and closer to the fundamental limit requires
a greater and greater expenditure of resources,
and thus in practice we expect there to be a gap between the achievable
power efficiency and the theoretical limit.
In the case of relativity, a non-zero mass particle accelerated
closer and closer to the speed of light requires a greater and greater
expenditure of energy.
In the case of channel coding, approaching closer and closer to the fundamental limit
on power efficiency requires a greater and greater expenditure of
bandwidth.
Thus, for any finite bandwidth, there will be a gap between the
achieved power efficiency and the fundamental limit,
but that gap can be reduced as the bandwidth is increased further.
\end{changea}

Shannon's result applies more generally than presented here,
because it can apply equally well to the constrained-bandwidth case.
Terrestrial communication overwhelmingly seeks to minimize signal bandwidth
as a priority over minimizing average power, due to the scarcity of spectral allocations
and the resulting high monetary value attached to spectrum.
In terrestrial wireless communication the radiated power at the transmitter is usually 
small relative to the power consumed in the
processing required in the receiver to deal with small signal bandwidth,
so substantial penalties in radiated power are normally quite acceptable as one price
paid for reducing the consumed bandwidth.
Appropriate channel codes for improving power efficiency with
unconstrained bandwidth are very
different from the approaches we are accustomed to in terrestrial communication.

\subsection{Guidance from the fundamental limit}

This report establishes the fundamental limit on power efficiency
for the interstellar channel, taking into account all known propagation
and motion impairments.
This establishes the maximum power efficiency that is theoretically
consistent with reliable communication in the interstellar channel.
The power efficiency that can be achieved with signal designs
 found in the interstellar communication literature are necessarily smaller,
and can be compared to this fundamental limit.
 A significantly more technologically advanced civilization
is subject to the same fundamental limit on power efficiency as are we, 
and if they pursue the
power-efficient design approach
we would expect them to arrive at similar design principles.
The notable exception to this occurs when a civilization is more concerned
about minimizing bandwidth at the expense of higher average power, in which case they will necessarily choose
a very different design approach.

While achieving the fundamental limit is
no more practically feasible than accelerating a particle with non-zero mass to the
speed of light,
a worthy goal of our design is to approach as close to the limit as we can.
It will be shown that a set of five simple principles of power efficient channel-coding design
can increase the power efficiency substantially.
In fact, for any information rate and average power falling outside the forbidden region,
it will be shown that
application of these principles in combination can asymptotically
achieve an error rate approaching zero.
By asymptotically, we mean that certain parameters of the design
grow without bound.
On the other hand, staying in the realm of what might be considered a practical choice of these parameters,
there remains a penalty in power efficiency relative to the fundamental limit, which we find is
on the order of one order of magnitude.

In addition to energy consumption, use of these five principles
of power efficient design can provide
a powerful form of implicit transmitter-receiver coordination.
The fundamental limit will presumably be known to both the
transmitter and receiver, and a design that seeks to approach that limit
will be a goal of any transmitter designer that considers average
power and the resulting energy consumption to be a more limiting resource than bandwidth.
The design of a channel code capable of approaching the limit is certainly not unique,
but in our judgement the channel coding we demonstrate is the
simplest way to achieve this goal.
All five principles are themselves simple, each solves a significant challenge posed
by impairments on the interstellar channel, and they are complementary to one another.
In addition, we show that it is relatively straightforward to discover signals
designed in accordance with these principles, whereas more complicated approaches
are likely to compromise discovery and render the reverse engineering of the
details of the channel code more difficult for the receiver.
Research into alternative approaches to power-efficient channel coding is nevertheless warranted.

\subsection{Structure of power-efficient signals}

The design principles we espouse 
result in a transmit signal with very specific characteristics.
Again, we can't say with certainty that this is the only way to approach the
fundamental limit, but in our judgement it is the simplest way.

A power-efficient signal can be expected to consist of
on-off patterns of energy (which we call ''energy bundles'').
\begin{changea}
These energy bundles are similar to the pulses of light that are
considered in optical SETI \citeref{655}, except that at radio wavelengths
heterodyne demodulation and matched filtering at baseband open up significant opportunities.
\end{changea}
An energy bundle consists of a bandlimited and time-limited waveform
which can be positioned in both time and frequency.
The frequency is controlled by choice of the carrier frequency used
to modulate the waveform.
The receiver estimates the energy in bundles 
by calculating the magnitude-squared of the output of a matched filter,
without seeking to recover any phase information.
This energy estimate can be applied to a threshold to decide
whether an energy bundle is present at a particular time and
frequency, or not.

Each codeword in a power-efficient channel code
consists of a specific pattern of energy bundles in time and frequency.
It is the job of the receiver to discern what pattern is present, and associate
that pattern with a recovered set of information bits.
The defining characteristic of this set of codewords is the
high sparsity (or low density) of energy bundles in both time and frequency.
Sparsity in frequency results in a large bandwidth relative
to the information rate, which is a mandatory characteristic of any channel code
capable of approaching the fundamental limit on power efficiency.
The purpose of this frequency sparsity is to convey \emph{multiple bits} 
of information by the location of a \emph{single} energy bundle,
thereby conserving energy.
For a fixed average power,
sparsity in time enables each bundle to contain a larger amount of energy than
would be possible if the transmitted signal were more continuously transmitted,
and this larger energy allows the presence of the bundle to be more reliably detected
during both communication and discovery.
The purpose of sparsity in time is thus to improve the reliability with which
an energy bundle can be distinguished from noise,
and this is how the error rate can be driven asymptotically towards zero.
A simple approach to discovery of the information signal is to
search for and detect individual energy bundles.
In fact this is the only way, since the frequency location of energy bundles
is random depending on the information being communicated.

\begin{changea}
Suppose the information rate, the average power, and the error rate are all fixed.
A subtle but important point is that
the rate at which energy bundles are transmitted can be changed.
To hold the information rate fixed, as that bundle rate is reduced the number
of candidate frequency locations must be increased to compensate, thereby increasing the total signal bandwidth.
To hold the average power fixed, as the bundle rate is reduced the 
energy per bundle can be increased to compensate.
Increasing the energy per bundle tends to render the detection of the bundle location more reliable, and
thus the target error rate can be achieved.
As the error rate target is made more stringent, the energy per bundle has to increase, and
thus the bandwidth also has to increase.
The sparsity of energy bundles increases in both time and in frequency.
In this fashion the fundamental limit is approached.
These neat tricks work only if the chosen information rate and average power
correspond to a power efficiency less than the fundamental limit.
The reason has to do with the false detection of energy bundles at locations where they were not transmitted.%
\end{changea}

\subsection{Interstellar coherence hole}

\begin{changea}
Radio wavelengths, especially at lower carrier frequencies,
have a poor reputation in the SETI community
due to the increasing prevalence of dispersion and scattering.
This reputation is not deserved, however, because through appropriate
design of the transmit signal these impairments can be avoided at all carrier
frequencies of interest and all wavelengths.
This avoidance strategy does not reduce the achievable
power efficiency, and for this reason none of the interstellar impairments
(except thermal noise) have an effect on
the fundamental limit on power efficiency, nor on the
achievable power efficiency in practice.
The same would not be true for design with constrained bandwidth,
so this is another indication of the desirable simplicity of power efficient design.%
\end{changea}

Transmitted energy bundles are optimally detected at the receiver
in the presence of noise by using a matched filter.
The waveform that is chosen by the transmitter designer as a ''carrier'' for energy bundles
can be specifically designed to avoid
most of the interstellar impairments, specifically by restricting the waveform's
time duration and bandwidth.
The resulting waveform falls within the coherence time and the coherence
frequency dictated by the interstellar propagation and transmitter-receiver motion.
When as many impairments as possible are avoided in this fashion,
the transmitted energy bundles are said to fall in an 
\emph{interstellar coherence hole} (ICH).

There are two remaining impairments that cannot be avoided
by reducing the energy bundle time duration and bandwidth: noise and scintillation.
At radio frequencies noise is due to thermal effects such as the
background radiation of the cosmos, star black-body radiation, and the finite temperature of the
receiver's circuitry.
Scintillation is due to scattering caused by the ionized nature of the interstellar medium,
the spatial inhomogeneities in ionization density,
and these two factors
working in combination with the lateral motion of the receiver and turbulence in the
interstellar clouds of ionized gas.
To the designer of the receiver,
scintillation is manifested by a time variation in the average signal flux arriving at the receive antenna.

Making use of the ICH in the design of energy bundles does not compromise
the available power efficiency.
In fact, it is a necessary step to approaching the fundamental limit on power efficiency,
or in other words
violating the ICH and thereby introducing impairments
in addition to noise and scintillation will prevent the fundamental limit
from being approached. 
The time duration and bandwidth of the resulting energy bundles are
constrained as to parameterization based on physical quantities
of the interstellar medium that can be observed and estimated by both transmitter and
receiver designers, providing another form of implicit coordination.
In fact, we should consider ourselves fortunate
because the very existence of interstellar and motion impairments
as our ability to observe and model them provides a wealth of information about the
nature of channel codes and signal waveforms designed specifically to avoid those impairments.

\subsection{Countering scintillation}

The basic building block of an energy bundle confined to the ICH followed
by a matched filter in the receiver is the optimum way to counter noise.
It does nothing to deal with scintillation, however.
The fundamental limit on power efficiency takes into account scintillation,
and therefore asserts that reliable recovery of information is possible in spite of scintillation.
\begin{changea}
Perhaps surprisingly, scintillation has no influence on the
power efficiency at the fundamental limit, because it causes a variation in the
received average power but not the statistical average of that average power.%
\end{changea}
The goal of a power-efficient channel code is to achieve reliable information
recovery at all times, even in the presence of scintillation.
\begin{changea}
The design of channel codes for interstellar communication is substantially
impacted by scintillation, even as the achievable power efficiency is not.%
\end{changea}
We show that scintillation does not compromise the ability to approach the fundamental
limit if \emph{diversity combining} is included among the five design principles.
This strategy divides the total energy of a single energy bundle into
a multiplicity of bundles which are transmitted with a large enough
spacing in time or frequency
that they experience statistically independent scintillation conditions.
The receiver averages the energy estimates for individual bundles
to obtain a more accurate estimate that is less affected by scintillation.

An alternative \emph{outage strategy} abandons the goal of reliable recovery
of all the information, but simply seeks reliable recovery of information
during periods of higher-than-average received signal flux and ignores
the received signal during outages, defined as periods of lower-than-average flux.
With this simpler strategy,
an additional burden is placed on a transmitter to encode the information in such a way that
massive losses of data during outages can be recovered at the receiver,
presumably through another layer of redundancy.
In interstellar communication, the same message is likely transmitted repeatedly,
since the transmitter has no idea when a receiver may be listening, 
and if so this redundancy
provides a natural way to recover the information lost due to outages,
and to a large extent information that is corrupted by random errors as well.

\subsection{Simplicity of power-efficient design}

One major benefit of power-efficient design in the context of interstellar communication
is its simplicity, which is quite beneficial when the transmitter and receiver
designs are not explicitly coordinated.
Removing the constraint on signal bandwidth simplifies design
and it is relatively easy to identify simple and practical design principles that
achieve high power efficiency.
This is in stark contrast to the types of channel coding schemes used in terrestrial communications,
which can be said without exaggeration to be vastly more complex and opaque than
what we have described for the power-efficient case.

One manifestation of this simplicity is due to the ICH.
The approach of the ICH is to design the transmit signal to \emph{avoid} as many interstellar
impairments as possible, rather than trying to \emph{compensate} for them in either
transmitter or receiver as would typically be attempted in terrestrial communication.
This implies that there is no processing required in the receiver related to interstellar
impairments, other than noise and scintillation, during either communication or discovery.
Bandwidth is increased modestly with this approach, but this is not an issue
when bandwidth is unconstrained.

This simplicity should allow
a receiver to readily ''reverse engineer''
a signal designed according to power efficiency principles
to recover the ''raw'' information content.
Of course inferring the ''meaning'' of that information entails a
whole other level of difficulty that is not addressed in this report.

\subsection{Discovery of power-efficient signals}

It serves no useful purpose to transmit a signal with impressive power efficiency, but which
 is very difficult for the receiver to discover in the first place.
It also serves little useful purpose to require another attention-attractive ''pilot'' or
''beacon'' signal to accompany the information-bearing signal,
if that beacon requires a relatively high average power, 
substantially increasing the power overall.
Thus, one significant design constraint in power efficient design is to
insure that the signal can be discovered with a reasonable level of resources
applied in the receiver.
This issue is made more critical by the
''needle in a haystack'' problem of lack of prior knowledge
as to the line of sight and frequency where a signal is to be found.
Thus the challenges posited by the discovery of signals designed
according to power-efficient principles is addressed in some detail
in this report.

\begin{changea}
Power-efficient signals are relatively simple to discover, because the
receiver's search strategy can focus on the detection of individual energy bundles.
Although the receiver can't be expected to know in advance the patterns of
energy bundles to be expected, using multi-channel spectral analysis the receiver can
simply look for energy bundles at a variety of carrier frequencies.
This search strategy is superficially similar to that employed in
optical SETI, which searches for ''pulses'' of light energy \citeref{655}.
Of course the details are considerably different, as optical SETI does not typically
employ multichannel spectral analysis, and uses energy criteria (photon counting)
rather than matching the received waveform against a specific bandlimited template.
What is in common with optical SETI is the lack of assumptions about persistence of
the signal.
A similar conclusion follows from cost considerations in high-power transmitter design
at radio wavelengths \citeref{569}.
\end{changea}

At first blush, one might expect that an information-bearing signal
specifically designed to have low average power relative to its information rate
would be difficult to discover, precisely
because that low-power makes the signal in some sense less ''visible'' to the receiver.
Surely, I hear you the reader say, a signal with lower average power will in general be more difficult to
discover than a signal with higher average power.
Generally that is true, but not in the way that one might expect.
It all goes back to the structure of power-efficient signals, which have
relatively low average power, but do \emph{not} have relatively low peak power.
In fact, they consist of energy bundles that are sparse in time and in frequency,
but which have \emph{individually} a sufficiently large energy to be readily detected.
Thus at lower average power levels the challenge of discovery is due to the increasing sparsity of the bundles,
rather than the energy of individual bundles.

Power-efficient information-bearing signals are amenable to being discovered by
an observation program that searches for individual energy bundles,
but is not cognizant of any other details of the signal structure or channel coding in use.
Although there is a challenge for the receiver in guessing the waveform used
to convey a single energy bundle, other than that they are relatively easy to detect
because of their high energy content (in relation to the noise) and
because they arrive with no appreciable impairments introduced by the
interstellar propagation (such as ''plasma dispersion'' or ''Doppler shift due to acceleration'')
that interfere with their detectability or require any related signal processing resources.
Following detection of one or more individual energy bundles, the receiver should monitor the ''vicinity''
(in both time \emph{and} in frequency) for more individual energy bundles.
Discovery is not a singular event, but rather emerges with a statistically
significant number of detections of individual energy bundles, and becomes
even more statistically significant as discernible and repeating patterns of energy bundles emerge
during extended observations.

The most significant challenge in discovery arises not only from the sparsity of energy bundles,
but in addition the variation in received signal flux (and thus observed bundle energy) wrought by scintillation.
Both these phenomena, individually and in conjunction with one another,
imply that the most efficient use of receiver resources is to make relatively short
observations looking for an energy bundle, and to repeat these short observations multiple
times in the same vicinity with intervening time intervals large relative to the coherence time of the scintillation.
Thus, not only is discovery search a long-term project because of the
large number of carrier frequencies and lines of sight to be searched, but also because discovery at a single location
(defined as carrier frequency and line of sight)
is a long-term project consisting of numerous short observations spread over a significant
period of time.
Fortunately, individual energy bundles in a power-efficient signal will by design
(in the interest of reliable information recovery) have sufficient
energy to be detected at high reliability whenever an observation happens to overlap an energy bundle
and simultaneously the scintillation is favorable.

In terms of the detection of individual energy bundles,
there is no fundamental lower limit on average power like that for communication of information.
This is because the receiver is extracting only a single bit of
information (one or more energy bundles is present, or not) and has in principle
an unlimited time (and hence an unlimited total available received energy) to work with.
Thus, as the receiver extends its total observation time in a given location, 
in principle perfect reliability
in detection of one or more energy bundles is approached regardless of the average power
of the signal and regardless of the strictness of the false alarm criterion.
The practical limitation on detection reliability follows from the
limited patience of a receiver to extend its observation time or its number of observations.
Patience is required for several reasons.
A transmitter may
be transmitting only sporadically along our common line of sight 
as part a targeted scanning strategy.
The transmitter can have no knowledge of the state of scintillation, and thus cannot
''cherry pick'' the right times to transmit.
Thus during those periods when a transmitter is active, the signal may or may not encounter
favorable scintillation.
Finally, when a transmission does encounter favorable scintillation,
energy bundles are still sparse in both time and frequency.
Fortunately sparsity in frequency is not an obstacle if the receiver employes multi-channel
spectral estimation as is the case with current SETI searches.

In design of a transmit signal, therefore, the challenges of discovery suggest the inclusion
of another objective in addition to power efficiency.
The power efficiency criterion means that we are seeking the lowest average signal power
consistent with a given information rate and a given objective for the reliability of information recovery
at the receiver.
In terms of discoverability, the transmitter has to assume that the receiver is making multiple
observations, but also evidently a finite number of observations.
The transmitter can make some assumption about the number of observations in place,
or in other words make an assumption as to the minimum resources devoted to discovery
by the receiver.
Subject to that assumption, as well as an objective for the reliability of discovery,
the transmitter seeks to minimize the average signal power.
We call this strategy \emph{power optimization} of the signal design.
The compatibility of power efficiency in conveying information reliably and power optimization
in enabling reliable discovery will be studied.

Fortunately these two complementary design objectives do not clash with one another.
\begin{changea}
Information-bearing signals designed according to power efficient principles are also close
to being power optimized for discovery, because in both cases the design objective to to endow
individual energy bundles with sufficient energy to be reliably detected.%
\end{changea}
There is a coupling between the number of observations
necessary at the receiver and the information rate.
Lower information rate results in lower average power and fewer energy bundles per unit time,
and hence a larger number of observations required of the receiver.
Another conclusion is that time diversity does interfere with discoverability, since it lowers the
amount of energy in each individual bundle, and this requires an appropriate modification
(with a resultant penalty in average power)
to the diversity strategy to maintain discoverability.
This additional factor favors the outage strategy for dealing with scintillation.

\subsection{Information-free beacons}

An alternative to an information-bearing signal is a \emph{beacon}, or
signal designed to be discovered but otherwise conveying no information.
\begin{changea}
The design of beacons and tradeoffs in the discovery of beacons is considered in detail
because this is a simpler context for modeling and studying discovery issues, and because
beacons may be important on their own.%
\end{changea}
Reliability of information recovery and power efficiency are not relevant criteria for beacon design,
so power optimization becomes the only relevant figure of merit.
A \emph{power-optimized} beacon design minimizes the average power of the beacon
for a fixed total observation time at the receiver.
Since it does not convey information, a beacon does not have to be
sparse in frequency or have a wide bandwidth in order to achieve power optimization.
At low average power, 
endowing individual energy bundles with a sufficient amount of energy
to enable reliable detection of individual energy bundles during discovery does imply sparsity of
energy bundles in time.
This structure as well as scintillation imply that the receiver must
make multiple observations in order to reliably detect 
a power-optimized beacon.
The transmitter can assume that the receiver is making
very few observations, in which case a relatively very high average power will be required
to insure reliable discovery of the beacon.

The Cyclops report prescribed a type of narrowband beacon in which energy is
transmitted continually, and this is the target of many current SETI searches \citeref{142}.
A comparison between power-optimized beacons and
a Cyclops beacon reveals that a substantial reduction in average power is available,
but this does require that the receiver make multiple short observations.
For example, assuming 1000 independent observations by the receiver,
an average power reduction by a factor of about three orders of magnitude
can be achieved.
In this example, an energy bundle is being transmitted and received about 3\% of the time.
As the number of receiver observations is increased, the required average power
is further reduced, with no lower limit.
Conversely, as the number of receiver observations is decreased the 
average power required for reliable discovery increases.

Such low-average-power beacons have very similar characteristics to
information-bearing signals with low information rate, and in fact offer
no particular advantage in terms of ease of discovery.
The major difference is that an information-bearing signal will consist of
energy bundles at multiple frequencies rather than just a single frequency, but multiple frequencies presents
no special challenge if we assume that the receiver performs a multichannel
spectral analysis.
Minor modification to the design of an information-bearing signal
entailing a small penalty in power efficiency may be required to enable
efficient discovery.
Both power-efficient information-bearing and power-optimized beacon signals can
be discovered using a common search strategy and algorithms, because they both
search for individual energy bundles.
Although neither type of signal is likely to be discovered by existing SETI observational strategies,
a roadmap is provided for how existing strategies can be initially enhanced,
and later optimized to search for power-efficient and power-optimized signals.

\subsection{The role of energy sparsity}

\begin{changea}
On the interstellar channel, its limited time and frequency coherence
dictates that energy bundles be restricted to the ICH to achieve the
maximum power efficiency.
A stream of information is embedded in the transmitted signal through the
location of these time- and bandwidth-restricted energy bundles.
In power-efficient signals most of the time and most of the bandwidth
is ''blank'', with no energy bundle present.
This is the major distinction between power-efficient signals and
the types of signals encountered in terrestrial communication
which are designed to be spectrally efficient and thus make
maximum usage of all the time available.

Sparsity has a different role in time as opposed to frequency.
Sparsity in time allows the energy of each bundle to be sufficiently
large to be reliably detected, regardless of the average power.
Sparsity in frequency allows a single energy bundle to convey
multiple bits of information, which is energy conserving.%
\end{changea}

\subsubsection{Sparsity in time}

Both information-bearing signals and information-free beacons
have a sparsity that grows as the average power of the signal is reduced.
Either no energy bundle or a single energy bundle is transmitted at any given time, and
the fraction of time devoted to transmission of an energy bundle is 
called the \emph{duty factor}.
There are many examples of low-duty-factor solutions to engineering challenges
where average power is a consideration.
A couple that have some similarity to interstellar communications are
the photographer's flash unit and a pulsed radar.
The flash unit is illuminating a target for the
purpose of capturing an image, the radar is illuminating a
target to monitor the reflection and estimate its position, and the
communications transmitter is illuminating the target conditionally 
based on some information it wishes to convey.
Examining the similarities and differences among these three applications offers some
insight into the benefit of low duty factor design.

The first motivation for low duty factor in all three applications is energy conservation.
In all three cases there is corruption by noise, and the intensity of the
electromagnetic illumination has to be sufficiently high to overwhelm any noise
that is present.
Briefly creating the necessary intensity requires much less average energy consumption
than would be necessary for continuous illumination at the same high intensity.
This is important for the flash unit because it is battery powered, and
reducing the energy required for each photograph allows more photographs
within the total battery capacity. 
Radar and interstellar communication are similar to one another in that reducing average
power reduces average energy consumption and eases the thermal management
challenges in the transmitter electronics.
The average power can be reduced by a reduction in the duty factor.


An astronomer might ask about the effect of ''integration time'', which is
averaging over longer periods of time.
This is an issue where the flash is significantly different from radar
and communications.
In the case of the flash, the illumination is essentially a thermal noise source,
having the same characteristics and statistics as the thermal noise introduced
in the sensor except (hopefully) with a higher noise temperature.
The sensor measures the temperature of the illumination plus sensor noise.
Averaging over longer periods of time reduces the spatial variation in the
estimate of temperature, but does not eliminate the need for the temperature of
the illumination to be high relative to the temperature of the sensor-introduced noise.
Thus, the main requirement is for the illumination to be high intensity,
with additional benefit from a longer duration of illumination
(at the expense of total energy expended)
in reducing the spatial statistical variation in the sensor's temperature estimate.

Integration time plays a very different role in radar and communications
because of the matched filter.
It would be possible to eschew the matched filter and operate like the
photographic flash and camera sensor, but the fundamental limit
on power efficiency could not be approached in this fashion.
This is because the sensor would respond to all the noise within the
bandwidth of the signal, whereas a matched filter responds only to
that fraction of this in-band noise that is perfectly correlated with a known 
deterministic signal waveform.
However, in order to implement a matched filter the receiver has to know
(or guess) the waveform of the signal.\footnote{
In principle a camera and its flash unit could use coherent detection,
thereby allowing a lower total-energy illumination for each photograph.
The reason it doesn't presumably has to do with the
technological difficulties of doing this at optical wavelengths,
where generating a non-monochromatic coherent waveform is difficult.}
That waveform can be anything, such as a sinusoid or a noise-like signal,
but the receiver has to know that exact waveform.
When the output of the matched filter is sampled at the right time, the ratio of signal
to noise depends only on the energy of the waveform, and not its
time duration or bandwidth or any other property.
This is a property of coherent detection, in contrast to the incoherent
estimation in the camera sensor.
With coherent detection, it is possible to increase the duration of the waveform
arbitrarily while keeping its total energy fixed without penalty, 
beneficially reducing the peak power.\footnote{
Actually one form of incoherence creeps in, specifically phase incoherence.
Even though the waveform may be known, there is an unknown phase factor
due to scintillation and unknown carrier phase.
The key property of this phase factor is that it is fixed for the duration of the
waveform, so the matched filter can perform is coherent integration
in spite of this unknown phase.
The output of the matched filter contains this unknown phase factor,
which can be eliminated by calculating the magnitude.
This is called phase-incoherent detection.}

Why then don't we fully exploit this integration time idea in radar and communications
in order to transmit continuously rather than in bursts?
With continuous transmission the fixed energy could be spread over greater time intervals,
reducing the peak power.
The answer has to do with motion, specifically the motion of the source and target and
(in the case of interstellar communication) motion of the turbulent clouds
of gasses in the interstellar medium.
This motion has the effect of modifying the waveform of the signal
in an uncontrolled way, potentially interfering with the proper role of integration
time in the operation of the matched filter.
The effect we see at the matched filter output is a reduction in signal level
(but no reduction in noise variance) due to the deleterious effects of motion.
The time duration of the signal waveform has to
be kept small enough to render this adverse effect of motion 
sufficiently small that it can be neglected, if we aspire to approach the
fundamental limit.

The low duty factor of low-power information-bearing signals or beacons is
a critical property, one that profoundly influences the strategies for discovery
of those signals.
This low duty factor is an inevitable consequence of three considerations.
First, the energy of individual bundles must be sufficiently large to
overwhelm the noise.
Second, the time duration of the energy bundles must be small enough
to render the coherence-destroying property of motion negligible.
Third, the only way to reduce the
average signal power and stay true to the first two considerations
is to maintain a fixed energy for each bundle
while reducing the rate at which bundles are transmitted, or in other
words reduce average power through a reduction in the duty factor.
The peak power of the signal stays constant even as the average
power is reduced.

\subsubsection{Sparsity in frequency}

\begin{changea}
A similar argument applies to frequency as well as time, in the case of
information-bearing signals.
Conveying information through the location of an energy bundle in frequency
is energy conserving because multiple bits of information can be represented
by a single energy bundle.
However, the coherence properties of the interstellar channel limit the
bandwidth of each of these energy bundles.
In this case incoherence is not due to motion, but rather due to wavelength-dependent
phase and amplitude shifts reflecting the effects of electromagnetic propagation
through an ionized conductive medium.
Combining a bandlimited energy bundle with a large number of frequency locations
for that energy bundle implies that the energy is sparse in frequency.

There there are two obvious ways to increase the 
quantity of information conveyed.\footnote{
Another obvious modification is to exchange the role of time and frequency,
conveying information through the location of an energy bundle in time
and employing diversity in frequency to combat scintillation.
This approach should be explored further.
}
One would be to locate two or more energy bundles (as opposed to one) at different frequencies, on in other words transmit more than a single energy bundle at one time.
The second would be to convey information by choosing between two or more possible energy
levels (rather than one).
It is straightforward to show that either of these modifications would reduce the
achievable power efficiency for a fixed reliability of information recovery.
Of course the ultimate validation of the efficacy of 
building up a signal through temporal-frequency patterns of energy bundles, each with the same energy level,
is the realization that this simple design approach is capable of approaching the fundamental limit
on power efficiency.%
\end{changea}

\section{Implications for SETI observation programs}

The end-to-end approach taken in this report yields relevant insights
into the design of both transmitter and receiver for interstellar communication.
Near term the most profound implications apply to receiver design,
which applies directly to the search for extraterrestrial intelligence (SETI).
Generally these searches have been targeted at beacons rather than
information-bearing signals, and thus concentrated on narrowband signals.
In addition,
most such searches at radio frequencies have assumed a continuously received
signal, exploiting that assumption to reject false alarms through a presumption
of signal persistence
and adding an expectation of extraterrestrial signatures 
such as Doppler-drift in frequency due to relative acceleration.
Implicit in these presumptions is a required
 average power for the beacon that is some orders of magnitude
larger than necessary, should the transmitter designer seek to reduce its energy consumption
and capital costs through a reduction in the duty factor of the signal.
\begin{changea}
There may of course be advanced civilizations to whom energy consumption
and conveying information are of little concern, 
and past and current SETI searches
have been seeking signals from such civilizations.

Fortunately these present-day search strategies are fairly easily modified
to search for power-efficient information-bearing signals and
power-optimized beacons.
This broadening of objectives would then include beacons
transmitted by civilizations who seek to reduce the average transmitted
power and energy consumption without harming the reliability with
which the beacon can be discovered.
This would also include information-bearing signals transmitted by
civilizations wishing to maximize the information rate achieved 
consistent with reliable extraction of that information at the receiver,
with whatever expenditure of average power and energy consumption is 
available.

The nature of an appropriate modification to search strategy is to 
substantially change the criterion for ruling out a given location.
With power-efficient and power-optimized of signals, 
short-term persistence of the signal is not expected,
and in the case of information-bearing signals the concentration of
energy at a single carrier frequency is also not expected.\footnote{
Existing searches may in fact have detected individual energy bundles from
a power-efficient information-bearing signal, or a power-optimized beacon,
but also rejected these detected events as ''false alarms''.
Such is the case for example for the most famous of these incidents, 
the so-called ''Wow!" signal \citeref{649}.}
The detection of an energy bundle is an indicator that
this location is worthy of further observation in the future, and not only the immediate
future, and in addition at the nearby frequencies, not only the same frequency.
Even locations where energy bundles are not detected need to
be revisited, gradually building a statistical case that no signal is present.
A permanent record should be kept of how often each location has been visited,
as well as the particulars of any energy bundles that have been detected
at that location, since all past and present observations are relevant to the
question of whether a signal of extraterrestrial origin is present.

A search for power-efficient and power-optimized signals
is appropriately staged over time.
Just a few observations at a given location without
detecting any energy bundles can rule out beacons
with a very large average power, or information-bearing signals
with a very large average power and commensurately large information rate.
Over time as more and more observations are conducted at a given
location it will be possible to statistically rule out lower and lower average power
signals, or conversely identify locations that are most promising
candidates for further observation.
To instrument such a staged search strategy, a database
must be maintained documenting all past observations and their outcomes.
\end{changea}

These searches can also be generalized to consider alternative waveforms
carrying an energy bundle, such as noise-like waveforms 
which are advantageous in their immunity to local sources of 
radio-frequency interference \citeref{211}.

A search for power-efficient and power-optimized signals most closely resembles
Astropulse, which is a search for isolated radio-transient events \citeref{647},
\begin{changea}
and optical SETI, which searches for isolated short-duration optical pulses \citeref{655}.%
\end{changea}
The main difference is both Astropulse and optical SETI assumes signal waveforms that 
are short in time duration and have wide bandwidth.
At radio wavelengths, this implies that the pulses are strongly
affected by the frequency incoherence of the interstellar medium, principally plasma dispersion.
As a result Astropulse requires a processing-intensive de-dispersion step to reverse
this effect.
Such a waveform is quite plausible for events of natural origin, but is not characteristic of 
explicitly designed power-efficient and power-optimized signals, which
 are necessarily and explicitly designed to avoid as many interstellar impairments as possible.\footnote{
 There is one good argument for transmitting a signal explicitly designed to
 be impaired, which is the ''interstellar signature'' imposed on the signal by that
 impairment, distinguishing it from signals of terrestrial or near-earth origin.
 }
 It is also not characteristic of a signal designed explicitly to be
 relatively easily discovered without a lot of processing resources, 
 given the ready alternative of avoiding these impairments.
A search for power-efficient signals fortunately requires no processing related to interstellar
impairments other than noise (in particular this is the role of the matched filter).
The parameters of the search, such as the presumed time duration and bandwidth of
the waveform supporting an energy bundle, are constrained by
 interstellar impairments subject to observation and modeling by
 both transmitter and receiver.


\chapter{Signal design for power efficiency}
\label{sec:powerefficient}

Most past treatises dealing with the search for extraterrestrial intelligence (SETI)
and interstellar communication at radio wavelengths have emphasized the importance of energy resources.
This is because the energy consumption associated with
transmission of the radio signal will be large relative to other communication
applications, largely because of the
large interstellar distances and the resulting large propagation loss.
\begin{changea}
Energy consumption can also be reduced by a reduction in information rate,
and thus the relevant concern is not the absolute energy consumption
but rather the energy consumption relative to a given information rate.
For a given information rate, the transmitter benefits by achieving the smallest average transmit power,
or rate of energy consumption.
Equivalently, for whatever energy resources may be available to the transmitter,
both the transmitter and receiver benefit from achieving the largest the attainable information rate.
\end{changea}

There are several ways of controlling the average power or energy consumption.
\begin{changea}
We have already mentioned the reduction in information rate, which then increases the
time required to transmit a message of a fixed size.%
\end{changea}
At a fixed information rate
energy consumption can be reduced by increasing the gain of the
transmit and/or receive antennas, 
which effectively reduces the end-to-end transmission loss.
This requires the transmit and/or receive antennas to be physically larger
(while no less dimensionally precise) and thus increases the capital expense \citereftwo{568}{569}.
The receiver sensitivity can be increaseed
by using more sophisticated signal processing techniques, but that has been largely
exhausted by using optimal or near-optimal signal detection techniques to counter
the inevitable noise of thermal origin.
Future improvements in the noise temperature of receiver
circuitry may well be possible, but the benefit of this is fairly modest.

There is another possibility for reducing energy consumption
without an impact on the information rate, and that is the
design of the transmit signals with the specific goal of reducing their average power
without affecting the reliability with which information can be recovered.
Such changes in transmit signals must be accompanied by modified receiver signal processing algorithms at
baseband to accommodate such signals.
This very challenge was addressed
in the 1950's and 1960's by communication engineers.
That decades-old work focused on maximizing the information rate achievable subject to
reliable recovery of that information by the receiver and subject to a constraint on the average power of the signal.
It was proved mathematically that there exists a fundamental limit, and signal design techniques that can
approach that fundamental limit asymptotically were uncovered.
This body of work was noted and some of its implications explored by
a couple of previous authors addressing interstellar communication  \citereftwo{196}{602}.
As shown here, there is a penalty of four to five orders of magnitude
in average power (relative to the fundamental limit) required for the types of signals typically assumed in
the SETI literature.
The remainder of this chapter and report is devoted to finding ways to reduce
that penalty substantially using relatively straightforward design techniques.

The models used in the earlier work assumed that the signal is affected by fading,
which is a random fluctuation of signal amplitude, as well as thermal noise.
Both of these phenomena are prominent in interstellar communication.
In astronomy scintillation is the same phenomenon as fading,
and noise of thermal origin is a universal feature of both the cosmos and receiver technology.
In this report the fundamental limit on the average power required
to achieve a certain information rate is derived taking into account the
specific impairments present the interstellar communication context.
Further, it is
shown that using a set of five simple signal-design principles accompanied by
appropriate receiver signal-processing algorithms this fundamental limit can be approached asymptotically.
No civilization, present or future, can bypass this fundamental limit
within the scope of the statistical channel models we use, which themselves invoke the classical physics of radio propagation and noise.

In order to accomplish this in the context of interstellar communication, 
we must address a couple of challenges not considered earlier in the communications
literature or in the interstellar communications literature.
First, there are a set of specific impairments caused by propagation effects through
the interstellar medium as well as by the relative motion of the transmitter and receiver,
and the interaction of these effects.
Earlier work has not modeled these impairments nor taken them into account, and hence
the results are incomplete.
Second,
in interstellar communication, the initial discovery of the signal is more challenging
than the communication of information once the signal is discovered, both because of the lack of coordination
between transmitter and receiver designers
and because of the ''needle in a haystack'' challenge of a large parameter space
over which to search.
The primary purpose of this report is to invoke the earlier theories relating to minimizing
average power (and hence energy consumption), coupled to an in-depth study of these two additional challenges
of interstellar impairments and signal discovery.

This chapter is an overview of our approach to power-efficient design.
While the fundamental limit on average power is universal,
concrete design approaches to reducing average power are less fundamental.
There is no theoretical basis to rule out the possibility that there may be other viable design approaches.
However, our approach does meet two important minimum requirements for power-efficient design.
First, it is able to approach the fundamental limit asymptotically.
Second, it is very straightforward, combining a set of five principles that individually
address one of the challenges posed by the interstellar communication channel.
Each of these is the simplest response we can think of to its respective challenge,
and none of these five principles can be abandoned without compromising power efficiency.
In our judgement, there could be no other simpler design approach that
is able to approach the fundamental limit.
Simplicity is important for interstellar communication because of the intrinsic lack of coordination.
Nevertheless, further research seeking possible alternative approaches to achieving power efficiency
should certainly be pursued.

\section{Efficiency measures for communication}
\label{sec:defpowere}

\begin{changea}
In communications, if all resources like energy and bandwidth are unlimited
then communication with perfect reliability is always feasible.
The more interesting question arises when resources are limited or constrained.
There are typically two resources that are limiting, one being the
acceptable energy consumption (or equivalently average power) and the
other the signal bandwidth.
The specific impact of these limited resources depend on the particulars of the
impairments encountered in the propagation of the signal.
Regardless of the impairments, however, as either the average power or bandwidth
of the signal increases then reliable communication can be achieved at higher rates.
This is obvious, since as we relax any constraint on either average power or bandwidth,
actually using less average power or less bandwidth is always an option, and thus
the achievable information rate must be at least as large.

More interesting and less obvious is a fundamental tradeoff that always exists
between average power and bandwidth.
These obey a rule analogous to the uncertainty principle of quantum mechanics;
namely; more of one permits less of the other.
In this report we argue that average power is significantly more important to
interstellar communication than is the bandwidth consumed by the signal.
In that event we would expect the bandwidth of an interstellar communication signal to be
large in the interest of reducing the average power for a given information rate,
or equivalently increasing the information rate for a given average power. 

Both the average power and the bandwidth are strongly affected by the
information rate.
Thus it is not the average power or bandwidth that matters in isolation,
but rather their values in relation to the information rate.
This reality is captured by two \emph{efficiency} measures, 
one for average power and one for bandwidth.
\end{changea}

\subsection{Power efficiency}

The average power of a signal, which we denote here by $\power$,
is obtained by accumulating the total energy of the signal over a very long time
and dividing by the time interval.
Average power is one metric, and another relevant
figure of merit is the information rate, which we denote here by $\rate$.
$\power$ has the units of watts, and $\rate$ has the units of bits per second.
As will be shown (and was illustrated in Figure \ref{fig:powerefficiencyconcept}),
if bandwidth is not constrained then for both practical realizations and at the fundamental limit
these two quantities are proportional to one another, or $\rate \propto \power$.
Given this reality, it is the ratio of the two quantities that is of greatest interest.
For an information-bearing signal, the \emph{power efficiency} $\powere$ is
defined as the ratio of the information rate $\rate$ and the average received power $\power$,
\begin{equation}
\label{eq:power}
\powere = \frac{\rate}{\power} 
\ \ \text{bits per joule}
\,.
\end{equation}
It is often convenient to work with the reciprocal of $\powere$,
\begin{equation}
\label{eq:energybit}
\energybit = \frac{1}{\powere}
\ \ \text{joules per bit}
 \,.
\end{equation}
The energy per bit
$\energybit$ is particularly easy to interpret, as it is the
energy required to reliably communicate exactly one bit of information.
If a message consists of $N$ bits,\footnote{
In this report our measure of message length is in bits,
and our measure of information rate is bits per second.
More fundamentally the ''size'' of a message can be
quantified by the number of possible distinct messages $M$,
With that measure, the size of the message in bits is
equivalently $\log_2 M$ bits.}
 then the total energy
required to convey that message is $N \energybit = N / \powere$.
The total energy required to convey this message is inversely
proportional to $\powere$, or proportional to $\energybit$.
If a sequence of bits is conveyed at information rate $\rate$,
the average power required for reliable
recovery of that information is $\power = \rate \energybit = \rate / \powere$.
Any way you look at it, $\powere$ is a valuable measure of the
energy efficiency with which information is conveyed.
Assuming that the transmitter considers energy as a limited resource
and considers more information to be desirable,
$\powere$ has the property that ''more is better'', as something
labeled as ''efficiency'' should.

The energy per bit $\energybit$ is one important figure of
merit, but there is also the issue of the reliability.
Inevitably some bits will be incorrectly decoded by the receiver
due to noise and other impairments introduced in the interstellar
signal propagation.
Such errors make a message more difficult to interpret,
and thus we want these bit errors to occur infrequently.
The reliability is measured by the bit error probability $\pe$.
For example, if $\pe = 10^{\,-4}$ then as a long-term average
one bit out of every 10,000 will be decoded incorrectly
(a ''0'' is turned into a ''1'' or a ''1'' is turned into a ''0'').
The power efficiency $\powere$ 
can only be stated in the context of a particular level of reliability.
If the receiver designer is willing to accept a poorer reliability
(larger $\pe$), then a higher $\powere$ can result.

\subsection{Spectral efficiency}

The major competitor to power-efficient design is spectrally efficient design,
where the major design objective is to minimize the bandwidth, which is
the range of frequencies occupied by the transmit signal.
Let $\Btotal$ be the total bandwidth required for an information-bearing signal.
The ratio of $\rate$ to $\Btotal$ is a good figure of merit
describing the total bandwidth requirements.
The \emph{spectral efficiency} $\spectrale$ is defined as
\begin{equation*}
\spectrale = \frac{\rate}{\Btotal}
\ \ \ \text{bits per second per Hz}
\,.
\end{equation*} 
The spectral efficiency is completely determined by the transmitter through its choice
of the structure and parameterization of the signal that bears information.
Assuming that the transmitter and receiver designers consider bandwidth to be a scarce resource,
then spectral efficiency also has the property that ''more is better''.

Two different measures of efficiency have been defined, and a natural
question is whether both types of efficiency can be achieved simultaneously.
As will be developed in the remainder of this chapter, the answer is
a definitive ''no''.
There is a fundamental and unavoidable tradeoff, in that high power efficiency
is associated with low spectral efficiency, and vice versa.
Thus, the designer of an interstellar communication system has to
choose sides.
Is that designer going to emphasize power efficiency, or spectral efficiency?
We believe that for interstellar communication the choice is obviously power efficiency.
This assertion will
now be justified by pointing out some implications of power efficiency
to interstellar communication.

\subsection{Transmitter energy consumption}

\begin{changea}
The primary practical implication of average power is the 
energy consumption that it imposes on the transmitter.
In terms of reliable recovery of information, it is the average power
delivered to the receiver's location that matters, and this is
by convention where we instrument and measure the power efficiency.
Due to the large propagation losses,
even a very small average power delivered to the receiver
can imply a very large radiated average power at the transmitter.
The relationship between transmit and receive average powers
depends on the directivity of the transmit antenna, 
the propagation loss over interstellar distances, and the collection area of the
receive antenna.
Physically the propagation loss is due to the spatial spreading of
the transmitted power, with the result that the average power intercepted
by the receive antenna decreases as the square of the distance.
In power-efficient design the average power is also proportional
to the information rate.
\end{changea}


\doctable
	{\small}
	{linkbudget}
	{Energy use and annual energy cost for interstellar communication
	at an information rate of 1 bit per second at a distance of 1000 light years, a carrier frequency of 5 GHz, and
	using prototype Arecibo antennas (305 meters) and electronics
	with an assumed 50\% antenna efficiency and a total system noise temperature of 8 degrees kelvin.
	The assumed price of energy is \$0.10 per kilowatt-hour, and all figures
	are for communication along a single line of sight.
	The fundamental limit is compared to an on-off keying (OOK)
	information-bearing signal following Figure \ref{fig:gapFundamental}
	at an assumed reliability of $\pe = 10^{-4}$.
	Noise only and noise plus scintillation are included to separate out the effect of scintillation,
	which does not influence the fundamental limit but has a substantial effect on OOK.
	}
	{p{5cm} l l l}
	{
	\toprule
	\textbf{Parameter} & \textbf{Fundamental limit} & \textbf{On-off keying} & \textbf{On-off keying} \\
	&& \textbf{Noise only} & \textbf{Noise plus scintillation} \\
	\addlinespace[3pt]
	\cmidrule{2-4}
	Transmit power consumption
	& 18.5 kilowatts & 664 kilowatts & 753 megawatts \\
	\addlinespace[3pt]
	\cmidrule{2-4}
	Annual cost of energy
	& \$16.2 thousand & \$581 thousand & \$659 million \\
	\bottomrule
	}

This situation is analyzed in Appendix \ref{sec:linkbudget},
where a link budget is developed for a typical interstellar communication
scenario, and a numerical example is displayed in Table \ref{tbl:linkbudget}.
A fundamental lower limit on the energy per bit $\energybit$
for the interstellar channel is developed later in this chapter
and in Chapter \ref{sec:fundamental},
and this lower limit on average power is compared to
on-off keying (OOK), which is described shortly and is typical of the
types of signals being targeted in current and past SETI searches.
In order to separate out the effects of noise and scintillation,
both the noise without scintillation and noise with scintillation cases are
shown (the fundamental limit is the same for both cases).
The latter radiated power requirement is about double the
 power consumption of the
Large Hadron Collider (LHC) at the European Organization for Nuclear Research (CERN),
which is about 300 megawatts.


We do not know the price of energy for another civilization, but we
do know that the price of electricity on Earth is roughly one
U.S. dollars per ten kilowatt-hours, and at that price
the annual energy cost of transmission at an information rate of one bit per second
increases from \$16 thousand at the fundamental limit to \$659 million for OOK
at 1 bps and 1000 light years.
This indicates that the cost of energy consumption, at least at
Earth pricing, is very significant for OOK, which suggests
great economic pressure to keep energy consumption
at a level as close to the fundamental limit as feasible.
Certainly this would be true if we were building a transmitter here on Earth,
and it is a reasonable assumption that this would be true in other worlds as well.
Of course we cannot rule out the possibility that another far more advanced
civilization has found far less costly energy sources.

There are several parameters that can be used to reduce the energy consumption of Table \ref{tbl:linkbudget}
and should be kept in mind:
\begin{description}
\item[Reduce information rate.] 
The average transmit power is proportional to the
information rate $\power \propto \rate$.
Accepting a lower information rate will lower the power requirement.
However, this approach may not be all it is cracked up to be.
If we wish to convey a message of a certain fixed length, then reducing the
information rate not only reduces the average power,
but it also extends the time duration for transmission of the message.
It does nothing to reduce the
total energy (or cost of energy) required to convey a fixed-length message.
\item[Increase carrier frequency.]
For fixed antennas, the average transmit power drops with an increase in carrier frequency $f_c$ as
$\power \propto f_c^{\,-2}$.
This is because the gain of an antenna is related to its physical size in relation to the wavelength.
However, larger $f_c$ does impose more stringent requirements on the
accuracy of antenna fabrication.
\item[Reduce distance.]
The average transmit power increases with distance $D$ as $\power \propto D^{\,2}$
in order to overcome the distance-related propagation loss.
For example, if the distance in the example of Table \ref{tbl:linkbudget} were halved to 500 light years,
the average transmit power would decrease by a factor of four.
\item[Accept outages and erasures.]
\begin{changea}
The ''apples and apples'' comparison in Table \ref{tbl:linkbudget} assumes the same
reliability of information recovery in all three cases.
However, we note that scintillation has a much bigger impact on the average power
required for OOK than does noise.
If we merely discard all information during low signal flux due to scintillation,
the average power decreases substantially to something more on the order of the noise-only case.
This is called the outage-erasure strategy, which is explored in Chapter \ref{sec:infosignals}.
It is an attractive approach for interstellar communication because typically a message
will be transmitted repeatedly, yielding an opportunity for the recovery of lost information.%
\end{changea}
\end{description}
As discussed in the remainder of this report, the transmit signal 
can be designed with the specific goal of increasing
the power efficiency,
thereby coming much closer to the fundamental
limit as was illustrated in Figure \ref{fig:channelcodingconcept}.
This is attractive because it allows the information rate to be increased or
the average power to be reduced.

\subsection{Thermal management}

Any transmitter will fundamentally suffer some inefficiency in conversion from
energy input to transmitted average power.
There are two implications to this.
First, the annual costs of energy consumption in Table \ref{tbl:linkbudget} must be
adjusted upward to account for this inefficiency.
Second, the energy lost to inefficiency is converted to heat, which must be dissipated
to (literally) prevent the transmitter electronics from melting.
The amount of heat to dissipate will be proportional to the average transmit
power, so OOK presents a significantly larger thermal management issue.
Heat dissipation at these power levels will consume a major portion of the capital
cost of building a transmitter.

\subsection{Multiple lines of sight}

The transmitter requires the expenditure of a very large total energy in what may prove to
be a hapless attempt to communicate with a receiver that never existed.
The transmitter probably also has little idea of where its signal may be observed,
and thus is likely to transmit signals along multiple lines of sight concurrently, just as the receiver
is observing along different lines of sight.
To have a reasonable expectation that a signal will be received, targeting thousands
or tens of thousands of relatively nearby stars will be necessary.
This multiple line-of-sight requirement magnifies by an equivalent factor the challenges relating to
the cost of energy consumption and heat dissipation.

\subsection{Peak power}

The peak power $\powerpeak$ is defined as the energy
of the signal averaged over short time periods.
Although the peak power does not relate directly to energy consumption or heat dissipation,
there will nevertheless be a penalty in the capital cost of building a transmitter for larger peak power.
It will be shown in the sequel that even as we reduce the average signal power there is little opportunity to 
simultaneously reduce the peak power.
Peak power must always remain large enough to overcome propagation losses and the noise floor at
the receiver, while average power is most efficiently reduced by
transmitting a signal intermittently.
The required peak power can be reduced by increasing the directivity of the
transmit antenna or increasing the collection area of the receive antenna,
so there is an opportunity to reduce both peak and
average power by expending more resources on the antennas.

\section{Basic building block: Energy bundles}
\label{eq:energybundles}

Our approach to power-efficient design uses the \emph{energy bundle} as a basic building block.
We now describe the concept of an energy bundle, how it is transmitted, and how it is detected.

Our basic approach is to transmit a bundle of energy and at the receiver
observe the bundle's excess energy over and above the noise.
There are two defining characteristics of an energy bundle.
First, at a given location (defined as carrier frequency and time)
an energy bundle is either present or absent.
Although this makes it appear like one bit of information is
conveyed by this
presence or absence of an energy bundle,
in fact the essence of power efficient communication is to
convey much more than one bit of information with a single
energy bundle, as will be seen.
\begin{changea}
The essence of the idea is to convey information not through the simple presence or absence,
but rather through the location of an energy bundle from among many alternatives.%
\end{changea}
Second, in its detection the receiver \emph{estimates} the amount of energy in a bundle
as the first step in \emph{detecting} the presence or absence of an energy bundle,
but is oblivious to any phase information.
The distinction between estimation and detection is significant.
Estimation seeks to answer the question ''how much energy is present
in a hypothetical bundle at this location'', while detection seeks to answer the question
''is there a bundle present or not at this location''.
A simple approach to detection is to apply a threshold
to an estimate of the energy,
but more refined approaches are often possible as will be seen.

In communications, a \emph{channel} is the physical entity that
conveys information from one physical location to another.
Over a physical channel based on radio propagation through the interstellar medium,
an energy bundle is associated with a  waveform $\sqrt{\energy_h} \, h(t)$,
which represents the electric field of the electromagnetic wave.
The waveform $h (t)$ is assumed to have unit energy, defined as
\begin{equation}
\label{eq:unitenergy}
\int_{0}^{T} | h(t) |^2 \, dt = 1 \,.
\end{equation}
The amount of energy in the bundle is $\energy_h$,
which is quite significant because $\energy_h$ determines how reliably
the energy bundle can be detected in the presence of noise.
Two other important characteristics of the energy bundle are the time duration $T$ and bandwidth $B$
of waveform $h(t)$.
We will see that these two quantities are constrained by the properties of the interstellar channel.

\subsection{Matched filtering}

Noise of thermal origin inevitably accompanies an energy bundle.
In the presence of this noise, the best method for detecting the excess
energy represented by $h(t)$ over and above the thermal noise is
a filter matched to $h(t)$.
This \emph{matched filter} is illustrated in Figure \ref{fig:mfillustrate} and discussed in greater detail in Chapters \ref{sec:channelmodel} and \ref{sec:detection}.
Detection of an energy bundle by
matched filtering is essential to power-efficient design, because any
other detection method will be less reliable and thus requires a larger
bundle energy $\energy_h$.
It is significant that in order to apply a matched filter the receiver must know (or be able to guess)
the waveform $h(t)$ chosen by the transmitter.
Alternatively, as part of a discovery search strategy the receiver
(at the expense of additional computation) can trial more than one waveform $h(t)$.

\incfig
	{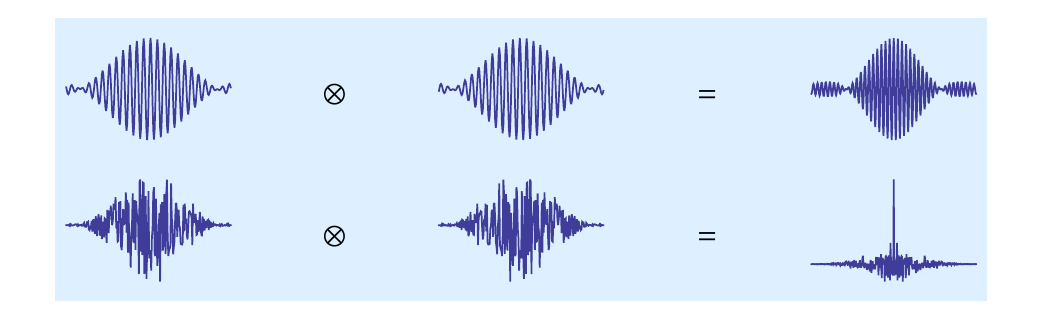}
	{1}
	{An illustration of an energy bundle and its detection using a matched filter.
	It consists of a waveform $h(t)$ shown on the left, which can be simple
	or complicated.
	The remainder of the
	figure illustrates how this bundle of energy can be detected using a matched
	filter, which consists of a convolution
	(represented by $\convolve$)
	 with a time-reversed and conjugated version $\cc h (-t)$ (illustrated in the middle),
	with the result equal to the autocorrelation of $h(t)$ (shown on the right). 
	This autocorrelation can be sampled at zero lag,
	 and the resulting value represents a filtered sample of the thermal noise.
	Various types of $h(t)$ can be used as the basis for an energy bundle, but arguably 
	the most useful are the segment of a sinusoid (on top) and the segment of
	a noise-like signal (shown on the bottom).}

When examining issues like the relative energy of an energy bundle and noise,
it is important to settle on a common place to observe and measure such things.
In radio communications it is conventional to do modeling and measurement of
the signal in the receiver, and at baseband following demodulation.
A slight complication is that a passband radio signal, which is associated with a physical
quantity like the electric field of the radio wave, is
intrinsically real-valued, but must
be represented at baseband by \emph{two} real-valued signals, called the in-phase
and quadrature signal components.
For mathematical convenience, as discussed further in Chapter \ref{sec:channelmodel},
it is conventional to represent these two signals all at once by a single complex-valued
signal, where the real-part represents the in-phase signal and the complex-part
represents the quadrature signal.
Thus, the baseband waveform $h(t)$ is actually assumed to be complex-valued, 
and is said to have
a magnitude $|h(t)|$ and a phase $\arg h(t)$.
The matched filter actually filters the reception 
by convolving it with an impulse response $\cc h (-t)$,  
a conjugated and time-reversed replica of $h(t)$.

In this report, $h(t)$ is assumed to be non-zero over the timescale $t \in [0,T]$.
The range of frequencies present in $h(t)$ depends on the frequency used in the
demodulation stage, but for simplicity and symmetry with time it is assumed that
the demodulation frequency is at the bottom of the band of passband signals.
That arbitrary convention makes $H(f)$, the Fourier transform of $h(t)$, non-zero for the range of
frequencies $f \in [0, B]$.
A waveform with only positive-frequency content is necessarily complex-valued,
but we already expected this.

\subsection{Phase-incoherent detection}

\begin{changea}
All known digital communication system designs 
at wavelengths where thermal noise is the dominant noise impairment
use something akin to the
energy bundle as their underlying building block.
In terrestrial systems it is common, however, to embed information
in the signal through using multiple amplitudes
(equivalent to multiple energies) and/or multiple phases.
These approaches are appropriate for achieving high spectral efficiency.
In contrast the greatest possible power efficiency can be achieved by
the simpler approach of embedding information in the
presence or absence of energy bundles, at the expense
of a penalty in spectral efficiency.
\end{changea}

The phase stability of an interstellar communication channel will be very poor
because of the great distances involved and also because of scattering and scintillation
phenomena as discussed in Chapter \ref{sec:channelmodel}.
Changes on the order of a fraction of a wavelength at the carrier frequency matter to phase,
and one wavelength is absolutely minuscule in comparison to the distances involved
in interstellar propagation.
In addition, the phase of the carrier frequency generated at the transmitter is unknown.
For both these reasons,
the receiver cannot depend on any phase information in a single energy bundle,
and must use phase-incoherent detection of the bundle.
This implies that the magnitude of the matched filter output is calculated, and
phase is discarded.
The square of this magnitude is an estimate of the single real-valued parameter $\energy_{h}$.
We call it an \emph{energy} bundle because all the receiver is able to measure is its energy.

\begin{changea}
This approach at radio wavelengths is still more coherent than the photon-counting approaches
used at optical frequencies \citeref{655}, but the difference is subtle.
Photon counting simple measures the total energy within a given bandwidth, without
distinguishing the ''shape'' of that energy vs. time as in a matched filter.
The phase-incoherent matched filter takes into account the time evolution of the waveform $h(t)$,
including both amplitude and phase,
but also assumes that the entire waveform is shifted by some constant-but-unknown phase.%
\end{changea}

\subsection{The interstellar coherence hole}

Impairments introduced in the interstellar medium (ISM) 
and due to the relative motion of the transmitter and receiver can have a
deleterious effect on the detection of energy bundles by modifying
the transmitted
$h(t)$ to something different before it appears in the baseband receiver processing.
However, by placing an upper bound on both $T \le T_c$ and $B \le B_c$, most of these
effects can be rendered negligible, as discussed in Chapter \ref{sec:incoherence}.
The largest values $T_c$ and $B_c$ are called respectively the
\emph{coherence time} and \emph{coherence bandwidth} of the interstellar channel,
and a waveform that meets these constraints is said to fall in an
\emph{interstellar coherence hole} (ICH).
The coherence time $T_c$ represents the time scale over which the
interstellar phenomenon can be approximated as time-invariant, or unchanging.
The coherence bandwidth $B_c$ is the bandwidth scale over which
any frequency-selective or frequency distortion phenomena can be
neglected.

One significance of the ICH is that as long as energy bundles conform to its
constraints, no processing is necessary in the transmitter or receiver relative to
the interstellar impairments.
However, the question arises as to whether the ICH constraint causes
some compromise in performance.
This is addressed in Chapter \ref{sec:fundamental}, where it is shown that
not only does limiting energy bundles to the ICH not compromise power efficiency, but
this constraint is actually necessary to approach the fundamental limits on power efficiency.
\begin{changea}
If $h(t)$ was allowed to wander outside the ICH but having too large a time duration or bandwidth,
matched filter detection will lose some efficiency in capturing the entire received energy,
and as a result the power efficiency will suffer.%
\end{changea}

\subsubsection{Two forms of coherence time}

There are actually two coherence times of significance.
The coherence time $T_c$ discussed thus far is the \emph{phase} coherence time,
and it measures the time interval over which the phase, while unknown, remains stable
enough that its effect on the energy estimate at the matched filter output can be neglected.
When observing two energy bundles at different times, there will also be a variation
in the observed energy contained within the bundles, even though each was
transmitted with the same energy.
This energy difference results from a scintillation-induced variation in the received flux
(the magnitude-squared of field intensity) impinging on the receiver's antenna.
A longer \emph{flux} coherence time $T_f$ is the time over which the flux, and hence
energy observed in each bundle, is stable.
Generally speaking $T_f \gg T_c$.

\subsubsection{Time-bandwidth product}

The number of \emph{degrees of freedom} of $h(t)$ is defined as 
\begin{equation*}
K = B \, T \,.
\end{equation*}
The degrees of freedom $K$ is an important characteristic of $h(t)$,
and is discussed further in Chapters \ref{sec:channelmodel} and \ref{sec:incoherence}.
Chapter \ref{sec:incoherence} estimates that for typical values of
the physical parameters of interstellar communication and motion,
$B_{c} \, T_{c} \gg 1$.
This is a physical property of the Milky Way of immense significance to communication, 
because if it were not true
then the interstellar propagation would be too unstable for reliable interstellar communication.
Nature has been kind to us and other civilizations that seek mutual contact.
In practical terms, it means that choosing an $h(t)$ that meets the constraint of the ICH
is feasible, and the assumption of simultaneous bandlimiting and time limiting
of $h(t)$ is a reasonable one when $K \gg 1$.\footnote{
There is an uncertainty relation for the Fourier transform which asserts that $h(t)$
cannot be strictly speaking simultaneously bandlimited and time-limited.
However, this assumption is a useful approximation for larger values of 
time-bandwidth product $K$,
but we can say for sure that $K \ge 1$.}

\subsubsection{Energy of a bundle and peak power}

\begin{changea}
The energy of a bundle $\energy_h$ is an important parameter in the design of the transmitter,
and one that has considerable consistency over different information rates
and between information-bearing signals and beacons.
As shown in Chapters \ref{sec:infosignals} and \ref{sec:discovery}
we must have at the receiver
\begin{equation*}
\frac{\energy_h}{N_0}  \approx 15 \ \text{dB}
\end{equation*}
in order that each energy bundle
can be detected with sufficient reliability.\footnote{
This conclusion does not apply when time diversity is employed
to counter scintillation for an information-bearing signal, as described in Chapter \ref{sec:infosignals}.
Time diversity divides the total energy into a set of redundant energy bundles
transmitted over time, so each bundle has correspondingly less energy.}
For example,
for the same scenario as in Table \ref{tbl:linkbudget}
(1000 light years, 8 degrees kelvin system temperature, Arecibo antennas),
the required energy of a bundle
at the transmitter is $\energy_h \approx 843$ kilojoules (see Appendix \ref{sec:linkbudget}).
\end{changea}

Together $\energy_h$ and $T$ determine the peak power,
\begin{equation*}
\powerpeak = \frac{\energy_h}{T} \,.
\end{equation*}
For the example just given, and assuming that $T = 1$ sec, the peak power would be 843 kilowatts.
Since a smaller $\powerpeak$ is generally advantageous, this suggests
choosing $T$ as large as possible, or $T \approx T_c$.
Radio astronomers call $T$ the ''integration time'', and seek to
maximize it in order to gather more energy from a source.

\begin{changea}
In summary, the energy and peak power of each transmitted bundle
is determined almost entirely by the antenna designs, the distance, and the receiver system noise temperature.
It does not depend to any significant degree on whether it is an information-bearing signal or beacon,
the information rate, or the average power.
The radiated average power at the transmitter is determined by the
energy per bundle $\energy_h$ (which is relatively fixed)
and by the rate at which energy bundles
are transmitted (which depends on other factors including the information rate).%
\end{changea}

\subsubsection{Impairments remaining in a coherence hole}

The impairments cannot \emph{all} be eliminated by sufficiently reducing
$T$ and $B$.
There are two impairments that persist no matter how small these parameters
are chosen, and the transmitter and receiver must deal with these two impairments.
These are additive noise (of thermal origin) and multiplicative
noise (called \emph{scintillation}).
They are analyzed in Chapter \ref{sec:channelmodel}, where a resulting model
for the interstellar channel is developed that can be employed in communication
system design.

The matched filter is the receiver's way of dealing with noise optimally.
When matched filter detection is used for this purpose, scintillation manifests itself as a variation in phase
and magnitude of the sampled output of the matched filter from one energy bundle to another.
The phase difference does not affect a
receiver that estimates only the magnitude or energy in the bundle.
Any difference in magnitude, however, is significant and a major consideration
in the design of both communications and discovery strategies.
Scintillation is an astronomical term,
and the same phenomenon is called \emph{fading} in the communications engineering literature.
Fading is a usual affliction for radio communication in the presence of relative motion
between transmitter, receiver, and the physical environment.

\subsubsection{Choice of $h(t)$}

The modeling of noise in the presence of matched filter estimation and
detection yields an important insight.
In terms of performance, the detailed shape of the waveform $h(t)$ does not matter,
as long as it conforms to the constraints $T \le T_c$ and $B \le B_c$.
The performance, as measured by detection reliability, is influenced only
by the energy $\energy_h$, and not the choice of waveform $h(t)$.

This lack of guidance from a performance standpoint makes the choice of $h(t)$ a potentially controversial issue.
Two cases of particular interest are illustrated in Figure \ref{fig:mfillustrate}.
The first is a time segment of a sinusoid, for which $B$ is on the order of $B \approx 1/T$,
and thus $K \approx 1$.\footnote{
For a complex-valued waveform, each degree of freedom is actually a complex-valued quantity,
so this one degree of freedom encompasses both a magnitude and a phase.}
The other interesting case is a segment of a noise-like waveform,
where $B \gg 1/T$ and $K \gg 1$.
The characteristic of a noise-like waveform that makes it particularly suitable is
that it can be designed to spread its energy more or less uniformly over the
entire range of times $t \in [0,T]$ and frequencies $f \in [0,B]$.
The considerations that enter the choice of $h(t)$ are discussed further
in Chapter \ref{sec:detection}.
The immunity to noise is not impacted by $h(t)$, but the
immunity to radio-frequency interference (RFI) that may be present at
the receiver is the major argument for choosing a noise-like waveform.
In particular, the sinusoid (although conceptually simple) is highly susceptible
to interference, whereas a properly designed noise-like waveform is maximally immune
to interference \citeref{211}.
Another conceptually simple choice is a narrow pulse, and this option has also been
pursued in SETI searches \citereftwo{631}{631}.
However, the narrow pulse violates the coherence bandwidth constraint $B \le B_c$ and thus is
not of interest for power-efficient design.

When $h(t)$ is chosen to match the constraints of an ICH,
almost all impairments introduced by radio propagation through the interstellar medium
and by the relative motion of transmitter and receiver are avoided.
The reason that an energy bundle should be designed to avoid these impairments
is that introducing impairments results in a reduction in the energy as it is observed at the
matched filter output, and this reduces the sensitivity of the detection
because it increases the energy necessary in each bundle to achieve reliable detection.
This in turn reduces the power efficiency.

\section{Information-bearing signals}
\label{sec:illustrations}

The average power $\power$ of a signal is defined as the
long-term average of the energy per unit time, and has the units of watts.
For an information-bearing signal communicating at an information rate $\rate$,
power-efficient design seeks to minimize power $\power$ for a signal
representing rate $\rate$.
If the information bearing signal has bandwidth $\Btotal$, then
the goal of spectrally efficient design is to minimize $\Btotal$
for a given $\rate$.
This section illustrates by example the major difference between power-efficient
and spectrally efficient design.
The significance of this difference is that terrestrial communication emphasizes
spectral efficiency, and this makes power-efficient design significantly distinctive and unfamiliar.

\subsection{Magnitude-based modulation}

When the desire is to achieve high spectral efficiency without concern for average power,
a good approach is to encode the information in the magnitude of an energy bundle.
This is efficient because increasing the number of possible magnitudes
increases the information content but has no impact on the bandwidth.

\subsubsection{On-off keying}

A simple baseline case, which is neither power-efficient nor spectrally efficient,
is the on-off keying (OOK) as illustrated in Figure \ref{fig:ookIllustrate}.
In OOK the presence or absence of a bundle of energy is used to represent
one bit of information.
In the $k$th time slot of width $T$, the signal is 
$A_{k} \cdot \sqrt{\energy_h}\, h(t - k T)$
where $A_{k} = 0$ or $A_{k} = 1$
depending on the information bit.
Although current and past SETI searches generally assume a beacon
rather than information-bearing signal, the signal they are searching
for strongly resembles OOK for $A_k \equiv 1$.

\incfig
	{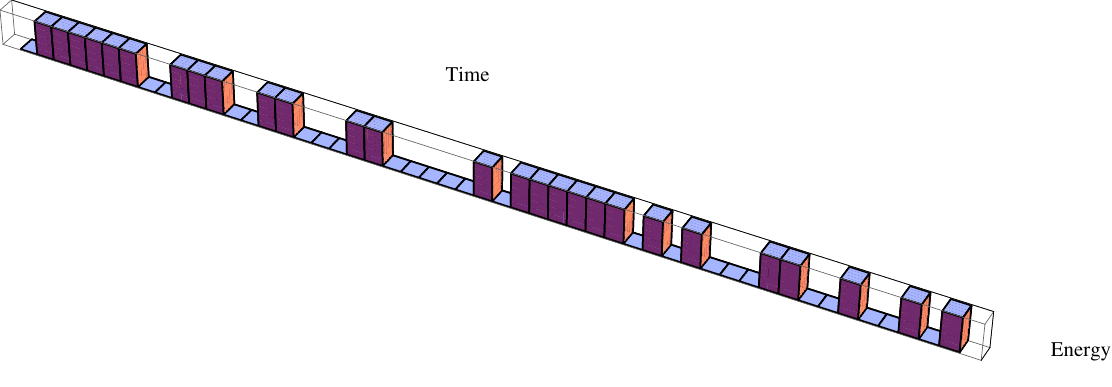}
	{1}
	{In on-off keying (OOK), in each time slot
	the presence or absence of a bundle of energy represents
	a single bit of information.}

\subsubsection{Multilevel keying}

 OOK is easily improved upon with respect to spectral efficiency
 by using multilevel keying,
as illustrated in Figure \ref{fig:mmkIllustrate}.
Rather than one information bit per energy bundle,
$\log_2 M$ bits of information can be conveyed
by transmitting one of $M$ different levels of energy.
(The case shown is $M = 8$, corresponding to $\log_2 8 = 3$ bits of information per bundle.)
This represents an improvement in spectral efficiency because each bundle of energy consumes the
same time and bandwidth as OOK, but conveys more information bits.

\incfig
	{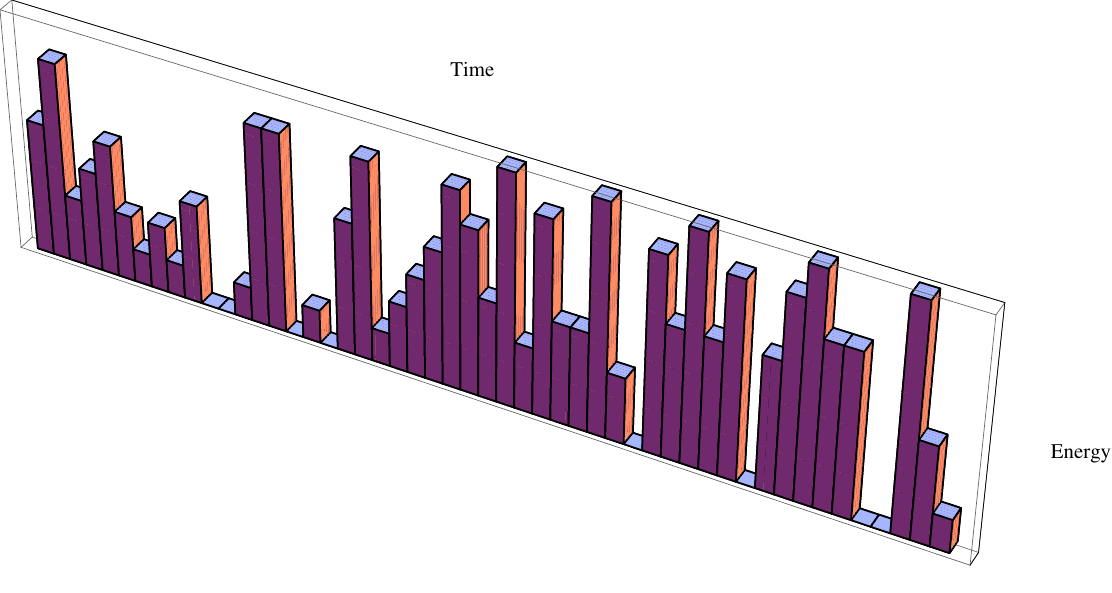}
	{1}
	{Spectrally efficient design is illustrated by multilevel keying.
	At each time slot, a
	bundle of energy assumes one of $M = 8$ values (including zero),
	and thus represents $\log_2 M = 3$ bits of information. }

Multilevel keying is inefficient in its use of power,
and thus represents a backward step in power efficiency.
In order to maintain fixed immunity to the noise that accompanies each
energy bundle, the average energy in each bundle
is increased by a factor proportional to $M^2$.
The number of bits communicated increases more slowly, 
by only a factor of $\log_2 M$, and thus the energy required to
communicate each bit of information is minimized for $M = 2$.
Suitably generalized,
this is a reason that
two values for the energy per bundle, zero and $\energy_{h}$,
is advantageous for power-efficient design.

\subsection{Location-based modulation}

If the desire is to improve power efficiency without concern
about bandwidth, a general approach is to convey information by the
\emph{location} of an energy bundle in frequency or time.
This way a single bundle with energy equal to $\energy_h$
can communicate multiple bits of information.
This potentially offers greater power efficiency than OOK
because the number of locations (and hence bits of information)
can be increased with less impact on the average power.\footnote{
Some increase in $\energy_h$ is necessary to maintain
fixed noise immunity because of the greater opportunity to make errors,
but it is modest.
This tradeoff is investigated in Chapters \ref{sec:fundamental} and \ref{sec:infosignals}.}

Suppose a single energy bundle is located at one of $M$ places.
These places should not overlap in time or frequency, but can be
located at either different times or different frequencies or both.
If the receiver is able to successfully ascertain the correct location
of the single energy bundle,
$\log_2 M$ bits of information has been conveyed from transmitter
to receiver.

\subsubsection{M-ary frequency-shift keying}
\label{sec:maryfsk}

In M-ary frequency-shift keying (FSK), a single energy bundle
is located at one of $M$ frequencies, thereby conveying $\log_2 M$
bits of information.
It is illustrated in Figure \ref{fig:fskIllustrate}
(the case shown is $M = 64$, corresponding to $\log_2 64 = 8$ bits of information).
The transmitted signal takes the form $\sqrt{\energy_h} \, h(t) \, e^{\,i\, 2 \pi A_{k} f_{0} t}$,
where $A_{k} \in \{ 0, 1, \dots \, M-1 \}$
and $f_{0}$ is the frequency spacing, typically equal to $f_{0} = B$
so that the energy bundle locations do not overlap in frequency.
Compared to multilevel keying, this requires $M$ times as much bandwidth,
so it is spectrally less efficient.
However, it is more power-efficient than either OOK or multilevel keying because the 
energy transmitted in each time slot is $\energy_{h}$
but each time slot communicates $\log_2 M$ times as much information for that fixed expenditure
of energy.

\incfig
	{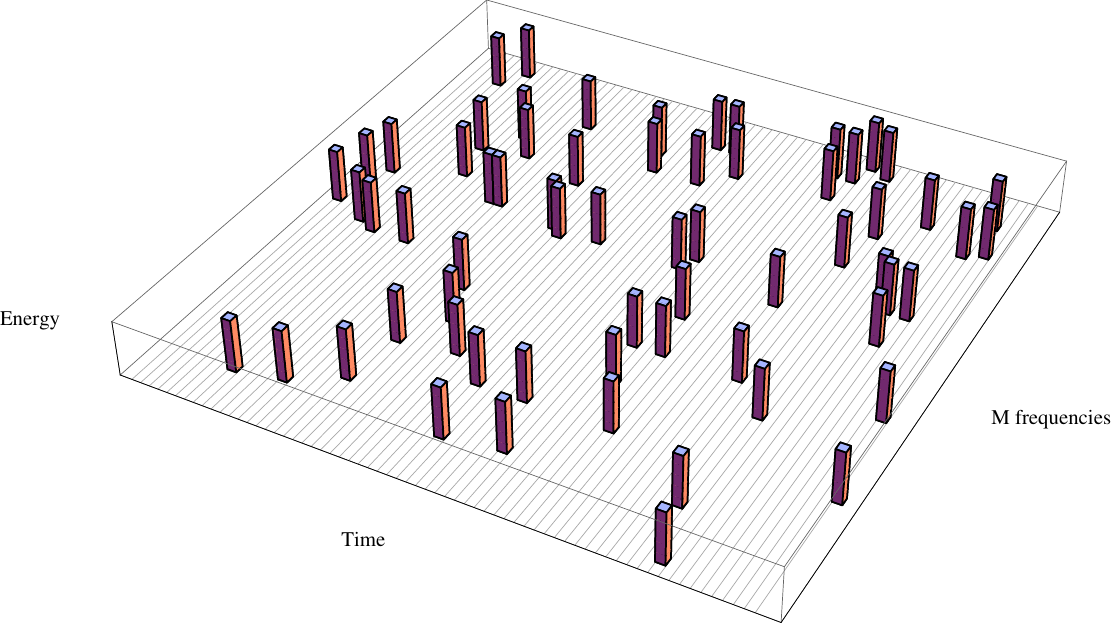}
	{1}
	{Power-efficient design is illustrated by M-ary FSK.
	In each time slot, a bundle of energy assumes one of $M=64$ different frequencies,
	representing $\log_{2} M = 8$ bits of information.}

It is extremely significant for interstellar communication that in location-based modulation
the bandwidth $B$ of each energy bundle can remain fixed, even as $M$ and the total signal bandwidth 
$\Btotal$ increases.
Increasing $\Btotal$ in this manner is totally consistent with confining each energy bundle to an ICH,
and thus maintaining the advantages of minimal impairment due to interstellar propagation.
In this manner,
increasing the total bandwidth $\Btotal$ need not invoke interstellar impairments
that we would normally associate with violation of the coherence bandwidth constraint.
This is because the signal has been broken down into smaller energy bundles that individually do not
trigger interstellar impairments other than noise and scintillation.
If those energy bundles do not overlap one another in time or frequency as transmitted,
the same is true as observed at the receiver because time-dispersive and frequency-dispersive
impairments introduced by the interstellar medium and motion have been rendered sufficiently
small that they can be ignored.

\subsubsection{M-ary pulse-position modulation}

While M-ary FSK represents information by the the location of
energy bundles in frequency, an alternative is
M-ary \emph{pulse-position modulation} (PPM), as illustrated in
Figure \ref{fig:ppmIllustrate}.
In PPM information is conveyed by the position of an energy bundle in time.
The transmitted signal is of the form $\sqrt{\energy_h} \, h(t - A_{k}/ M T)$,
where $A_{k} \in \{ 0, 1, \dots \, M-1 \}$.

\incfig
	{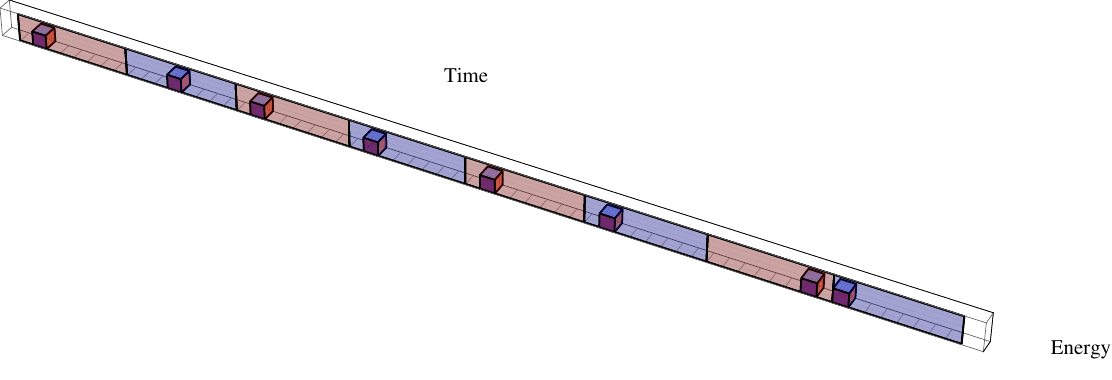}
	{1}
	{
	Power-efficient design is also illustrated by
	pulse-position modulation (PPM).
	Time is divided into codewords with $M=8$ positions each, and $\log_{2} M = 3$ bits
	of information are conveyed by the location of a single pulse within
	each codeword.
	}

Because of the presence of frequency-dependent dispersion in the interstellar channel,
FSK is a more natural choice than PPM.
As we attempt to increase the power efficiency in PPM, each bundle assumes a shorter and shorter
time duration and this expands its bandwidth.
Eventually that bandwidth will exceed the frequency coherence of the propagation, and the
energy bundles will suffer significant distortion.
In FSK, in contrast, power efficiency improves as bundles are transmitted on more frequencies,
but the parameterization of each individual bundle remains unchanged.
For this reason, M-ary FSK is adopted in our design approach, and PPM will receive no further consideration.

\subsection{Choosing an information rate and power}
\label{sec:irAndPower}

In M-ary FSK $\log_{2} M$ bits of information are conveyed by a single energy bundle.
This grouping of $M$ possible energy bundle locations and the associated $\log_2 M$ bits of information
is called a \emph{codeword}.
This terminology comes from channel coding, which will be described shortly.
The information rate $\rate$ is easily adjusted by changing the rate at which these
codewords are transmitted, and this choice also affects the average power $\power$.
Suppose that codewords are transmitted at rate $F$ codewords per second.
The $\rate$ and $\power$ then scale as
\begin{align}
\label{eq:ratepower1}
&\rate = F \cdot \log_{2} M \\
\label{eq:ratepower2}
&\power = F \cdot \energy_{h} \,.
\end{align}
Assuming that $M$ and $\energy_h$ are held fixed,\footnote{
The value of $\energy_{h}$ is determined by how much energy in the signal
is necessary to overcome the noise.
Thus, the minimum value of $\energy_{h}$ is determined by the reliability with which
the $\log_{2} M$ information bits are to be extracted from the signal.
This required value of $\energy_h$ actually depends on $M$, as
follows from more detailed modeling in Chapters \ref{sec:fundamental} and \ref{sec:infosignals}.}
both $\rate$ and $\power$
can be manipulated by the choice of $F$, with a energy per bit
\begin{equation*}
\energybit = \frac{\power}{\rate} = \frac{\energy_h}{\log_2 M} \,.
\end{equation*}
For fixed $M$ and $\energy_h$,
the relationship $\rate \propto \power$ is maintained as $F$ is varied.
The rate $F$ is simply a way to dial in any rate $\rate$ and
average power $\power$ desired, with the constraint that $\power \propto \rate$.

There is an upper limit on the codeword rate $F$ dictated by the coherence bandwidth $B_c$
in conjunction with the uncertainty principle $B T \ge 1$.
Together these two constraints imply that
\begin{equation*}
T \ge \frac{1}{B} \ge \frac{1}{B_c} \,.
\end{equation*}
The time duration $T$ of each energy bundle  is constrained by the coherence properties of the interstellar channel
at the carrier frequency in use, and this places an upper limit on the codeword rate $F \le B_c$.
This limit on $F$ does not, however, constrain the information rate $\rate$ in \eqref{eq:ratepower1}
since $M$ can be arbitrarily large as long as the total bandwidth $\Btotal$ is not constrained.

\subsubsection{Duty factor}

For any value of codeword rate $F$, the \emph{duty factor} is defined as
\begin{equation*}
\delta = F \, T \,.
\end{equation*}
Since any $F \ge 0$ is suitable, the only constraint on $\delta$ is $\delta \ge 0$.
The rate $\rate$ and average power $\power$ can be expressed in terms of duty factor as 
\begin{align}
\label{eq:ratepowerdelta1}
&\rate = \delta \cdot \frac{\log_{2} M}{T} \\
\label{eq:ratepowerdelta2}
&\power = \delta \cdot \frac{\energy_{h}}{T} = \delta \cdot \powerpeak \,.
\end{align}
The concept of duty factor is most useful when $0 \le \delta \le 1$.
In this case $\delta$ has the interpretation as the fraction of the time that is occupied by
an active energy bundle, with the
remaining fraction of time $(1-\delta)$ consisting of blank intervals during which no 
energy bundle is being transmitted.
When $\delta = 1$ (or equivalently $F = 1/T$) an energy bundle is being transmitted
at all times.
Adjusting the duty factor is an equivalent way to control the rate $\rate$
and average power $\power$.
An M-ary FSK signal with a duty factor of $\delta = 1/4$ is illustrated in Figure \ref{fig:fskIllustrateDutyCycle}.

\incfig
	{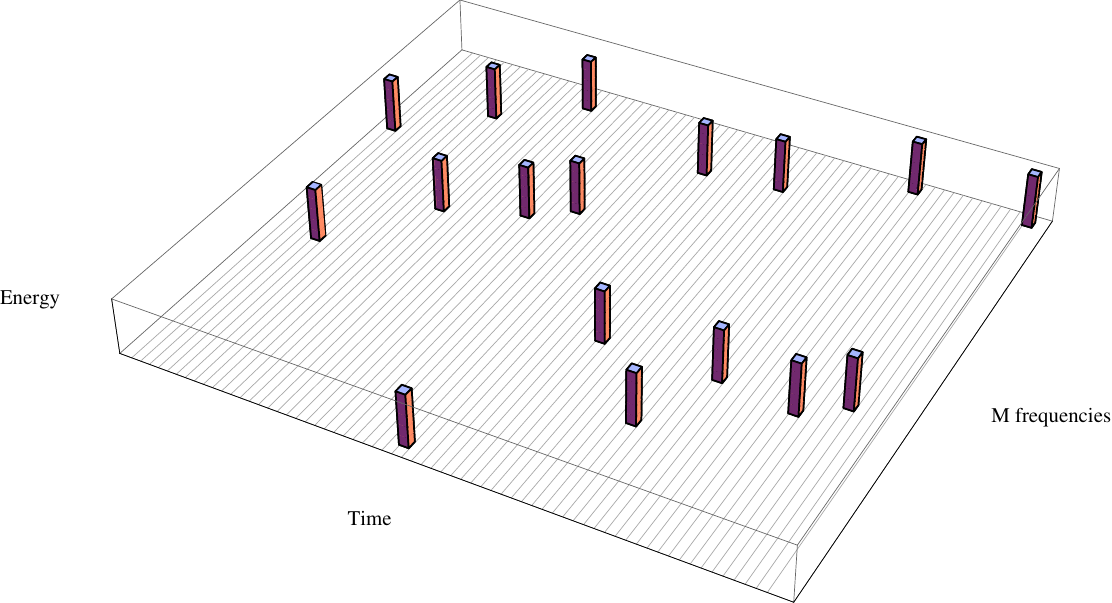}
	{1}
	{
	Figure \ref{fig:fskIllustrate} is repeated with a duty factor $\delta = 1/4$.
	Both the
	average power $\power$ and information rate $\rate$ are reduced by $\delta$.
	}

\section{Power efficiency vs. spectral efficiency}
\label{sec:tradeoff}

In the last section we saw an interesting divide.
There seemed to be two distinct sets of techniques,
one (using multilevel modulation) conserving bandwidth but wasting power,
and the other (using M-ary FSK) conserving power but wasting bandwidth.
Is this just an accident, or clever choice of examples, or is something more fundamental going on?
It is fundamental.
Recall the definitions of power efficiency and spectral efficiency
in Section \ref{sec:defpowere}.
We will now show that at the fundamental limit higher power efficiency
inevitably results in poorer spectral efficiency, and vice versa.
One consequence is that there is a global maximum
power efficiency $\powere \le \powerenoise$ for interstellar communications
that cannot be exceeded.
Since $\powerenoise$ is a constant independent of $\power$ and $\rate$,
it follows that the fundamental limit obeys
the relationship $\rate \propto \power$ that was illustrated in Figure \ref{fig:powerefficiencyconcept},
and also characterizes the rate-power tradeoff for practical communication
as was also illustrated in that same figure.

\subsection{Channel models}

There is no single fundamental limit on communications, but rather
it depends on the nature of impairments encountered by the signal
as it passes from transmitter to receiver,
a model of which is called a communications channel.
It is helpful to use the simplest model that captures the
essential impairments, those impairments that place limits on the ability to communicate reliably.
The impairments relevant to interstellar communication are covered in
some detail in Chapter \ref{sec:incoherence}, but it is also shown there
that most of them can be rendered negligible by appropriate design of the
transmit signal.
The impairments that cannot be avoided are modeled in Chapter \ref{sec:fundamental}
and the fundamental limit on power efficiency that follows
from this model is determined in Chapter \ref{sec:fundamental}.

To get us started, and to capture the flavor of the fundamental limit,
a simpler case of additive noise in isolation is considered here.
This is modeled as 
\begin{equation}
\label{eq:awgnchannel}
Y(t) = X (t) + N(t) \,,
\end{equation}
where $Y(t)$ is the received signal corrupted by noise, 
$X(t)$ is the portion of $Y(t)$ attributable to the transmitted signal, and
$N(t)$ is  \emph{additive white Gaussian noise} (AWGN).
AWGN is modeled as a real-valued zero-mean random process $N(t)$ with
a Gaussian amplitude distribution, and is wide-sense stationarity
with a constant power spectral density $N_{0}$.
Since the power spectral density is flat, not a function of frequency,
the noise is said to be \emph{white}.

The significance of the whiteness property is that any two samples $N(t_{1})$ and $N(t_{2})$
are statistically independent for $t_{1} \ne t_{2}$.
$N_{0}$ has the interpretation of the average power of the noise within a one Hz bandwidth,
and it has the units of watts per Hz, or equivalently joules.
As a consequence of the unit energy in $h(t)$ of \eqref{eq:unitenergy}, it turns out
that the variance of the noise at sampled output of the matched filter is also $N_{0}$.
This can get a little confusing, because $N_{0}$ plays two roles, one as the
power spectral density of the white noise at the input to the matched filter,
and another as the variance (or energy) of the noise component of one sample at the output of the matched filter.

Scintillation is another property introduced by scattering phenomena in the
interstellar medium.
The model of this (not included in \eqref{eq:awgnchannel}) is a complex-valued Gaussian random variable
that multiplies the signal with a single
parameter, its variance $\vs$.
As discussed in Chapter \ref{sec:channelmodel}, due to the physics of scattering
it is always the case that $\vs = 1$.
However, for generality, and to highlight the role that scintillation plays, we always
carry the $\vs$ parameter without substituting its numerical value.

\subsection{Energy contrast ratio}

The matched filter output sampled at correct point in time
(aligned with an input energy bundle) is characterized by three
parameters $\{ \energy_{h}, \, \vs , \, N_{0} \}$.
The signal component, ignoring any unknown phase shift, is the product
of $\sqrt{\energy_{h}}$ and a scintillation multiplicative noise, with has zero mean and variance $\vs$.
Of course, in the absence of any energy bundle, the signal component is identically zero.
In addition, there is an additive noise sample which is Gaussian distributed with zero mean and variance $N_{0}$.

The interpretation of a given energy $\energy_{h}$ at the output of the
matched filter depends on the size of the scintillation and the level of the noise $N_{0}$.
In particular, when we characterize the reliability of algorithms for detecting an energy bundle,
it turns out that it is the ratio of $\energy_{h}$ and $N_{0}$ that matters, rather than the
values of each parameter alone.
When the noise variance gets larger, the signal energy must get larger in proportion to maintain the same reliability.
For that reason, we define the \emph{energy contrast ratio} as
\begin{equation}
\label{sec:ecrdefinition}
\xi = \frac{\text{Signal energy}}{\text{Average noise energy}}
\end{equation}
in a single sample at the output of the matched filter.\footnote{
The terminology ''energy contrast ratio'' is adopted from \citeref{650}.
There is a resemblance to the concept of signal-to-noise ratio (SNR), except that SNR
is always defined as a ratio of powers rather than energies.
The use of SNR can be confusing since its numerical value depends on several assumptions,
like the bandwidth in use and the point in the system that it is measured.
The energy contrast ratio does not depend on factors like integration time or bandwidth,
and is measured in one standard place, in one sample at the matched filter output.
}
For the simplest case where scintillation is absent,
at the matched filter output the size of the signal is $\sqrt{\energy_h}$ so that the numerator
in \eqref{sec:ecrdefinition} equals the square of the signal, while the denominator equals the variance of the noise term. 

A couple of different definitions of ''signal energy'' in the numerator will prove useful.
When scintillation is present the signal term is actually random, so $\xi$ is defined
with the signal-squared in the numerator replaced by the variance of the signal,
\begin{equation}
\label{eq:ecrs}
\ecrs = \frac{\energy_{h} \, \vs}{N_{0}} \,.
\end{equation}
For some purposes, the signal energy $\energy_h$ needs to be replaced by
the signal energy per bit $\energybit$, resulting in an alternative definition of energy contrast ratio,
\begin{equation}
\label{eq:ecrb}
\ecrb = \frac{\energybit \, \vs}{N_{0}} \,.
\end{equation}

\subsection{Tradeoff between power efficiency and spectral efficiency}
\label{sec:tradeoffpowerspectral}

The Shannon limit
takes the form of a quantity called the \emph{channel capacity} $\capacity$
that can be quantified for any given channel model \citeref{164}.
The significance of $\capacity$ is that whenever the information $\rate$ is
less than or equal to capacity $\rate \le \capacity$ it is mathematically
possible to communicate with perfect reliability through the channel.
Conversely, whenever $\rate > \capacity$ it is mathematically impossible
to communicate reliably.
Perfect reliability is an asymptotic result, applying in the limit of unbounded
complexity and time delay.

\textbox
	{Existence of fundamental limits was a revelation}
		{
This idea that communication is subject to fundamental physical limits
was a revelation when Shannon introduced it in 1948 \citeref{164}.
Prior to that time, it was expected that more and more complicated and sophisticated
techniques could achieve higher and higher information rates.
As it turns out there is a hard limit on the information rate that can be achieved.
Also, it was assumed that completely reliable communication
would be impossible at any rate due to the deleterious effect of noise.
In fact, it is theoretically possible to achieve arbitrarily high reliability for any information
rate, as long as that rate is less than capacity.
It was also found in the late 1940's that practical communication systems at the
time performed poorly relative to this fundamental limit, and the research that
followed seeking to reduce that penalty has remarkably improved the performance
of communication systems.
\boxbreak
Similarly, the existence of a fundamental limit is a great asset for interstellar communication.
This limit alerts us to the fact that relative to the types of signals typically considered
for interstellar communication to date,
a substantial improvement in power efficiency is feasible.
The fundamental limit also constrains what any civilization can achieve,
no matter how advanced.
If we can achieve a practical design that approaches the fundamental
limit, we know that no other civilization can do significantly better.
We also find that a design that approaches the fundamental limit becomes
strongly constrained, reducing design freedom and 
providing specific guidance on signal structure to both transmitter and receiver.
Thus, the fundamental limit proves valuable as a form of implicit coordination
between transmitter and receiver \citeref{583}.
	}

One of the easiest channel models for determining capacity is the AWGN
model of \eqref{eq:awgnchannel}.
The impact of scintillation on the achievable power efficiency
is considered in Chapter \ref{sec:fundamental} and was accounted for
in the example of Table \ref{tbl:linkbudget}, but in the interest of simplicity let us neglect
scintillation and the other impairments on the interstellar channel for now.
Shannon showed that the capacity of an AWGN channel is given by
\begin{align}
\label{eq:capacity}
&\rate \le \capacity =   \Btotal \cdot \log_{\,2} \left( 1 + \snrin \right) \text{   bits/sec} \\
\label{eq:snrin}
&\snrin = \frac{\power}{N_{0} \, \Btotal} 
\,.
\end{align}
This famous formula applies to the case where the average signal power is constrained to
be $\power$, and the total bandwidth of the signal is constrained to be $\Btotal$.
In this case $\snrin$ is the signal-to-noise ratio at the \emph{input} to the receiver, since
$N_{0} \, \Btotal$ is the total noise power within the signal bandwidth
and it is advantageous for the front end of the receiver to remove all noise outside the signal bandwidth
before doing anything else.

Although it isn't immediately obvious, \eqref{eq:capacity} harbors within it
a fundamental tradeoff between power efficiency and spectral efficiency.
Specifically, the bound of \eqref{eq:capacity} is easily algebraically manipulated
to create an equivalent bound on power efficiency $\powere$
expressed in terms of spectral efficiency $\spectrale$,
\begin{equation}
\label{eq:powerVsSpectral}
\powere \le \powerenoise \cdot g( \spectrale )
\end{equation}
where the two factors are
\begin{equation}
\label{eq:powerenoise}
\powerenoise =  \frac{1}{N_{0} \, \log 2}
\ \ \ \ \text{and} \ \ \ \ \
g( x ) =  \frac{x \, \log 2}{2^{\,x}-1} \,.
\end{equation}
In this bound, $\powerenoise$ is the maximum possible power efficiency $\powere$
that can be achieved for the AWGN channel, and
the factor $ g( \spectrale )$ represents a reduction in power efficiency $\powere$
that must be sacrificed to achieve a given spectral efficiency $\spectrale$.
The region represented by
\eqref{eq:powerVsSpectral} is illustrated in Figure \ref{fig:powerVsSpectral}
by plotting the function $g(\spectrale)$.

\incfig
	{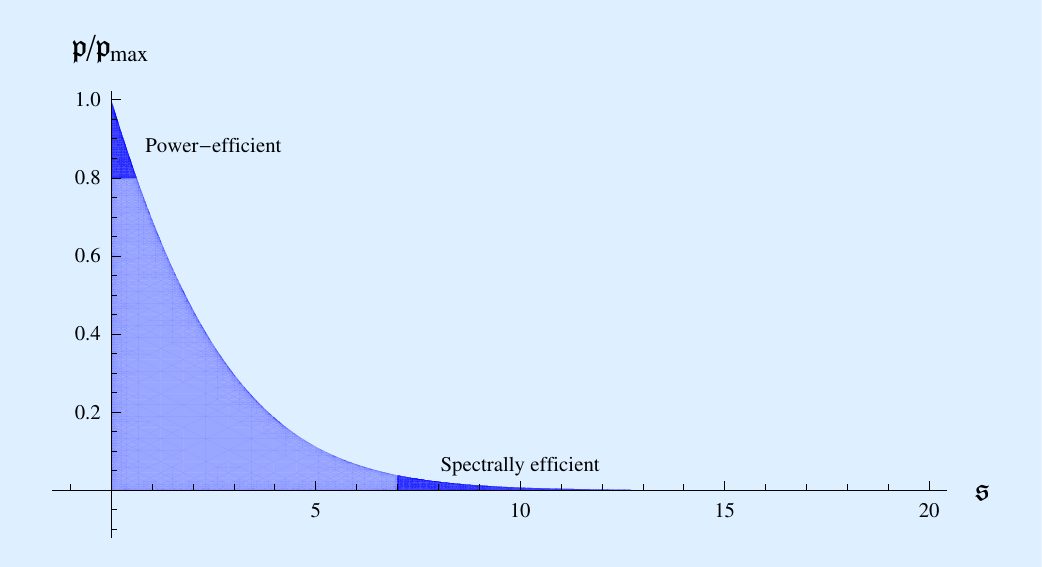}
	{1}
	{In the presence of white Gaussian noise (characteristic of natural
	sources of thermal noise), Shannon proved that reliable
	communication is possible within the shaded region of 
	$\{ \powere/\powerenoise , \, \spectrale \}$ and not possible outside that region.
	The horizontal axis is the spectral efficiency $\spectrale$ and the
	vertical axis the power efficiency $\powere$ normalized by the global
	maximum possible power efficiency $\powerenoise$.
	Commercial terrestrial wireless systems invariably operate in the
	spectrally efficient regime where $\spectrale$ is favored at the expense of $\powere$.
	Interstellar communication should operate in the power-efficient regime where
	$\powere$ is favored at the expense of of $\spectrale$.
	As a result, we would expect an interstellar communication system to use
	a relatively wide bandwidth relative to the information rate.
	 The global maximum power efficiency $\powerenoise$ can only be achieved
	 as $\spectrale \to 0$, or equivalently the total signal bandwidth $\Btotal$ is unbounded.}

Figure \ref{fig:powerVsSpectral} harbors several important insights.
Substantial increases in power efficiency $\powere$ are feasible only
by forcing spectral efficiency $\spectrale$ to get small, or equivalently
for a given information rate $\rate$ reducing average power $\power$
to its lowest levels dictates increasing bandwidth $\Btotal$.
There is, however, a global maximum $\powerenoise$ that can be achieved,
no matter how small $\spectrale$.
Another insight is that spectral efficiency $\spectrale$ can be increased
without limit -- there is no global maximum -- but only at the expense of reducing $\powere$.
High spectral efficiency pays a price in terms of the required $\power$.

As shown this tradeoff between the two types of efficiency only applies to a channel
model that includes AWGN, but not scintillation.
In Chapter \ref{sec:fundamental} the issue of scintillation is addressed.
While scintillation has a material effect on the channel capacity $\capacity$
in general, it is shown there that
scintillation has no effect on the global maximum power efficiency $\powerenoise$.
Thus scintillation does not reduce (or increase) the maximum achievable 
power efficiency.

\subsubsection{Terrestrial vs interstellar communication}

Terrestrial communication usually emphasizes high spectral efficiency $\spectrale$ 
due to the high commercial value attached to spectrum allocations.
Each government allocates a specific band of frequencies for specific purposes,
and that band is typically bought and sold by different companies as they seek
commercial gain from offering wireless services based on that bandwidth.
Further, there is considerable pressure to match information rates between
these wireless services and pre-existing copper wire-based networking,
such as ethernet, which places a premium on high information rate $\rate$.
Thus, there is an ever-increasing economic value placed on the spectrum,
and increasing pressure to reduce the bandwidth $\Btotal$
by using more and more sophisticated and processing-intensive signal
processing that improves the spectral efficiency.
In addition, in terrestrial services, particularly over relatively short range,
the average transmit powers are typically not an issue from the perspective
of energy costs.
There is pressure to increase battery life in mobile terminals, but the power
devoted to radio transmission is typically small relative to the power consumed
by the aforementioned signal processing.
Thus, an acceptable tradeoff is to increase power $\power$ and reduce
power efficiency $\powere$ in the interest of less bandwidth $\Btotal$
and higher spectral efficiency $\spectrale$.
Putting these considerations together, the motivation to operate in the
''spectrally efficient'' region shown in Figure \ref{fig:powerVsSpectral} is overwhelming.
In contrast, as argued earlier in this chapter there is strong pressure to
operate in the ''power efficient'' region in interstellar communication.

\subsubsection{Fundamental limit on power efficiency}

Fom \eqref{eq:powerenoise} there is a fundamental global limit on the
power efficiency $\powere$ that can be achieved for the AWGN channel
regardless of the total bandwidth $\Btotal$ consumed,
\begin{equation}
\label{eq:poweremaxAWGN}
\powere \le \powerenoise = \frac{1}{N_{0} \, \log 2} \,.
\end{equation}
Keep in mind that $\power$ is the average power (not peak power)
measured in the receiver at the same point the total noise $N_{0}$ is measured,
typically in the receiver baseband processing.
While $\power$ is directly related to the transmit average power and energy consumption,
it is also influenced by the propagation distance, transmit antenna directivity or gain, and receive antenna
collection area.
Expressed in terms of the energy contrast ratio per bit defined in \eqref{eq:ecrb}, 
\eqref{eq:poweremaxAWGN} becomes
\begin{equation}
\label{eq:ecrawgn}
\ecrb \ge \log 2 = 0.693 \ \ (- 1.59 \ \text{dB} )\,.
\end{equation}
This is a remarkably simple and universal result, which will be shown
in Chapter \ref{sec:fundamental} to apply to the interstellar channel, even with
its numerous propagation and motion impairments including scintillation.
It simply asserts that each bit of information has to be accompanied
by a certain energy delivered to the receiver, that energy being proportional to the
size of the noise $N_{0}$ power density.
Surely this is one of the important and fundamental properties of our Galaxy, as it
governs the minimum energy that must be delivered to the receiver for each bit of information
when the goal is reliable communication
among civilizations separated at interstellar distances.

\textbox
	{Power-efficient design is conceptually simple}
	{
One of the strongest arguments for using power-efficiency
(as opposed to spectral efficiency) as the primary design goal
is the conceptual simplicity of power-efficient design.
The reason is that both power and spectral efficiency place a constraint
on power and energy, but spectral efficiency places an additional bandwidth constraint.
This additional constraint in the design makes it conceptually and practically much harder.
The penalty in average power for practical techniques relative to the fundamental limit on
power efficiency was essentially closed by 1967.
A similar accomplishment in approaching the fundamental limit including a bandwidth constraint
did not occur until 1994.
There is a vast difference in conceptual and practical complexity of the techniques
in the two cases.
\boxbreak
The practical significance of this to interstellar communication follows
from the lack of coordination between transmit and receive designers.
In interstellar communication, it can be argued convincingly that the
design techniques must be conceptually simple to have hope of
transmitter and receiver landing in a compatible zone.
This places a practical obstacle in the way of achieving high spectral
efficiency, but far less of an obstacle to achieving high power efficiency.
The relatively low processing requirements for power-efficient designs are
also a considerable advantage during discovery, where
our experience in SETI observation programs has been that the speed of search is practically
limited by the technology, budgetary, and energy consumption requirements of the
receive signal processing.
	}

In summary, there are numerous reasons in interstellar communication to put
power efficiency at the front of the line, making it a higher priority than spectral efficiency.
This places operation in the ''power efficient'' regime in Figure \ref{fig:powerVsSpectral},
rather than the ``spectrally efficient'' regime.
This makes interstellar communication a very different animal from terrestrial
communication, as will be seen.
It also implies that one expects a signal for interstellar communication to have relatively poor spectral efficiency,
or consume relatively wide bandwidths relative to its information rate.
On the face of it this would seem to contradict the need to avoid interstellar impairments,
but in fact it does not.
Typically in power-efficient design (as illustrated by M-ary FSK) the bandwidth of an
energy bundle $B$ is much less than the total bandwidth of the signal $\Btotal$,
and these parameters can be chosen independently.
It is $B$ that is governed by the ICH, and it is $\Btotal$ that determines the
spectral efficiency $\spectrale$.

\begin{changea}
One striking feature of the fundamental limit on power efficiency is that it is
not affected by the various interstellar impairments and is not affected
by carrier frequency.
This is true as long as the noise is accurately modeled by AWGN, which
is true for carrier frequencies below about $f_c =$ 60 GHz.
We will find that the concrete designs
considered in Chapter \ref{sec:infosignals}, are also
not affected by carrier frequency.
Keep in mind, however, that the fundamental limit has been defined in the
receiver baseband signal processing.
Referencing this to the required transmit average power,
there are some wavelength-dependent effects, mainly the
transmit antenna gain or directivity.
Thus, higher carrier frequencies remain attractive because they
allow higher transmit antenna gain for a given transmit antenna geometry,
but only for this reason.
Interstellar impairments like dispersion and scattering generally become
more severe for lower carrier frequencies, but this has no
effect on the achievable power efficiency.
\end{changea}

\subsection{Channel coding}

In proving his famous theorem, Shannon introduced channel coding.
In general channel coding takes a group of $\log_2 M$ bits and associates
a different waveform with each of the $M$ alternatives represented by those bits.
For the particular case where the channel code is structured around
energy bundles, that waveform consists of a collection of one or more
energy bundles at different times and translated by different frequencies.
M-ary FSK is a simple example.
Given a ``base'' waveform $h(t)$, 
the waveform actually transmitted over time duration $T$ is equal to $h(t)$
shifted by one of $M$ uniformly spaced frequencies.
The frequency offset is determined by all $\log_{2} M$ bits, 
and observing the frequency of the energy bundle maps into the
entire set of $M$ bits.
There is no way to infer one bit without inferring all $M$.
As $\log_2 M$ grows, this mapping becomes exponentially more complicated.

\subsection{Wake-up call from the fundamental limit}

To gain some perspective on the significance of \eqref{eq:poweremaxAWGN}, it is
useful to relate it to a simple but concrete modulation scheme, OOK as illustrated
in Figure \ref{fig:ookIllustrate}.
Recall that OOK is a ''middle of the road'' option,
one that is neither particularly spectally efficient nor power efficient.
It is characteristic of the type of signals that have been the primary target for
SETI discovery searches over the past five decades.

OOK is analyzed in Appendix \ref{sec:peook}, where its probability of error
$\pe$ in the presence of AWGN with or without scintillation is determined as a function of 
the energy contrast ratio $\ecrb$, and the result is
plotted in Figure \ref{fig:gapFundamental}.
The AWGN-only case is included even though it is not directly relevant to the
interstellar channel because it allows us to quantify how much of the penalty in power
efficiency is due to noise, and what additional penalty is due to scintillation.
As expected, $\pe$ decreases rapidly as the signal energy $\energybit$
increases in relation to the noise energy $N_0$.
Also shown, as a vertical line, is the fundamental limit
on $\ecrb$ at minus 1.59 dB, as given by \eqref{eq:ecrawgn}.
This plot reveals a substantial penalty for OOK relative to the fundamental
limit.
That penalty depends on the reliability that we demand, but at a moderate
reliability of $\pe = 10^{-4}$ it  is 15.5 dB for AWGN and 46.1 dB for AWGN in
combination with scintillation, corresponding to linear factors of 35.9 and 40,740.
Recall that the average power required is proportional to the information rate, or $\power = \rate$.
The conclusion of Figure \ref{fig:gapFundamental}, is that the 
required energy per bit $\energybit$ is
about 40,000 times larger for OOK than the fundamental limit,
and this value was assumed in Table \ref{tbl:linkbudget}.

In Chapter \ref{sec:fundamental} a simple channel code is demonstrated that can
approach the fundamental limit of \eqref{eq:ecrawgn} asymptotically
as the bandwidth is increased.
In Chapter \ref{sec:infosignals} it is shown that a constrained-complexity
version of that channel code can substantially increase the power efficiency,
although never eliminate the penalty entirely.
This channel code combines
five principles of power-efficient design principles that will now be described.
A simpler outage/erasure strategy is also analyzed in Chapter \ref{sec:infosignals}, 
and shown to also be effective in dealing with scintillation
as long as it is acceptable to suffer a major loss of recovered information 
during periods of low received signal flux due to scintillation.

\incfig
	{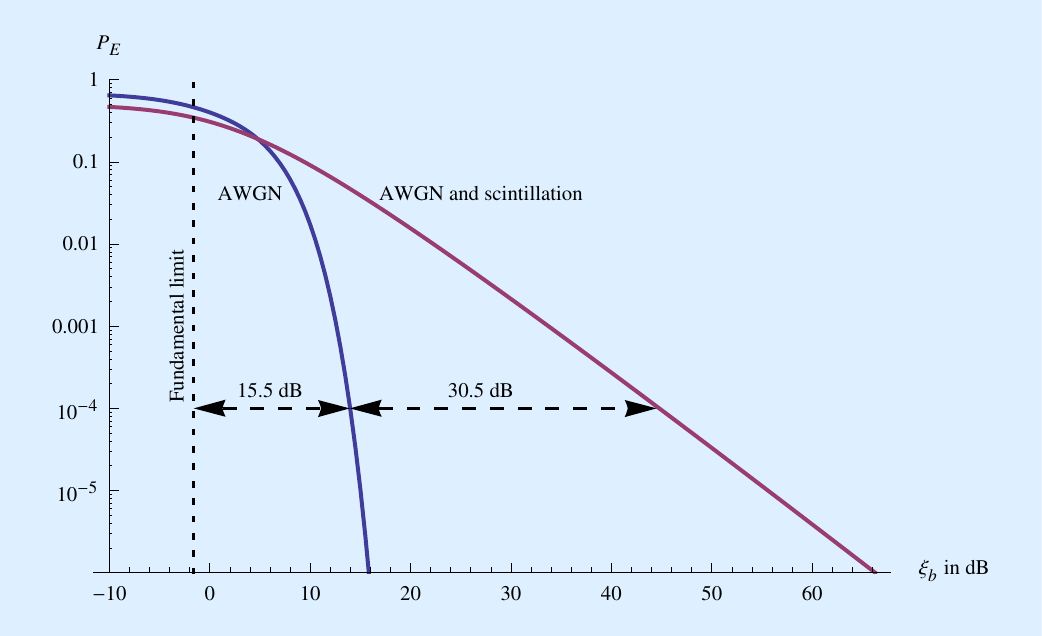}
	{1}
	{
	A log plot of the error probability $\pe$ vs the energy contrast ratio pr bit $\ecrb$
	for the baseline on-off keying (OOK) modulation of Figure \ref{fig:ookIllustrate}.
	Two cases are considered, average white Gaussian noise (AWGN) alone
	and AWGN combined with scintillation.
	The fundamental limit is $\ecrb = - 1.6$ dB, and is shown as the vertical line on the left.
	The penalty for the two cases relative to the fundamental limit is arbitrarily defined at
	a bit error probability of $\pe = 10^{-4}$ (an average of one bit out of every 10,000 in error)
	although it gets larger for smaller $\pe$ and gets smaller for larger $\pe$.
	With scintillation, the penalty in average power is 46.1 dB, or a linear factor of 40,740.
	For AWGN alone, the penalty is 15.4 dB, or a linear factor of 35.9.
	Scintillation adds a penalty of 30.5 dB, or a linear factor of 1135.
	}

\section{Five principles of power-efficient design}
\label{sec:principles}

The bound on energy contrast ratio of \eqref{eq:ecrawgn} is 
a provable mathematical result.
If this bound is satisfied, reliable communication is possible,
and if it is not satisfied then reliable communication is not possible.
As shown in Chapter \ref{sec:fundamental}, this limit applies to the
interstellar channel with all its impairments, including but not limited
to noise and scintillation.
This provides a measure of how well we are doing relative to
an absolute standard, the same standard that will constrain any
other civilization, no matter how advanced.
The techniques for approaching the fundamental limit
not only improve the power efficiency, which we have argued is
a compelling objective for interstellar communication, but they also
act as a form of implicit coordination between transmit and receive designs.

In this section, five design principles that collectively approach the
fundamental limit are listed and described, referring back to earlier
examples for illustration.
It is demonstrated in Chapter \ref{sec:fundamental} that by combining
these principles in a concrete channel code the fundamental limit can be approached
asymptotically. 
Four of these are drawn from R.S. Kennedy's research in the 1960's
into power-efficient design for the fading noisy channel \citeref{637}.
Kennedy's work of course did not address the specific impairments of the interstellar
channel, so we have added a fifth principle, which is to restrict
the energy bundles to fall within the parameters of the interstellar coherence hole (ICH).
It is shown in Chapter \ref{sec:fundamental} that this added principle does not affect
the fundamental limit, and in fact is necessary to approach the fundamental limit.
Fortuitously, it converts the interstellar channel into 
something similar to that analyzed by Kennedy
in the 1960's.

Any set of practical techniques for improving power efficiency $\powere$
and for approaching the fundamental limit are not as fundamental
as the limit itself.
It is possible that a designer somewhere else in the Milky Way
could come up with a different collection of design techniques that
works as well.
What gives us confidence that these five principles do have universal appeal
is that they are each very straightforward, and each has a compelling and complementary
purpose in increasing power efficiency and approaching closer to the fundamental limit.
They are simple enough to be practical for the uncoordinated interstellar channel.
Our own judgement is that it would not be possible to approach the fundamental
limit in any simpler way.
Further, the existence of a fundamental limit provides an absolute
and universal standard against which to evaluate the
utility of these principles.
Recognizing, however, that there is nothing absolute and
universal about these five design principles, it is certainly appropriate to put further thought
and research into alternative ways of
approaching the fundamental limit on power efficiency.

The five principles will now be described in order. 
Section \ref{sec:illustrations} and Chapters \ref{sec:infosignals}
and \ref{sec:discovery} illustrate the application of these principles to concrete designs.
In particular, the implication of constraining the complexity of the channel code
is addressed in Chapters \ref{sec:infosignals}, where the resulting penalty relative to the fundamental
limit is quantified.

\subsection{Use energy bundles}

The presence of a signal of technological origin along a given line of sight can
be detected by bundles of excess energy conveyed by some carrier waveform $h(t)$.
The transmitter should use only two values for the energy of an energy bundle, zero and $\energy_{h}$,
because to use any more values while maintaining detection reliability increases the power
requirement more than it increases the information rate.
The receiver, knowing or guessing $h(t)$, can determine the
 portion of the energy in the reception that is aligned with $h(t)$
by applying a matched filter to the reception followed by sampling and taking the magnitude squared.
Due to phase instability in the interstellar channel combined with motion of transmitter
and receiver, no phase information can be trusted, and this is why the receiver should estimate
magnitude (total energy) only.
That total energy aligned with $h(t)$ includes both signal energy of technological origin (if present)
and inevitably noise energy as well.
When the total energy aligned with $h(t)$ is significantly larger than what would be
expected of noise alone, then it can be inferred with confidence (high probability)
that excess signal energy of technological origin carried by $h(t)$ is present.

If the power spectral density of the noise is unknown or uncertain, the receiver
may not have full visibility into how much noise energy is typically aligned with $h(t)$.
This may occur, for example, when there are astronomical sources of thermal noise
along and near the line of sight.
In that case, the energy aligned with $h(t)$ can be compared to the total energy of the
signal within the time duration $T$ and bandwidth $B$ occupied by $h(t)$.
Regardless of the noise level, when a larger fraction of the total energy is aligned
with $h(t)$ this is indicative of a signal of technological origin carried by $h(t)$.
This simple approach is described and analyzed in Chapter \ref{sec:detection}
and shown to be very effective, especially as the number of degrees of freedom $K = B \, T$
in $h(t)$ grows.

The ability to distinguish energy of technological origin from thermal noise of natural
origin depends on the energy in each  bundle $\energy_{h}$ arriving in the baseband circuitry of the receiver.
Other parameters of $h(t)$, including bandwidth $B$, time duration $T$, and its detailed waveform
do not influence the receiver's ability to distinguish $h(t)$ from noise.
This signal energy depends on the total transmitted energy in the bundle,
the propagation loss (which depends on distance),
and the transmit and receive antenna gains (which depends on the size and precision of the antennas).
Since this baseband bundle energy has to be larger than the noise energy, there is
little opportunity to reduce the average signal power by the specific means of reducing the amount of energy
in individual energy bundles.
This reality is established by modeling the discovery process in Chapter \ref{sec:discovery}
and the post-discovery communication process in Chapter \ref{sec:infosignals}.
This implies that the requirements on peak transmitted power and antenna gains in conjunction with propagation loss
cannot be relaxed without harming the reliability of communication.
Reductions in average receive and transmit power must be obtained by reducing the duty factor
rather than reducing the amount of energy in each bundle.

\subsection{Choose transmit signal to avoid channel impairments}

A variety of impairments introduced by radio propagation through the interstellar medium,
as well as by the relative motion of transmitter, receiver, and interstellar medium,
are listed, described, and modeled in Chapter \ref{sec:incoherence}.
One approach to dealing with these impairments would be to compensate for
them in the receiver.
This is the normal approach in spectrally efficient design, but is inappropriate
for power-efficient design because such
processing would inevitably carry a penalty
in enhancement of noise, and
thereby reduce the receiver sensitivity and necessitate an increase in signal power to compensate.
The alternative is to avoid the impairments in the first place by the appropriate choice
of the transmit signal $h(t)$ serving as a carrier of energy, at a modest penalty in bandwidth.
It is shown in Chapter \ref{sec:incoherence} that by restricting the time duration $T \le T_{c}$ and
bandwidth $B \le B_{c}$ of a transmitted waveform $h(t)$,
the impairments reflected in the energy estimate at a receive matched filter output can be
rendered negligible.
The parameters $\{ T_{c} ,\, B_{c} \}$ define the interstellar coherence hole (ICH),
and depend on parameters of the interstellar propagation, parameters of motion, and carrier frequency.

There are two exceptions, noise and scintillation, which cannot be circumvented 
by restrictions on $T$ and $B$.
The integrity of individual energy bundles are not affected by scintillation
(because $T$ is chosen to be short relative to the coherence time $T_c$ of the scintillation),
but scintillation will result in a variation and uncertainty in the energy of individual
energy bundles arriving at the receiver. 
These remaining impairments are an important feature of the resulting model of
the interstellar channel, and they are described and modeled in Chapter \ref{sec:channelmodel}.

\subsection{Convey information by location rather than amplitude or phase}

In communication engineering, there are three common ways to convey one or more bits of
information to a receiver: by adjusting the amplitudes and/or phases of
a carrier waveform $h(t)$, by adjusting the location of $h(t)$ in time and/or frequency,
and by transmitting one from among a set of waveforms $\{ h_{n} (t) , \, 1 \le n \le N \}$.
Amplitude/phase modulation is commonly used in designs seeking spectral
efficiency because increasing the number of alternative amplitudes and/or phases
does not increase the required bandwidth.
However, this makes inefficient use of energy because for fixed noise immunity the
required energy per bit $\energybit$ increases faster than the number of bits
conveyed by modulating the amplitude or phase of a single waveform.

For power-efficient design, the second and third techniques are available, because
multiple bits of information can be conveyed with the energy of a single waveform.
This is illustrated by M-ary FSK, which is also an example of the third technique for the special
case where $N = M$ and $\{ h_{n} , \, 1 \le n \le N \}$ are orthogonal waveforms,
called orthogonal modulation.
M-ary FSK is also an example of a channel code, where the location
of a single energy bundle is determined not by an individual bit, but by
a collection of $\log_2 M$ bits.
Finally, approaching the fundamental limit on power efficiency requires a bandwidth
that is growing without bound, and M-ary FSK is consistent with this requirement.

\subsection{Make energy bundles sparse in time as well as frequency}

Delivering an energy per bundle large enough to overwhelm the noise at the receiver
allows individual energy bundles to be reliably detected.
This is useful for reliable information recovery, and also for discovery of the
signal in the first place.
With this constraint, the only way to reduce the average signal power is to have
fewer energy bundles per unit time.
The time duration of each energy bundle is also bounded by the
coherence time of the interstellar channel, $T \le T_c$.
This implies that at low average power levels energy bundles will be sparse in time,
by which we mean that periods of transmitted power are interspersed with
large blank intervals with no transmitted signal.
Combining that with location-based information recovery, 
which results in energy bundles that are sparse in frequency,
the conclusion is that bundles are sparse in both time and frequency.

In this report, the same basic approach of bundle sparsity
in time is encountered consistently,
including the following contexts:
\begin{description}
\item[Reduce information rate $\rate$.] 
As discussed in Section \ref{sec:irAndPower}, to maintain the
reliability of energy recovery, $\rate$ should be reduced by reducing
the duty factor $\delta$.
Low duty factor is synonmous with energy bundles that are sparse in time.
\item[Increasing power efficiency $\powere$.]
Suppose we wish to minimize the average power $\power$
required for a given fixed information rate $\rate$, or in other words
approach the fundamental limit in power efficiency.
As shown in Chapter \ref{sec:fundamental}, this can be accomplished
by using M-ary FSK as $M \to \infty$, which encodes more and more bits
of information using the location of a single energy bundle, implying increasing sparsity of energy
bundles in frequency.
For fixed $\rate$, $M \to \infty$ also implies that duty factor $\delta \to 0$,
 implying increasing the sparsity of energy bundles in time as well.
\item[Minimizing number of observations in discovery.]
If the transmitter wishes to reduce its energy consumption by reducing average power $\power$,
then during discovery of that signal the receiver must increase the number of observations
required to reliably detect that signal.
It is shown in Chapter \ref{sec:discovery} that the best tradeoff between transmitter energy
consumption and receiver observations is achieved when the transmitter reduces its power
by reducing the duty factor $\delta$, implying that the energy bundles are increasingly sparse in time.
\end{description}

One surprising result from Chapter \ref{sec:fundamental} is that
the fundamental limit for the interstellar channel is not affected by scintillation.
This does not mean that scintillation can be ignored in practice, but merely that
the presence of scintillation does not 
increase or reduce the maximum power efficiency that
can be obtained.
The sparsity of energy bundles is the reason for this.
As the fundamental limit is approached, the density of energy bundles in time
and frequency approaches zero.
As a result the error probability becomes dominated by the receiver's
ability to accurately detect the \emph{absence} of energy bundles, or in other words
the primary source of unreliability is false alarms in the detection of individual energy bundles.
These false alarms are a noise-only phenomenon, and are not influenced by scintillation.

\subsection{Combat scintillation with time-diversity combining}

While a channel code like M-ary FSK
can approach the fundamental limit for power efficiency on the
AWGN channel without scintillation,
it does nothing to counter scintillation.
Scintillation causes a relatively slow variation in the signal magnitude, as well
as relatively rapid variations in phase.
For a signal consisting of intermittent energy bundles,
like M-ary FSK,
scintillation causes a variation in observed energy per bundle with time.
As it turns out, scintillation (discussed in Chapter \ref{sec:channelmodel})
does not affect the \emph{average} energy per bundle,
and it also does not affect the fundamental limit on power efficiency
(see Chapter \ref{sec:fundamental}).
It does cause the \emph{received} energy per bundle to sometimes be greater than
average and at other times less than average.
If no other measures are taken, this will cause the reliability of
information extraction to vary with time.
In particular, there will be periods during which essentially no
information is extracted, and the result of that is poor overall reliability.

The transmitter can counter scintillation by introducing \emph{time diversity},
which then allows the receiver to do \emph{diversity combining}.
The idea is simple, and widely used in terrestrial wireless systems.
Instead of transmitting a single bundle with energy $\energy_h$ to represent $M$ bits,
the energy is divided among $J$ bundles with energy $\energy_h$ spaced out in time
at intervals, so the total energy becomes $J \energy_h$.
Crucially, these energy bundles are spaced at time intervals large relative to the coherence time.
Statistically, some of those bundles will be attenuated by scintillation,
and some will be amplified, but these effects will be statistically independent.
In diversity combining, the receiver combines these bundles
by averaging the per-bundle received energy estimates, that averaging being
a much more reliable estimate of the transmitter's intention because it averages
out the statistical variations in bundle energy due to scintillation.
Because of this averaging, the energy in each bundle $\energy_h$ can be reduced,
perhaps by as much as a factor of $J$.

An example of a time-diversity channel code is the ''time-diversity FSK''
that is illustrated in Figure \ref{fig:timeDiversityCode}.
It uses the M-ary FSK code described in Section \ref{sec:maryfsk} as a starting point
by defining $M$ frequencies where a single energy bundle may be present,
communicating $\log_2 M$ bits of information with a single energy bundle.
For time diversity, within each codeword the energy bundle is repeated $J$ times.
The $J$ exact replicas are spaced at intervals larger than the flux coherence time $T_f$,
so that they each experience a different 
(and hopefully independent) state of scintillation.
We would like this spacing to be large enough that the $J$ values of scintillation
are statistically independent.
The receiver can estimate the energy at each individual time and
frequency, but accumulate that energy over the $J$ replica locations to average out
the variation in flux $\flux$ due to scintillation.
This averaging is called \emph{phase-incoherent time diversity combining}.

\incfig
	{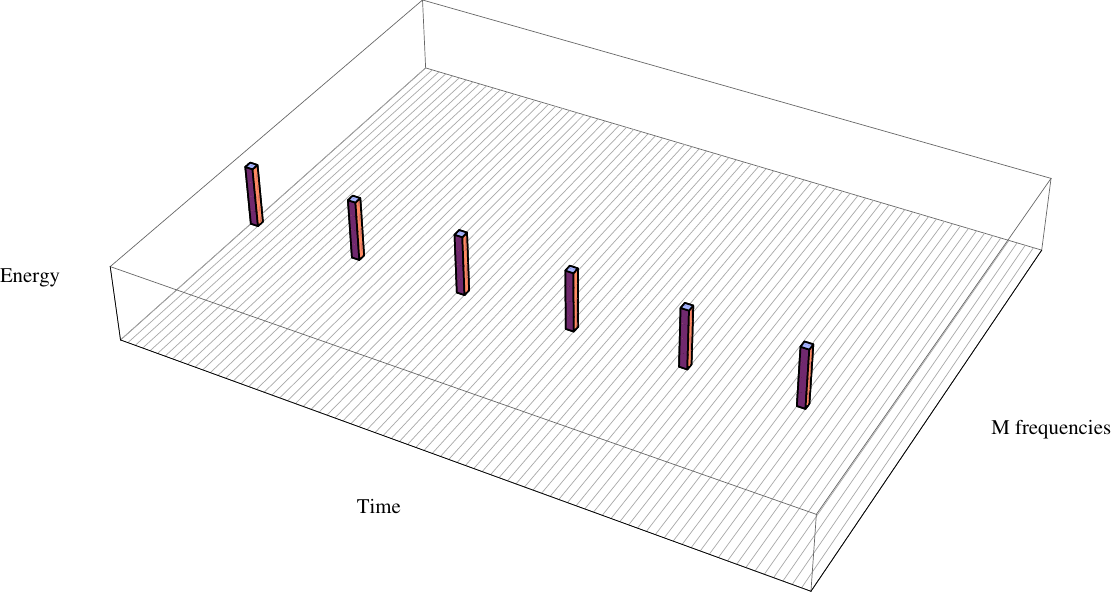}
	{1}
	{
	Example of one codeword from a time-diversity channel code for the interstellar 
	channel with noise and scintillation.
	This code uses M-ary FSK ($M = 64$ is illustrated) to communicate $\log_2 M$ bits of
	information ($\log_2 64 = 8$ is illustrated).
	In this case, the frequency chosen is 25 (out of 64 possibilities).
	Each energy bundle at one FSK frequency is repeated $J$ times ($J = 6$ is illustrated) for time diversity,
	with a spacing between repetitions equal to $12 \times T$ in this case.
	This spacing between repetitions should be larger than the flux coherence time $T_f$ of the channel
	(which is likely to be much larger than shown in this example).
	There are various ways to position codewords relative to one another, as
	discussed further in Chapter \ref{sec:infosignals}).
	}

The utility of a combination of M-ary FSK and time diversity to reduce the penalty
relative to the fundamental limit is illustrated in Figure \ref{fig:PeBFSK}, drawing on
results from Chapter \ref{sec:infosignals}.
Increasing $M$ from $M=2$ to $M=2^{20}$
(to combat AWGN) and increasing $J$ from $J = 1$ to
$J= 49$ (to combat scintillation) reduces that penalty by 37.2 dB,
which corresponds to a reduction in average power by a linear factor of 5,300.
This reduction in power is obtained at the expense of increasing the total bandwidth $\Btotal$
by a factor of $2^{20} \approx 10^6$.
For example, if the bandwidth of each energy bundle is one Hz, then $\Btotal$
is approximately 1 MHz.
This is consistent with what was seen in Figure \ref{fig:gapFundamental},  in that without the benefits of
M-ary FSK and time diversity a very large average signal power
is necessary to overcome scintillation and still achieve a consistently low
error probability $\pe$.
In Chapter \ref{sec:fundamental}
it is shown that a channel code like Figure \ref{fig:timeDiversityCode}
can approach the fundamental limit on power efficiency
for the interstellar channel asymptotically, in the sense that for any power efficiency
below the fundamental limit the error probability can be driven to zero $\pe \to 0$ as
$J \to \infty$ and $M \to \infty$.
Figure \ref{fig:PeBFSK} illustrates what portion of that job
 can be accomplished with finite $J$ and $M$, 
 but also emphasizes that a penalty remains due to using finite values.\footnote{
Some other factors than finite $J$ and $M$ contribute to this penalty.
The method used for detection of energy bundles is modified to be insensitive to
the bundle energy and noise power, at some expense in power efficiency.
Phase-incoherent detection of energy bundles also contributes.
}
 That penalty is 10.4 dB, or a linear factor of 10.9 in average power.

\incfig
	{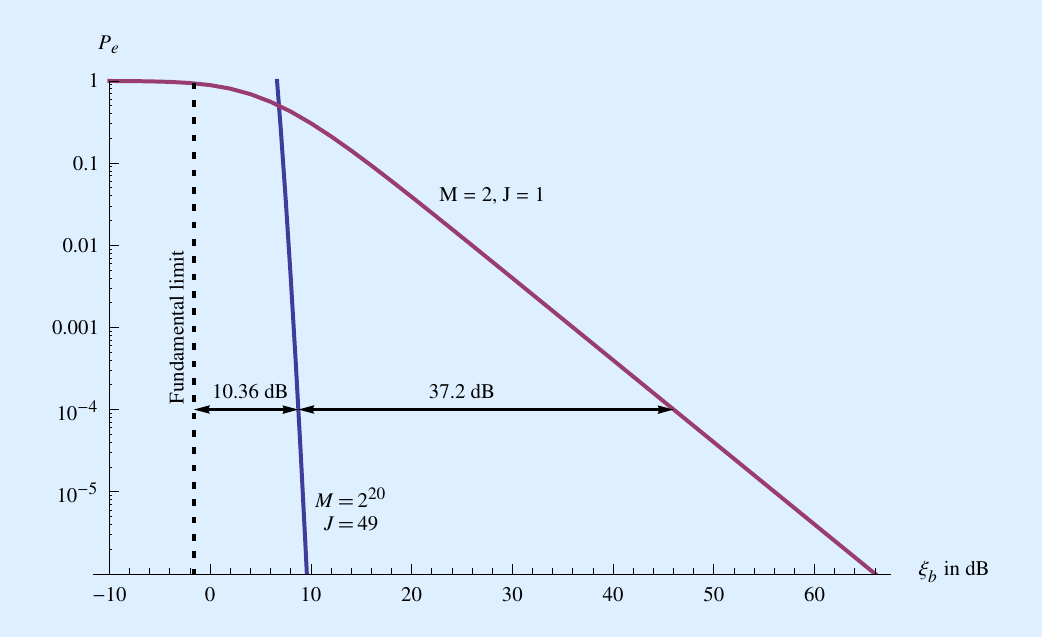}
	{1}
	{A comparison of M-ary FSK with time diversity
	and binary FSK without time diversity,
	made by plotting the bit error probability $\pe$
	against the energy contrast ratio per bit $\ecrb$.
	When $M = 2^{20} \approx 10^6$, the degree of time diversity
	$J = 49$ is chosen to minimize $\ecrb$ as will be shown in Chapter \ref{sec:infosignals}.
	By increasing $M$ from $M=2$ to $M=2^{20}$ and increasing $J$ from $J = 1$
	to $J = 49$, the average power required to achieve an error
	probability $\pe = 10^{-4}$ is reduced by 37.2 dB, or a linear factor of 5,300.
	Nonetheless there remains a penalty relative to the fundamental limit
	of $\ecrb = -1.6$ dB of 10.36 dB, or a linear factor of 10.9 in average power.
	}

In Chapter \ref{sec:infosignals} a simpler strategy is also explored,
based on the observation that M-ary FSK alone can approach to within 3 dB the fundamental limit
on the AWGN channel without scintillation (that 3 dB is due to phase-incoherent detection of
energy bundles).
If the designers abandon the objective of obtaining \emph{consistently} small $\pe$,
then a simpler approach is feasible.
The receiver can monitor the state of the scintillation (ideally at each of the
$M$ frequencies), and
declare an \emph{erasure} whenever an energy bundle is not detected
at a frequency where the received signal flux is relatively high.
(When the total bandwidth $\Btotal$ is small enough that all frequencies
are affected in essentially the same way by scintillation, then
an erasure becomes an \emph{outage}.)
Any information that might have been conveyed by an erasure or outage is
discarded.
Knowing explicitly what information has been discarded
is easier to deal with than random errors in unknown locations.
By deliberately throwing away some information, the information $\rate$
is effectively reduced, which reduces the effective power efficiency.
The discarded information must be recovered by 
taking advantage of redundancy in the received message.

\section{Discovery}
\label{eq:multipleobservations} 

The preceding has emphasized the reliable recovery of information from a signal.
However, this
extraction of information from an information-bearing signal is only possible once
the presence of that signal has been established.
That is the role of \emph{discovery},
which is addressed in Chapter \ref{sec:discovery}.
Discovery involves a different set of challenges from communication,
even when dealing with the same information-bearing signal.
Discovery is made difficult due to the ''needle in a haystack'' challenge of searching a large parameter
space consisting of spatial, frequency, and time variables.
Also it lacks prior knowledge of signal
characteristics and parameterization, such as power level, repetition time,
channel coding structure, duty factor, etc.

If the signal is assumed to consist of energy bundles
arranged in time and in frequency, as strongly suggested by power-efficient design principles,
then the challenge becomes simpler.
The initial stage of discovery can focus on detecting individual energy bundles,
a challenge addressed in Chapter \ref{sec:detection}.
Whereas in communication the receiver knows fairly precisely where and when
to look for energy bundles, during discovery there is no prior guidance based on prior observations.
However, there is substantial guidance based on fundamental limits
and the design principles that they suggest.
For power-efficient signals energy bundles are sparse,
meaning most observations will not overlap an energy bundle.
For the case of power-efficient information-bearing signals, that sparsity
is in frequency as well as time.
When the power and rate of the signal are reduced by lowering the duty factor,
this becomes even more extreme.
Scintillation is another factor that affects all signals, power efficient or not,
and results in periods when the per-bundle energy is attenuated.

As a result of signal intermittency by design and due to scintillation,
a search for individual energy bundles
cannot hope to be successful with a single observation at each frequency.
Multiple observations in the same location 
(line of sight and carrier frequency) are necessary to have a reasonable
chance of an observation interval overlapping an actual energy bundle with sufficiently large power to detect reliably.
To take account of scintillation, multiple observations are most efficient when
they are spaced at intervals larger than the coherence time of the ISM.
This critical role of multiple observations in combating scintillation
was pointed out by Cordes, Lazio, and Sagan \citereftwo{122}{121},
although their modeling focuses exclusively on scintillation and does not take
account of the low duty factor characteristic of power-efficient signals.

After one energy bundle is detected, discovery can
shift to a pattern recognition mode, in which additional energy bundles
are sought at the same carrier frequency or nearby frequencies,
and with an expectation that they will be separated by ''blank'' intervals.
This is another area where the discovery of power-efficient signals differs from present and past
practice in SETI observation programs, which have assumed
that energy is continuously transmitted, and confined to a single frequency
or a frequency that is varying linearly with time due to the effects of acceleration.

\subsection{Effective energy}

Power efficiency is not relevant as a metric of performance in the discovery stage, 
because there is no concept of
information rate $\rate$ to maximize.
Rather we use another metric called \emph{effective energy}
that captures the operational resources devoted at transmitter and receiver, and the
possible tradeoffs between those resources.
There are two primary operational costs in discovery.
For the transmitter,
the primary operational cost is energy consumption,
which relates directly to the average transmitted signal power.
As a proxy for this transmit power, we use the average receive power $\power$,
which is proportional to transmit power if the transmit and receive antenna gains and
the propagation distance are held fixed.
At the receiver, the primary operational cost during discovery is the signal processing
devoted to multiple observations, and the energy consumed by this processing.
The total processing devoted to a single line of sight is proportional to $L \, T_{o}$,
where $L$ is the number of observations and $T_{o}$ is the time interval devoted
to each individual observation.
The total effective energy of those energy bundles that overlap the receiver's observations is
the product of the average power and the observation time, or
\begin{equation}
\label{eq:energyeff}
\energyeff = \power \, L \, T_{o} \,.
\end{equation}

\subsubsection{Power optimization for discovery}

The design of the transmitted signal has a profound
impact on the effective energy required for discovery of that signal at a given level of reliability.
That reliability is measured by the false alarm probability $\pfa$ and the
detection probability $\pd$.
In Chapter \ref{sec:discovery} a ''detection event'' is defined as a decision that one or more
energy bundles is present during the total observation period.
Thus, $\pfa$ is defined as the probability of a detection event
when in fact the reception consists of noise alone,
and $\pd$ is the probability of a detection event when the reception contains a signal
of technological origin corrupted by scintillation and noise.
The design of a signal has been \emph{power-optimized for discovery} when
at a given reliability $\{ \pfa , \, \pd \}$ the value of effective energy $\energyeff$ is minimized.
This is equivalent to minimizing the total resources devoted by transmitter and receiver.

\subsubsection{Total effective energy hypothesis}

It is not surprising that, as shown in Chapter \ref{sec:discovery},
 if the signal design has been power-optimized for discovery,
then at a given level of detection reliability
the effective energy $\energyeff$ stays almost constant as
$\power$, $L$, and $T_{o}$ are varied (subject to some constraints on $T_{o}$ discussed shortly).
This \emph{total effective energy hypothesis} quantifies an important
tradeoff between transmitter and receiver resources.
According to our principles of power-efficient design, the average power
is reduced by decreasing the duty factor $\delta$, and not by reducing the
energy per bundle.
Thus individual energy bundles can be detected reliably, but only when the receiver's observations
overlap an energy bundle.
With these assumptions,
the necessary total observation time $L \, T_{o} \propto 1/\power$.
This suggests (and it is in fact true) that a signal with arbitrarily small $\power$
can be detected at any level of reliability if the receiver is willing to increase its observation time
sufficiently.
In practical terms, the receiver may not be able or motivated to extend observations
at a given location (line of sight and frequency) indefinitely, and any such practical limitation will
place a lower bound on the average received power $\power$
of a signal that can be detected reliably. 

As the transmitter reduces its average transmit power, the receiver
can make up for this by increasing the total observation time $L \, T_{o}$,
thus transferring cost from transmitter (in the form of energy consumption for
transmission) to receiver (in the form of energy consumption for processing).
Of course a similar transfer in the opposite direction, from receiver to transmitter, is also possible.
Several economic arguments point in different directions:
\begin{enumerate}
\item
The \emph{value} of discovery accrues
to the receiver alone, at least in the timeframe of the generation conducting the
observations.
This would suggest a high expectation on the part of the transmitter designer
that the receiver should make a disproportionate investment, allowing the transmitter
 to reduce its energy consumption.
\item
The receiver processing power is disproportionately large because
of the ''needle in a haystack'' issue of having to search a large parameter space
consisting of different frequencies and different lines of sight,
and the transmitter should be sympathetic to this and reduce the receiver's energy consumption
required for observations in a single location
by increasing the transmitter's own energy consumption.
\item
It is likely that the transmitter will have multiple targets in different
lines of sight.
This raises the energy costs for the transmitter, and suggests transferring those
costs to the receiver.
\end{enumerate}
Overall it seems as if considerations 2. and 3. tend to neutralize one another,
leaving 1. as the dominant consideration.
If so, this favors using low-power signals at the transmitter, and accompanying this
by a relatively large number of observations $L$ at the receiver.

There is another tradeoff internal to the receiver between the number of observations
and the duration of each observation.
In particular,
for a given average receive power $\power$ the number of observations $L \propto 1/T_{o}$.
Thus, the number of observations can be reduced if each observation has proportionally greater duration.
However, this exact tradeoff holds only when duration of each observation $T_{o}$ contains no more than
one energy bundle.
The reason is that two or more bundles occurring during a single observation 
have a statistical correlation in the amount of energy they carry if (as it turns out is likely)
these energy bundles are spaced at an interval that is small relative to the coherence time of the scintillation.
In the presence of scintillation, it is more efficient to space observations far enough apart
that the energy of any energy bundles that overlaps those observations is statistically independent, and to
keep the observation duration $T_{o}$ small enough that there is no more than one energy
bundle overlapping a single observation.
This places an upper bound on $T_{o}$.
On the other end, in order to capture the total effective energy $\energyeff$ within a bundle,
the observation time $T_{o}$ must be greater than the duration $T$ of a single energy bundle.
This places a lower bound on $T_{o}$.

\subsubsection{Bundle energy hypothesis}

The total effective energy hypothesis comes with an important caveat.
It is valid only if the design of the signal by the transmitter has been
power-optimized for discovery.
This hypothesis in and of itself does not say anything about how to
perform this power optimization.
For a signal that is structured as a repeated transmission of
energy bundles, such as the power-efficient information-bearing
signals considered earlier,
the design question becomes a simple one.
For a total average received power $\power$, what should be the
rate at which energy bundles are transmitted and received, and what amount of
energy $\energy_h$ should be contained within each energy bundle?
For a given average power $\power$,
the designer of the transmit signal has a choice between a larger amount of energy per bundle
and a lower average rate of transmission of bundles, or the reverse.

A second \emph{bundle energy hypothesis} is verified in Chapter \ref{sec:discovery}.
This hypothesis asserts that as average power $\power$ is varied,
and the noise power spectral density $N_{0}$ is held fixed,
the energy per bundle $\energy_{h}$ 
(or equivalently the energy contrast ratio $\ecrs$) should remain approximately constant.
The logic behind this hypothesis simply notes that when
an energy bundle overlaps a single observation in the receiver,
the detection probability is determined by the energy contrast ratio $\ecrs$.
Thus the bundle energy hypothesis is equivalent to hypothesizing that
the detection probability for this individual energy bundle should remain
 fixed, even as the rate of transmitted bundles and the average power $\power$
is varied.
It turns out that a detection probability of 57\% is about optimum,
corresponding to $\ecrs \approx$ 14 to 15 dB.
The practical basis for this hypothesis is that if the designer attempts
to reduce $\energy_{h}$ in the interest of reducing $\power$,
then the detection probability falls fast and dramatically.
Making up for this by increasing the number of observations $L$ is possible, but this
is an inefficient tradeoff because it results in an $\energyeff$ much larger than necessary.

\textbox
	{The pervasive use of low duty factor to reduce energy consumption}
	{
It seems as if current and past SETI discovery search strategies are strongly
influenced by receiver design issues, but have not paid sufficient attention to
transmitter design, and in particular the issues of energy consumption.
Not only is the propagation loss that has to be overcome very great, but a transmitter is likely to transmit
along multiple lines of sight, magnifying the total energy consumption.
Many engineering applications have used low duty factor as a tool for
reducing energy consumption while retaining high peak powers for functional purposes.
In view of this engineering history, the conclusions here are hardly surprising.
\boxbreak
A premier example is radar, where high energy is necessary due to
low target reflectivity.
Radars invariably transmit short bursts of energy.
This has nothing to do with delay and distance estimation
(transmitting white noise and cross-correlating transmission and reception would work well),
but is done to reduce energy consumption.
Other examples abound.
Lighthouses concentrate the light in one direction in order to be reliably detected (seen), and then
rotate the whole assembly for 360 degree coverage without the energy consumption
that would be required for isotropic illumination at similar powers.
Mobile photography employs light flashes rather than continuous illumination
to increase battery life.
Pneumatic drills use high-energy impacts to fracture rock,
but space those impacts over time to reduce energy consumption.
	}

\subsection{Design modularity}

As illustrated in Table \ref{tbl:modularity}, the two energy hypotheses create
a desirable modularity in the design.
By this, we mean that there are two distinct sets of design parameters, one associated with the total effective energy hypothesis, and the other associated with the bundle energy hypothesis.
Within each of these two sets, the design parameters can be traded off against one another while
remaining faithful to the respective hypothesis.
However, as each set of design parameters is manipulated, this does not interact with the
parameters associated with the other set.
Engineers appreciate modularity because it results in a separation of concerns,
which in turn reduces the complexity of the design process.
This separation of concerns is particularly significant in the context of interstellar communication with
its lack of explicit coordination between the transmitter and receiver designs.

\doctable
	{\normalsize}
	{modularity}
	{
	From the perspective of discovery, two hypotheses separate the design
	parameters into two sets.
	The first set captures those concerns that affect the amount of energy per
	bundle at the receiver baseband.
	The second set captures those concerns that affect the tradeoff between
	average receive power and the number of observations performed by the
	receiver in each location (line of sight and frequency band).
	These concerns do not overlap one another, and thus can be addressed independently.
	The hypotheses are verified in Chapter \ref{sec:discovery}.}
	{p{2cm} p{5cm} p{7cm}}
	{
	\toprule
	\textbf{Hypothesis} & \textbf{Design invariant} & \textbf{Contributing design parameters}
	\\
	\addlinespace[3pt]
	\cmidrule{1-3}
	\textbf{Bundle \newline energy} &
	$\energy_{h} =$ received energy \newline per bundle  &
	Transmit energy per energy bundle
	\newline
	Peak transmitted power
	\newline
	Transmit antenna gain
	\newline
	Receive antenna collection area
	\\
	\addlinespace[3pt]
	\cmidrule{1-3}
	\textbf{Total \newline effective \newline energy} &
	$\energyeff =$ total effective \newline received energy \newline overlapping all observations &
	Average transmit power
	\newline
	$\power$ = average received power
	\newline
	$L$ = number of independent observations
	\newline
	$T_{o}$ = duration of each observation
	\\
	\bottomrule
	}

We have argued that a major consideration in interstellar communication
is energy consumption.
In the transmitter most energy consumption is related to the average transmit power $\power$,
and in the receiver to the baseband signal processing, which is related to the total observation time
$L \, T_{o}$ in each location.
The bundle energy hypothesis is consequential, because it tells us that the average
power should be reduced not by reducing the amount of received energy $\energy_{h}$ in each bundle, but
rather by reducing the rate at which energy bundles are transmitted.
This implies that reaching low average powers is achieved primarily by reducing
the duty factor $\delta$ of the signal, making it more and more intermittent in character.
While reducing $\delta$ necessitates increasing the number of independent observations $L$ or the duration of each observation $T_{o}$,
this increase in observation time is far smaller than would
be necessary to compensate for the alternative strategy of reducing $\power$ by reducing $\energy_{h}$.

Energy consumption applies to the operational cost, but also to that portion of the
capital costs related to heat dissipation due to inefficiencies in energy conversion 
to radio frequencies.
Other capital costs are influenced by the peak transmit power,
the size of the transmit and receive antenna, and the signal processing hardware at baseband.
To a degree the peak transmit power can be reduced by extending the duration $T$
of each energy bundle, but this opportunity is limited by the coherence time $T_{c}$ of the interstellar channel.
In Chapter \ref{sec:incoherence} it is shown that the coherence time is dominated by
uncompensated acceleration due to orbital dynamics,
although this source of incoherence can be compensated by the transmitter and receiver
in which case the dominant source of time incoherence becomes scintillation.
The strategy of reducing average transmitted power by reducing duty factor will cause
the peak transmitted power to be larger than the average transmitted power.
Within the constraint of a fixed energy per bundle $\energy_{h}$,
tradeoffs can be made among peak transmit power, transmit antenna 
directivity or gain, and receive antenna collection area.
However, reductions in average transmit power in the interest of reducing the
transmitter's operational costs will not yield an overall benefit in these
other sources of capital expense.
	
\subsection{Beacons}

Many SETI observation programs have drawn on the
Cyclops study published in 1970 \citeref{142}.
Cyclops did not emphasize information-bearing signals, but
rather proposed a search for beacons.
An interstellar beacon is designed to attract attention, but not embed
within it any other information.
The primary argument for beacons is simplicity, since there
is no need to guess or make assumptions about the structure of 
an information-bearing signal.
The type of beacon proposed by Cyclops is simply an unmodulated carrier,
which can be detected by multichannel spectral analysis.
Cyclops also assumed that relative acceleration of source and
observer would cause the observed frequency of the unmodulated carrier
to vary linearly with time.

The relative simplicity of the beacon makes it a good place to
begin in addressing discovery issues.
A simple generalization of this Cyclops beacon is a
\emph{periodic beacon} consisting of the regular transmission of
energy bundles.
This approach is motivated by the bundle energy hypothesis,
which addresses how to perform power-optimization of a beacon for discovery.
If the transmitter wishes to reduce the average receive power $\power$
(and hence the average transmit power) while minimizing the
impact of this power reduction on the resources allocated by the receiver, then it will do so
by keeping the energy of each bundle $\energy_h$ fixed
and reducing the duty factor.
Adhering to this hypothesis,
suppose the energy bundles are spaced at intervals of $T_p$,
so that the period of the beacon is $T_p$.
If each energy bundle has duration $T$, then the duty factor of the beacon is
\begin{equation*}
\delta_p = \frac{T}{T_p} \,.
\end{equation*}
A periodic beacon is illustrated in Figure \ref{fig:beaconIllustrate}
for the case of a 12.5\% duty factor.

\incfig
	{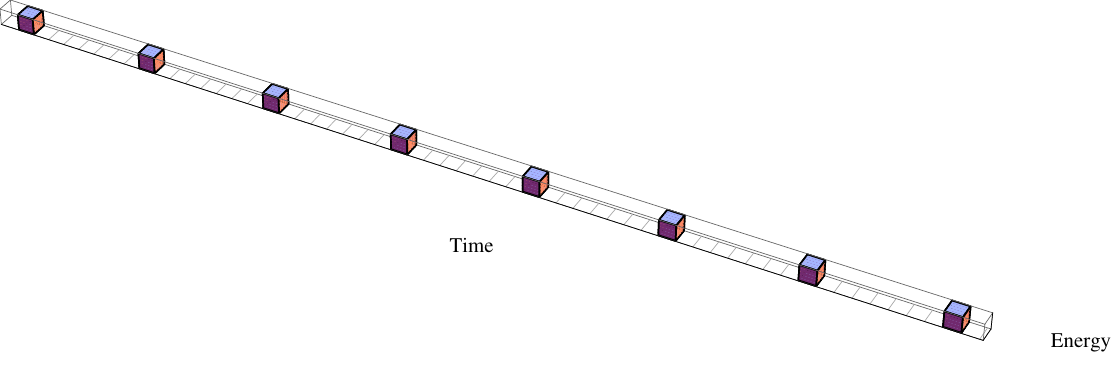}
	{.8}
	{
	Illustration of a periodic beacon consisting of periodic energy bundles.
	The duty factor of this example is $\delta_p = 1/8$ (12.5\%).
	}

The Cyclops beacon can be considered a special case of
a periodic beacon with a particular choice of waveform
and with 100\% duty factor,
\begin{align*}
&h(t) = \frac{1}{\sqrt{T}} \,, \ \ \ \ 0 \le t < T \\
&T_p = T \ \ \ \text{or} \ \ \ \
\delta_p = 1 \,.
\end{align*}
Within the class of periodic beacons, the Cyclops beacon design
maximizes the operational cost to the
transmitter by maximizing the average power $\power$
and thus its energy consumption, and
minimizes the cost to the receiver by minimizing the number
of observations $L$.
In fact, if $\energy_h$ is large enough to overcome scintillation, 
even $L = 1$ observations will be adequate for reliable detection.
On the other hand, if we assume that the transmitter designer
is interested in reducing its energy consumption, it is unlikely to
transmit a 100\% duty factor Cyclops beacon.
If the transmitter does reduce $\power$ by choosing a lower
duty factor, then the search strategies pursued over the past
four decades generally fail to discover the beacon.

\textbox{Beacons can be narrowband or broadband}
{
The tradeoff between power efficiency and spectral efficiency
of Figure \ref{fig:powerVsSpectral} applies to information-bearing signals,
predicting that interstellar information-bearing signals will
have a relatively large bandwidth $\Btotal$ relative to their information rate.
This is not because energy bundles are necessarily broadband
(they can have bandwidth $B$ between $1/T$ and the
coherence bandwidth $B_c$), but because information is conveyed
by the position of energy bundles in time and frequency, which consumes
extra bandwidth.
Since a beacon is an information-bearing signal in the limit of
zero information rate $\rate$, this conclusion does not have much to say
about the bandwidth of a beacon.
In fact, a beacon can be low power and at the same time be narrowband.
\boxbreak
For example, the bandwidth of a periodic beacon is
equal to the bandwidth $1/T < B \le B_c$ of its constituent
energy bundles.
A periodic beacon can have the same bandwidth as a Cyclops beacon
with the choice $B \approx 1/T$,
even as its average power is much lower due to its low duty factor.
Alternatively, a periodic beacon can be relatively broadband if $B \approx B_c$
as permitted by the coherence bandwidth of the interstellar channel
(which is highly dependent on the carrier frequency).
As its average power $\power$ is changed,
the bandwidth $\Btotal$ of a beacon does not need to change.
}

A more detailed analysis of a periodic beacon design is
performed in Chapter \ref{sec:discovery}, and in particular
the considerations that go into the choice of duty factor.
Only a fraction $\delta_p$ of observations overlap a given observation,
assuming that the duration of each observation is $T_o = 2 \, T$.
The number of observations 
$L$ required to achieve a given reliability of discovery has to be
increased to compensate for this lower probability of overlap.
By the bundle energy hypothesis, however, the alternative of reducing $\power$
by reducing the bungle energy $\energy_h$ would have a much more dramatic effect on $L$.
The value of $\delta_p$ can be optimized for each value of $\power$,
and this confirms the bundle energy hypothesis to good approximation.
In particular, the regime $1 \le L \le 10^{\,3}$ is explored.
At the $\power$ for which the extreme value $L = 10^{\,3}$ is optimized
for reliable detection, for example, it is found that the optimum duty factor
is $\delta_p \approx 3$\%, and a Cyclops beacon ($\delta_p = 1$)
would require $L \approx 10^{15}$ to achieve the same detection reliability.
That is about twelve orders of magnitude greater, demonstrating that
the Cyclops beacon is clearly impractical at an average power this low.

In practice there cannot be a joint optimization of $\power$
(chosen by the transmitter) and $L$ 
(chosen by the receiver) due to the lack of coordination.
Without coordination, the transmitter chooses a value for $\delta_p$ and the
receiver must choose values for $L$ and $T_{o}$.
Once the value of $\delta_{p}$ is chosen and fixed by the trasnsmitter,
any variation in average received power $\power$ is due to
a change in the amount of energy per bundle $\energy_{h}$,
which in turn is due to uncertainty or change in transmitted energy per bundle,
transmit antenna gain, propagation distance, and receive antenna collection area.
Choosing a smaller $\delta_{p}$ benefits the transmitter designer by reducing
the energy consumption if $\energy_{h}$ is kept fixed, or alternatively
allows a larger transmitted energy per bundle if energy consumption is kept fixed.
Generally the transmitter will want to conservatively choose a sufficiently large
energy per bundle so that $\energy_{h}$ is sufficiently large to account for the
expected worst-case conditions (distance, receive antenna collection area, and noise level).
From this anchor point, the transmitter designer can adjust $\delta_{p}$
to trade off energy consumption against its expectations about the total observation
time at the receiver.

The receiver designer can render the receiver more sensitive,
meaning able to reliably detect the beacon at lower average power levels,
by increasing the receive antenna collection area, the number of
observations $L$, and the duration of each observation $T_{o}$.
The collection area is based on an assumption as to the transmitted
energy per bundle, the transmit antenna gain, and the propagation distance.
$L$ and $T_{o}$ are based on an assumption as to the
minimum duty factor $\delta_{p}$ of the beacon, or in other words the energy consumption
the transmitter is willing to allocate.
For a give choice of $L$ and $T_{o}$, the detection will become unreliable if $\delta_{p}$
is too small and the probability of overlap between an energy bundle and the observation shrinks.

The transmit and receiver designs are both based on worst-case assumptions.
If any of those assumptions are violated, the reliability of detection will suffer.
If any of the parameters are larger than the worst-case assumption,
then the choices made are not resource-optimizing but nevertheless the
reliability of detection is not harmed. In fact, it is enhanced.
For example, Figure \ref{fig:beaconCompare} shows how the
reliability of detection (as represented by the miss probability ($1-\pd$)
varies with average received power $\power$ for two sets of design choices made by the
transmitter and receiver.
For fixed $\delta_p$ the power is varied by a change in energy
per bundle $\energy_h$, which will occur with a variation in
propagation distance or transmitter and receiver antenna gains.
Two cases are compared, a Cyclops beacon
with a small number of observations
($\delta_p = 1$ and $L=3$)
and an intermittent beacon with low-duty-factor and
a large number of observations
($\delta_p = 3$\% and $L = 1000$).\footnote{
The values of $\delta_p$ and $L$ were power-optimized at $\xi = 1$ (0 dB) corresponding to the
vertical line on the left to achieve a miss probability of $1 - \pd =10^{-4}$.}
The intermittent beacon can achieve high reliability of discovery
at a considerably lower average power.
For example, at a miss probability of 
$1 - \pd =10^{-4}$ the average power of the intermittent beacon is 28 dB lower,
corresponding to a reduction in average power $\power$ by a linear factor of $708$.
Of course smaller $L$ enables smaller reductions in power, while larger $L$
can accommodate smaller average power $\power$, without limit.
As predicted by the total effective energy hypothesis, the relationship is
$\power \propto 1/L$ for approximately fixed detection reliability.

\incfig
	{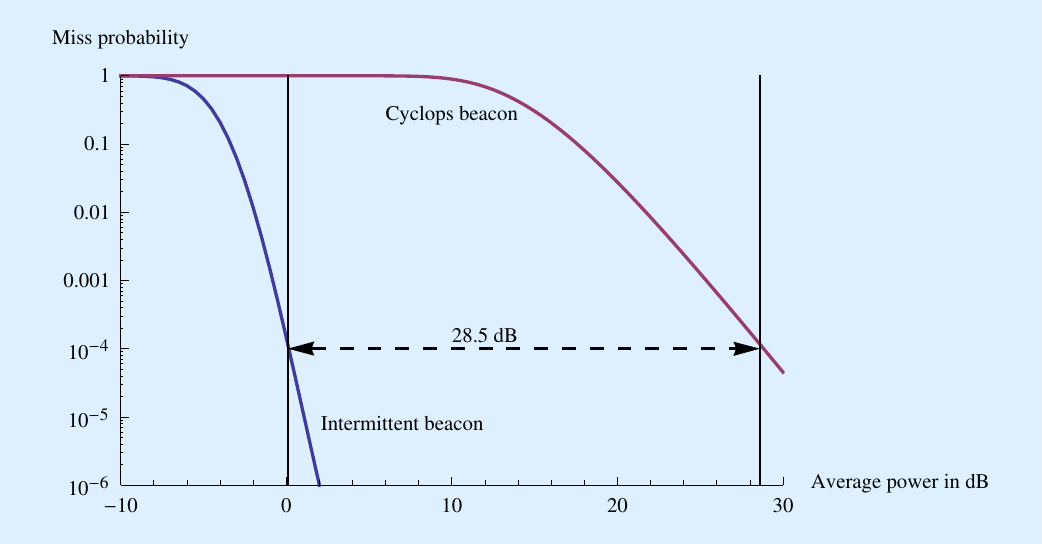}
	{1}
	{A log plot of the miss probability $(1 - \pd)$ as a function of the	
	 of the average received power in dB in the presence of AWGN and scintillation.
	The absolute power scale is not specified, and is irrelevant to the comparison.
	The two cases shown correspond to a Cyclops beacon
	($\delta_p = 1$ and $L=3$)
	and an intermittent periodic beacon
	($\delta_p = 3$\% and $L = 1000$).
	These design choices remain fixed as the average received power
	(and average transmit power) are varied.
	See Chapter \ref{sec:discovery} for a more detailed comparison of beacon designs.
	}

\textbox
	{Why transmit a beacon?}
	{
One basic question that SETI researchers have always faced is
whether to expect to find a beacon rather than an information-bearing signal.
Generally this question has been resolved by searching for a beacon,
the logic being that beacons are simpler and thus easier to discover.
This logic has some validity in the regime of spectrally efficient signals,
because achieving high spectral efficiency requires very complex
channel coding techniques.
At minimum this makes such information-bearing signals difficult to reverse engineer to extract
the embedded information, but it may also interfere with their discovery.
In any case, all this complexity does not contribute to discoverability
in any constructive way, and discovery is the foremost objective.
\boxbreak
Within the regime of power-efficient design, in contrast, there is no
big distinction between a beacon and an information-bearing signal.
Both have the same basic structure consisting of a sparse assembly
of energy bundles, the only difference being a greater randomness in the
location of the bundles for the information-bearing signal.
The way to achieve lower average power is the same:
Keep the amount of energy in each energy bundle $\energy_{h}$ constant while making
them more sparse in time by reducing the duty factor $\delta$.
Thus, both can be discovered in the same way, by first detecting an
energy bundle and then continuing the observations looking for more energy bundles in the
near vicinity in time and frequency.
Essentially the distinction between the two types of signals disappears,
and it seems unlikely that a civilization going to all the trouble of
transmitting a signal would neglect to embed information within
that signal at a rate $\rate$ consistent with the average power $\power$ that it can afford.
Embedding that information does not help discovery, but does not harm it either.
Further, once that information has been successfully decoded the statistical
argument for technological and intelligent origin of that signal is vastly improved,
and the receiving civilization is informed by any information content that is successfully decoded.
\boxbreak
While the transmitter has to make a choice between a beacon and an information-bearing
signal, fortunately the receiver can hedge its bets by searching for both simultaneously.
The search for an information-bearing signal follows the same prescription as a beacon,
but has to be a bit more flexible at the post-detection pattern-recognition phase.
In particular, it has to hold open the possibility of detecting energy bundles at nearby frequencies.
	}

\subsection{Information-bearing signals}

A difference between a beacon and an information-bearing signal
as described thus far
is that the former is designed for reliable discovery, and the latter for
reliable recovery of information.
In spite of difference in objectives, the design objectives for the beacon 
and for the information-bearing signal are well aligned.
The periodic beacon is structured around energy bundles, as is the
information-bearing signal.
The impact on the receiver when lowering the average received power of a beacon
is minimized by keeping the energy per bundle $\energy_h$ fixed and reducing the
duty factor $\delta$.
The same design principle applies to an information-bearing signal:
To keep the reliability of information extraction fixed,
as the information rate $\rate$ (and thus average power $\power$)
is reduced the energy per bundle $\energy_h$ and energy per bit $\energybit$
are kept fixed, and the duty factor $\delta$ is reduced.

An information-bearing signal also has to be discovered, and thus
its design has to consider the dual purpose of reliable information extraction
and reliable discovery.
Addressing for a moment the discovery challenge,
the only difference between a periodic beacon and an
information-bearing signal following the power-efficiency design
rules is that the information-bearing signal has some additional
randomness in the location of energy bundles in frequency.

In the power-optimized and power-efficient regime, discovery of a beacon
and an information-bearing signal are essentially the same.
Both concentrate on the detection of an individual energy bundle
by making a relatively large number of observations in the same location.
Following the detection of one or more energy bundles, those observations should
be extended at that location, including nearby frequencies.
A pattern-recognition algorithm operating over an extended period of time
can then distinguish a false alarm from an actual signal of non-natural or technological origin,
and also discern from the pattern of locations of energy bundles the structure and objective of the signal.
For example, an M-ary FSK signal consists of periodic energy bundles,
typically with less than 100\% duty factor,
just like the beacon.
The difference is that these bundles are randomly situated at
$M$ uniformly spaced frequencies, rather than all appearing at the
same frequency.
In the context of multi-channel spectral analysis combined with
matched filtering,
the detection of individual energy bundles is anticipated,
but the receiver must be prepared for additional randomness
in the time-frequency locations where those detections occur.

Since during discovery the receiver has no idea whether a non-natural signal is present or not,
the timing of its observations have no relationship to the actual location of energy bundles.
Thus, any additional randomness in location will not affect the detection process.
In a typical discovery search scenario the receiver is performing a multichannel spectral analysis,
so it will observe energy bundles that occur at nearby frequencies as easily as bundles
that occur at a single frequency.
The existence of these multiple frequencies increases the number of locations
that must be processed in looking for energy bundles, and this greater number of
locations increases the false alarm probability $\pfa$.
However, this increase is a consequence of the multichannel spectral analysis
and not the randomness of the information-bearing signal, and the same
increase in $\pfa$ will occur with either a beacon or an information-bearing signal.
The conclusion is that if a beacon and an information-bearing signal possess the
same average rate of energy bundle transmission, and those energy bundles
have the same energy $\energy_{h}$, then there is essentially no difference in the reliability
with which a beacon and an information-bearing signal can be detected.

A transmitter designer choosing a beacon rather than an information-bearing signal
is likely to be motivated by achieving a low average power $\power$ using a low
duty factor $\delta_{p}$.
If the transmitter designer chooses instead to transmit an information-bearing
signal, there is an additional benefit in increasing $\power$ and $\delta_{p}$,
which is a higher information rate $\rate$.
If that is the choice, then the information-bearing signal can be discovered
while employing less resources (in the form of observation time) at the receiver than
can the beacon.
This is because the rate at which energy bundles are arriving is greater
for the information-bearing signal,
and thus fewer observations are required for reliable detection.
This conclusion is valid, however, only if the energy per bundle $\energy_{h}$
is comparable for the beacon and the information-bearing signal.
This question is investigated in Chapter \ref{sec:discovery}, where it is found
that the value of $\energy_{h}$ is approximately the same between a beacon designed
in accordance with the bundle energy hypothesis and an information-bearing
signal with sufficiently large $\energy_{h}$ to enable reliable extraction of information.
This is not surprising, since in both cases $\energy_{h}$ is chosen to enable
a reliable decision between the presence or absence of an energy bundle in
a given location.
Thus, a beacon and an information-bearing signal of the same average power $\power$ can
be detected with the same reliability, and that information-bearing signal will
also have sufficiently high $\power$ to reliably extract information from it.

The foregoing applies to an information-bearing signal such as M-ary FSK
without time diversity.
When time diversity is introduced, the discovery issue is complicated
by the lower energy per bundle $\energy_h$.
This makes the signal intrinsically harder to discover, although an
adjustment to the signal design to account for discovery is needed, as pursued
in Chapter \ref{sec:discovery}.

\section{History and further reading}

Relevant past work can be found in the literature both from
terrestrial communication and from the past research on SETI,
and also from the astrophysics and astronomy literature related to
the interstellar medium.

\subsection{Cyclops report}

\begin{changea}
A required starting point for any study of interstellar communication
is the Cyclops report \citeref{142},
which is the result of an intensive collaborative study in the summer 1969.
The last four decades of SETI observations have 
in no small part been based on the original
prescription of this report of 1970 \citeref{142}.
This inspirational effort comprehensively studied various aspects of interstellar communication and discovery,
and remains valuable reading today.
Its discussion of important and relevant issues like antenna design
and motion remain the best sources, in part because they integrated
the perspectives of engineering and astronomy in a unique and valuable way.

Of course there has been much progress in areas relevant to
interstellar communications in the intervening more than four decades,
which are covered in the more recent articles referenced below.
The biggest advancement has been in the understanding of the
ISM flowing from the new science of pulsar astronomy
and related astrophysics modeling.
Cyclops emphasizes beacons rather than information-bearing signals, 
so only narrowband signals are considered.
Issues surrounding average signal power,
which are emphasized of this report, fall outside the scope of the Cyclops study.%
\end{changea}

\subsection{Modeling the interstellar medium}

One principle of power-efficient design is to choose a transmit signal that avoids
as many interstellar propagation impairments as possible.
Given the current state of spacecraft and propulsion technology \citeref{617}, there is no possibility of our own direct experimentation with interstellar communication except with the cooperation of intelligent civilizations elsewhere in our galaxy. Fortunately astrophysical models and astronomical observations (especially from pulsar astronomy \citeref{212}) provide relevant models that can be used as a basis of design and predictions of performance.
This methodology is available to both the transmitter and receiver designers.

It is a well established principle in astrophysics that over relatively short periods of time the integrity of a monochromatic wave is preserved by the ISM, with the exceptions of a frequency shift (due to Doppler) and scintillation (time variations in phase and amplitude due to scattering in the interstellar medium) \citeref{132}.
This concept is generalized here to waveforms with finite bandwidth, establishing
the concept of an interstellar coherence hole (ICH), which turns out to be
critically important to both power efficiency and ease of discovery.
The time-bandwidth product of the ICH is a critical parameter, and estimates
establish that it is large enough to support power-efficient communication
at all carrier frequencies of interest.
It is also large enough to support interesting spread-spectrum options in the choice of waveforms \citeref{211}.

\subsection{Terrestrial wireless communication}

While the details and parameterization are different, both scattering
and motion impairments have much in common with terrestrial wireless communication.
Thus there is 
a wealth of theory, experience, and practice with terrestrial wireless
impairments that can be tapped \citeref{574}.
The major difference, as emphasized earlier, is the emphasis on
power efficiency as opposed to minimizing bandwidth, which drives the
design of channel codes in very different directions.

 \subsection{Optical communication}
 
 It is interesting that an especially relevant parallel to interstellar communication at \emph{radio}
 wavelengths is terrestrial communication at \emph{optical} wavelengths.
 Power-efficient techniques at the sacrifice of bandwidth are more commonplace in
 optical communication, which is rarely limited by bandwidth resources \citeref{640}.
 The search for power-efficient radio signals also has similarities to pulsed optical SETI \citereftwo{652}{655},
 which interestingly also assumes unconstrained bandwidth and the conservation of energy
 through pulsed signals.
 In addition, the scale of physical features for propagation through the ISM are
 very large relative to a wavelength, even at radio frequencies, 
 and are thus best understood in terms of geometrical optics \citeref{598}.
 Thus,
 the literature of optical as well as radio communication is an aid to understanding interstellar radio communication.
 
 \subsection{High-power microwave transmitters}
 
 \begin{changea}
 Part of the argument for limitations that may exist in the peak and average power
 is the capital cost of building a high-power radio transmitter.
 While this report does not consider this technological issue,
 the current state of the art in transmitter design and the implications
 to interstellar communication are summarized by the Benford's \citereftwo{568}{569}.
 They arrive at similar conclusions to optical SETI \citeref{655}, namely that
 pulsed signals are highly beneficial in terms of average power and the capital costs
 of building a transmitter.
 
 The type of signals required for power-efficient design as in this report are
 different, however, in that our results assume coherent matched filtering
 of a specific transmitted waveform $h(t)$.
 The highest power pulsed microwave signals are typically generated by
 storing and releasing energy from a cavity, which will result in a waveform
 with limited coherence; that is, while it may have limited bandwidth and time duration, it may also be more
noise-like and unpredictable over that bandwidth.
Appropriately designed, such signals may be able to avoid dispersion and motion impairments.
However, they will not be amenable to the full benefits of matched filtering, as assumed in this report.
Thus more study of linear amplification of microwave signals at high power levels would be appropriate
before we can understand the full technological issues surrounding power efficient signals
transmitted at high power levels.
Of particular interest would be any fundamental limitations that exist, since a transmitting civilization
is presumably more technologically advanced than us.
 \end{changea}

\subsection{Power-efficient design}

The origin of power-efficient communication is the insight of
Shannon into fundamental limits to communication \citeref{164}.
First applied to the AWGN, this theory clearly demonstrated the tradeoff
between power efficiency and spectral efficiency.
 Further, this theory highlighted the fact that substantial progress was
 possible in both regimes because of a substantial penalty for the best practices at the time
 relative to the fundamental limits.
 While the proof of Shannon's theorem is not constructive
 and gave no specific prescription, it
 provided a general direction by showing that channel coding is necessary to close that penalty.
 
Early and rapid progress in the power-efficient regime came relatively early
because it is conceptually and practically simpler.
It was shown in 1961 that orthogonal coding
 could achieve capacity asymptotically on the AWGN channel
 \citeref{639}.
 M-ary FSK and PPM are examples of  the more general
 class of orthogonal codes, all of which exhibit identical
 $\pe$ performance.
 Also directly relevant to interstellar communication and discussed in
 Chapter \ref{sec:channelmodel},
 the capacity of a Rayleigh fading channel (equivalent to scintillation in the ISM)
 was established to be
 the same as the AWGN by R.S. Kennedy during the 1960's \citeref{637}.
 The proof was constructive, based on demonstrating a combination of
 techniques that asymptotically achieved the AWGN capacity.
 
 This flurry of activity in power-efficient design was primarily of theoretical
 interest driven by the the relative conceptual simplicity of the challenge.
 It is always better to solve the easier problem before moving on to the harder problem.
 Little research on power-efficient design has appeared in
 the literature since 1970, as the problem was essentially solved by then.
 Rather, attention has focused on the much more conceptually challenging
 (and commercially important) spectrally efficient domain, which is also more technically challenging.
Approaching the fundamental limit with a severe
bandwidth constraint on the Gaussian channel took several additional decades of research,
culminating with success in the mid-1990's with the identification of
turbo codes \citeref{618} and low-density parity check codes \citeref{651}.
 On the other hand, power-efficient design has been used in
 optical communications, usually in the form of PPM \citeref{640},
 because spectral efficiency is hardly a consideration.
 Terrestrial optical communication thus serves as a better practical validation of power-efficient design
 than terrestrial wireless communication.
 
 \subsection{Broadband interstellar communication}
 
The famous Shannon capacity formula for the AWGN channel appears in several papers on SETI, 
most notably in the work of S. Shostak in which the Shannon formula
was applied to characterize the fundamental limit on the interstellar channel \citeref{616}.
The work of Jones \citeref{196} and Fridman \citeref{602} implicitly consider the fundamental limit on
power efficiency for the Gaussian channel.
These authors do not consider the impact of scintillation 
and other impairments on the fundamental limit.
\begin{changea}
Jones noted the abundance of bandwidth in the microwave window, and observed that orthogonal coding
(M-ary FSK is a special case of orthogonal coding)
is well suited to minimizing average power 
and is capable of approaching the fundamental limit on the Gaussian channel \citeref{196}.
Fridman includes a good discussion of the tradeoff between bandwidth and average power,
and argues that average power is more important \citeref{602}.
Like this report he considers the end-to-end communication system design issue,
considering the relative merits of PSK and FSK and advocating M-ary FSK because of
its more favorable power characteristics.
\end{changea}

Numerous authors have expressly considered those interstellar channel
impairments without explicitly designing a communication system taking them into account.
Of particular interest here is past work on using wider bandwidth in interstellar communication.
Clancy is notable as the first author to suggest the combination of a broadband signal
(in his case a narrow pulse) taking some account of the dispersion in the ISM \citeref{578}.
This is a very different approach to expanding bandwidth than dictated by
power-efficient principles because it unnecessarily triggers interstellar impairments.

A couple of authors have proposed, as we do here, transmit signals specifically tailored to circumvent
effects introduced in interstellar radio propagation, including
Cordes and Sullivan \citeref{123} and Shostak \citeref{127}.
In particular,
Shostak notes the significance of dispersion and scattering and suggests that it may be necessary to limit the per-carrier bandwidth, and suggested working around that limitation using multiple carriers, each modulated with a narrower bandwidth . 
This is called \emph{multicarrier modulation} (MCM) \citeref{612}.
MCM is a technique widely used in terrestrial wireless, with the same motivation,
and is a meritorious way to achieve low bandwidth (although
it cannot achieve high power efficiency).
However, the common thread between Shostak's observations and this report is to divide the signal into narrower
bandwidth slices to avoid interstellar dispersion.
Shostak proposes doing that by transmitting independent information-bearing signals, whereas
our power-efficient approach is to transmit energy bundles at different frequencies
(not simultaneously!) as part of a channel coding scheme for a single information-bearing signal.

Harp et. al. propose a receiver autocorrelation technique supported by the transmit signal that can limit the
effect of ISM impairments \citeref{581}.
Morrison addresses the discovery of broadband signals
by highlighting a specific example \citeref{577}.
Other authors have suggested some benefits for using wider bandwidth.
In addition to those previously mentioned \citerefthree{196}{578}{602}, 
Cohen and Charlton advocated the use of
wideband signals to overcome frequency-selective fading in improving the chances of discovery \citeref{140},
as did Cordes and Sullivan \citeref{123}.
In our own work, the ability of broadband signals to improve immunity to radio-frequency
interference was advocated \citeref{211}, and the spread-spectrum signals
proposed there are fully compatible with the design of energy bundles for power-efficient design.

\subsection{Interplanetary internet}

\begin{changea}
One characteristic of deep-space communication is large delays.
Researchers interested in data collection and preservation in deep-space missions
have been developing new networking protocols that are data-tolerant \citeref{654}.
This effort is called the \emph{interplanetary internet} \citeref{653}, and deals concretely with
some of the issues addressed more abstractly in \citeref{583}.
The applications addressed by the interplanetary internet
involve space vehicles launched from earth, implying that their distance is confined to our
solar system or (in the future) nearby stars, and within that context the coordinated design of
the communications and networking infrastructure.
This report focuses
on communication with other stars in the Milky Way, and deals with
communication at potentially much greater distances, usually with other intelligent civilizations
rather than spacecraft that are launched by our own civilization.
Some of the phenomena addressed here, such
as impairments due to the interstellar medium, are not significant for the 
interplanetary internet because of the shorter distances involved.
It should also be noted that the interplanetary internet's use of the term
''bundle'' refers to a higher-level protocol element, and has no connection
to that same term as used in this report.
\end{changea}
 
 \subsection{Energy-efficient wireless communication}
 
 Due to the emergence of battery-operated mobile terminals,
 there has been much recent attention to ''energy-efficient''
 wireless communication \citeref{641}.
 In this case energy efficiency refers to extending the battery life.
 Because the radio communications demands spectral efficiency,
 battery life is dominated by the extensive processing required
 in the terminals for signal processing and protocol implementation,
 and the power devoted to radio-frequency transmission is small
 by comparison.
 Thus, this recent work focuses mostly on protocol design for
 lower power consumption in the receiver computation, and has little relevance to
 interstellar communication, where discovery rather than communication
 is the major contributor to the receiver's processing burden and energy consumption.


\chapter{The interstellar coherence hole}
\label{sec:channelmodel}

In Chapter \ref{sec:powerefficient} a fundamental limit was established on
the power efficiency obtainable on the 
additive white Gaussian noise (AWGN) channel.
However, in interstellar communication noise is only one of a number of impairments due to the ISM
and due to relative motion between source and observer.
In this chapter we first review the impairments that are encountered generally,
together with a more detailed description of the impairments encountered
by energy bundles that fall within an interstellar coherence hole (ICH).
The remaining impairments, which determine the parameters of the ICH,
are described in Chapter \ref{sec:incoherence}.
In power-efficient design, these other impairments are avoided by the
appropriate design of the transmit signal.

\begin{changea}

\section{Review of all impairments}
\label{sec:impairmentreview}

Although the primary goal of this chapter is to discuss impairments that affect
an energy bundle waveform $h(t)$ that falls in an ICH, it is useful to list and review
briefly all the interstellar channel impairments, including those that determine the size of the ICH,
those that affect signals within a single ICH, and those that impact the relationship between
two ICH's that are non-overlapping in time or frequency.

\subsection{Determining the size of the ICH}

There are four classes of impairments that determine the size of the ICH
 illustrated in Figure \ref{fig:ICH-illustration}.
These impairments are caused by four distinct physical phenomena,
two of which are caused by propagation effects through the
interstellar medium (ISM) and two of which are caused by the
relative motion of the transmitter, receiver, and turbulent gasses
within the ISM.
Three of the four impairments can be traced back to the presence of inhomogeneous and turbulent
clouds of ionized gasses in the ISM, which are conductive and thus
influence the propagation of radio waves over very long distances.

\incfig
	{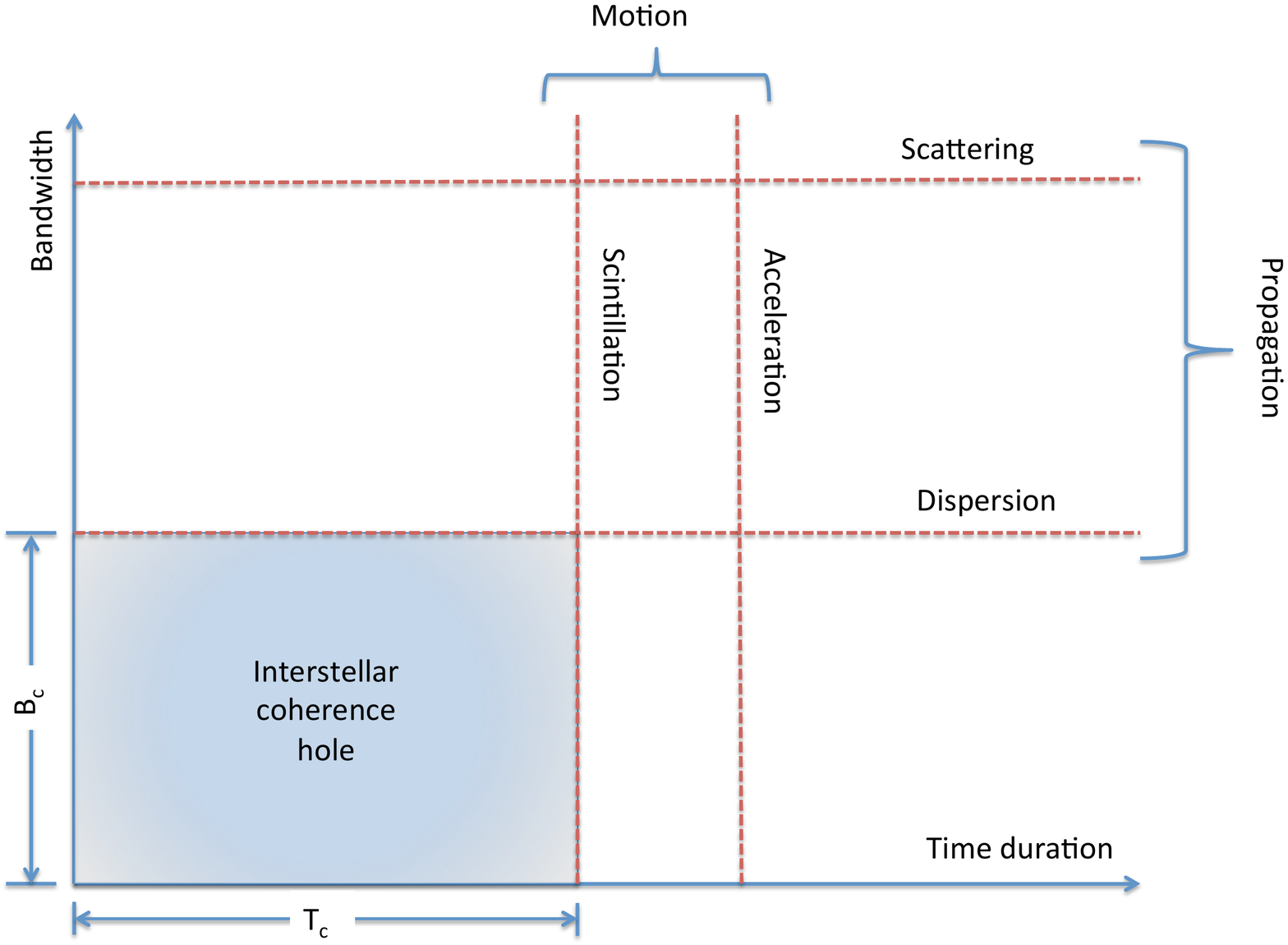}
	{.9}
	{Illustration of the interstellar channel impairments affecting the size interstellar coherence hole (ICH),
	including the coherence time $T_c$ and coherence bandwidth $B_c$.
	$T_c$ is determined by the more stringent of the scintillation coherence time 
	and the acceleration coherence time.
	$B_c$ is determined by the more stringent of the dispersion coherence bandwidth
	and the scattering coherence bandwidth.
	}

The propagation effects are due to an interaction between radio waves
and the clouds of ionized gasses present in the ISM, and are time-invariant effects that result in
a frequency-dependent amplitude and phase shift.
There are two phenomena to be concerned with:
\begin{description}
\item [Plasma dispersion.]
The interaction of radio waves with free electrons
that have been stripped from hydrogen atoms results in a wavelength-dependent
group delay.
\item [Scattering]
The inhomogeneity in these ionized gas clouds causes
diffraction and refraction of the radio waves (analogous to lensing effects in optics), and this results
(under some conditions) in multiple paths for radio waves propagating between
transmitter and receiver.
Those multiple paths can be associated with distinct group delays, and the
resulting delay spread can result in a frequency-dependent amplitude and phase shift. 
\end{description}
Both of these phenomena can be neglected when the signal $h(t)$ has a sufficiently small bandwidth,
called respectively the dispersion coherence bandwidth and the scattering coherence bandwidth.
Under typical conditions the dispersion coherence bandwidth is numerically the smaller of the two, and thus
determines the overall coherence bandwidth $B_c$ associated with the ICH.

Typically the transmitter and receiver are in relative motion relative to one another,
and it is sufficiently accurate to model this motion in terms of its velocity and acceleration.
The motion impairments cause a changing path length from transmitter to receiver
as well as a changing view of the scattering phenomenon along the line of sight.
This can result in changes to $h(t)$ that are time-varying rather than
frequency-dependent. 
These two motion impairments are:
\begin{description}
\item [Acceleration.]
The component of acceleration between the transmitter and receiver in the direction of the
line of sight results in a time-varying phase shift over the time duration of baseband signal $h(t)$.
\item [Scintillation.]
The component  of the transmitter's and receiver's velocity transverse (perpendicular) to the
line of sight interact with any multiple propagation paths due to scattering,
the result of which is
a time variation in the constructive and destructive interference of
signals arriving along these distinct paths.
This is manifested as a multiplicative time variation in both the magnitude and phase of the signal
waveform $h(t)$.
\end{description}
Similar to propagation effects,
both of these phenomena can be neglected when the signal $h(t)$ has a sufficiently small time duration,
called respectively the acceleration coherence time and the scintillation coherence time.
Which of these limitations to time duration is numerically the smaller depends on whether the
transmitter and receiver undertake a correction to their respective contributions to acceleration, which is
relatively easy to do.
Absent such a correction, the overall coherence time $T_c$ associated with the ICH
is typically determined by acceleration, and when both corrections are performed
$T_c$ typically increases substantially and
is determined by the scintillation rather than the acceleration.

A transmitted signal $h(t)$ avoids all four of these impairments
if its time duration is constrained to $T \le T_c$ and its bandwidth to $B \le B_c$.
The coherence parameters of the interstellar channel are modeled in Chapter \ref{sec:incoherence}.
Avoiding impairments by transmit signal design is one of the principles of
power-efficient design, since compensating for them at the receiver,
even if that is feasible, typically invokes a noise penalty and thus reduces the achieved power efficiency.

\textbox
	{Simultaneous bandlimit and time-limit}
	{
Strictly speaking a waveform cannot be simultaneously bandlimited
and time limited.
However, a classical uncertainty principle of Fourier transforms \citereftwo{599}{625}
states that if $K = B T \gg 1$ then waveforms exist with
most of their energy concentrated in $f \in [0,B]$ and $t \in [0,T]$,
and ''most'' approaches ''all'' as  $K  \to \infty$.
This should not be surprising in light of the sampling theory, which says
that an $h(t)$ precisely confined to $f \in [0,B]$ can be sampled
at rate $1/B$ without losing any information.
If only $K$ consecutive samples are non-zero (the remaining infinity of samples are all zero)
then the waveform is
approximately (although not precisely) confined to time interval $T = K/B$.
	}

\subsection{Impairments within in a coherence hole}

Restricting the time duration and bandwidth of baseband signal waveform $h(t)$ cannot
eliminate all impairments.
In particular two remain, both of them modeled as random phenomena:
\begin{description}
\item[Noise.]
Additive white Gaussian noise (AWGN) of thermal origin from
	background radiation, emission from stars, and due to the non-zero receiver temperature
	is present at the input and output of the matched filter
	(see Section \ref{sec:noiseICH}).
	The average energy of this noise at the output of the matched filter is not
	influenced by either the time duration $T$ nor the bandwidth $B$ of waveform $h(t)$.
\item[Scintillation.]
While the \emph{time variation} in the multiplicative amplitude and phase
factors of scintillation can be neglected if the time duration $T$ of $h(t)$ is sufficiently small,
there remains a \emph{fixed} multiplicative amplitude and phase.
This is modeled as a random multiplicative scintillation factor
	(see Section \ref{sec:scintillation}).
\end{description}

\subsection{Relationship between impairments for distinct coherence holes}
\label{sec:jointtreatment}

The channel codes discussed in Chapters \ref{sec:fundamental}
and \ref{sec:infosignals} are consistently structured as a pattern of energy bundles
arrayed in time and in frequency.
These energy bundles are non-overlapping in time, in frequency, or both,
and each energy bundle is assumed to fall within the parameterization of the ICH.
Thus each of these energy bundles is \emph{individually} impaired by additive noise and multiplicative scintillation.
In understanding the effects of these impairments on the channel code as a whole,
it is critical to model the \emph{relationship} of these impairments for distinct non-overlapping  energy bundles.
For example, if the multiplicative scintillation is statistically independent as it affects different energy bundles,
the impact on the performance of the resulting channel code is very different 
than if it is statistically dependent (or correlated) for those same energy bundles.

The relevant relationships among distinct non-overlapping energy bundles are:
\begin{description}
\item [Noise.]
The noise samples at the output of two matched filters, each matched to non-overlapping energy bundles,
are always statistically independent.
\item [Scintillation.]
The relationship between the scintillation multipliers experienced by two non-overlapping energy bundles
depends on whether they are separated in time or in frequency.
\begin{description}
\item [Separated in time.]
If the time separation is significantly greater than the scintillation coherence time, then the
random scintillation multipliers can be considered statistically independent.
\item [Separated in frequency.]
If the separation in frequency is less than the scattering coherence bandwidth, the scintillation
multipliers can be considered identical, but if that separation is significantly greater 
then the respective scintillation multipliers can be considered
to be statistically independent.
This distinction is significant since (from Figure \ref{fig:ICH-illustration}) the
scattering coherence bandwidth is typically quite a bit larger than the coherence bandwidth $B_c$,
which is itself dominated by the typically smaller dispersion coherence bandwidth.
\end{description} 
\item [Acceleration.]
Inevitably there is a relative velocity between the transmitter and receiver, and this causes a Doppler shift
in the carrier frequency $f_c$.
A \emph{constant} velocity is not significant because the receiver possesses no absolute reference for
$f_c$, but any \emph{change} in relative velocity at two times corresponding to two energy bundles separated in time
(reflecting the presence of acceleration)
will be manifested as a differential shift in carrier frequency associated with those energy bundles.
If the transmitter compensates the transmit signal for its 
component of acceleration along the line of sight,
 and the receiver has done the same for its receive signal, then
this differential frequency shift is avoided.
\end{description}

\end{changea}

\section{Definition of the coherence hole}

\begin{changea}

Thus far the concept of ICH is vague, but it can be made precise by considering
a model for the passband and baseband transmission through the interstellar channel.

\subsection{Passband and baseband equivalent channels}

Interstellar radio propagation is a passband phenomenon, meaning the
radio signal as seen in the interstellar medium (ISM) is located at some carrier frequency $f_c$,
where typically $f_c$ falls in a range of frequencies between 0.5 and 60 GHz where both
the atmosphere and the ISM are relatively transparent.
For purposes of analysis, it is often convenient to work with a baseband-equivalent channel
as illustrated in Figure \ref{fig:basebandequivalent}.
This diagrams what happens to a single energy bundle waveform $h(t)$ as it passes
through the modulator, the passband channel, and the demodulator.
When nothing happens in the passband channel to modify the signal,
the input to the final lowpass filter (LPF) is
\begin{equation*}
\sqrt{\energy_h} \, h(t) \, e^{\,i\,\theta}
\left( 1+ e^{\, - i \, 2 \pi \, (2 \, f_c - 2 \, \theta) } \right) \,.
\end{equation*}
The role of the final LPF is to eliminate the double-frequency term located at carrier frequency $2 \, f_c$,
in which case the only signal remaining at the output of the LPF is $\sqrt{\energy_h} \, h(t) \, e^{\,i\,\theta}$.
The result is an equivalent baseband channel as pictured in Figure \ref{fig:basebandequivalent},
in which all signals are complex-valued.
The real part of these baseband signals correspond to the in-phase component of the passband signal,
and the imaginary parts correspond to the quadrature components.
In practice things will happen in the passband channel, but it is usually possible with some
minor algebraic manipulation to find a baseband channel model that is equivalent.

\incxy
	[@C=.6cm] 
	{basebandequivalent}
	{
	Illustrated above is a passband channel accompanied by modulation and demodulation to baseband.
	The notation $\Re \{ \cdot \}$ denotes taking the real part of a complex-valued signal.
	The carrier frequency $f_c$ Hz is accompanied by an unknown phase shift $\theta$ radians
	in the transmitter.
	The transmitted signal has energy $\energy_h$ and unit-energy waveform $h(t)$
	that is limited to time $t \in [0,T]$ and frequency $f \in [0,B]$.
	Thus the time duration of $h(t)$ is $T$ and the bandwidth is $B$.
	Illustrated below is an equivalent baseband channel model that is valid in the frequency range $f \in [0,B]$.	
	All signals represented by a single arrow ($\longrightarrow$) are real-valued,
	and all signals represented by a double arrow ($\Longrightarrow$) are complex-valued.
	}
	{
 	\sqrt{\recenergy} \, h(t) \ar@{=>}[r] 
	&
	*{\bigotimes} \ar@{=>}[r] 
	&
	*+<1pc>[F-,]{2 \,\Re \{ \cdot \}} \ar@{->}[r] 
	&
	*+<1pc>[F-,]{
	\begin{matrix}
	\text{Passband} \\
	\text{channel}
	\end{matrix}
	} \ar@{->}[r] 
	&
	*{\bigotimes} \ar@{=>}[r] 
	&
	*+<1pc>[F-,]{\begin{matrix}\text{LPF} \\ f \in [0,B] \end{matrix}} \ar@{=>}[r] 
	&
	r(t)
	\\
	& 
	\displaystyle e^{\, i \, ( 2 \pi \, f_{c} \, t + \theta)} \ar@{=>}[u] 
	&&&
	\displaystyle e^{ - \, i \, 2 \pi \, f_{c} \, t } \ar@{=>}[u]
	\\
	&
	\sqrt{\recenergy} \, h(t) \ar@{=>}[r] 
	&
	*+<1pc>[F-,]{
	\begin{matrix}
	\text{Phase shift} \ \theta \\
	e^{\,i\,\theta}
	\end{matrix}
	} \ar@{=>}[r] 
	&
	*+<1pc>[F-,]{
	\begin{matrix}
	\text{Baseband equivalent} \\
	\text{channel}
	\end{matrix}
	} \ar@{=>}[r] 
	&
	r(t)
	}

\subsection{A passband channel absent impairments}

Under what conditions does the passband channel not introduce any impairments of note?
At first blush, we might say that this corresponds to replacing the passband channel
and the baseband equivalent channel in Figure \ref{fig:basebandequivalent} by ''nothing''.
However, this is far too stringent, because there are some happenings in
the passband channel that are totally benign.
One of these is a phase shift, which is benign since it introduces no \emph{additional} uncertainty about the
phase beyond the confusion already introduced by the carrier phase $\theta$.
Another is a constant group delay, which is also benign since there
is always a large (and not accurately known) propagation delay over long distances.
A passband channel which introduces only a phase shift $\theta_0$ and a constant group delay $\tau_0$
can be modeled by a frequency transfer function $G(f)$ given by
\begin{equation}
\label{eq:ICHfreqRes}
G(f_c + \Delta f ) =  \exp \{ \, i \, \phi_0 - i \, 2 \pi \, \Delta f \, \tau_0 \} \,, \ \ 
0 \le \Delta f \le B
\,.
\end{equation}
The two parameters $\{ \phi_0 , \, \tau_0 \} $ of this benign passband channel are real-valued.
It is important to note that to be benign, this passband channel need only possess this
transfer function $G(t)$ over the range of frequencies $f \in [f_c , \, f_c + B ]$ occupied
by the signal $h(t)$, and it doesn't matter what happens at other frequencies.

\incxy
	[@C=.6cm] 
	{noimpairments}
	{
	The baseband equivalent of a passband channel that introduces no impairments.
	}
	{
	\sqrt{\recenergy} \, h(t) \ar@{=>}[r] 
	&
	*+<1pc>[F-,]{
	\begin{matrix}
	\text{Phase shift} \ \theta \\
	e^{\,i\,\theta}
	\end{matrix}
	} \ar@{=>}[r] 
	&
	*+<1pc>[F-,]{
	\begin{matrix}
	\text{Phase shift} \ \phi_0 \\
	e^{\,i\,\phi_0}
	\end{matrix}
	} \ar@{=>}[r] 
	&
	*+<1pc>[F-,]{
	\begin{matrix}
	\text{Group delay} \ \tau_0 \\
	e^{\,- i\, 2 \pi \, f \, \tau_0}
	\end{matrix}
	} \ar@{=>}[r] 
	&
	r(t)
	}

\subsection{Idea of the coherence hole}

The interstellar channel propagation together with impairments introduced by the
motion of the transmitter and receiver are definitely not benign  in the sense
just discussed.
Fortunately there aren't any nonlinearities to worry about.
However, as shown in as shown in Figure \ref{fig:ICH-illustration}
there are time-varying effects due to motion to worry about (acceleration and scintillation).
As also shown in Figure \ref{fig:ICH-illustration}, 
there are two frequency-response effects to worry about (dispersion and scattering).
The concept of the ICH is that these nasty effects can be rendered negligible
for a restricted group of signal waveforms $h(t)$, those waveforms said to fall into a coherence hole.
First, if the time duration $T$ of $h(t)$ is made sufficiently small, the time-varying effects can
be rendered negligible.
The result of ignoring the time variations is that the remaining impairments can be modeled
as a frequency transfer function $G(f)$.
Second, if the bandwidth $B$ is made sufficiently small, the effects in $G(f)$ that
are not captured in the benign transfer function of \eqref{eq:ICHfreqRes} can also be
neglected.

If $h(t)$ is restricted to an ICH in this manner, does the benign model of
Figure \ref{fig:noimpairments} then apply as a way of predicting the baseband output?
No, it definitely does not because not all interstellar impairments go away for small $T$ and $B$.
Scintillation does not go away completely, it merely becomes manifested as a \emph{fixed}
(as opposed to time-varying) amplitude for a single energy bundle.
Another ubiquitous phenomenon is additive noise, at radio frequencies principally
due to thermal effects in stars, in the cosmic background radiation, and the non-zero
temperature of the receiver circuitry.
When these two remaining impairments are added to the benign channel of
Figure \ref{fig:noimpairments}, the resulting baseband channel model is shown in
Figure \ref{fig:ICHmodel}.
In this model the unknown phases $\theta$ and $\phi_0$ have been left out
because the scintillation factor $S$ always has a uniformly-distributed phase
and thus any other sources of phase shift have no added effect.

\incxy
	[@C=.6cm] 
	{ICHmodel}
	{
	The baseband equivalent approximation to the interstellar channel
	with the input waveform $h(t)$ is restricted to an interstellar coherence hole,
	or its time duration is restricted to $T \le T_c$ and $B \le B_c$.
	There are other impairments not included in the model, but they are all
	minimal enough that they can be neglected.
	In addition to an unknown group delay $\tau_0$, there is a complex-valued
	multiplier $S$ that encompasses the effect of scintillation (and any other sources of unknown phase)
	and an additive noise $N(t)$.
	$S$ is modeled as a random variable and $N(t)$ as a random process, and thus
	the baseband reception $R(t)$ is also a random process
	(random variables and random processes are consistently represented by capital letters).
	}
	{
	\sqrt{\recenergy} \, h(t - \tau_0) \ar@{=>}[r] 
	&
	*{\bigotimes} \ar@{=>}[r] 
	&
	*{\bigoplus} \ar@{=>}[r] 
	&
	R(t)
	\\
	& 
	S \ar@{=>}[u] 
	&
	N(t) \ar@{=>}[u] 
	}

\subsection{Formal definition of the coherence hole}

As $T \to 0$ and $B \to 0$ the simple model for the interstellar channel's response
to a single energy bundle shown in Figure \ref{fig:ICHmodel} becomes increasingly accurate.
There are, however, reasons uncovered later why we might not want to go overboard
and choose $T$ and $B$ any smaller than necessary.
The question then arises ''how small is small enough''?
The answer of course is ''small enough not to affect the sensitivity or performance of the receiver''.
This answer is not completely satisfactory, however, since it depends on the design of the receiver,
in addition to the transmit signal.
The way to address this issue is to design the receiver \emph{assuming} that the
baseband channel model of Figure \ref{fig:ICHmodel}, and then for each
interstellar impairment that is neglected in the model of Figure \ref{fig:ICHmodel} ask
how the impact of that impairment on receiver sensitivity can be rendered insignificant.
This is the methodology that is followed in Chapter \ref{sec:incoherence} for each of the
four impairments shown in Figure \ref{fig:ICH-illustration}.

The goal of the receiver is to detect whether an energy bundle is present in a given reception $R(t)$
or not, where ''not'' specifically means $\energy_h = 0$.
When $N(t)$ is AWGN, then the best means of detection by a variety of criteria is a matched
filter followed by a sampler.\footnote{
Even if the noise arriving from the passband channel is AWGN, the lowpass filter in Figure \ref{fig:basebandequivalent}
will render this noise non-white.
However, this LPF is actually not necessary when the receiver implements a matched filter,
because the latter also eliminates all signal components outside the bandwidth $f \in [0,B]$.}
A filter matched to $h(t)$ has impulse response $\cc h (-t)$, so the outcome of applying this matched filter
is shown in Figure \ref{fig:ICHidea}.
In equation form, this figure becomes
\begin{equation}
\label{eq:ICHmodel}
Z = \begin{cases}
\sqrt{\energy_h} \, S + M \,, &\ \ \text{energy bundle present} \\
M \,, &\ \ \text{noise only} \,,
\end{cases}
\end{equation}
where the notation has been simplifed to $Z(\tau_0) = Z$.
For additional understanding and justification of the matched filter
see Chapter \ref{sec:detection}.

\incxy
	[@C=.6cm] 
	{ICHidea}
	{
	The ideal receiver structure for detecting the presence or absence of an energy
	bundle based on the ICH model of Figure \ref{fig:ICHmodel}.
	In the receiver baseband processing, the output of a matched filter with
	impulse response $\cc h(-t)$ is sampled at time $t = \tau_0$ to
	obtain a decision variable $Z ( \tau_0 )$.
	Since $h(t)$ is assumed to be a unit energy waveform,
	this can be transformed into the discrete-time model below (with only one sample),
	where $M$ is a complex-valued noise random variable.
	}
	{
	\sqrt{\recenergy} \, h(t - \tau_0) \ar@{=>}[r] 
	&
	*{\bigotimes} \ar@{=>}[r] 
	&
	*{\bigoplus} \ar@{=>}[r] 
	&
	*+<1pc>[F-,]{
	\begin{matrix}
	\text{Matched filter} \\
	\cc h (-t)
	\end{matrix}
	} \ar@{=>}[r]
	&
	*+<1pc>[F-,]{
	\begin{matrix}
	\text{Sampler} \\
	t = \tau_0
	\end{matrix}
	} \ar@{=>}[r]
	&
	z(\tau_0)
	\\
	& 
	S \ar@{=>}[u] 
	&
	N(t) \ar@{=>}[u] 
	\\
	&
	\sqrt{\recenergy} \ar@{->}[r] 
	&
	*{\bigotimes} \ar@{=>}[r] 
	&
	*{\bigoplus} \ar@{=>}[r] 
	&
	z(\tau_0)
	\\
	&& 
	S \ar@{=>}[u] 
	& 
	M \ar@{=>}[u] 
	}

In Chapter \ref{sec:incoherence} the different interstellar impairments are added
to the model of Figure \ref{fig:ICHidea} and \eqref{eq:ICHmodel} to gauge under
what conditions the magnitude of $Z$
becomes an accurate model, yielding typical values for the coherence times and
bandwidths promised in Figure \ref{fig:ICH-illustration}.
This methodology of examining the effect of impairments at the \emph{output} of the
matched filter is notable, because in some cases impairments that might appear
to be significant at the channel output (input to the matched filter) turn out to be not so
significant after all.
\end{changea}
The significant physical parameters for interstellar communication
are listed in Table \ref{tbl:assumptions}, including values
that will be used for numerical examples.
The nature and significance of these parameters will be explained as we proceed.

\textbox
	{A word about terminology and nomenclature}
	{
This paper deals with both physical models and communications.
Generally we use a nomenclature consistent with the astrophysics literature
when dealing with physical models, and a notation consistent with the communications
engineering literature when dealing with communications.
For example, the dispersion measure is $\dm$, but the $K$-factor for a Rician
fading model is $\kappa$.
A similar division is used for terminology, for example in dealing with
scintillation and fading (which are similar phenomena with different
names in astrophysics and communications).

Generally deterministic signals are denoted by a lower-case letter and
random signals by upper case (like $z(t)$ vs $Z(t)$).
Physical constants are usually denoted by a Greek letter, unless the
convention in the physics or communications literature is different
(for example the dispersion measure $\dm$ or noise power spectral density $N_{0}$).
The probability of an event $A$ is denoted by $\prob{A}$.
For a random variable $X$, the $f_{X} (x)$ is the probability density function
and $F_{X} (x)$ is the cumulative distribution function,  so the two are related by
\begin{equation*}
F_{X} (x) = \prob{X \le x} = \int_{- \infty}^{x} f_{X} (u) \, du 
\ \ \ \ \text{and} \ \ \ \ 
f_{X} (x) = \frac{d}{d x} F_{X} (x)
\,.
\end{equation*}
The averaging or expectation operator is denoted as $\expec{\cdot}$
(whereas in the physics literature the alternative notation $< \cdot > \equiv \expec{\cdot}$
is often used).
For example
\begin{equation*}
\expec{X^{2}} = \int x^{2} \, f_{X} (x) \, dx \,.
\end{equation*}
The variance of a random variable $X$ is denoted as
$\var{X} = \expec{X^{2}} - ( \expec{X} )^{2}$,
and the standard deviation of $X$ is $\sqrt{\var{X}}$.
	}

\doctable
	{\small}
	{assumptions}
	{All the significant physical parameters determining the
	strength of impairments in
	interstellar communication, together with
	numerical values used in examples.
	Generally the values are chosen to be worst-case,
	in the sense that they minimize the ICH size.}
	{l c p{12cm}}
	{
	\textbf{Source} & \textbf{Parameter} & \textbf{Description}\\
	\toprule
	\addlinespace[3pt]
	ISM & $\dm$ &
	The dispersion measure equals the total column density of
	free electrons along the line of sight (Section \ref{sec:dispersion}).
	\\
	&& $\dm = 1000$ is
	approximately the largest $\dm$ observed in pulsar astronomy;
	$\dm$ will be much smaller for most lines of sight.
	\\
	&& Pre-equalization in either transmitter or receiver can
	reduce the effective value to the uncertainty $\Delta \dm$ 
	(Appendix \ref{sec:equalization}).
	\\
	\addlinespace[3pt]
	\cmidrule{2-3}
	& $D$ &
	A distance-of-propagation measure. 
	In Figure \ref{fig:scattering-iD-illustration},
	$D$ is the distance from screen to observer, and
	more generally it is defined by \eqref{eq:d12}.
	\\
	&&The coherence bandwidth of scattering decreases with $D$
	and the scattering becomes stronger,
	so we use a relatively large value of
	 $D = 250$ light years
	 (this corresponds to $D_{1} = D_{2} = 500$ light years)
	\\
	\addlinespace[3pt]
	\cmidrule{2-3}
	& $\rdiff$&
	The diffraction scale characterizes the turbulent clouds of 
	interstellar plasma, and equals the range of $x$ at the scattering screen
	over which the phase introduced by the screen is approximately constant.
	\\
	&&It is wavelength-dependent, with $\rdiff \propto f_{c}^{\,6/5}$,
	with a typical value of $\rdiff = 10^{\,7}$ meters at $f_{c} = 0.3$ GHz \citeref{598}.
	\\
	\addlinespace[3pt]
	\cmidrule{1-3}
	Motion & $P$ &
	Period of circular motion, either diurnal rotation about axis or orbital motion around star.
	\\
	&& $P= 1$ day ($8.6 \times 10^{4}$ sec)  for the diurnal rotation of the Earth.
	\\
	\addlinespace[3pt]
	\cmidrule{2-3}
	& $R$ &
	Radius of circular motion, either diurnal rotation about axis or orbital motion around star.
	\\
	&& $R = 6.38 \times 10^{6}$ meters for the diurnal rotation of the Earth.
	\\
	\addlinespace[3pt]
	\cmidrule{2-3}
	& $v_{\bot}$&
	The velocity of the observer transverse to the line of sight.
	\\
	&& Although \citeref{122} estimates a typical value $v_{\bot} = 10$ km per sec,
	we conservatively adopt $v_{\bot} = 50$ km per sec. 
	\\
	\addlinespace[3pt]
	\bottomrule
	}

\section{Model of a coherence hole}
\label{sec:modelICH}

Once the groundwork of establishing the validity of the ICH is complete,
the model of \eqref{eq:ICHmodel} for the response of the interstellar channel
and receiver matched filter is simple.
There are only two sources of randomness, the scintillation magnitude $S$
and the additive noise $M$.
Both of these are complex-valued random variables.
$S$ will often be written in terms of its magnitude and phase as
\begin{equation}
S = \sqrt{\flux} \, e^{\,i\,\Phi}
\end{equation}
where $\flux$ is called the received signal flux 
(normalized so that $\flux = 1$ in the absence of scintillation)
and $\Phi$ is a random phase that is uniformly distributed on $[ 0 , \, 2 \pi ]$.
Like all the other parameters of the model,
the signal energy $\energy_h$ is measured at the receiver, and thus
does not take into account propagation loss.

The magnitude of the signal is affected by both $\recenergy$ and $\flux$.
These are both energy measures, so their contribution to the signal magnitude
is the square root.
The received signal energy is multiplied by $\flux$,
becoming $\recenergy \cdot \flux$,
and it is possible to have $\flux > 1$ or $\flux < 1$ implying that
scintillation results in either an increase or a decrease in signal level.
The presence of the random phase term $e^{\,i\,\Phi}$ 
in the signal renders  the phase of $Z$ useless, but the magnitude $|Z|^{2}$
contains useful information about the presence of a signal ($\recenergy > 0$)
or the absence of a signal ($\recenergy = 0$).

In the model of \eqref{eq:ICHmodel}, by definition
$\expec{\flux} = \vs$ and $\expec{|M|^{2}} = N_{0}$.
The energy contrast ratio $\ecrs$ which was defined earlier in \eqref{eq:ecrs}
can be expressed directly in terms of the flux parameters as
\begin{equation}
\ecrs =  \frac{\recenergy \cdot \expec{\flux}}{N_{0}} \,.
\end{equation}
Another useful parameter of the scintillation is the \emph{modulation index},
defined as
\begin{equation}
\label{eq:modulationindex}
m_{\flux} = \frac{\sqrt{\var{\flux}}}{\expec{\flux}} \,.
\end{equation}
This index is a measure of the typical variation in the flux $\flux$
(as measured by standard deviation)
in relation to its mean-value.
Thus when $m_{\flux} \approx 0$ there is a very small variation in
signal level due to scintillation (at least in relation to the average signal level)
and when $m_{\flux} \approx 1$ the standard deviation is
equal to the mean value.

This simple model of \eqref{eq:ICHmodel} is now related to the physics of scattering \citeref{598}
followed by noise of thermal origin.

\subsection{Scattering within a coherence hole}
\label{sec:scintillation}

Scattering is due to variations in electron density within the turbulent interstellar clouds.
Because the distance scales are so large, even at radio frequencies
the phenomenon can be modeled by geometrical optics
using a ray approximation to the radio propagation.
This is illustrated by a simple one-dimensional model in Figure \ref{fig:scattering-iD-illustration}.
Although the conductive medium in turbulent clouds is actually
spread through three dimensions, this simplified model captures the essential features.
A thin one-dimensional \emph{scattering screen} is placed in the path of a plane wave,
and an observer views the resulting intensity at distance $D$.
The essential feature that causes scattering is the non-homogeneity of the screen
which has a group delay $\tau (x)$ that depends on the position $x$ on the screen.
The resulting diffractive and refractive effects at the observer result from
constructive or destructive interference among rays passing through different $x$
that influence the signal level.

\incfig
	{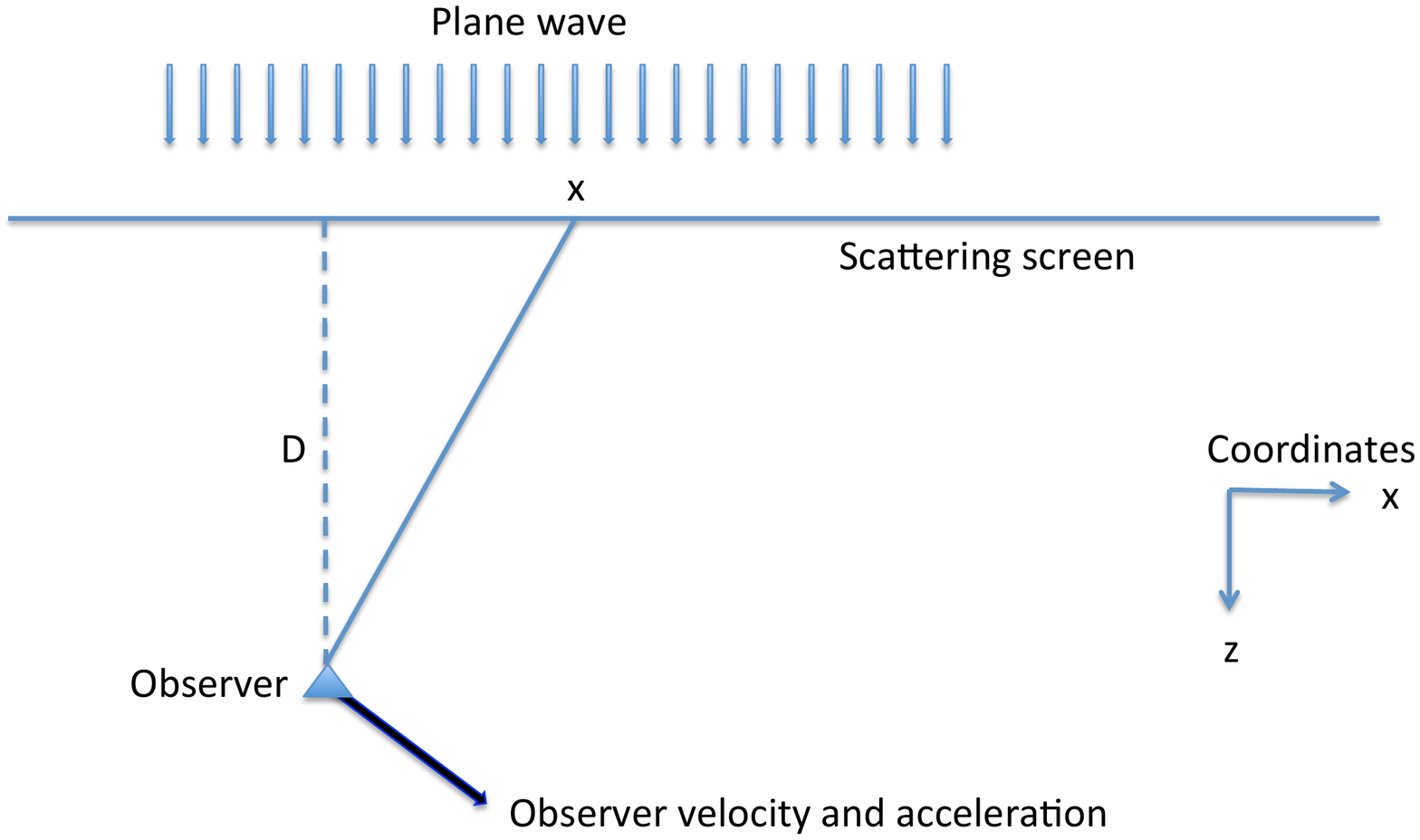}
	{.8}
	{Illustration of the geometry of scattering using a simplified one-dimensional
	scattering screen. Rays pass through the screen
	(located at distance $D$ from the observer) at different lateral distances $x$ and
	encounter different group delays and phase shifts due to small variations in electron density.}

The phase shift encountered by a given ray is affected by (a) the phase shift encountered at the
screen, (b) the phase shift due to propagation from point $x$ on the screen to the observer,
and (c) changes in the latter phase shift with time due to the observer's motion.
Although (b) and (c) are deterministic and easily modeled, (a) is unknown and subject only to
statistical analysis based on a model of turbulent motion of interstellar clouds.
An essential characteristic of this model is the \emph{diffraction scale} $\rdiff$,
which measures the range of $x$ over which the phase is statistically likely to remain fairly constant.
It is these ''coherent patches'' of stationary phase that contribute most strongly to the
energy that arrives at the observer.
Other areas on the screen without coherent phase tend not to contribute much net energy,
since they result in a jumble of phases that destructively interfere.

A second important geometric feature of the scattering is the
\emph{Fresnel scale} $\rf$,
given by
\begin{equation}
\label{eq:rf}
\rf = \sqrt{\lambda_{c} D } =  \sqrt{\frac{c \, D}{f_{c}}}
\end{equation}
where $\lambda_{c}$ is the wavelength and $f_{c}$ is the frequency of the carrier,
 $c$ is the speed of light.
 The Fresnel scale
 $\rf$ is the distance over which the phase shift due to the
 variation in distance with $x$ is small,
and has no relationship to the phase shift of  the scattering screen.
 The phase shifts as a function of $x$ are small over a scale $|x| \le \rf$
 (called the first Fresnel zone)
 because over this scale the variation in propagation distance is smaller than
 one wavelength.
 Along with coherence of phase introduced by the screen, coherence of phase
 due to the geometry is another necessary condition for the constructive interference
 of rays and hence a significant contribution to the energy arriving at the observer.
 
There is an important interplay between $\rdiff$ and $\rf$ that determines the
character of the resulting signal impairment.
The regime where $\rdiff \gg \rf$ is called \emph{weak scattering},
and $\rdiff \ll \rf$ is called \emph{strong scattering}.
Interstellar communication can encounter both types of scattering
as determined primarily by the interaction between the parameters $D$ and $f_{c}$.
The scales follow $\rdiff \propto  f_{c}^{\,6/5}$ and $\rf \propto f_{c}^{\,- 1/2}$,
and hence $\rdiff / \rf \propto f_{c}^{\, 17/10}$, and in addition $\rdiff /\rf \propto D^{-1/2}$ \citeref{598}.
Adopting these models,
the boundary between these regimes is illustrated in Figure \ref{fig:weakVsStrong}
for a typical value for $\rdiff$.
Strong scattering dominates for lower carrier frequency $f_{c}$ 
and larger distance $D$.

The character of scintillation is different for strong and weak scattering,
so they are now considered separately.

\incfig
	{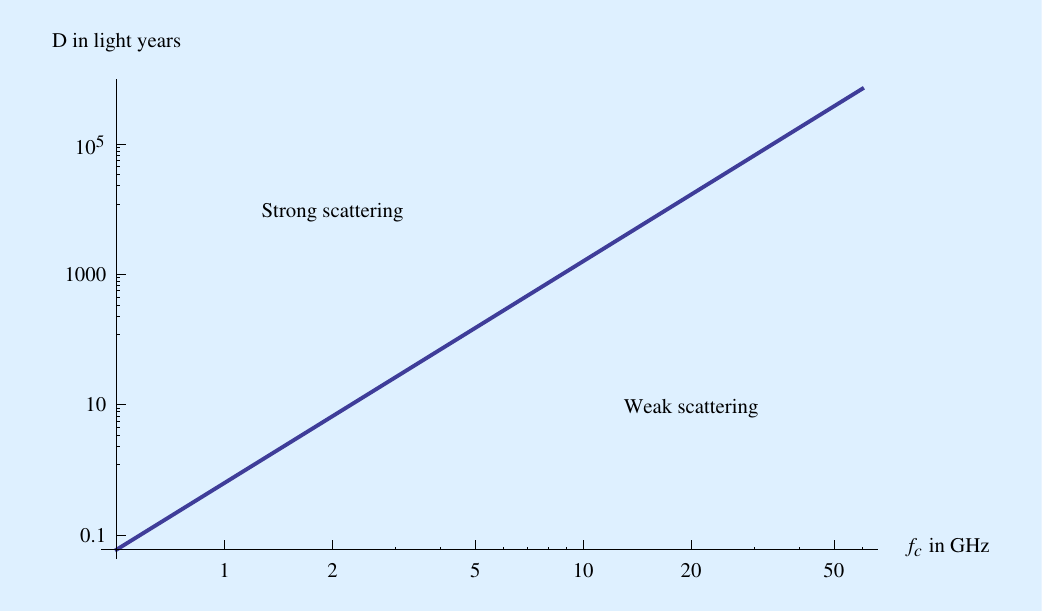}
	{.8}
	{A log-log plot showing the boundary between weak and strong scattering for variation of the
	distance $D$ in light years (according to the definition of \eqref{eq:d12}) and the carrier frequency $f_{c}$.
	The numerical value of $\rdiff$ given in Table \ref{tbl:assumptions} is assumed.}

\subsubsection{Strong scattering regime}

See Figure \ref{fig:scattering-strongvsweak} for an illustration of the distinctive
geometries that separate strong vs. weak scattering.
In the strong scattering regime of Figure \ref{fig:scattering-strongvsweak}a, 
$\rdiff \ll \rf$, and thus the typical coherent patch
of stationary phase is at a scale much shorter than $\rf$.
Within the first Fresnel zone the geometric variation in phase
is not much of a factor, but
even coherent patches occurring outside the first Fresnel zone are significant contributors
to the received flux,
albeit at a reduced energy, because the geometric contribution to phase variation can
still be small over the smaller $\rdiff \ll \rf$ scale.
In fact, there is a longer \emph{refraction scale} $\rref > \rf$ over which coherence patches
make a significant contribution (it happens that $\rref = \rf^{\,2} / \rdiff$).
Within the refraction scale there are multiple distinct coherent patches,
each contributing energy with a propagation delay that increases with $x$.
The maximum delay that is a significant contributor roughly corresponds to $x = \rref$.
In strong scattering there is \emph{multipath distortion} due to the multiple
pathways that arrive at the observer, each corresponding to a distinct coherent
patch, and each having a distinct group delay.

\incfig
	{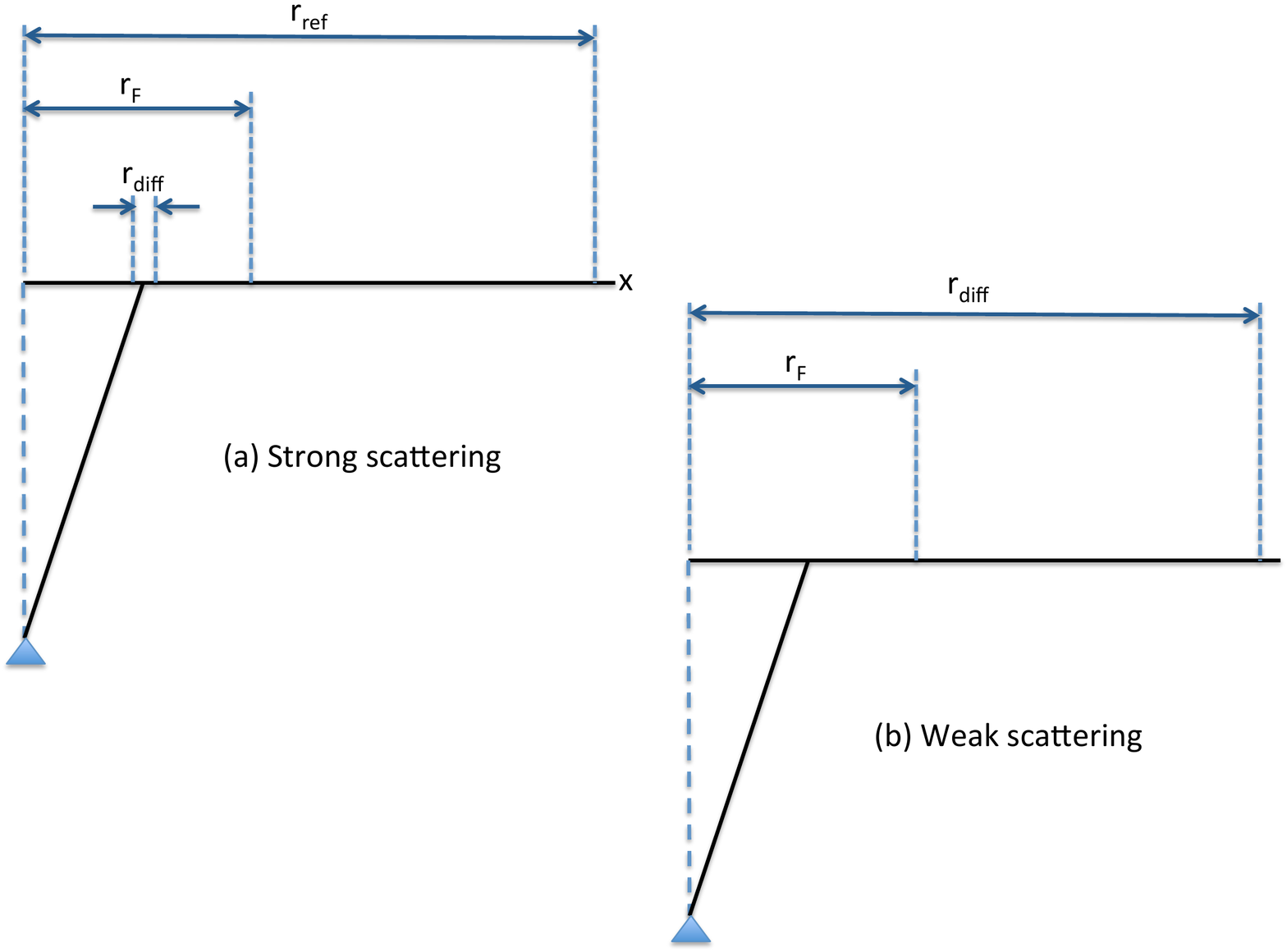}
	{.8}
	{An illustration of the difference in geometries between (a) strong scattering
	and (b) weak scattering. The $x$ scale is exaggerated relative to the $z$ scale
	to magnify the distinction.}

Thus far the discussion has assumed a monochromatic (sinusoidal) signal
is propagating through the scattering screen.
To model the ICH, the question is what the effect of this multipath distortion
is on a finite bandwidth waveform $h(t)$.
This question is considered in more depth in Chapter \ref{sec:incoherence},
but the results can be summarized as follows.
Assume that there are $N$ paths
corresponding to distinct coherent patches, where
the $n$th path is characterized by an amplitude
$r_k$ and a phase shift $\phi_{n}$.
Although $h(t)$ will also be distorted by plasma dispersion (see Section \ref{sec:dispersion}),
within an ICH this effect can be neglected.
Although there is some smearing of energy over time due to the
geometric variation of propagation delay within a single coherent path,
within the ICH due to the choice of a maximum bandwidth for $h(t)$
this effect can be neglected to lump all this energy into a single discrete delay $\tau_{n}$.
Although the exact amplitudes are unknown, it is the
phases that are very random.
Phase is very sensitive to slight perturbations
of conditions over such long distances and those perturbations can easily be large
relative to one wavelength.

In our simple model,
the observation consists of a superposition of these paths,
\begin{equation}
\label{eq:multipathsum}
r(t) = \sqrt{\recenergy} \cdot \sum_{n=1}^{N}  r_k \, e^{\,i\,\Phi_{n}} \, h( t - \tau_{n} ) \,.
\end{equation}
This basic model is
very familiar in terrestrial wireless communication,
where the multipath distortion is due to radio waves bouncing over
various obstacles like buildings and hills \citeref{574}.
There are two distinct effects to consider: the delays $\{ \tau_{n} \}$
and the phases $\{ \phi_{n} \}$.

Consider first the delays  $\{ \tau_{n} \}$.
As seen from geometric reasoning, all $\tau_{n} < \infty$
since as $x$ increases (and equivalently propagation delay increases)
eventually a point is reached where the geometric variation in phase
across a coherent patch yields an insignificant observed net flux.
Thus, define the \emph{delay spread} $\tau_{s}$ of the scattering as
the maximum of the $\{ \tau_{n}, \, 1 \le \tau \le N \}$.
Recall that in the model of the ICH of Figure \ref{fig:ICHidea},
what counts is the response of a matched filter to the reception $r(t)$.
The response of the matched filter to a single term $h(t - \tau_{n})$ is
\begin{equation}
\label{eq:multipath}
\int_{0}^{T} \, h(t - \tau_{n}) \, \cc h (t) \, dt
=  \int_{0}^{B} | H(f) |^{2} \, e^{-\,i\,2 \pi f \, \tau_{n}} \, d f \,.
\end{equation}
The effect of delay $\tau_{n}$ is reflected
in the frequency-domain by the phase term $e^{-\,i\,2 \pi f \, \tau_{n}}$.
This term can be neglected (and hence the group delay $\tau_{n}$ can be neglected) if
the total phase shift variation across $f \in [0,B]$ is less than $\alpha \pi$, or
\begin{equation}
\label{eq:delaycbw}
2 \, B \, \tau_{s} \ll \alpha \,,
\end{equation}
where $0 \le \alpha \le 1$ is a parameter chosen to adjust the
strictness of the criterion.
Intuitively, the ''width'' or rate of change of $h(t)$ is limited by $B$; the larger
$B$, the faster can be the variations in $h(t)$.
As long as $\tau_{s}$ is less than $\alpha/2B$, the effect on $h(t)$ is
hardly noticeable because the waveform of $h(t)$ is changing too slowly to
see the effects of that delay.

This establishes delay spread as a frequency coherence issue, in that
its effects are mitigated and eventually can be neglected as bandwidth $B$ is decreased.
Thus, within an ICH it can be assumed that \eqref{eq:delaycbw} is satisfied
and \eqref{eq:multipathsum} simplifies to
\begin{equation}
\label{eq:scinmulti}
Z(0) = \left. r(t) \convolve \cc h (-t) \right|_{t = 0} 
= \sqrt{\recenergy} \cdot \sum_{n=1}^{N} r_{n} \, e^{\,i\,\Phi_{n}}
\,,
\end{equation}
which is consistent with \eqref{eq:ICHmodel} without the noise.
In \eqref{eq:scinmulti} the magnitude and phase of the scintillation are,
\begin{equation}
\label{eq:S}
 \sqrt{\flux} \, e^{\,i\,\Phi} = S = \sum_{n=1}^{N} r_{n} e^{\,i\,\Phi_{n}}\,.
\end{equation}
A new complex-valued random variable $S$ has been introduced, capturing
both the magnitude and phase of the scintillation.

The flux $\flux$ is the average observed energy associated with unit-energy waveform $h(t)$ after it is 
intercepted by the screen.
No reduction in $B$ and/or $T$ can rid us of $\flux$,
and thus scintillation is a characteristic of the ICH model.
It affects even the limiting case of a monochromatic signal ($B \to 0$) \citerefthree{121}{122}{132}.
The phase shifts of different paths are a very sensitive function of
the physical parameters, including plasma density and path length.
For example, differences in the $\{ \tau_{n} \}$ too small to cause frequency incoherence
will still have a big effect on the $\{ \Phi_{n} \}$.
On the other hand, the $\{ r_n \}$ can be expected to vary reasonably slowly with time,
since they are dependent on the large-scale geometry of multiple paths of propagation
through the ISM.
Thus, in \eqref{eq:scinmulti} the liberty has been taken to make the $\{ r_n \}$
constants and the $\{ \Phi_n \}$ random variables.

Working back from these assumed statistics of the $\{ \Phi_{n} \}$ it is straightforward to show
(Appendix \ref{sec:withoutmod}) that $S$ has zero mean
($\expec{S} = 0$), and by a Central Limit Theorem argument it is a Gaussian
random variable with second order statistics
\begin{equation}
\label{eq:strongsecorder}
\expec{S^{2}} = 0
\ \ \ \text{and} \ \ \ \ 
\vs = \expec{|S|^{2}} = \sum_{n=1}^{N} r_n^{\,2} \,.
\end{equation}
The condition $\expec{S^{2}} = 0$ implies that the real and imaginary parts
$\Re \{ S \}$ and $\Im \{ S \}$ 
are statistically independent and each has variance $\vs/2$.
In different ICH's that do not overlap in time or in frequency, the values of $S$
may have some statistical dependency, more so in magnitude than in phase
because the $\{ r_{n} \}$ are determined primarily by turbulence in the interstellar
clouds and the motion of the receiver.
The effect of motion is considered in more detail in Section \ref{sec:fading},
where it is shown that it results in time incoherence that can be
neglected for sufficiently small $T$.

From the Gaussian distribution of $S$ and \eqref{eq:strongsecorder}, it follows that
$\expec{\flux} = \vs$ and $\var{\flux} = \sigma_{s}^{\,4}$,
and thus from \eqref{eq:modulationindex}
the modulation index is $m_{\flux} = 1$.
Thus, the standard deviation of flux equals its mean, independent of $\vs$.
However, a deeper examination of the physics reveals that \citeref{598}
\begin{equation}
\label{eq:sigmasone}
\expec{\flux} = \vs = 1 \,.
\end{equation}
This is valid for both weak and strong scattering, and thus
neither weak nor strong scattering affects the numerical value of the
energy contrast ratio $\ecrs$ that was defined in \eqref{eq:ecrs}.

There is a simple physical argument behind \eqref{eq:sigmasone}.
\begin{changea}
Consider a sphere centered at the source and intersecting the observer.
Scattering is a lossless phenomenon, 
and thus while it redistributes the energy as it propagates through the
sphere boundary it does not affect the total energy propagating through that sphere boundary.
Thus, while an observer moving along the sphere boundary will see a variation in signal flux
due to changes due to turbulence in the scattering medium internal to the 
sphere and also due to the observer's own motion,
the average value of the signal flux observed over a relatively long period of time
is not affected by scattering.
Even though a typical observer is not moving along the sphere boundary,
consider the effect of the components of observer velocity transverse and parallel
to the line of sight.
The transverse component of this motion is approximately along the sphere
boundary since the radius of the sphere is quite large,
and thus it will not affect the average flux noticably.
The parallel component of velocity will affect the average of the observed
flux due to changing propagation loss, but this will only be noticeable over very long
time intervals at typical observer velocities.
\end{changea}

For strong scattering $\sqrt{\flux}$ has a Rayleigh
distribution, and thus this type of scintillation is
called \emph{Rayleigh fading} in the communications
engineering literature.
There is a wealth of relevant experience in designing wireless mobile communication
systems that encounter Rayleigh fading.

\subsubsection{Weak scattering regime}

Returning to Figure \ref{fig:scattering-strongvsweak}b,
in weak scattering (where $\rdiff \gg \rf$) the scale of coherent patches
is large relative to the Fresnel scale.
This implies that geometric effects are dominant in the sense that
the delay spread and received flux are determined by the geometry
more than the coherence patches, although the latter remain very
significant in determining whether the flux is high (a coherence patch
overlaps the observer's first Fresnel zone) or low (the observer's first Fresnel
zone falls in a region of chaotic phase or transition).
For weak scattering, the modulation index is less than unity \citeref{598},
\begin{equation}
\label{eq:mfluxRician}
m_{\flux} = \left( \frac{\rf}{\rdiff} \right)^{\,5/6} < 1 \,,
\end{equation}
or the variations in flux $\flux$ are smaller than the mean.
Since $m_{\flux} = 1$ for a completely random scintillation, this implies
that $\flux$ has a deterministic component that 
contributes to $\expec{\flux}$ but does not contribute to $\sqrt{\var{\flux}}$.
In communications engineering this called a \emph{specular} component,
and the case $m_{\flux} < 1$ is called \emph{Rician fading},
because $\sqrt{\flux}$ has a Rician distribution.
The model used in communications engineering for Rician flux is \citeref{574}
\begin{align}
\label{eq:ricianflux}
&\sqrt{\flux} \, e^{\,i\,\Phi} =\sqrt{\gamma\,} \, \sigma_{s} \, e^{\,i\,\Psi}
+ \sqrt{1-\gamma\,} \, \sigma_s \, S \\
&\gamma = \frac{\kappa}{1 + \kappa}
 \,,
\end{align}
where $S$ has the same statistics as in Rayleigh fading.
Comparing with \eqref{eq:S}, there is the addition of
a specular term which has a constant magnitude
and a random phase $\Psi$ uniformly
distributed on $[0, 2 \pi ]$ and independent of $S$.
A parameter
$0 \le \kappa < \infty$ called the $K$-factor controls
how much of the energy is in the constant-magnitude term vs the random term,
all the while maintaining a constant average flux $\expec{\flux} = \vs$.
The modulation index
\begin{equation}
\label{eq:mfluxRicianK}
m_{\flux} = \frac{\sqrt{1 + 2 \, \kappa}}{1 + \kappa} \,,
\end{equation}
is plotted in Figure \ref{fig:scinMod}.
The Rayleigh fading case (strong scattering) corresponds to $\kappa = 0$,
while $m_{\flux}$ decreases as $\kappa$ increases.
Rayleigh fading is therefore ''worst case'' in the sense of ''largest modulation index''.

Rician fading is characteristic of wireless propagation when there is an unobstructed specular path
from source to observer that contributes a constant flux
(albeit with random phase $\Psi$), accompanied by multiple random
paths that contribute to a variation in flux.
In order to get proper articulation between weak and strong scattering at the
transition ($\rdiff = \rf$), we must have $\vs = 1$ in \eqref{eq:ricianflux}.
For a given value of $m_{\flux}$ from \eqref{eq:mfluxRician},
the appropriate $\kappa$ can be read off of Figure \ref{fig:scinMod}
or calculated from \eqref{eq:mfluxRicianK}.

\incfig
	{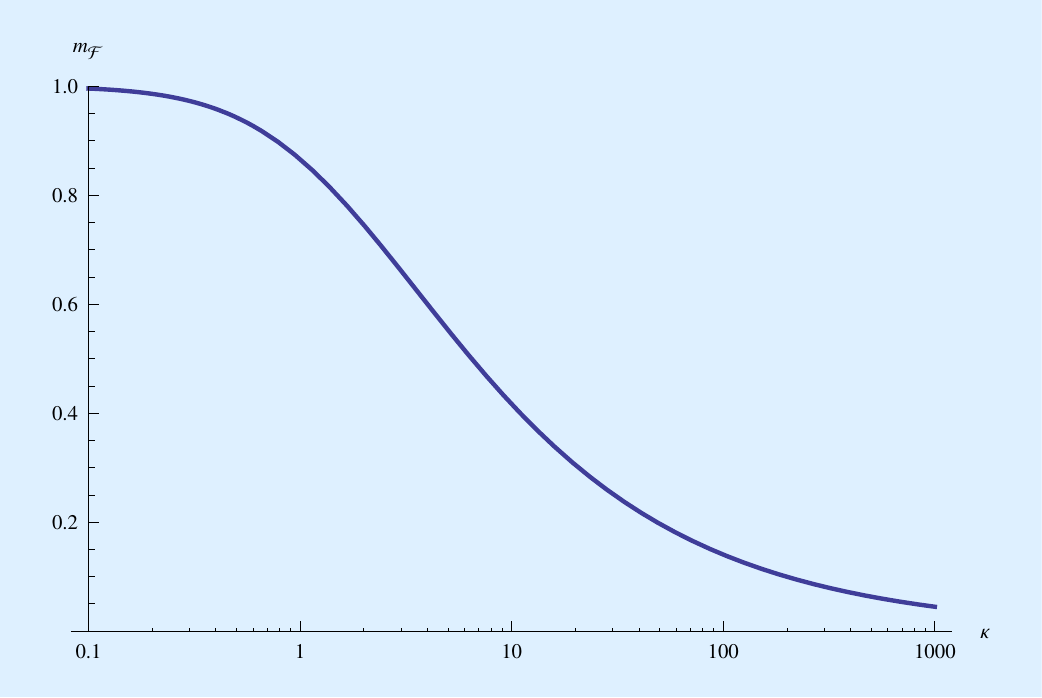}
	{.6}
	{The modulation index $m_{\flux}$ of Rician fading as a function of the
	$K$-factor $\kappa$.}

\subsection{Noise within a coherence hole}
\label{sec:noiseICH}

Suppose that the ISM introduces AWGN $N(t)$
with power spectral density $N_{0}$ into $r(t)$.
Then the noise at the matched filter output in Figure \ref{fig:ICHidea} is
\begin{equation}
\label{eq:noiseICH}
M = \int_{0}^{T} e^{-\,i\, 2 \pi \, f_{c} \, t} \, \cc h (t) \, N(t) \, dt \,.
\end{equation}
This random variable is (Appendix \ref{sec:modelICH}) zero-mean Gaussian with
variance $E|M|^{2} = N_{0}$, and
further $\expec{M^2} = 0$ and thus  $\Re \{ M \}$ and $\Im \{ M \}$ are independent and
have the same variance $N_{0}/2$.

It is significant that neither $B$ nor $T$ affects the statistics of $M$.
Thus, the statistics of the noise at the matched filter output do not change as $B$ and $T$ decrease, and thus
 noise remains an important impairment within an ICH.
The intuitive explanation for this is that within bandwidth $B$ the
total noise power is $N_0 B$, and thus the total noise energy within
time $T$ is $N_0 \, B \, T = N_0 \, K$.
Although this \emph{total} noise energy increases with $K$, the matched filter
extracts the noise in only one degree of freedom and rejects the 
noise in the remaining $K-1$ degrees of freedom,
leaving a total noise energy aligned with $h(t)$ equal to $N_0$
and independent of $K$.

\section{Time-bandwidth product and degrees of freedom}
\label{sec:channelmodeldof}

An important secondary parameter of the ICH is its time-bandwidth product
$K_{c} = B_c T_c$, which is called the \emph{degrees of freedom} (DOF).
$K_{c}$ is the number of complex-valued numbers
that are required to fully specify any waveform $h(t)$ that falls in an ICH.
$K_{c}$ depends
on the specific line of sight (LOS) and the carrier frequency $f_{c}$.
If $K_{c} \le 1$, then interstellar communication becomes very difficult because
for all practical purposes the ICH does not exist, or putting it another way
the inherent incoherence of interstellar propagation between a source and observer
in relative motion severely distorts virtually any waveform $h(t)$ that we attempt to transmit.
This would be particularly problematic for discovery, since there is no opportunity
to estimate $h(t)$ in advance nor even observe the current ISM ''weather'' conditions
to infer some of its likely properties.
If on the other hand, if $K_{c} \gg 1$, then interstellar communication becomes
relatively straightforward, because the transmit signal can be chosen to
circumvent all impairments save noise and scintillation.
In Chapter \ref{sec:incoherence},
$K_{c}$ is estimated using the numerical values of Table \ref{tbl:assumptions}
and the result is shown in Figure \ref{fig:dofICH}.
This estimate varies over a range $K_c \approx 1000$ to $10^5$ 
for the range $0.5 \le f_c \le 60$ GHz.
In addition, if both the transmitter and receiver correct for their respective accelerations
in the direction of the line of sight, then $T_{c}$ (and hence $K_{c}$) increases substantially.

\incfig
	{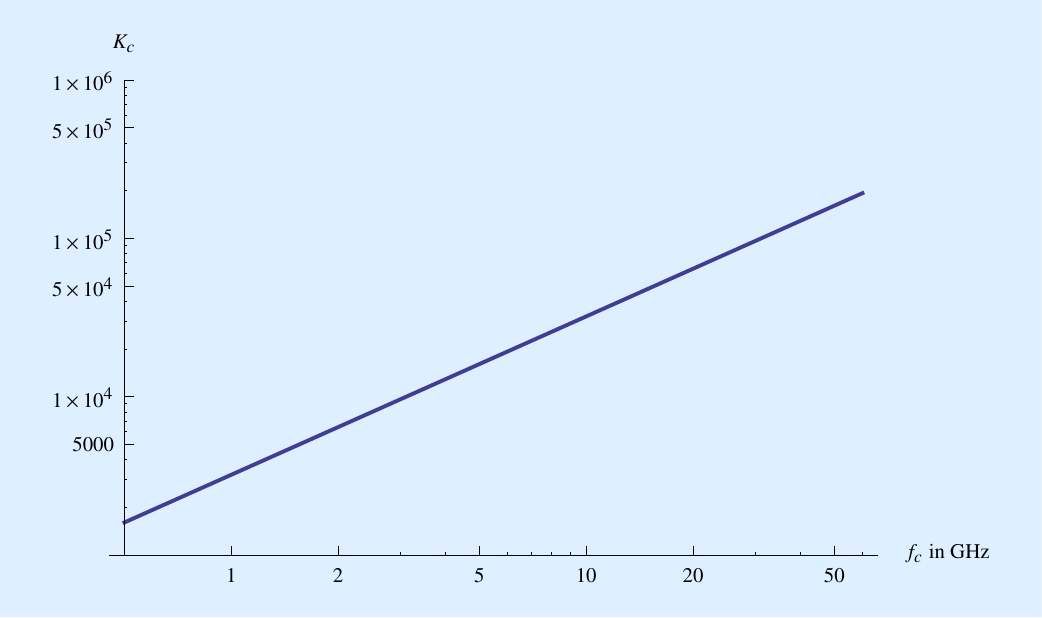}
	{.8}
	{A log-log plot of $K_c = B_c T_c$, the degrees of freedom for an ICH,
	vs the carrier frequency $f_c$ for the numerical values of
	Table \ref{tbl:assumptions}.
	A safety factor of $\alpha = 0.1$ is assumed, except for acceleration
	where the safety factor is $\alpha = 0.01$.
	The components $B_c$ and $T_c$ are shown in Figure \ref{fig:btCoherence}.}

The choice of $h(t)$ is of considerable importance from two perspectives.
First, the choice should be a waveform that the receiver can guess
in order to apply a matched filter, because
there is no possibility of conveying the precise $h(t)$ to the receiver prior to discovery.
Second, while the choice of $h(t)$ has no relevance to the
power efficiency of the system,
it does have considerable significance to the RFI immunity of the signal,
which argues for choosing as large a $K$ as possible \citeref{211}.
Of course, the largest value is $K = K_{c}$ consistent with the ICH.

\section{Two energy bundles}
\label{sec:twobundles}

\begin{changea}
The idea behind the ICH is to avoid as many interstellar impairments
as possible through choice of a transmit signal.
The remaining impairments at the sampled output of a matched filter
are an additive Gaussian noise and a multiplicative scintillation.
Of considerable interest in the evaluation of channel codes
in Chapters \ref{sec:fundamental} and \ref{sec:infosignals}
and discovery algorithms in Chapter \ref{sec:discovery} is the relationship
between two or more energy bundles that are
non-overlapping in time or frequency. 
This question is considered in Chapter \ref{sec:incoherence}, and
the conclusions there will now be summarized.

The noise accompanying two non-overlapping energy bundles is always
statistically independent, but there may be a dependence in the
scintillation fiux and there may be a differential frequency offset
due to acceleration.

\subsection{Scintillation}

For two energy bundles non-overlapping in time, but overlapping in
freqency, if the time offset is greater than the scattering coherence
time pictured in Figure \ref{fig:ICH-illustration}, the received scintillation flux
cannot be counted on to be the same or nearly the same.
As the time offset is increased, the coherence in
the scintillation phase is the first to break down, since it is sensitive to a fractional wavelength
shift in the propagation.
Thereafter the coherence in the received scintillation flux will also break down.
At time offsets considerably greater than the scattering coherence
time the two scintillation multipliers
(amplitude and phase) can be considered to be statistically independent.
For the time diversity combining strategies 
considered in Chapter \ref{sec:infosignals}
this independence is desirable.

For two energy bundles overlapping in time but non-overlapping in
frequency, the same statement applies with ''scattering coherence bandwidth''
substituted for ''scintillation coherence time''.
This scattering coherence bandwidth is typically larger than the
dispersion coherence bandwidth, and the latter dominates the bandwidth of the ICH.
Therefore, in M-ary FSK it is perfectly feasible to see both cases,
coherence and incoherence in the scintillation between two frequencies.
This is an important consideration in the design of M-ary FSK
receivers in the absence of time diversity in Chapter \ref{sec:infosignals}.

\end{changea}

\subsection{Differential frequency offset}
\label{sec:freqoffset}

Going back to the case of non-overlapping energy bundles in time,
any uncorrected acceleration will cause a differential frequency offset
between the two bundles.
The significant sources of acceleration are the diurnal rotation of
a planet or moon about its axes and the orbital motion of a planet about its
host star or a moon about its host planet.
A transmitter and a receiver can readily correct the signal to remove the effects of
this acceleration based on their knowledge of the local orbital dynamics.
However, if either fails to make this correction, acceleration typically becomes more
important than scintillation in determining the coherence time $T_{c}$.
Any frequency offset affects the extraction of information from the
received signal considered in Chapter \ref{sec:infosignals},
since information is conveyed by the location of an energy bundle in frequency.

\section{Some observations about the ICH}

We can consider ourselves fortunate that the ISM and source/observer motion
introduce impairments from the perspective of implicit coordination of transmitter and
receiver \citeref{583}.
The source and observer can independently
estimate $\{ T_{c} , B_{c} \}$ and the resulting $K_{c}$ based on common observable physical
characteristics of line of sight, and they should arrive at similar (although not identical) estimates.
This provides source and observer a common reference in choosing some free parameters
in the transmit signal $h(t)$, and reduces the ambiguity and the size of a discovery search
at the receiver.
In addition to RFI immunity, this is another argument for choosing $K = K_{c}$,
as this choice minimizes the ambiguity in $K$ between transmitter and receiver.
At the same time as ambiguity is reduced, ISM and motion impairments
(with the exceptions of noise and scintillation) do not adversely affect 
the feasible power efficiency,
our ability
to reliably communicate, or introduce any added processing requirements
in either transmitter or receiver.

The estimate of $\{ T_{c} , B_{c} \}$ depends on several circumstances:
\begin{description}
\item[Definition of coherence.]
Coherence improves gradually as $\{ T , B \}$ are decreased,
but perfection is not attainable.
How much coherence is enough depends to some extent on the usage; for example,
some information-representation schemes are more sensitive to incoherence 
(and to particular
types of incoherence) than others.
Thus the transmitter and receiver will adjust their definitions of coherence
somewhat (fortunately in compatible ways) based on other aspects of the system design.
\item[The line of sight and observer motion.]
Dispersion and scattering
and source/observer motion depend on the line of sight.
In particular, the relevant parameters of the line of sight from Table \ref{tbl:assumptions}
are the dispersion measure $\dm$, the diffraction scale $\rdiff$, and the distance $D$.
Astronomical observations yield considerable information \citeref{214}, but with some residual uncertainty.
A lot is also known about the motion impairments, which depend on the
radial acceleration $a_{\parallel}$ and the transverse velocity $v_{\bot}$.
This is particularly true of $a_{\parallel}$ because acceleration can be inferred from
an accelerometer or equivalently from modeling of the dominant sources of acceleration,
which are the orbit of a planet about its star as well as the revolution about its axis.
\item[Carrier frequency.]
All impairments depend on the wavelength (carrier frequency) in use.
\end{description}
Inevitably there will be some discrepancy between the source and observer estimates,
particularly in reference to the density of electrons along their line of sight
and the nature of their inhomogeneity.
Information about this is presumably due primarily to pulsar observations,
and the source and observer must base their estimates on pulsars that typically
have a different orientation relative to the line of sight.

There are also two forms of compensation that can be performed at the transmitter
and that have the effect of increasing the size of the ICH:
one is for acceleration and the other equalizes for the best-case dispersion
along the line of sight (Chapter \ref{sec:incoherence}).
The receiver designer must therefore account for the possibility that the
transmitter may have compensated for one or both of these effects, or not.

\section{Summary}

While the interstellar medium and relative motion of source and observer introduce impairments
to signals that can be severe, most of these can be avoided by reducing the time duration and bandwidth of the
waveform $h(t)$ that is used as the basis of an energy bundle.
The resulting interstellar coherence hole (ICH) serves as the foundation of a
power-efficient design for an interstellar communication system.
Two impairments that cannot be avoided in this way are
additive white Gaussian noise (AWGN) and scintillation.
AWGN is due to cosmological and stellar sources, as well as the non-zero temperature
of the receiver circuitry.
Scintillation is due to the constructive and destructive interference of multiple paths
through the interstellar medium due to the presence of inhomogeneous clouds of
ionized gasses.
The ICH is characterized by a coherence time and a coherence bandwidth for the
interstellar channel, both of which can be estimated for typical physical parameters
as shown in Chapter \ref{sec:incoherence}.


\chapter{The fundamental limit}
\label{sec:fundamental}

In Section \ref{sec:tradeoff} the fundamental limit on power efficiency through the
channel model with additive white Gaussian noise (AWGN) and assuming an average power constraint
was displayed, and used to demonstrate that there is a tradeoff between power efficiency
and spectral efficiency.
However, as shown in Chapter \ref{sec:channelmodel} and \ref{sec:incoherence}, although AWGN is
one of the impairments on the interstellar channel, there are others.
This chapter develops a fundamental limit for the interstellar channel based
on the results of Chapters \ref{sec:channelmodel} and \ref{sec:incoherence}.
Generally, it considers only the case of the power efficiency limit,
in the absence of any constraint on the total signal bandwidth $\Btotal$,
because this unconstrained bandwidth case is far easier.
That is, this chapter explores the fundamental limit applying to power-efficient design, rather than
spectrally efficient design.

The results of this chapter can be summarized in the single equation
\begin{equation}
\label{eq:fundlimit}
\ecrb = \frac{\energybit \, \vs}{N_0}  > \log 2 \,,
\end{equation}
where $\ecrb$ is the energy contrast ratio per bit defined in \eqref{eq:ecrb}.
This is the fundamental limit on power efficiency that can be obtained on
the interstellar channel.
In particular it establishes a minimum energy that the transmitter must deliver
to the receiver for each bit of information that is conveyed reliably, meaning
with arbitrarily small probability of error.
This chapter establishes this relationship, and
Chapter \ref{sec:infosignals} describes concrete and practical designs for channel coding
on the interstellar channel and compares the power efficiency
that they achieve to the fundamental limit of \eqref{eq:fundlimit}.

The most significant result of this chapter is a concrete channel code that can approach the
fundamental limit asymptotically.
By that we mean that as long as $\energybit$ satisfies \eqref{eq:fundlimit}
for this channel code, then the probability of error $\pe$ can be driven
arbitrarily close to zero as some parameters of the channel code become
unbounded.
This channel code combines the five principles of
power-efficient design outlined in Section \ref{sec:principles}.
Both the channel code and its proof of asymptotic optimality provides considerable insight
into what is necessary to approach the fundamental limit.
It also provides practical guidance for the concrete designs in Chapter \ref{sec:infosignals}.

\section{Role of large bandwidth}

The Shannon \emph{channel capacity}, published in 1948,
is a mathematically provable upper bound on the information
rate $\rate$ that can be achieved reliably for a given channel model \citeref{164}.
More modern developments of Shannon's results are found in \citereftwo{171}{172}.
Although channel capacity is a general concept,
for the AWGN channel of \eqref{eq:awgnchannel} the result is
particularly simple to derive.
Unless the average power of $X(t)$ is constrained, the capacity becomes
infinite, so it is assumed that $\expec{X^{2} (t)} \le \power$.
It is important to note that the average signal power $\power$ is referenced
to the same point as the noise in \eqref{eq:awgnchannel}, which is usually
chosen to be at the receive antenna output and before baseband signal processing.
To understand the implications of this to the average transmit power,
we have to work backwards, taking into account the transmit and receive antenna gains
and the propagation loss.

The Shannon capacity for the AWGN channel was given in \eqref{eq:capacity},
and from that result the fundamental limit on power efficiency $\powere$ 
was inferred.
That limit, when expressed in terms of the energy contrast ratio per bit $\ecrb$ is
given by \eqref{eq:fundlimit}.
At this fundamental limit,
$\rate \propto \power$, or every increase in $\rate$ demands a proportional
increase in $\power$.
This is exactly the same tradeoff that was seen in practical designs
like M-ary FSK in Section \ref{sec:illustrations}, 
although the constant of proportionality is certain to
be smaller reflecting a penalty that must be incurred
 by practical outcomes.

Perhaps the most significant observation was the tradeoff between spectral efficiency
and power efficiency, and in particular the observation that high power efficiency implies
poor spectral efficiency, or equivalently a large total bandwidth $\Btotal$ relative to the information rate $\rate$.
It is important to understand the critical role of large bandwidth, and that is a theme of the
remainder of this chapter.
A different but insightful interpretation follows from
the observation that the input signal-to-noise ratio $\snrin$ of \eqref{eq:snrin} is very small
in the regime where $\Btotal$ is allowed to be large \citeref{602}.
When $\snrin \approx 0$,
 \eqref{eq:capacity} can be expanded in a Taylor series as
 \begin{equation}
 \label{eq:capproxBlarge}
\rate \le \frac{\power}{N_{0} \, \log 2} \cdot
\left( 1 - \frac{\snrin}{2} + \frac{\snrin^{\,2}}{3} - \dots \right) 
\underset{\snrin \to 0}{\longrightarrow}  
\frac{\power}{N_{0} \, \log 2} 
\,.
\end{equation}
Thus, another characteristic of power-efficient communication is
a very small $\snrin$ because of a large $\Btotal$,
and in this special case \eqref{eq:capacity} simplifies
to \eqref{eq:capproxBlarge}, in which $\rate \propto \power$ at the limit.

To approach the fundamental limit on
power efficiency $\powere$ the bandwidth $\Btotal$ must be
\emph{unconstrained}.
It is important to note that unconstrained bandwidth is different than \emph{unlimited} bandwidth, because
the bandwidth actually utilized in any practical design
(such as those in Section \ref{sec:illustrations} or Chapter \ref{sec:infosignals}) will always be finite.
Unconstrained bandwidth merely assumes
that bandwidth is not a consideration in the transmitter's signal design,
which focuses exclusively on achieving high power efficiency.
However, one implication of \eqref{eq:capproxBlarge} is that any approach that
can asymptotically approach the fundamental limit must consume a total bandwidth
$\Btotal$ that approaches infinity.
Putting it another way, any practical approach (such as those discussed in Chapter \ref{sec:infosignals}
must use a finite bandwidth, and this reality alone inevitably creates a penalty in power
efficiency relative to the fundamental limit.

\textbox
{How does large bandwidth square with interstellar impairments?}
{
One possible source of misunderstanding arises from the seeming
contradiction between two observations.
In this chapter, it is shown in several ways that large total signal bandwidth $\Btotal$ is necessary
to asymptotically approach the fundamental limit on power efficiency.
In Chapter \ref{sec:incoherence} it is shown that a large bandwidth is problematic
in that some interstellar impairments become increasingly significant with increasing bandwidth.
By restricting the bandwidth $B$ of an energy bundle to $B \le B_c$, where
$B_c$ is the coherence bandwidth, a decrease in receiver sensitivity
due to these impairments is avoided.
Thus, a large bandwidth and a restricted bandwidth are simultaneously necessary to approach the fundamental limit.
Are these two observations incompatible with one another?
\boxbreak
No, they are completely compatible!
In this chapter, it is shown that a channel code composed of energy bundles
can approach the fundamental limit asymptotically.
Basing this channel code on energy bundles makes it impervious to all
interstellar impairments other than noise and scintillation, as long as
the constraint $B \le B_c$ is adhered to (as well as a similar constraint on time duration).
By spreading these energy bundles out in frequency as well as time,
the bandwidth of each codeword $\Btotal$ is increased accordingly.
Any channel code that approaches the fundamental limit must asymptotically
have codewords that
obey the condition $\Btotal \to \infty$, even as each energy bundle constituent of the
codewords must obey the constraint $B \le B_c$.
}

\section{Approaching the fundamental limit: channel coding}

Channel coding is the most general way to think of transmitting
a sequence of information bits over a physical channel.
Shannon's channel capacity is based on this general model,
deriving conditions under which channel coding can
or cannot achieve reliable communication of information.

Channel coding can be described in a general way as follows.
Supposing the information rate is $\rate$, then the number of bits to be communicated
during a time period with duration $t_{0}$ is $t_{0} \rate$, and if the average power is constrained to $\power$
then the total energy available during this period is $t_{0} \power$.
This can be accomplished by defining a \emph{codebook} of $2^{\, t_{0} \rate}$ distinct waveforms
on this interval,
where each \emph{codeword} in this codebook meets the energy constraint.\footnote{
More accurately, individual codewords can violate the energy constraint as long as the average
over all codewords does not.}
Depending on the actual $t_{0} \rate$ information bits presented to the transmitter, the appropriate codeword is transmitted.
Shannon's channel capacity theorem states that if $\rate \le \capacity$ then there exists a codebook for each $t_{0}$
with the property that as $t_{0} \to \infty$ the probability of the receiver choosing
the wrong codeword (due to noise or other impairments) approaches zero.
Conversely, if $\rate > \capacity$ then the probability of choosing the wrong codeword is bounded
away from zero.

The proof of Shannon's theorem demonstrates that a sequence of codebooks
exists such that as $t_{0} \to \infty$ the error probability goes to zero.
However, the theorem is not constructive and doesn't say how to design actual codebooks.
Construction of a codebook is usually challenging, in part because its membership is
growing exponentially with $t_{0}$.
In the spectrally efficient regime, the additional constraint on the bandwidth of
each codeword is challenging.
But the unconstrained bandwidth case is much simpler.
Although the existence of a fundamental limit inspired immediate and substantial progress,
it took less than two decades of research by applied mathematicians
to find codebooks that approach capacity in the easier power-efficient case, 
and about three \emph{additional} decades for the spectrally efficient case.

The M-ary FSK of Section \ref{sec:illustrations}
is a simple examples of channel coding.
The characteristic that makes them a channel code is that it
maps $M$ information bits onto a signal, rather than doing that mapping
one bit at a time.
The resulting signal is a function of all $M$ bits, and there is no feasible way to
determine a single bit without determining all $M$ of them.

\section{A channel code that approaches the fundamental limit}
\label{sec:asymptcode}

M-ary FSK is a very useful weapon in the power-efficient design arsenal,
and for this reason is included among the five design principles of Section \ref{sec:principles}.
In fact, the first case attacked and solved was the AWGN channel
with unconstrained bandwidth $\Btotal \to \infty$, and M-ary FSK was
shown to fit the bill in this case, achieving an error probability that approaches zero
as $M \to \infty$.
This applies directly only to the case of phase-coherent detection
on an AWGN channel without scintillation.

When scintillation is added to AWGN, as well as the numerous other impairments
due to interstellar propagation and motion of the source and observer,
we have to work harder.
This is attributable to the necessity of phase-incoherent detection, and also due to the
statistical variation in the received energy per bundle due to scintillation.
We now demonstrate a concrete channel code that overcomes both of these challenges and approaches
the fundamental limit of \eqref{eq:fundlimit} for the AWGN channel both with and without
scintillation, and in the scintillation case in the absence of any side information about the scintillation
in the receiver.
This will be a key step in the primary result of this chapter, which is to show that
\eqref{eq:fundlimit} is the fundamental limit for the interstellar channel, implying that
the impairments on the interstellar channel other than noise do not influence the fundamental limit
on power efficiency.
This also gives us valuable guidance on how to improve the power efficiency
for the interstellar channel in practice, which is pursued in Chapter \ref{sec:infosignals}.

\begin{changea}
It is important to realize that approaching the fundamental limit requires an explosion of
bandwidth toward infinity, and as we will see in addition an explosion in energy per bundle.
Neither of these is practical, so we expect in practice (and this is confirmed in Chapter \ref{sec:infosignals})
that practical designs will display a residual penalty relative to the fundamental limit. 
The situation is rather like the role of the speed of light in special relativity.
While the speed of light represents the maximum attainable speed achievable by any particle,
actually accelerating a particle to the speed of light is hard, and some remaining gap is inevitable.
The growing mass of the particle as its speed increases and the growing energy required
to increase its speed further is analogous to the growing bandwidth and energy per bundle
of a signal as it approaches the fundamental limit on power efficiency.
\end{changea}

\subsection{Structure of the channel code}
\label{sec:noside}

We now demonstrate a concrete channel code
that can asymptotically approach the fundamental limit on the interstellar channel.
This channel code is conceptually simple, and thus provides considerable guidance
on how to efficiently communicate through the interstellar channel in practice.
Further, this channel code is unbeatable.
More advanced civilizations might be able to come up with a different channel code,
but they could never improve on its power efficiency.
To design the code, the five principles of power-efficient
design listed in Section \ref{sec:principles} are combined.

\subsubsection{Use energy bundles}

The channel code is based on transmitting energy bundles, and constraining those bundles
to conform to the coherence bandwidth $B \le B_c$
and the coherence time $T \le T_c$.
Each codeword will communicate $\log_2 M$ bits of information, or in other words
the codebook counts $2^M$ member codewords.
The transmitter chooses one of those  $2^M$ codewords, and the receiver
detects which codeword was transmitted.
We will show that as long as \eqref{eq:fundlimit} is satisfied,
the probability of error $\pe$ in the detection of a single bit approaches $\pe \to 0$
as $M \to \infty$.
Of course for any finite value of $M$ a more stringent criterion than  \eqref{eq:fundlimit}
will have to be satisfied to achieve $\pe \approx 0$, and the resulting penalty relative to
the fundamental limit is explored in Chapter \ref{sec:infosignals}.

\subsubsection{Use M-ary FSK}

The $\log_2 M$ bits of information are conveyed by the location of energy bundles
at one of $M$ different frequencies.
These frequencies are chosen to be sufficiently far apart that there is no
frequency-overlap among the $M$ possible locations for energy bundles.
An arbitrarily large amount of information can be communicated by each
codeword simply by choosing a large $M$.

It is interesting that
we do not need to resort to multicarrier modulation (MCM) to 
achieve any $\rate$ we desire.
In fact, we \emph{cannot} approach the fundamental limit using MCM.
In practice, of course, a finite $M$ (number of frequencies in FSK)
and $J$ (number of repetitions of an energy bundle in the time diversity)
must be chosen
with a resulting penalty relative to the fundamental limit, and in that case
MCM is a valuable practical technique for increasing the information $\rate$
without resorting to impractically large values of $M$.

\subsubsection{Use time diversity}

Each actual energy bundle is redundantly repeated $J$ times with an intervening interval larger
than the coherence time of the scintillation
and time diversity combining is utilized by the receiver to counter the effects of scintillation.
An example of a channel code 
that combines M-ary FSK with time diversity
was illustrated in Figure \ref{fig:timeDiversityCode}.
A similar illustration of a single codeword (showing the locations of energy bundles in time and frequency)
is given in Figure \ref{fig:oneCodeword}.
The basic idea is that the total energy is divided between the $J$ replicas,
which experience the scintillation independently.
Averaging the $J$ estimates gives a more reliable indication of the
transmitter intention.
The $J$ energy bundles are located at one frequency out of $M$ possibilities,
conveying $\log_2 M$ bits of information.

\incfig
	{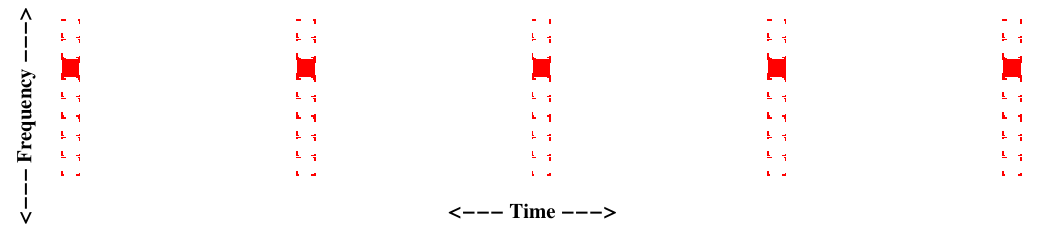}
	{1}
	{
	A two-dimensional representation of a single codeword for
	 M-ary FSK ($M=8$ is illustrated) combined with time diversity using $J$ identical repetitions of
	 each FSK symbol ($J=6$ is illustrated).
	 Each energy bundle has time duration $T$, and
	 the spacing between repetitions is  $11 \times T$.
	 Although the parameterization is different, this is similar to the codeword illustrated in
	 Figure \ref{fig:timeDiversityCode}.
	 }

Interestingly, if $M$ is sufficiently large each codeword will encounter significant
variation in the received signal flux at different frequencies.
Fortunately this is not an issue because each codeword uses only a single frequency
to transmit its $J$ energy bundles.
Those $J$ bundles beneficially experience the statistical variation of energy per bundle
with time.
The details of the received signal flux at other frequencies may be different, but
no energy bundles are being transmitted at those frequencies so that variation is
not observed by the receiver.

Figure \ref{fig:oneCodeword} shows only a single codeword out of a codebook
containing $M$ codewords.
An issue which turns out to be important is how a multiplicity of these codewords
are positioned in time to convey a sequence of information bits.
One possibility, illustrated in Figure \ref{fig:sequentialCodewords}, is to
transmit codewords sequentially.
Another possibility, illustrated in Figure \ref{fig:interleavedCodewords},
is to interleave the codewords in time.
Anywhere between two (the example shown) and $J$ codewords can be interleaved in this
fashion.

\incfig
	{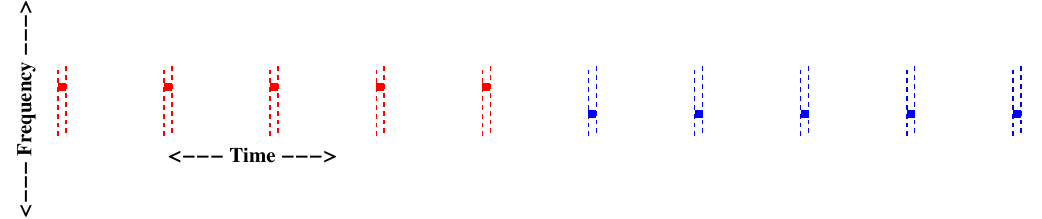}
	{1}
	{
	An illustration of how the individual codewords of Figure \ref{fig:oneCodeword}
	can be transmitted sequentially to communicate a stream of information.
	Shown are two codewords communicating a total of $6$ bits of information
	in twice the time as would be required for a single codeword.
	The duty cycle of this scheme is $\delta = 1/12$, because transmission occurs
	during $T$ seconds out of every $12 \times T$ seconds.
	}

\incfig
	{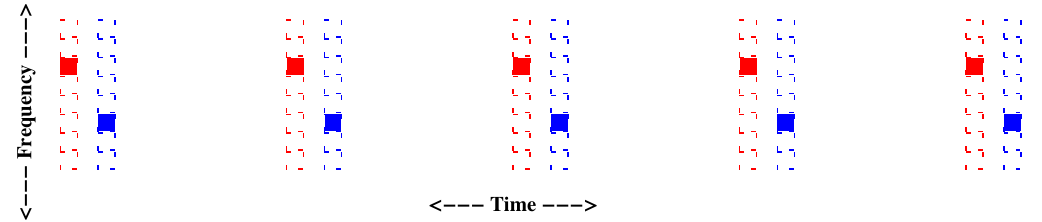}
	{1}
	{An illustration of how the individual codewords of Figure \ref{fig:oneCodeword}
	can be interleaved in time to communicate a stream of information.
	This interleaving would be combined with sequential transmission
	(which was illustrated in Figure \ref{fig:sequentialCodewords}).
	Shown are two codewords communicating a total of $6$ bits of information
	in slightly more time than would be required for a single codeword.
	The duty cycle of this scheme is $\delta = 1/6$, because transmission occurs
	during $2 \times T$ seconds out of every $12 \times T$ seconds.
	This duty cycle is double that of
	Figure \ref{fig:sequentialCodewords} because twice as many codewords are packed in the same overall time.
	}

It will turn out that the fundamental limit is approached as $J \to \infty$.
This allows the law of large numbers to kick in and give the receiver an
increasingly accurate estimate of the transmitted energy per bundle even in the
face of the statistical variations in the energy of each bundle due to scintillation.
Together, the two conditions $M \to \infty$ and $J \to \infty$ have profound
implications.
A single code word grows in both bandwidth and in time duration, in both
cases without bound.
The number of information bits communicated by a single codeword also
grows, so for a fixed total information rate it follows that individual codewords are transmitted at a lower and lower rate.
In summary, as we approach the fundamental limit, more and more information is
crammed in a single codeword, and each codeword grows in size
in both time and frequency with a correspondingly lower rate of
transmitting codewords.

\subsubsection{Use low duty cycle}

For any fixed information rate $\rate$ and average power $\power$ that we choose as design goals,
there is a tradeoff between using a larger $M$ and lower duty cycle $\delta$, or a smaller $M$
and larger duty cycle.
It will turn out that the fundamental limit is only approached as $M \to \infty$,
or equivalently $\delta \to 0$.
It will also turn out that the reason for this is that as $\delta \to 0$ the energy in
each bundle $\energy_h$ grows, and this is necessary to achieve a vanishing
error probability $\pe \to 0$.
Thus a reduction in duty cycle is used as a way to ''punch through''
the scintillation, since it is also true that for a constant information rate $\rate$ and average power $\power$ the 
energy in each bundle $\energy_h$ increases as the duty cycle $\delta$ is reduced.

Together these characteristics imply that each codeword becomes increasingly sparse
as the fundamental limit is approached.
Each codeword has $M \cdot J$ possible locations for energy bundles,
but only $J$ of these locations are used.
Out of the $M$ frequencies in total, only one is used.
Similarly, the duty cycle $\delta$ is small, and only a fraction $\delta$
of the time is there an energy bundle currently being transmitted at \emph{any} frequency.
For a fixed average power $\power$, this sparsity allows the energy of each
bundle that is transmitted to be as large as necessary to insure reliable detection of energy bundles.
	
\subsection{Parameters of the codebook}

The parameters of the codebook as we have defined it
are $\{ \rate ,\, \power ,\, T ,\, M ,\, J ,\, \energy_h ,\, \delta \}$.
The two primary parameters $\{ \rate ,\, \power \}$
are fixed as a design objective, and our challenge is to
adjust the secondary parameters $\{ T ,\, M ,\, J ,\, \energy_h \, \delta \}$
in a way that reduces the error probability $\pe$.
In the following, it is extremely important to keep the mindset
that $\{ \rate ,\, \power \}$ are fixed and unassailable.
The question then is under what conditions on $\{ \rate ,\, \power \}$
it is possible to drive $\pe \to 0$.
The answer will be that this is possible if $\{ \rate ,\, \power \}$ satisfy
\begin{equation}
\label{eq:fundlimit2}
\frac{\rate}{\power} < \frac{\vs}{N_0 \, \log 2} \,,
\end{equation}
which is equivalent to the fundamental limit of \eqref{eq:fundlimit}.

The other question is how the secondary parameters $\{ T ,\, M ,\, J ,\, \energy_h \, \delta \}$
should be adjusted, under the condition that $\{ \rate ,\, \power \}$ are fixed,
to achieve $\pe \to 0$ when \eqref{eq:fundlimit2} is satisfied.
The parameter $T$, the duration of each energy bundle, is relatively
simple to deal with since it is constrained by the coherence time $T \le T_c$.
Generally it is advantageous to choose $T$ as large as possible,
since that reduces the transmitted peak power.
Otherwise $T$ is not particularly significant.
Regarding the remaining parameters,
we must allow $J \to \infty$ and $\delta \to 0$,
which implies also that $M \to \infty$.
Roughly speaking choosing a large $M$ counters the additive noise
while a large $J$ counters the scintillation.
The purpose of choosing a small $\delta$ is to increase $\energy_h$, which allows more
reliable decisions on the presence or absence of an energy bundle to be made.

\subsubsection{Information rate}

In terms of the other parameters, the net information rate is
\begin{equation}
\label{eq:rate0}
\rate = \delta \cdot \frac{\log_2 M}{J \, T} \,.
\end{equation}
This is because $\log_2 M$ bits of information are conveyed
by each codeword, each of which consumes a total transmission time of $J \, T$.
For fixed $M$ and $J$, the highest information rate occurs with a 100\% duty cycle,
meaning an energy bundle is being transmitted at some frequency at all times.
This can be achieved,
for example, by interleaving codewords to avoid any ''blank'' intervals,
and then transmitting the resulting fully-interleaved codewords sequentially with time.
More generally, when $\delta < 1$ the result is a reduction in $\rate$.

Remember however that $\rate$ is actually fixed, and also assume that $T$ has been
chosen according to time coherence criteria.
In that case \eqref{eq:rate0} reveals a tradeoff between the parameters
$\{ M ,\, J ,\, \delta \}$.
Generally as $J$ is increased and $\delta$ is reduced, which is
necessary to approach the fundamental limit, then $M$ has to be
increased to compensate.
Increasing $J$ is consuming more transmission time, so achieving the
desired rate $\rate$ requires that more information be conveyed
during that transmission time by increasing $M$.
Similarly, reducing $\delta$ reduces the transmission time,
requiring that the remaining transmission time be used more efficiently
by increasing the ''information density'' of the transmission time
actually in use, again by increasing $M$.

The duty cycle is determined by how successive codewords
are positioned in time.
For example, $\delta$ is different in Figures \ref{fig:sequentialCodewords}
and \ref{fig:interleavedCodewords}.
The only significant parameter of a scheme for transmitting codewords
sequentially or interleaving them, or some combination, is $\delta$,
which affects the information rate $\rate$ by the factor $\delta$ in \eqref{eq:rate0}.
Driving $\delta$ to zero can be accomplished, for example, by adding
a ''blank'' interval of arbitrary length between transmitted codewords.

\subsubsection{Average power}

In terms of the parameters of the channel code,
the average power is
\begin{equation}
\label{eq:power0}
\power = \delta \cdot \frac{\energy_h}{T} \,.
\end{equation}
When the
energy of each constituent bundle is $\energy_h$,
each codeword requires a total energy of $J \, \energy_h$.
Since the cumulative transmission time is $J \, T$,
the power during the codeword transmission is  $J \, \energy_h$
divided by $J \, T$.
This would equal the average power if codewords were being
transmitted with a 100\% duty cycle.
The actual average power is reduced by a factor of $\delta$.

The energy per bit $\energybit$ of this channel code is
\begin{equation}
\label{eq:deltavseh}
\energybit = \frac{J \, \energy_h}{\log_2 M} = \frac{\delta \, \energy_h}{\rate \, T} 
\ \ \ \text{or} \ \ \ \ 
\ecrb = \frac{\delta}{\rate \, T} \cdot \ecrs
\,,
\end{equation}
where $\ecrb$ is the energy contrast ratio per bit defined in \eqref{eq:ecrb}
and $\ecrs$ is the energy contrast ratio per energy bundle defined in \eqref{eq:ecrs}.

Now we can understand the reason that $\delta \to 0$ is necessary to approach the fundamental limit.
When operating near the fundamental limit, $\ecrb \approx \log 2$ is approximately fixed.
Holding $\ecrb$, $\rate$, and $T$ fixed, from \eqref{eq:deltavseh} we have that $\ecrs \propto 1/\delta$.
The duty cycle $\delta$ is a means of controlling $\ecrs$, and thus controlling the
reliability of the decision on the presence or absence of each individual energy bundle.
A minimum value of $\ecrs$ is required to make a reliable decision on the presence
or absence of an energy bundle at each candidate location.
By choosing $\delta$ appropriately, we can insure that $\ecrs$ is sufficiently large.
In fact, as $\delta \to 0$ it happens that $\energy_h \to \infty$, which is precisely how
the code achieves an error probability $\pe \to 0$.
Putting it another way,
at a constant power $\power$, reducing the duty cycle $\delta$ allows the
energy in each bundle to increase, and the channel code is able to do this
while holding the information rate $\rate$ fixed by increasing $M$.
Thus, the code has effectively traded increased $M$ (and bandwidth $\Btotal$)
for increased $\ecrs$ so as to overwhelm the noise and scintillation with a very large
amount of energy in each energy bundle.
As $M$ increases and $\delta$ decreases, energy bundles become more and more
sparse in both frequency and in time, and hence each bundle can have a larger energy $\energy_h$
while maintaining constant average power $\power$.

In effect a lower duty cycle $\delta$ is used to ''punch through'' the scintillation
by jacking up the energy in each bundle, even as the number of bundles
is reduced to compensate for this and keep the average power fixed.
This is an important lesson to remember in designing practical channel codes
for the interstellar channel, as is undertaken in Chapter \ref{sec:infosignals}.
Increasing the energy per bundle by reducing the duty cycle is also a technique
that makes signals easier to discover, as discussed in Chapter \ref{sec:discovery}.

\subsection{Receiver processing for time diversity}

The receiver makes productive use of time diversity by averaging the
received energy over time, thus averaging the variations in scintillation
and yielding a more reliable indication of whether an energy bundle is present or not.
In particular, assume that there are $M \cdot J$ matched filters, one matched
to each candidate location.
Let the complex-valued sampled outputs of these filters be 
$\{ Z_{k,m}, \, 1 \le k \le J , \, 1 \le m \le M \}$,
where the $k$ index refers to different times and the $m$ index
refers to different frequencies.
If the actual energy bundle is located at frequency $m=n$,
\begin{equation}
Z_{k,m} = \begin{cases}
\sqrt{\, \energy_h \,} \, S_{k,n} + M_{k,n} \,, & m = n \\
M_{k,m} \,, & 1 \le m \le M-1 \,, \ m \ne n \,.
\end{cases} \,.
\end{equation}
where $S_{k,m}$ is a complex-valued random scintillation and $M_{k,m}$ is a complex-valued additive noise.
By assumption, the $\{ S_{k,m}, \, 1 \le k \le J , \, 1 \le m \le M \}$ and 
$\{ M_{k,m}, \, 1 \le k \le J , \, 1 \le m \le M \}$ are Gaussian and mutually independent.
The receiver can average over the $J$ energy estimates formed by
the matched filter outputs at each index $m$,
forming $M$ decision variables
\begin{equation}
\label{eq:diversitydecisionvar}
Q_m =  \frac{1}{J}  \sum_{k=1}^{J} | Z_{k,m} |^2 \,, \ \ 1 \le m \le M
\,.
\end{equation}
The decision algorithm processes $\{ Q_m , \, 1 \le m \le M \}$ to choose
a value of $m$ as the location of the single energy bundle.

\subsection{The result}

It is shown in Appendix \ref{sec:capacitynoknowledge} that when
\eqref{eq:fundlimit2} is satisfied (and hence \eqref{eq:fundlimit} is satisfied),
the bit error probability  $\pe \to 0$ as $J \to \infty$ and $\delta \to 0$.
The proof assumes that the receiver has precise knowledge of the noise power spectral density $N_0$,
the energy per bundle $\energy_h$, and the scintillation variance $\vs$,
and uses a threshold $\lambda$ applied to the  $\{ Q_m , \, 1 \le m \le M \}$
to choose a value of $m$.
On the other hand,
the receiver needs no knowledge of the scintillation $S_k$
(including the current flux $\flux_k = |S_k|^2$) or noise $M_k$.

The proof in Appendix \ref{sec:capacitynoknowledge}  
applies equally to the AWGN channel with scintillation and the
AWGN channel without scintillation.
This implies that the channel code defined here will work equally
well for weak scattering (Ricean statistics) and strong scattering (Rayleigh statistics).
Communication engineers will undoubtably find this observation surprising,
because we would not normally consider using time diversity in the absence of scintillation,
and because phase-coherent detection is an option in the absence of scintillation and
this would presumably improve the performance.
The reason for this goes back to the observation that $\energy_h \to \infty$,
which swamps out the noise in the interest of driving $\pe \to 0$ and makes the
form of detection irrelevant in the limit.

A related surprising observation is that scintillation does not affect the
fundamental limit.
The reason for this becomes evident in the proof of Appendix \ref{sec:capacitynoknowledge}.
The M-ary FSK plus time diversity channel code circumvents scintillation by making
energy bundles sparse in both frequency and time.
Thus it neatly avoids scintillation almost all of the time, by transmitting nothing.
The error probability $\pe$ then becomes dominated by the probability of a false
alarm in detection of an energy bundle where none exists, and that probability
is not affected by scintillation.

It is also significant that the fundamental limit on power efficiency
is \emph{not} affected by the size of the ICH represented by parameters $\{ B_{c}, \, T_{c}, \, K_{c} \}$,
assuming of course that $K_{c} \gg 1$.
Intuitively this is because the only impact of the ICH size is
the duration of energy bundles $T$ and their bandwidth, neither of which
influences the fundamental limit on power efficiency.

Remarkably, a simple channel code has been found that approaches the
fundamental limit asymptotically as its complexity (as measured by the
$M \cdot J$ locations in each codeword) grows without bound.
A constrained-complexity version of this same channel code (with finite $M$ and $J$)
is a good candidate for practical application to the interstellar channel,
and this is explored further in Chapter \ref{sec:infosignals}.

\textbox{Characteristics of power-efficient information-bearing signals}
	{
	By displaying a channel code that asymptotically approaches the fundamental
	limit on the interstellar channel, some characteristics of such signals are revealed.
	First of all, the critical and complementary role of each of the five 
	principles of power efficient design of Section \eqref{sec:principles} is revealed.
	\boxbreak
	Stepping back and examining the coding scheme, the most defining characteristic is
	the sparsity of energy bundles, in both time and frequency.
	The channel code achieves power efficiency by conserving on the number of energy bundles
	that are transmitted, beneficially increasing the amount of energy in each bundle.
	This becomes evident in both the frequency and time dimensions:
	\boxbreak
	\begin{itemize}
	\item
	Asymptotically, the bandwidth of each individual codeword 
	becomes unlimited, because $M \to \infty$.
	This unlimited bandwidth is characteristic of any channel code that approaches
	the fundamental limit, because the spectral efficiency must asymptotically approach zero.
	\item
	At any given time there is either zero or one energy bundle being transmitted,
	and taken together with the large bandwidth the density of
	 energy bundles becomes sparse in the frequency dimension. 
	\item
	The total time duration of each codeword becomes very large as the
	time diversity grows, or $J \to \infty$.
	This growing and unlimited time duration is also 
	characteristic of any channel code that approaches the fundamental limit.
	\item
	Asymptotically the fraction of the time that an energy bundle is being transmitted,
	which is the duty cycle, shrinks as $\delta \to 0$.
	Thus, energy bundles become very sparse in the time dimension as well.
	\item
	Asymptotically the amount of energy in each bundle grows, also without bound.
	This is critical to achieving $\pe \to 0$.
	\end{itemize}
	These are only asymptotic properties. 
	Any practical and concrete code will use fixed values of $M < \infty$, $J < \infty$, and $\delta > 0$
	and as a result the energy per bundle will also be finite.
As a result of these practical choices, there must be a penalty in power efficiency
(represented by the energy contrast ratio per bit $\ecrb$ that is actually achieved) relative to the fundamental limit of \eqref{eq:fundlimit}.
Choosing $J$ larger helps to close that penalty because the law of large numbers,
and choosing $\delta$ smaller helps to close the penalty because it increases the energy per bundle.
In either case the consequence is a larger $M$ and total bandwidth $\Btotal$.
Practical approaches to improving the power efficiency and the consequences are considered in Chapter \ref{sec:infosignals}.
	}

\subsection{Argument for uniqueness of this channel code}

Basing a practical channel code on one that is capable of
approaching the fundamental limit on power efficiency is attractive.
Not only can high power efficiency be achieved this way, but
it potentially also provides an indirect window into
the thought processes of a transmitter designer who follows
this same approach.
An important question, however, is the uniqueness of such
a channel code.
There is no mathematical basis for claiming uniqueness.
As an example of this, we need look no farther than spectrally efficient
design, where there are two distinct (and very different) classes
of codes that have been shown to approach the Shannon limit,
turbo codes \citeref{618} and low-density parity check codes \citeref{651}.

The relative simplicity of power-efficient design is a significant
help here.
That we are able to so directly and easily find a code with the
desired properties, based on simple intuition no less, is encouraging.
While we could never make a mathematical argument for the
uniqueness of this solution, we can make a logical argument.
The reader will have to judge how compelling this argument is,
and of course the challenge goes out to readers to identify a different but
equally simple channel code that also possesses the asymptotic property.
We have tried and failed.

First, our code is structured around the energy bundle.
All known digital communication systems use a transmitted signal
that is similarly structured about the ''modulation'' of some
continuous-time analog waveform $h(t)$.
The first specific characteristic of the energy bundle idea,
as opposed to the most general use of modulation, is that there
is no attempt to embed information in the phase, only the
magnitude.
This is necessary because of the intrinsically unstable phase
due to scintillation, where natural phenomena and motion need modify the
effective distance by a mere wavelength
(out of multiple light years) before the phase becomes randomized.
The second specific characteristic is the on-off nature of the magnitude.
A couple of arguments can be made for this.
One is Occam's razor,
since two levels is surely the simplest choice, 
and it is sufficient to achieve the asymptotic property,
The other argument is the inherent
power inefficiency in using more than two levels.
At least if those levels are uniformly spaced for fixed noise immunity,
the energy increases more rapidly than the increase in information conveyed.
The additional ICH constraints on time duration and bandwidth for an energy bundle
are necessary to achieve the maximum power efficiency as shown
later in Section \ref{sec:otherimpairments}. 

It is notable that the crucial discovery phase can be based on the
detection of individual energy bundles.
Thus from the perspective of discovery the only characteristic of
the channel code that matters is its reliance on energy bundles, and
an assumption that each bundle has sufficient energy
to overwhelm the noise during favorable scintillation conditions.
Other features of the channel code can be inferred following discovery from observation
of the time-frequency pattern of detected energy bundles.

Second, our code is structured around the idea of using the location of an energy
bundle in time or frequency to encode multiple bits of information.
An arbitrary amount of information (specifically $\log_2 M$ bits)
can be embedded in using one of $M$ alternative locations.
We know that there could be more energy-efficient mechanism for encoding information
(this one can approach the fundamental limit, after all),
and finding a functionally simpler approach seems out of the question.
For example, if we transmit $L >1$ energy bundles in $M$ locations, then
it is easy to show that the power efficiency is monotonically decreasing in $L$.

Third, our code specifically uses frequency as the locator
(frequency-shift keying).
Here there is clearly a viable competitor, which is
using $M$ locations in time (this is pulse-position
modulation).
By using frequency as a locator, there is no need to change the
time duration or bandwidth of each energy bundle as the fundamental
limit is approached, or in other words the limited frequency coherence
property of the ICH need never be violated.
Consuming bandwidth is a necessary element of power efficiency, while
using time as the locator would necessarily increase the total
bandwidth $\Btotal$ by requiring the bandwidth $B = \Btotal$ of
individual bundles to expand, but this is incompatible with other
impairments on the interstellar channel since these bundles would
eventually violate the limits on ICH frequency coherence.

Countering the effects of scintillation
is a challenge where there is greater flexibility.
At the expense of consuming even more bandwidth there is
the option to use frequency diversity rather than time diversity.
There may also be some utility in using more than one level for
the redundant energy bundles, and in fact we will see an example of
that in Chapter \ref{sec:discovery} in the interest of improving
discovery reliability.
There is also the option explored in Chapter \ref{sec:infosignals}
of relaxing our reliability objective and adopting a simpler outage strategy
to deal with scintillation.
This approach trades simplicity for periods of information loss
during periods of low received signal flux.

\section{Fundamental limit for the interstellar channel}

To summarize the results thus far, \eqref{eq:fundlimit} quantifies
the greatest power efficiency that can be obtained on the AWGN channel
without scintillation, and absent a bandwidth constraint.
This is a straightforward consequence of Shannon's channel capacity formula
for that channel, as shown in Section \ref{sec:tradeoffpowerspectral}.
In addition, a concrete channel code based on M-ary FSK and time diversity
was shown in Section \ref{sec:asymptcode} to asymptotically approach the
fundamental limit of  \eqref{eq:fundlimit}.
This was established in the absence of any side information available to the
receiver as to the state of the received signal flux.
This channel code was based on energy bundles falling in the ICH,
and therefore is compatible with the interstellar channel.

Together, these results suggest (but do not prove) that the fundamental limit
on the power efficiency that can be obtained on the interstellar channel
is given by \eqref{eq:fundlimit}.
Why are we not quite there yet?
All that has been shown thus far is that scintillation 
and the lack of knowledge of the scintillation state in the receiver
do not \emph{decrease} the attainable power efficiency.
But is it possible that scintillation could
\emph{increase} the available power efficiency.
The same question applies to all those
other impairments described in Chapter \ref{sec:incoherence}.
They cannot decrease the available power efficiency, since
a channel code based on the ICH is able to asymptotically approach  \eqref{eq:fundlimit},
but is it possible that they could relax the fundamental limit on power efficiency?

The answer is that \eqref{eq:fundlimit} is the fundamental limit on power efficiency
for the interstellar channel.
To establish this, we now tie up these two loose ends.
Readers interested primarily in interstellar communication can safely skip to
Chapter \ref{sec:infosignals}, as the following does not offer
significant new insights on how to communicate more reliably.

\subsection{Lower bound on power efficiency}

To show that power efficiency cannot be improved by scintillation,
it suffices to find a lower bound on power efficiency, and show that
this lower bound equals the fundamental limit of \eqref{eq:fundlimit}.
The lower bound we use is based on the strong assumption that
the receiver has complete knowledge of the current state of
the scintillation.
This added knowledge can only relax the fundamental limit,
allowing reliable communication at equal or higher power efficiencies,
since the receiver always has the option of ignoring this side information.

To derive this result, a discrete-time channel model that characterizes
a communication system constrained to transmit a
sequence of energy bundles conforming to the ICH is defined.
Then a fundamental
limit for this channel assuming receiver knowledge of the state of
the scintillation is determined, and shown to equal \eqref{eq:fundlimit}.

\subsubsection{Discrete-time channel model}

The basic idea behind structuring the signal around the ICH is to
transmit a sequence of energy bundles, the only constraint being that
these bundles do not overlap in time or frequency.
Since the total bandwidth $\Btotal$ is considered to be unconstrained,
we don't have to worry about the details of where these
bundles are placed.
These locations have no impact on the power efficiency $\powere$
that can be achieved nor on the derivation of the results to follow.

A channel model appropriate for studying fundamental
limits requires a generalization of the input-output model of Figure \ref{fig:ICHidea}.
That model is linear in $\sqrt{\energy_{h}}$, so that the sequence of
root-energies can be replaced by a general discrete-time complex-valued input signal $X_{k}$, resulting in a relationship between input sequence $\{ X_k \}$ and
sampled matched filter output sequence $\{ Z_k \}$,
\begin{equation}
\label{eq:discretechannelmodel}
Z_{k} = S_{k} \cdot X_{k} + M_{k} \,.
\end{equation}
For the channel code considered in Section \ref{sec:asymptcode},
$\{ X_k \}$ is always chosen to be drawn from the
alphabet $X_k \in \{ \sqrt{\energy_h}, 0 \}$,
but for purposes of determining fundamental limits
it is important not to constrain the values of $\{ X_k \}$ in any other manner
than average power.
The index $k$ does not necessarily represent time only, as energy bundles
may be transmitted at different frequencies as well as times.

The total rate at which energy bundles are transmitted and received
is assumed to be $F \ \text{sec}^{-1}$.
Since the bandwidth $\Btotal$ is unconstrained, the value of $F$
can be chosen freely.
The average power constraint will be met if
\begin{equation}
\label{eq:powerFsigmaX}
F \cdot \sigma_X^{\,2} \le \power \,,
\end{equation}
where $\sigma_X^{\,2} = \expec{|X_k|^2}$.
Note that there is a tradeoff between $F$ and $\sigma_X$,
and in particular as $F \to \infty$ we must have $\sigma_X \to 0$.

\subsubsection{Side information}

For the channel model of \eqref{eq:discretechannelmodel}, 
in order to obtain a lower bound assume that the
receiver knows the scintillation sequence $\{ S_k \}$.
This strong assumption can only
maintain or improve the attainable power efficiency
since the receiver has the option of ignoring this knowledge.
The first action the receiver should perform is to equalize the
phase of the scintillation multiplier by forming the sequence
\begin{equation}
Z_{k} \cdot \frac{\cc S_k}{| S_k |} = 
\sqrt{ \flux_{k}} \cdot X_{k} + M_{k} \cdot \frac{\cc S_k}{| S_k |}\,.
\end{equation}
This processing is reversible, and thus cannot affect the
fundamental limit.
The channel model can thus be replaced by
\begin{equation}
\grave{Z}_{k} = 
\sqrt{ \flux_{k}} \cdot X_{k} + \grave{M}_{k} \,,
\end{equation}
where $\{ \grave{Z}_{k} \}$ and $\{ \grave{M}_{k} \}$
are phase-shifted versions of $\{ Z_{k} \}$ and $\{ M_{k} \}$.
In particular, the distribution of noise  $\{ \grave{M}_{k} \}$
is identical to the original noise $\{ M_{k} \}$.
Of course the flux $\flux_k = |S_k|^2$ is a real-valued quantity, and
is assumed to be known to the receiver but not the transmitter.

\subsubsection{Ergotic capacity}
\label{sec:ergoticcapacity}

Suppose for a moment that $\flux_k = \flux$
is constant.
Then the capacity of a discrete-time channel with
additive complex-valued zero-mean Gaussian noise is
well known \citerefWloc{43}{Section 4.2.3},
\begin{equation}
\label{eq:capfixedflux}
\rate \le \capacity = F \cdot
\log_{\,2} \left( 1 + \frac{\flux \, \sigma_X^{\,2}}{N_0} \right)
\,.
\end{equation}
This is a discrete-time version of the AWGN channel capacity \eqref{eq:capacity}.
The ergotic capacity merely averages the capacity over the flux,
\begin{equation}
\label{eq:capavgflux}
\rate \le \capacity =
\expec{
F \cdot
\log_{\,2} \left( 1 + \frac{\flux \, \sigma_X^{\,2}}{N_0} \right)
}
\end{equation}
where the expectation is over the random ensemble for $\flux$.

The assumptions behind the ergotic capacity
 can be summarized as follows  \citerefWloc{574}{Sec. 5.4.5}.
For any fixed value of flux $\flux$, \eqref{eq:capfixedflux} applies.
This ''instantaneous'' capacity is increased relative to the noise-only case if $\flux >1$
or decreased if $\flux < 1$.
As the actual flux $\flux_k$ changes with time, this ''instantaneous'' capacity also changes,
and \eqref{eq:capavgflux} is the average of these instantaneous capacities.\footnote{
The term ''ergotic'' refers to an equality between a time average
and ensemble average.
This definition of capacity assumes that the time-average capacity is equal to the
ensemble-average capacity, which will be valid for a wide range of conditions.}
The transmitter cannot tailor the channel code at different times according to the
instantaneous conditions.
The transmitter can, however, use a codebook that spans a time duration that is very large
relative to the known time scale of the fluctuations in $\flux$,
and also tailor the code to the known distribution of the scintillation.
In this case each code word statistically experiences the full range of $\flux_k$ for which it is designed.
Thus, predicated on the assumption that the receiver knows $\{ S_k \}$, the ergotic capacity is
the fundamental limit on the rate $\rate$ that can be achieved.

\subsubsection{Power efficiency with side information}

Expressing the ergotic capacity of \eqref{eq:capavgflux} in terms of
 the average power $\power$ from \eqref{eq:powerFsigmaX},
\begin{equation}
\rate \le \capacity =
\expec{
F \cdot
\log_{\,2} \left( 1 + \frac{\flux \, \power}{F \, N_0} \right)
} \,.
\end{equation}
The argument of $\expec{\cdot}$ is monotonically increasing in
rate $F$ (for every value of $\flux$), so by taking advantage of unconstrained
bandwidth and letting $F \to \infty$ (the unconstrained bandwidth case) the
maximum information rate is achieved.
This maximum rate is
\begin{equation} 
\rate \le \capacity
\underset{F \to \infty}{\longrightarrow}
\expec{ 
\frac{ \flux \, \power}{N_0 \, \log 2}
}
= 
\frac{ \power \, \vs}{N_{0} \, \log 2} \,.
\end{equation}
This is equivalent to \eqref{eq:fundlimit}, establishing that
knowledge of the scintillation by the receiver does not
increase the achievable power efficiency or, in other words, relax the fundamental limit.

\subsection{Effect of other impairments on power efficiency}
\label{sec:otherimpairments}

In fact, this result is stronger that it appears because any transmitter design that
violates ICH integrity introduces additional ISM and motion impairments into the
received signal, and we would expect that dealing with these would reduce
(and could not increase) the power efficiency.
A reasonable hypothesis is that a design based on the ICH structure can achieve greater power efficiency
than any design that deliberately violates the integrity of the ICH.

To investigate this further,
assume that the transmitter uses a unit-energy waveform $h(t)$ as a carrier for an energy bundle
and the receiver implements a filter matched to $h(t)$, but the actual received
waveform has been changed to $S \cdot g(t)$ by the effects of interstellar impairments
including a scintillation multiplier $S$.
Then the magnitude of the sampled matched filter output $Z$ can be bounded by the Schwartz inequality,
\begin{equation}
| \, Z \, |^{\,2} = 
\left| \, \int_0^T S \cdot g (t) \cdot \cc h (t) \, dt \, \right|^{\,2} \le
\int_0^T | \, S \cdot g (t) |^{\,2} \, dt = \flux \cdot \int_0^T | \, g (t) \, |^{\,2} \, dt 
\,,
\end{equation}
with equality if and only if $g(t) = z \cdot h(t)$ for some complex-valued constant $z \ne 0$.
On the other hand, if there is no interstellar impairment other than scintillation,
or in other words $h(t)$ falls in the ICH, then the
same output is $| \, Z \, |^{\,2} = \flux$.
Thus, the only way that non-scintillation impairments could increase the signal level,
and hence improve on the receiver sensitivity, is if they increase the signal energy, or
\begin{equation}
1 < \int_0^T | \, g (t) \, |^{\,2} \, dt  \,.
\end{equation}
While we know that scintillation can increase signal energy temporarily
(although not on average),
an examination of the other impairments in Chapter \ref{sec:incoherence}
reveals none that are capable of increasing the signal energy.
For example, plasma dispersion is an all-pass (phase-only)
filtering function, and thus neither reduces nor increases the total signal energy.

As a proof of the fundamental limit on power efficiency for the interstellar channel,
this argument does have one shortcoming.
We have shown than when communication is based on the energy bundle, giving us the
option of avoiding all interstellar impairments save noise and scintillation,
then it has been shown that the best power efficiency results when the energy bundle adheres to the ICH.
This does not rule out the possibility that deliberately violating the ICH constraints
would allow even higher power efficiency than the fundamental limit of \eqref{eq:fundlimit}
by using some other approach than the energy bundle.
Since none of the ICH-excluded impairments would permit more energy to arrive
at the receiver, it is our judgment that this could not be the case.
Under these conditions generally impairments result in reduced (rather than increased)
power efficiency due to the enhancement of noise and increase in false alarm probability
resulting from processing to encounter impairments.
This possibility should, however, be explored 
with greater mathematical rigor before ruling it out permanently.

\section{History and further reading}

The publication of Shannon's channel capacity theorem in 1948 resulted
in a flurry of activity pursuing concrete channel codes that could
approach the fundamental limit.
The easier unconstrained bandwidth case was addressed first.
Fano showed in 1961 that
M-ary FSK and PPM both approach fundamental limits on 
power efficiency on the AWGN channel as $M \to \infty$
and using phase-coherent detection \citeref{639}.
These are special cases of a more general class of \emph{orthogonal codes} that
all have this nice property.
For spectrally efficient design, where $\Btotal$ is constrained, fundamental limits
were approached by very complex turbo codes in 1993 \citeref{618}.
Later it was shown that low-density parity check codes
(first described by Gallager in 1960) can also approach
the fundamental limit \citeref{651}.

Because scintillation (fading) is an important impairment in terrestrial radio
communications, there is an extensive literature.
There is no single definition of capacity,
because it is dependent on the assumptions
about knowledge of the current scintillation conditions at the
transmitter and at the receiver \citeref{632}.
For an intuitive explanation of these alternative sets of assumptions
and their consequences, 
\citerefWloc{574}{Sec. 5.4} is recommended.

With the exception of the use of the ICH and the role of interstellar impairments more
generally, most of the results of this chapter can be traced to the work of
R.S. Kennedy in 1969 \citeref{637},
who determined the fundamental limit on power efficiency for a
continuous-time Rayleigh fading channel in the absence of
any such knowledge in the receiver.
For a more readily available and concise derivation see \citerefWloc{638}{Section II},
and Kennedy's results are also summarized in Gallager's 1968 textbook on
information theory \citerefWloc{635}{Section 8.6}.

\section{Summary}

The fundamental limit on power efficiency for the interstellar channel
has been determined for unconstrained bandwidth and in the absence of
any side information about the state of the scintillation in the receiver.
A concrete channel coding based on M-ary FSK combined with time diversity
has been shown capable of approaching the fundamental limit asymptotically,
and will serve as the basis for a practical constrained-complexity design in
Chapter \ref{sec:infosignals}.
That channel coding makes use of all five of the principles of
power-efficient design described in Section \ref{sec:principles},
and concretely illustrates the necessary and complementary role of each of those principles in achieving
high power efficiency.
The fundamental limit is not relaxed if the receiver possesses side information about the scintillation.
To obtain the maximum possible power efficiency, bandwidth must be unconstrained
and allowed to increase arbitrarily,
so very poor spectral efficiency is a necessary feature of any interstellar communication
system operating near the power efficiency limit.

The fundamental limit depends on the two physical parameters $\{ N_0 , \, \vs \}$ alone,
and is not affected by the ICH parameters $\{ T_c, \, B_c \}$ or the carrier frequency $f_c$.
Its numerical value is not affected by the presence or absence of scintillation,
since under all conditions of scattering (both weak and strong)
the phenomenon is lossless and thus $\vs =1$.
It is also shown that a signal design structured about the interstellar coherence hole (ICH)
developed in Chapter \ref{sec:channelmodel} possesses the same fundamental limit, implying
that nothing need be lost in power efficiency by constraining the design in this way.
Indeed any design using signals violating the
time duration and bandwidth constraints of the ICH will necessarily sacrifice some 
signal level at the matched filter output, and thus operate at a reduced power efficiency
for equivalent reliability.

Perhaps the most significant outcome of this chapter is a fundamental limit that constrains
all civilizations, no matter how experienced or advanced they may be.
In demonstrating a channel coding that asymptotically approaches this limit,
a great deal of insight into how both we and other civilizations might approach the
fundamental limit is developed.
This should help us anticipate the types of signals to expect, and thereby aid in the
discovery of such signals in an uncoordinated environment.


\chapter{Communication}
\label{sec:infosignals}

Having established the fundamental limit on power efficiency
in Chapter \ref{sec:fundamental},
the question is how close this limit can be approached by a concrete design.
We have seen that actually approaching the limit requires a channel code with
codewords that grow with time duration and expand in bandwidth, and
an energy per bundle that grows, all without bounds.
In contrast any concrete channel code must necessarily
have a finite duration and finite bandwidth, a finite energy in
each bundle, and a probability of error $\pe > 0$.
Putting in this ''complexity'' constraint will limit the available power efficiency
to something less that permitted by the fundamental limit.
The one aspect of the channel code of Chapter \ref{sec:fundamental}
that carries over without modification
is the restriction of energy bundles to the interstellar coherence hole (ICH).

The question addressed in this chapter is how close to the fundamental limit we can reach with
such a parameterization while staying in the realm of what might be considered
practical.
This is considered for the M-ary FSK channel code with time diversity
considered in Chapter \ref{sec:fundamental}, but restricting $M$ and $J$ to be
fixed and finite values rather than allowing them to grow without bound.
A simpler approached based on M-ary FSK (to counter noise)
combined with an outage approach 
(in which information is lost during periods of low received signal flux)
is then considered and compared.
The outage approach looks attractive for the interstellar channel.

According to the classification of Figure \ref{fig:threelayers} 
proposed in \citeref{583}, this design
only addresses the signal layer, omitting the reliability and message layers. 
The signal layer is intrinsically unreliable (due to $\pe > 0$)
due to noise and scintillation, and thus the
design of the reliability and message layers will be affected by the
results described here.
In particular, the outage strategy changes the reliability layer
requirements profoundly.

\incxy
	{threelayers}
	{
	Three distinct layers of functionality
	in the interstellar communication of messages \citeref{583}.
	These are the signal layer (including generation and detection),
	the reliability layer (including redundancy and assertion),
	and the message layer (including creation and interpretation).
	Our concrete design in this chapter considers only the signal layer,
	but demonstrates that the reliability layer
	must deal with not only random dispersed errors but also,
	in the case of an outage strategy,
	substantial periods with no recovered data due to 
	low received signal flux.
	}
	{
		*+<1pc>[F-,]{
			\begin{matrix}
			\text{Message} \\
			\text{creation}
			\end{matrix}
			}
		\ar@{=>}[r] &
		*+<1pc>[F-,]{
			\begin{matrix}
			\text{Reliability} \\
			\text{redundancy}
			\end{matrix}
			}
		 \ar@{=>}[r] &
		 *+<1pc>[F-,]{
		 	\begin{matrix}
			\text{Signal} \\
			\text{generation}
			\end{matrix}
			}
		  \ar@{=>}[r] &
		  	{
		  	\begin{matrix}
			\text{Radio} \\
			\text{channel}
			\end{matrix}
			}
		  \\
			{
			\begin{matrix}
			\text{Radio} \\
			\text{channel}
			\end{matrix}
			}
		 \ar@{=>}[r] &
		*+<1pc>[F-,]{
			\begin{matrix}
			\text{Signal} \\
			\text{detection}
			\end{matrix}
			}
		\ar@{=>}[r] &
		*+<1pc>[F-,]{
			\begin{matrix}
			\text{Reliability} \\
			\text{assertion}
			\end{matrix}
			}
		 \ar@{=>}[r] &
		 *+<1pc>[F-,]{
			 \begin{matrix}
			\text{Message} \\
			\text{interpretation}
			\end{matrix}
		 } & \\
      }

\section{Time diversity strategy}
\label{sec:diversitystrategy}

Diversity combining is one of the
principles of power-efficient design of Section \ref{sec:principles}.
The idea behind time diversity is to transmit each
energy bundle multiple times with a time spacing large
relative to the coherence time of the channel, or a frequency spacing
large relative to the coherence bandwidth of the channel.
At the receiver, diversity combining averages the signal energy
from these redundant versions of the bundle to form an
estimate that asymptotically averages out the received flux
and thus allows reliable extraction of the information in spite of scintillation.
To make it practical, finite values for $M$ and $J$ are used,
creating a penalty in the achieved energy contrast ratio per bit
$\ecrb$ relative to the fundamental limit.
Our primary interest is in the quantifying this penalty, which also depends
on the objective for error probability $\pe$.

\subsection{Parameters and tradeoffs}

The channel code of Section \ref{sec:noside} can asymptotically approach
the fundamental limit on power efficiency
on the interstellar channel, so our concrete design uses a constrained-complexity
version of this code.
As illustrated in Figure \ref{fig:oneCodeword},
one codeword consists of an energy bundle at one frequency out of $M$
and is repeated $J$ times to counter the time-varying effects of scintillation.
To convey a stream of information, codewords are merged
by some combination of interleaving 
(as illustrated in Figure \ref{fig:sequentialCodewords})
or sequential transmission (Figure \ref{fig:interleavedCodewords}).

The parameters of the channel code are listed in
Table \ref{tbl:fskTDparameters}, using the same notation as
Section \ref{sec:noside}.
The duty factor $\delta$ is now a secondary parameter, determined by the
choice of  $\{ \rate , \, T , \, J, \, M \}$.
However, not all combinations of these parameters are feasible, corresponding
to the constraint $0 < \delta \le 1$.
Although any $\rate$ can be accommodated in principle by making $M$ sufficiently
large, for given values of $\{ \rate , \, T ,\, M \}$, the available time diversity is
limited by
\begin{equation}
\label{eq:Jmax}
J \le \frac{\log_2 M}{\rate \, T} \,.
\end{equation}
This bound arises because there is limited transmission time available for redundant energy bundles.
Choosing a larger $M$ has the beneficial effect of increasing the available transmission time,
thus allowing a larger $J$.

On-off keying (OOK) with 100\% duty factor has rate $\rate = 1/T$,
so $\rate \, T = 1$ corresponds to this baseline OOK case.
When $\rate \, T < 1$, this is equivalent in information rate to OOK with
less than 100\% duty factor, and  $\rate \, T > 1$ has a higher
information rate than can be achieved by single-carrier OOK
(this is possible because $M$ can be as large as we choose).
It appears that a smaller information rate $\rate$ is generally helpful,
because from \eqref{eq:Jmax} this allows a larger $J$ for a given $M$.
However, it will be now be shown that choosing a larger $J$ is not always beneficial.

\doctable
	{\small}
	{fskTDparameters}
	{A list of parameters of a constrained-complexity channel
	code based on FSK and time diversity.
	The design parameters describe all relevant choices.
	The secondary parameters describe the consequences of
	those choices in determining the information rate and power efficiency.
	}
	{p{2.5cm} p{3cm} p{9cm}}
	{
	\toprule
	\textbf{Category} &\textbf{Parameter} & \textbf{Description}
	\\
	\cmidrule{1-3}
	\textbf{Design \newline parameters}
	& $J$ & Number of time repetitions of each energy bundle
	\\
	\cmidrule{2-3}
	& $M$ & Number of frequency offsets for energy bundles
	\newline Each codeword conveys $\log_2 M$ bits of information
	\newline The minimum is $M \ge 2^{\,J \, \ratenorm}$
	\\
	\cmidrule{2-3}
	& $\rate \, T$ & Normalized information rate
	\newline $\rate \ T = 1$ equals the rate of 100\% duty factor on-off keying (OOK)
	\\
	\cmidrule{2-3}
	& $\energy_h$ & Received energy in each bundle
	\\
	\cmidrule{1-3}
	\textbf{Physical \newline parameters}
	& $N_0$ & Power spectral density of noise
	\\
	\cmidrule{2-3}
	& $\vs$ & Average of the received signal flux $\flux$
	\newline $\vs = 1$ based on the physics of scattering
	\\
	\cmidrule{1-3}
	\textbf{Secondary \newline parameter}
	& $\delta$ & Duty factor
	\newline Faction of time an energy bundle is transmitted
	\\
	\cmidrule{1-3}
	\textbf{Reliability}
	& $\pe$ & Probability of error in each decoded bit
	\\
	\bottomrule
	}

\subsection{Reliability}
\label{sec:reliabilityFSKdiversity}

In Section \ref{sec:noside} it was established that the bit error probability
$\pe \to 0$ is possible using this channel code at power efficiency $\powere$ less than the fundamental limit
(energy per bit $\energybit$ above the fundamental limit).
A threshold detection algorithm in the context of M-ary FSK in combination
with time diversity was used (and is necessary) to prove this result.
The threshold algorithm
required precise knowledge of the parameters $\{ \energy_h , \, N_0 \}$.
For practical application, however, a \emph{maximum energy} detection algorithm is
more desirable.
This algorithm chooses the value of $m$ meeting the criterion 
\begin{equation}\
\label{eq:maxAlgorithm}
\max_{1 \le m \le M} Q_m
\end{equation}
as the location of an energy bundle,
where $Q_m$ is the decision variable given by \eqref{eq:diversitydecisionvar}.
A desirable characteristic of this maximum algorithm, which is actually a feature of M-ary FSK as
a channel coding technique, is that knowledge of $\{ \energy_h , \, N_0 \}$ is not needed.
We know that an energy bundle is present at exactly one frequency, and the only issue
is to decide which one.
The maximum energy algorithm simply chooses the location where the estimated energy
is highest, assuming that this ''excess'' energy is due to the presence of an energy bundle.

When $M$ and $J$ are finite we must have $\pe >0$.
A tight upper bound on the error probability $\pe$
for the maximum algorithm is determined in Appendix \ref{sec:pefsktd},
\begin{equation}
\label{eq:peVecrsFSKtd}
\pe \le \frac{M}{2} \cdot \left( \frac{4 \, ( 1 + \ecrs )}{( 2 + \ecrs )^{\,2}} \right)^J \,.
\end{equation}
Note that $\pe$ depends only on the energy contrast ratio $\ecrs$, and not on the
individual parameters $\{ \energy_h , \, N_0 \}$.
The maximum algorithm adjusts automatically to wherever $\ecrs$ may be,
always yielding the most reliable detection (lowest $\pe$) dictated by whatever conditions it encounters.

In Figure \ref{fig:PeBFSK} the tradeoff between the energy contrast ratio per bit $\ecrb$
and $\pe$ was illustrated by plotting \eqref{eq:peVecrsFSKtd} for two cases,
$M = 2$ and $M = 2^{20} \approx 10^6$.
For the second case $J = 49$ is chosen because that choice minimizes $\ecrb$, as
will be shown shortly.
This figure illustrates that there is no single value for the penalty in average power
relative to the fundamental limit, but rather this penalty depends on our objective for $\pe$.
In numerical examples we choose $\pe = 10^{-4}$, or an average of one error
out of every 10,000 bits.

\subsection{Total bandwidth}

In the following numerical comparisons, the value $M = 2^{20} \approx 10^6$ is assumed.
This corresponds to a total bandwidth $\Btotal = 1$ MHz  if the energy bundle bandwidth is $B = 1 Hz$,
or correspondingly greater $\Btotal$ for larger values of $B$.
For comparison purposes, an $M$ and $\Btotal$ three orders of magnitude smaller and larger
are also calculated.

One issue that should be kept in mind is the frequency stability of the received transmission.
If there is any frequency drift due to acceleration, then this will interfere with the accurate
recovery of the information.
Thus, this small $B$ and large $\Btotal$ assumes that both transmitter and receiver
have corrected for their respective acceleration as discussed in Section \ref{sec:freqoffset}.

\subsection{Total energy hypothesis}

One question is how the energy per bundle $\energy_h$
is affected by the transmission of $J$ bundles in time diversity.
This question is best addressed by fixing the error probability $\pe$
and then varying $J$ for a fixed value of $M$.
A \emph{total energy hypothesis} states that the
reliability as measured by $\pe$ is approximately fixed
if the total energy $J \, \energy_h$ 
(or equivalently $J \, \ecrs$)
is held fixed as $J$ increases.
As shown in Figure \ref{fig:ecrsVsJ}, $\ecrs$ does decrease
steadily as time diversity $J$ increases, but the total energy $J \ecrs$
does remain relatively fixed.\footnote{
This is in contrast to the asymptotic results of Section \ref{sec:noside},
where $\ecrs \to \infty$ as the fundamental limit is approached.
The difference is that there $M \to \infty$, and here $M$ is held fixed.
}
There is a minimum value of total energy at $J \approx 2^5$,
and beyond that there is a slow increase in the total energy.
Subsequent calculations will reveal that $J$ should not be
chosen too large, so most we expect to operate near this minimum.

Based on the total energy hypothesis, $\ecrs \approx 1/J$ maintains
approximately fixed error probability $\pe$ as $J$ increases.
Time diversity allows the energy per individual bundle to decrease, and
it also desirably decreases the peak transmitted power by spreading the
total energy of each codeword over a greater total transmission time $J T$.
On the other hand, as discussed in Chapter \ref{sec:discovery}, it
also makes discovery more difficult because the detection of each
individual energy bundle less reliable.
It is shown there that a small adjustment to the channel code can mitigate this adverse effect
at a small penally in power efficiency.

\incfig
	{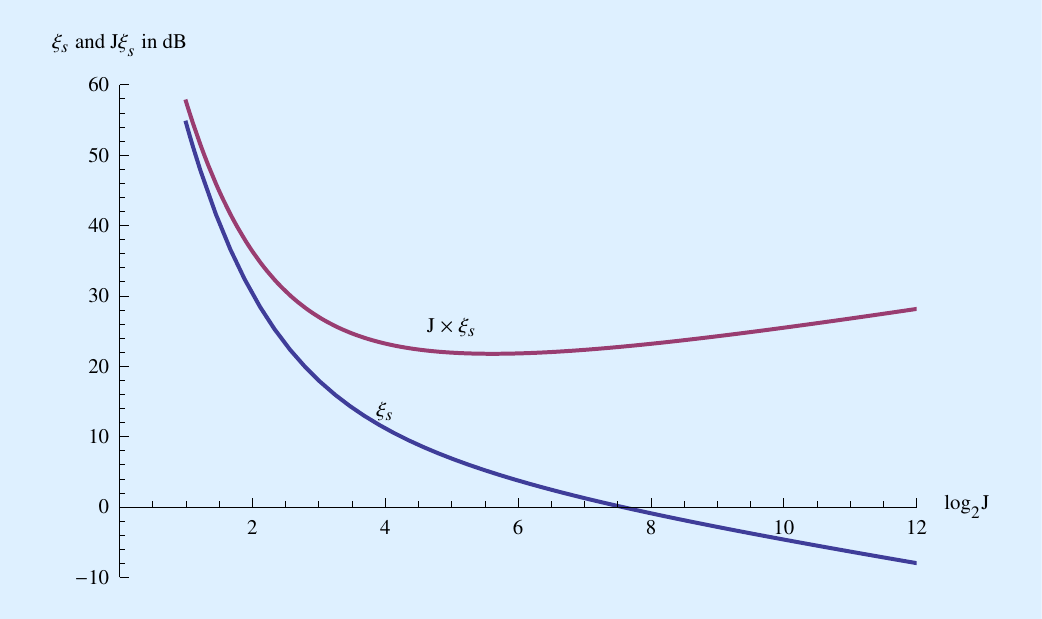}
	{1}
	{
	A plot of the energy contrast ratio $\ecrs$ in dB for each energy
	bundle required to achieve a reliability $\pe = 10^{-4}$
	vs. the time diversity $J$.
	The number of frequencies is kept fixed at $M=2^{20}$.
	There is a steady decrease in the required $\ecrs$
	as a benefit of the law of large numbers.
	The total energy $J \, \ecrs$ in all bundles
	remains fairly constant, confirming the total energy hypothesis.
	}

\subsection{Power efficiency}

To see how the power efficiency $\powere$ is affected by a change in the
parameterization of the channel code, it is convenient to work instead with
the energy per bit $\energybit = 1/\powere$ or equivalently the
energy contrast ratio per bit $\ecrb$.
This quantity is plotted in Figure \ref{fig:ecrbVsJdifferentM} for three
different values of $M$ and as a function of $J$.
The values of $M$ illustrated are $\{ 2^{10} \approx 10^3 , \, 2^{20} \approx 10^6 , \, 2^{20} \approx 10^9 \}$.
The total bandwidth $\Btotal$ for these cases depend on the bandwidth $B$ of each energy bundle
and the frequency spacing of the bundle locations, which is at least $B$.
The three choices correspond to about one kHz, one MHz, and one GHz for $B = 1$ Hz.

This figure illustrates three different observations:
\begin{itemize}
\item
Consistent with \eqref{eq:Jmax}, for any value of $\{ \rate , \, M \}$ there is a maximum feasible
value of $J$.
This maximum value is proportional to $\log_2 M$, or $J \propto \log_2 M$.
\item
The energy contrast ratio $\ecrb$ does not decrease monotonically with $J$.
There is an optimum value of $J$ that gets us the closest to the fundamental limit,
and fortunately (from an implementation perspective) that optimum value is modest in size.
The reason for this behavior is a tradeoff between two completing effects.
On the one hand, as $J$ increases the receiver's estimate of bundle energy becomes more accurate,
and that is reflected in the $J$ exponent in \eqref{eq:peVecrsFSKtd}.
On the other hand, as $J$ increases, for fixed $M$ the energy contrast ratio
decreases as $\ecrs \propto 1/J$ for fixed $\energybit$.
This is because there are $J$ energy bundles
contributing to the total received energy, and if $J$ is larger there is less energy to
allocate to each of those $J$ energy bundles.
\item
If the optimum value of $J$ is chosen,
increasing $M$ is beneficial in getting us closer to the fundamental limit on $\ecrb$
for the optimum choice of $J$.
\end{itemize}
The direct coupling between $M$ and $J$ (the latter at the optimum point) is helpful
to implicit coordination between transmitter and receiver, because it indicates that for
a transmitter intent on reducing its transmitted power one of these variables is dependent
on the other.
Reflecting that, in the subsequent comparisons we can eliminate the parameter $J$
by substituting its optimum value.
That optimum value is shown in Figure \ref{fig:JvsMconstantRate} for a fixed information rate,
demonstrating a linear relationship.

\incfig
	{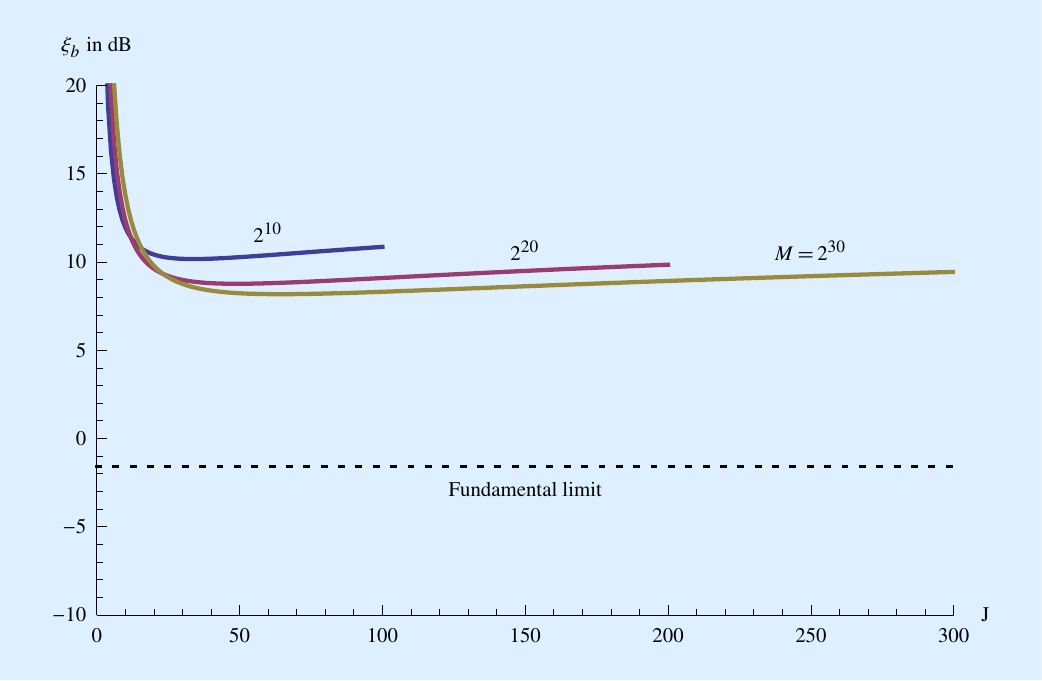}
	{1}
	{The energy contrast ratio $\ecrb$ in dB is plotted against the degree of
	time diversity $J$ for $\log_2 M = 10$, 20, and 30.
	All curves correspond to information rate $\rate \, T = 0.1$, or an information
	rate equal to 10\% of an OOK signal with 100\% duty factor,
	and an error probability $\pe = 10^{-4}$.
	The range of $J$ is upper bounded by \eqref{eq:Jmax}.
	The fundamental limit on power efficiency at $\ecrb = -1.6$ dB is shown by the dotted line.}

\incfig
	{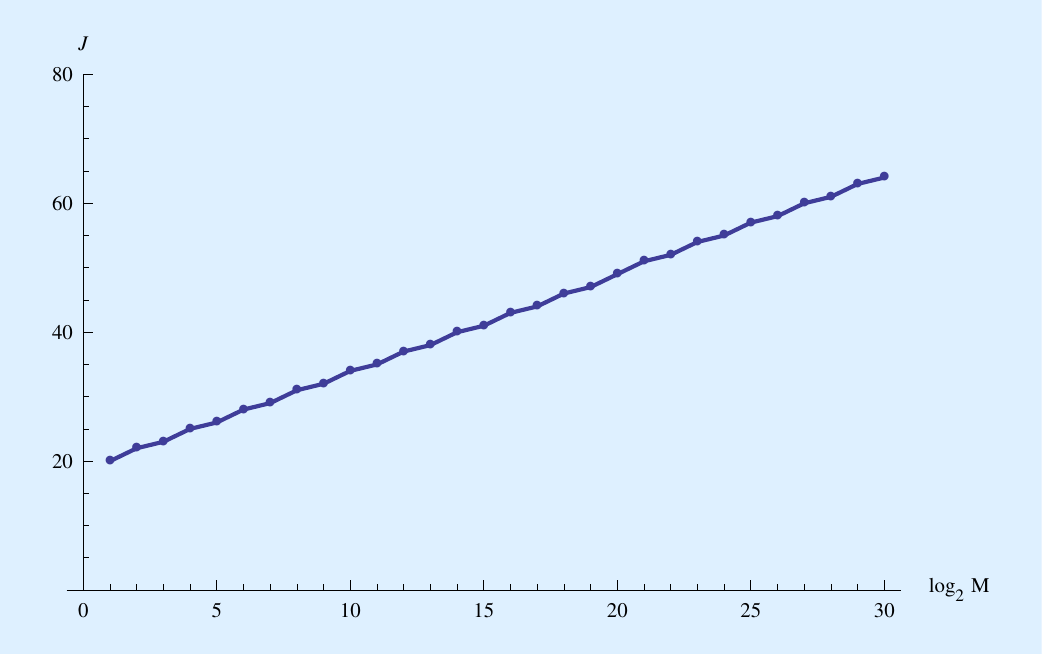}
	{1}
	{The degree of time diversity $J$ that minimizes the energy contrast ratio per bit $\ecrb$
	(and hence maximizes the power efficiency) is plotted as a function of $\log_2 M$.
	The information rate is held fixed at $\rate \, T = 0.1$,
	and the error probability is held fixed at $\pe = 10^{-4}$}

Assuming that the transmitter always chooses a value for $J$
that maximizes the power efficiency, the relationship of the achieved energy contrast ratio
per bit $\ecrb$ to $\log_2 M$ is illustrated in Figure \ref{fig:ecrbVsMdifferentRate}
for several information rates $\rate$.
The channel code performs well for all rates less than OOK with 100\% duty factor
($\rate \, T < 1$) because there is sufficient room for the optimum degree of time diversity $J$
at larger values of $\log_2 M$.
However, at information rates higher than OOK, the channel code performs poorly
because the constraint on $J$ given by \eqref{eq:Jmax} is too small to allow
the optimum choice of $J$ to be used.
There is simply too little transmission time available to achieve the needed
time diversity (which consumes transmission time since energy bundles are
repeated $J$ times) at higher rates, where we want to transmit codewords more frequently.
	
\incfig
	{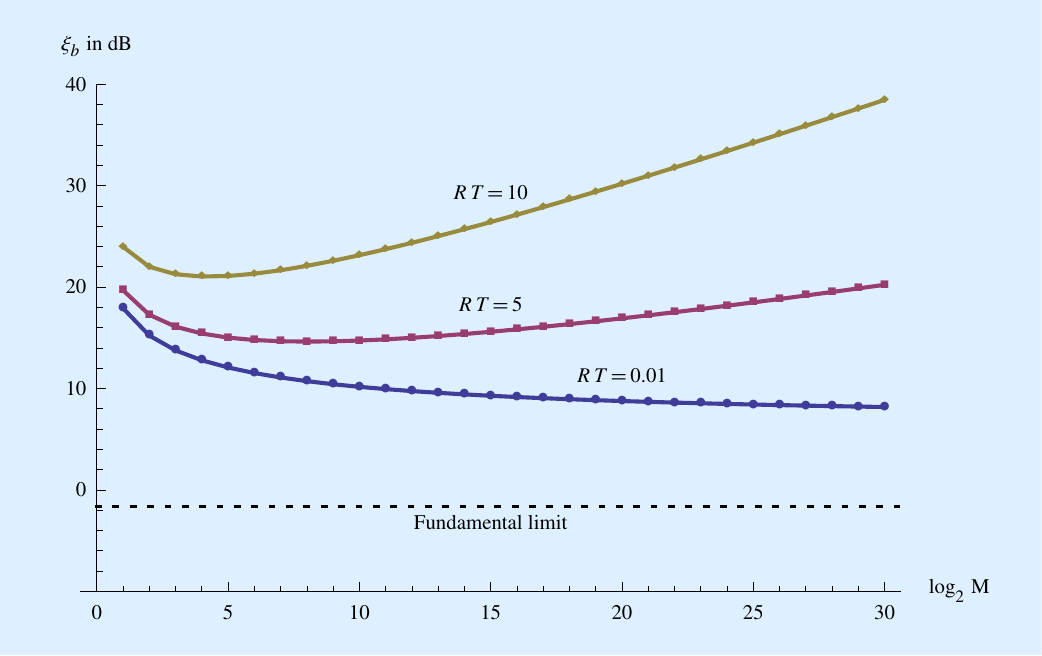}
	{1}
	{A plot of the minimum energy contrast ratio per bit $\ecrb$ that can be achieved
	when the optimum value of $J$ is used against $\log_2 M$, for three different information rates $\rate$.
	The error probability is held fixed at $\pe = 10^{-4}$
	The case $\rate T = 0.01$ is typical of all low rates, where
	the achievable $\ecrb$ does not depend on $\rate$.
	The other two cases show that this channel code performs poorly at rates higher
	than OOK with 100\% duty factor ($\rate \, T \gg 1$), the reason being that the
	degree of time diversity is too constrained by \eqref{eq:Jmax} to achieve the
	greatest possible power efficiency for a given value of $\log_2 M$. }

The M-ary FSK with time diversity channel code, after the optimizing choice of $J$,
suffers a material penalty relative to the fundamental limit due to the choice of finite
values for $M$ and $J$.
At $M = 2^{20}$ the penalty is 10.4 dB (a linear factor in average power of 10.9).
The penalty decreases slowly with $M$, becoming 9.8 dB (a linear factor of 9.5)
at $M = 2^{30}$.
We expect that the penalty will not approach zero as $M \to \infty$ for two reasons.
One is the use of phase-incoherent detection, and the other is the use of the
maximum algorithm of \eqref{eq:maxAlgorithm} as a concession to practical concerns about
knowledge of bundle energy $\energy_h$ and noise power spectral density $N_0$.

\subsection{Multicarrier modulation}
\label{sec:mcm}

\textbox
	{Multicarrier modulation}
	{
Multicarrier modulation (MCM) is widely used because it is recognized to have a number of advantages
other than immunity to dispersion and trading spectral efficiency for power efficiency. 
If the distinct carrier frequencies deliberately violate frequency coherence,
the scintillation experienced by
different carriers will be unrelated to one another (this is called \emph{frequency-selective fading}).
This observation can be used to improve the reliability for communication as well as discovery,
for example by creating data redundancy across different carriers.
The relative powers allocated to different carriers can be adjusted
to accommodate frequency-dependent phenomena such as a planet's
larger atmospheric noise due to water vapor (concentrated at $f_c \gg 15$ GHz).
The power allocated to one (and only one) subcarrier can be artificially increased
to make discovery easier with only a small power penalty overall.
MCM also lends itself to parallelism in transmitter implementation.
For example, any collection of subcarrier signals can be transmitted independently
of the others using separate
RF electronics and antennas, each optimized for its own carrier frequency.
A transmitter designer on a limited budget can 
expand total information rate over time by increasing $M$, 
with a capital cost that scales no faster than $\rate$.
Both transmit and receive implementations lend themselves to
joint subcarrier manipulation at baseband using discrete Fourier
transform techniques \citeref{612}.
	}

We have seen in Figure \ref{fig:ecrbVsMdifferentRate}
that a finite-complexity channel code becomes inefficient
at information rates rates higher than OOK with 100\% duty cycle.
The problem is our attempt to use a single carrier, which at higher rates
does not afford sufficient time to transmit our redundant time diversification energy bundles
when $M$ is constrained to reasonable values.
There is an easy way to circumvent this limitation, and that is to use multiple carriers.
A sequence of codewords can be transmitted on each of $L$ carriers,
taking care to keep the information rate on each carrier sufficiently low to achieve
the best power efficiency,
and also taking care to space the carriers far enough apart to avoid frequency overlap.
The resulting information rate is increased by a factor of $L$, and the average
power is increased by the same factor of $L$, so the power efficiency is not affected.
This approach is called \emph{multicarrier modulation} MCM \citeref{612}.

In 1995, S. Shostak anticipated the importance of
MCM (although he didn't call it that), motivated (as here) by its immunity to ISM impairments \citeref{127}.
Similar to M-ary FSK, MCM is a way to trade additional bandwidth and average power
for higher information rate.
MCM merely maintains the same power efficiency as the bandwidth expands,
whereas M-ary FSK improves the power efficiency.
On the other hand, MCM can achieve higher information rates for the same
increase in bandwidth, so it is more spectrally efficient.
MCM demonstrates one approach to trading higher spectral efficiency
for poorer power efficiency.
As such, it is widely used in terrestrial wireless systems, where spectral efficiency
is usually a dominant consideration.

\section{Outage/erasure strategy}
\label{sec:outageerasure}

A channel code based on M-ary FSK with time diversity
pursues the ambitious goal of maintaining consistently reliable communication
in spite of scintillation.
An alternative  \emph{outage/erasure} strategy relaxes this goal in the interest of simplifying the channel coding.
This strategy combines three ideas:
\begin{changea}
\begin{enumerate}
\item 
This strategy eschews time diversity entirely, so
the channel code consist of single M-ary FSK codewords without redundancy.
This implies that there are codewords that suffer a large probability of error
(during periods and frequencies of low scintillation flux)
and codewords that are interpreted reliably
(during periods and frequencies with high flux).
\item
This strategy may rely on the capability of the receiver to identify
times and frequencies with low scintillation flux, in which case it can ignore or put
less reliance on information extracted from the signal during these periods.
\item
This strategy relies on redundancy in the transmitted information
in order to recover information that is inevitably lost to scintillation.
If there is sufficient redundancy, 
it may be possible to recover the original information reliably
even without knowing times or frequencies with low scintillation flux, and
the second part of the strategy may be unnecessary.
\end{enumerate}
\end{changea}

This outage strategy places a burden on the reliability and message layers 
of Figure \ref{fig:threelayers} to accommodate a substantial loss of
information while retaining message integrity.
In terrestrial systems this would be dealt with in the reliability layer by combining
interleaving with another layer of redundant channel coding.
\begin{changea}
In interstellar communication there may be a simpler approach.
 The transmitter presumably has no idea when a communication is being received
 or anybody is even listening.
 A transmitted message also has to be carefully composed to be
 understandable in the absence of any shared language, and this would
 require considerable resources.
 In response to this issue, the transmitter is likely to transmit an single message
 over and over, rather than attempt to communicate an open-ended stream of information.
If this is the case, a reasonable way for the receiver to deal with the loss
of unreliable information is to
accumulate multiple received copies of the transmitter's intended message, 
and gradually interpolate missing or unreliable information.

Even without explicit knowledge of which part of the message is corrupted,
with sufficient redundant copies of the message a reliable representation
can be recovered by \emph{majority logic}.
If a particular segment of the message is identical in two or more copies, 
then this segment is almost surely correct.
This relies on the statistical reality that two corrupted segments
are very unlikely to be identical.

\subsection{Impact of incoherence on M-ary FSK}

Our non-time-diversity strategy can still be based on M-ary FSK,
a channel code that can asymptotically approach the fundamental limit on
power efficiency if (a) scintillation is absent and (b) phase-coherent detection is used.
In the presence of scintillation there is no alternative to using phase-incoherent
detection, and this will inevitably impose a 3 dB penalty in sensitivity.
To understand the impact of scintillation on M-ary FSK,
recall that an energy packet may be transmitted at $M$ different  frequencies.
The dependence of scintillation flux on frequency is quite relevant.
As illustrated in Figure \ref{fig:threeBandwidths}, 
there are three frequency scales to take into consideration:
\begin{itemize}
\item 
The bandwidth $B$ of an energy bundle must be less than or equal to
both the dispersion and scattering coherence bandwidths.
The smaller of these two coherence bandwidths determines the
ICH coherence bandwidth $B_c$.
\item
The dispersion coherence bandwidth is typically smaller
and thus determines $B_c$, the
bandwidth of a coherence hole and the upper bound on $B$.
\item
The scattering coherence bandwidth is typically larger
than the dispersion coherence bandwidth.
Significantly for M-ary FSK, the scattering coherence bandwidth
is a determinant of scintillation as it varies with frequency.
\end{itemize}
In Chapter \ref{sec:incoherence} the dependence of scintillation on
both time and frequency is discussed.
The observation most relevant to M-ary FSK is that
any two energy bundles with frequency spacing
within the scattering coherence bandwidth will experience nearly the
same scintillation (amplitude and phase).
Conversely, when the frequency spacing becomes larger than the
scattering coherence bandwidth the scintillation experienced
by the two energy bundles will
begin to diverge, eventually becoming statistically independent for a
frequency spacing large relative to the scattering coherence bandwidth.

\incfig
	{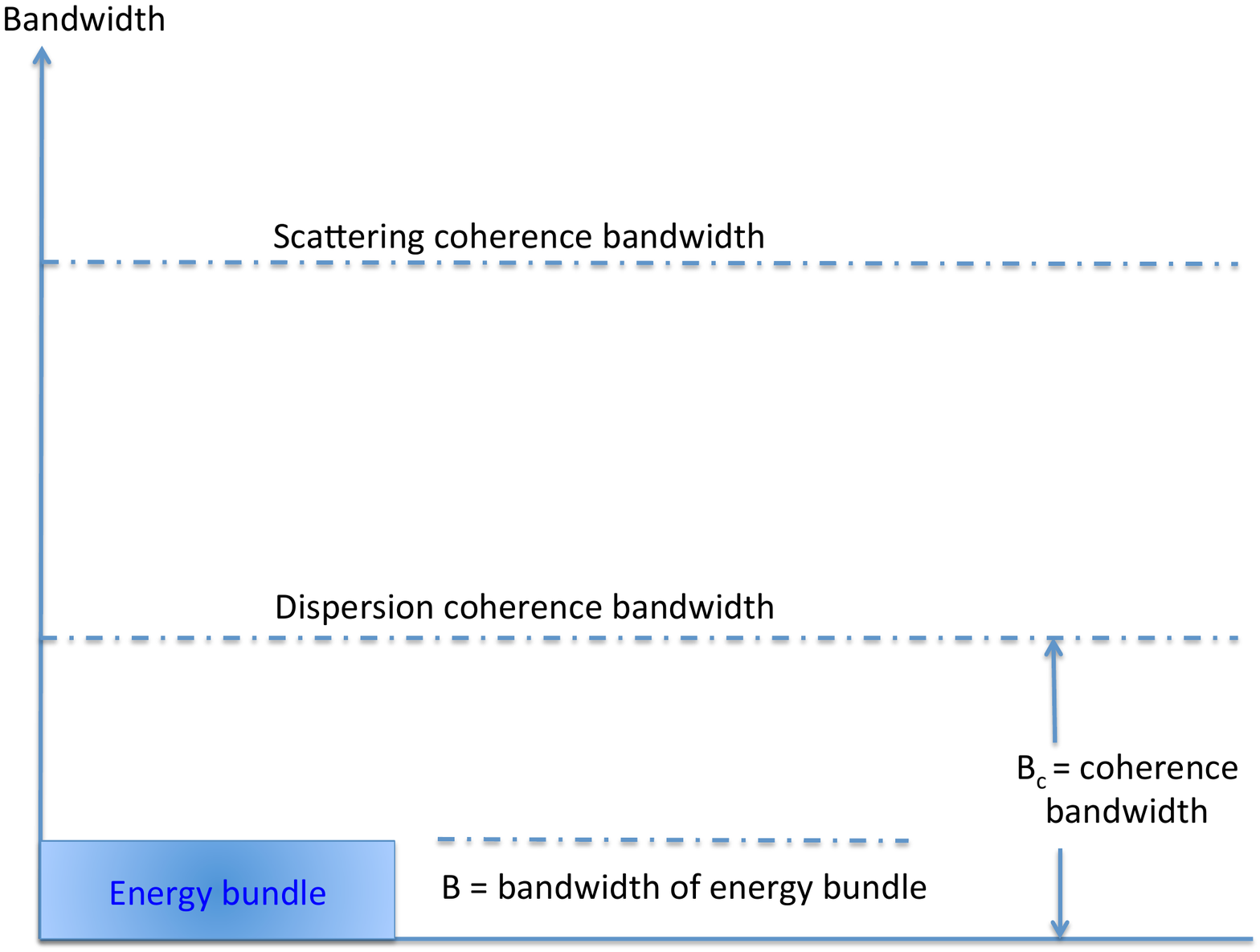}
	{.9}
	{
	Three bandwidth scales of interest to M-ary FSK without time diversity.
	}

One way to think about this is that the time evolution of scintillation has to
be considered separately for each of the $M$ frequencies that are
candidate locations for an energy bundle in M-ary FSK.
The time evolution of scintillation will generally be the same for
closely spaced frequencies (relative to the scattering coherence bandwidth)
and statistically independent processes for widely spaced frequencies.
This latter phenomenon is called \emph{frequency-selective} scintillation.
Cohen and Charlton noted the relevance of this phenomenon to SETI in 1995,
and proposed to exploit it for the purpose of reducing the effects of scintillation
by transmitting a ''polychromatic'' SETI beacon \citeref{140}.

As to whether M-ary FSK codewords will experience frequency-selective
fading depends on the relationship between the total bandwidth $\Btotal$
and the scattering coherence bandwidth.
This relationship depends on the carrier frequency $f_c$, as well
as the value of $M$ and the bandwidth $B$ of each energy bundle
(where $\Btotal \approx M \, B$).
The dispersion and scattering coherence bandwidths typically diverge
at low carrier frequencies $f_c$ and then parallel one another, as illustrated in
Figure \ref{fig:scatVsdispCB}.
The ratio of these two bandwidths, as illustrated in Figure \ref{fig:ratioCB},
varies over about four orders of magnitude for this case.
This ratio governs the largest $M$ for which M-ary FSK avoids
frequency-selective scintillation assuming that each energy bundle
has the largest bandwidth $B = B_c$ permitted by the dispersion coherence
bandwidth.
When $B$ is chosen to be smaller than $B_c$,
this transition value of $M$ can be correspondingly larger.
Earlier in this chapter the value $M = 2^{20} \approx 10^{6}$ was used
as a baseline in numerical examples.
For this relatively large value of $M$ frequency-selective
fading will be a reality under most conditions.

\incfig
	{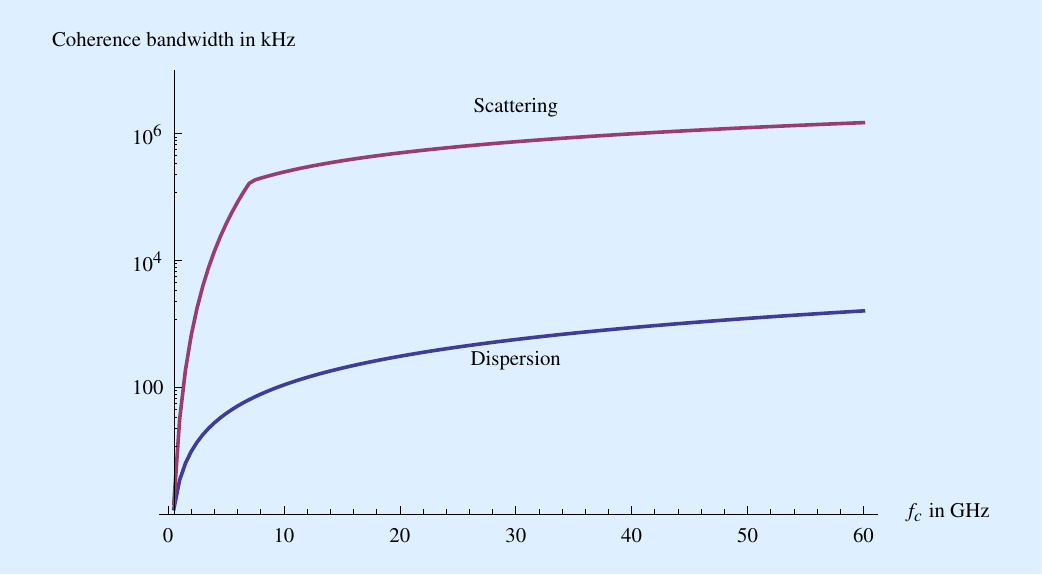}
	{1}
	{A log plot of the dispersion coherence bandwidth and the scattering coherence bandwidth
	(in kHz) vs the carrier frequency $f_c$ (in GHz)
	for the typical physical parameters of Table \ref{tbl:assumptions}
	and with a safety parameter $\alpha = 0.1$.
	The discontinuity in the scattering plot is where the strong and
	weak scattering approximations intersect; reality will be more continuous
	at the transition.}

\incfig
	{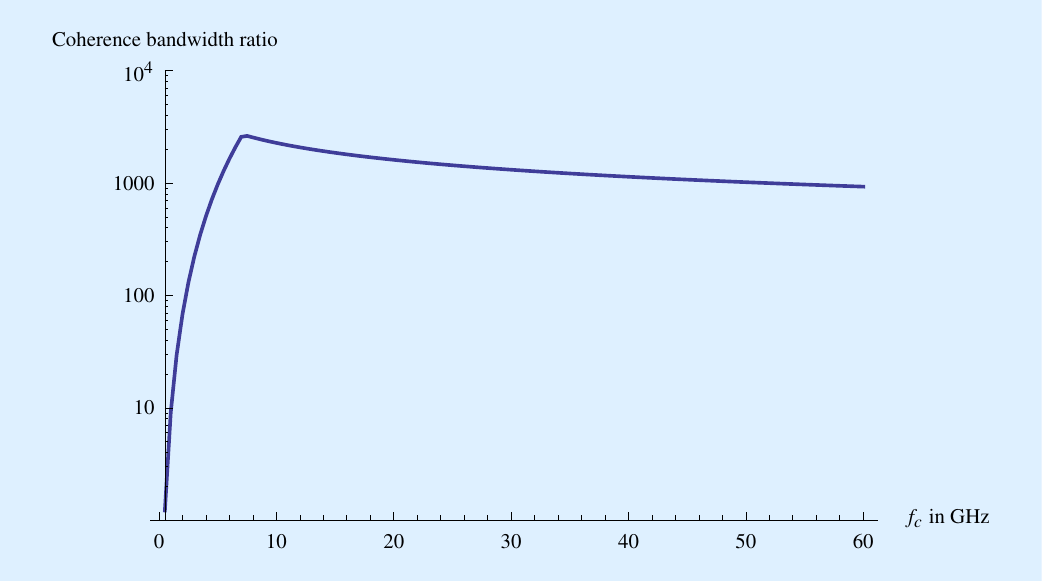}
	{1}
	{A log plot of the ratio of the larger scattering 
	coherence bandwidth to the smaller dispersion coherence bandwidth
	for the values plotted in Figure \ref{fig:scatVsdispCB}.
	}

\subsection{Pattern of corrupted information}

At times and frequencies with low scintillation flux the recovered information will
be corrupted.
Depending on the choice of parameters $\{ f_c, \, B, \, M \}$,
 the typical patterns of corruption will be different.
In the absence of frequency-selective fading (large $f_c$, smaller $M$)
the corruption will be on the scale of entire codewords, and
typically affect a collection of codewords grouped in time.
This phenomenon is called an \emph{outage}, which is defined as a period of time
during which information recovery is totally and consistently corrupted.
During outage periods, the entire $M$ bits recovered from
each codeword can be considered to be corrupted.

The opposite circumstance occurs when the total bandwidth $\Btotal$
is much larger than the scattering coherence bandwidth (small $f_c$, larger $M$).
In this case the scintillation will be frequency-selective.
At any given time energy bundles at some bands of frequencies can be detected
reliably, and there are other bands of frequencies where the energy of the
bundles is too low to be detected reliably.
This implies that some combinations of the $M$ bits recovered from a codeword
are likely to be correct, and some combinations likely to be are corrupted.
This phenomenon is called an \emph{erasure}, and those combinations
of bits which cannot be relied upon are called erasures.

The decoder that decides which of the $M$ frequencies
harbors the energy bundle should pass on to the reliability layer in
Figure \ref{fig:threelayers} all the information that it possesses,
to help that layer more reliably recover the entire message.
For example, if there is an available estimate of scintillation flux
at a time or frequency accompanying a particular pattern of
$M$ bits, then that side information should be passed on.

\subsection{Available single-carrier information rate}

In the absence of time diversity, or $J = 1$, the rate becomes
\begin{equation}
\rate = \delta \, \frac{\log_2 M}{T} \,.
\end{equation}
For a single carrier the available normalized information rate is bounded by
\begin{equation}
0 < \rate \, T \le \log_2 M \,.
\end{equation}
Not having to reserve time for the $J >1$ redundant copies of each
energy bundle allows much higher information rates $\rate$
on a single carrier, reducing the need for relying on multicarrier modulation (MCM).
For example, with $M = 2^{20} \approx 10^6$, the information rate $\rate$
can be as much as 20 times higher than OOK, 
at the expense of an average power $\power$ correspondingly large.

\end{changea}

\subsection{Choice of an outage/erasure threshold}

In order compare an outage/erasure strategy with
the time-diversity alternative, it is helpful to develop an
estimate of power efficiency or some equivalent measure.
For this purpose it is necessary to make further assumptions
on the operation of this strategy.

A conventional outage strategy would maintain a
running estimate of the received signal flux based on
a time average of the past detected energy bundles.
In an erasure strategy, the running estimate would be
repeated for different frequency bands, with a frequency granularity
determined by the scattering coherence bandwidth.
If such an estimate is available, the receiver can declare an outage
or erasure whenever the received signal flux is below a threshold $\oot$,
or $\flux < \oot$.
We will now define an effective power efficiency based on the
assumption that all information declared to be in an outage or subject
to an erasure is useless and is discarded, and the remaining information is
reliable and is usable.\footnote{
A better performing approach would be to pass on a 
finer granularity measure of
reliability with each segment of recovered information,
rather than the crude ''reliable'' or ''discarded'' measure.
In this case the effective power efficiency becomes more difficult to determine.
}
The following description assumes only outages, but the existence of
erasures does not modify the results.

Suppose the probability of an outage is $\po$.
Then the effective rate at which information is reliably recovered is not
the transmitted rate $\rate$, but rather $(1 - \po ) \cdot \rate$.
Then a reasonable definition of effective power efficiency is
\begin{equation}
\label{eq:powereoutage}
\powere = \frac{(1 - \po) \cdot \rate }{\power} \,.
\end{equation}
The $(1-\po)$ factor takes into account the reduction in actual recovered
information rate due to the loss of all information during outages.
$\rate$ is the information rate embedded in the information-bearing signal,
and $\power$ is the average received power with or without scintillation
(which are the same since $\vs = 1$).

A principled way to choose $\oot$ is to maximize the effective
power efficiency.
For a chosen $\oot$ and for Rayleigh fading (scintillation due to strong scattering), the outage probability is
\begin{equation}
\label{eq:poutage}
\po = \prob{ \flux < \oot } = 1 - \exp \left\{ - \frac{\oot}{\vs} \right\} \,.
\end{equation}
During non-outages, the reliability of information recovery still varies
widely depending on the actual flux $\flux$.
Our criterion for choosing $\oot$ is to insure that $\pe \approx 0$
in the worst case, which is $\flux = \oot$.
At all other times during non-outages the reliability is even better, and usually \emph{much} better.
It is useful to remember that when $\pe$ is calculated for $\flux = \oot$,
this is not representative of the overall average error probability during non-outages,
which is much smaller, nor the overall average error rate which is strongly influenced
by outages.

Zeroing in on the case $\flux = \oot$, and ignoring the actual variation in $\flux$
due to scintillation, the received energy per bundle is changed from $\energy_h$
to $\oot \, \energy_h$.
The receiver can add a flat gain (or loss) $1/\sqrt{\oot}$ without affecting the
available power efficiency or the $\pe$ that can be achieved.
After this receiver adjustment,
the energy per bundle becomes $\energy_h$, and the additive noise
power spectral density is changed from $N_0$ to $N_0 / \oot$.
The resulting channel model, which applies only at the threshold of outage,
is an AWGN channel with noise power spectral density equal to $N_0 / \oot$.
The fundamental limit on power efficiency for this channel is
\begin{equation}
\frac{\rate}{\power} < \frac{1}{(N_0/\oot) \, \log 2}
\end{equation}
and thus the fundamental limit on effective power efficiency
becomes
\begin{equation}
\frac{(1 - \po) \, \rate}{\power} < \frac{\oot \, (1 - \po) }{N_0 \, \log 2} \,.
\end{equation}
The choice of $\oot$ that maximizes this effective power efficiency
is $\oot = \vs = 1$, in which case $\oot \, (1 - \po) = 1/e$.
The resulting effective power efficiency is
\begin{equation}
\frac{(1 - \po) \, \rate}{\power} < \frac{1}{N_0 \, e \, \log 2} \,,
\end{equation}
and the resulting outage probability is $\po = 63\%$.

At the maximum effective power efficiency, the threshold
of outage is at the point where the received signal flux $\flux$ equals its average value $\vs = 1$.
The resulting channel model at the threshold of outage is an AWGN channel
with scintillation absent.
The resulting effective power efficiency is reduced relative to that AWGN channel
by a factor of $1/e=0.368$ (or 4.3 dB) due to the deliberate loss of 63\% of the embedded information
during periods of outage.
For any specific channel code, we can determine the penalty in power efficiency
for the AWGN channel absent scintillation, and then add 4.3 dB to that penalty to
find the penalty in effective power efficiency.

We will now compare OOK and M-ary FSK as two channel coding
approaches that could be combined with an outage strategy.

\subsection{Comparison of channel codes}

OOK is representative of the type of signal sought by current SETI searches,
so it represents a useful benchmark.
M-ary FSK is known from our prior analysis to be a good channel code for
maximizing power efficiency on the AWGN channel.
There is no reason to include time diversity with FSK since we are using the alternative outage
strategy to counter scintillation.
These channel codes can be analyzed for the AWGN channel absent scintillation,
which is appropriate at the optimum threshold of outage.

OOK conveys one bit of information per codeword by the presence of absence
of an energy bundle with energy $\energy_h$.
The resulting energy per bit $\energybit = \energy_h /2$ if the two choices are equally likely.
The resulting $\pe$ is determined in Appendix \ref{sec:OOKawgnOnly}.

The M-ary FSK channel code is the same as studied in Section \ref{sec:diversitystrategy}, and uses the
maximum algorithm of \eqref{eq:maxAlgorithm} which (in contrast to OOK)
requires no knowledge of $\energy_h$ or $N_0$.
There are two differences from Section \ref{sec:diversitystrategy}, however.
There is no time diversity needed, so $J = 1$.
Also, the analysis of $\pe$ is different from that leading to \eqref{eq:peVecrsFSKtd} because
the model for the channel at the threshold of outage is absent scintillation.
A tight upper bound on the bit error probability $\pe$ is determined in Appendix \ref{sec:peppm},
\begin{equation}
\label{eq:peFSKbound}
\pe \le \frac{1}{4} \, M^{1 - \frac{\ecrb}{2 \, \log 2}} \,.
\end{equation}
$\pe$ is compared for OOK and M-ary FSK in Figure \ref{fig:peOOKvsFSK} for 
a typical value $M = 2^{20}$.
This shows that at $\pe = 10^{}$ at the threshold of outage,
M-ary FSK with $M=2^{20}$ can reduce the penalty in effective power efficiency 
relative to the fundamental limit from a linear factor of $96.6$ for OOK
down to a factor of $8.3$.

\incfig
	{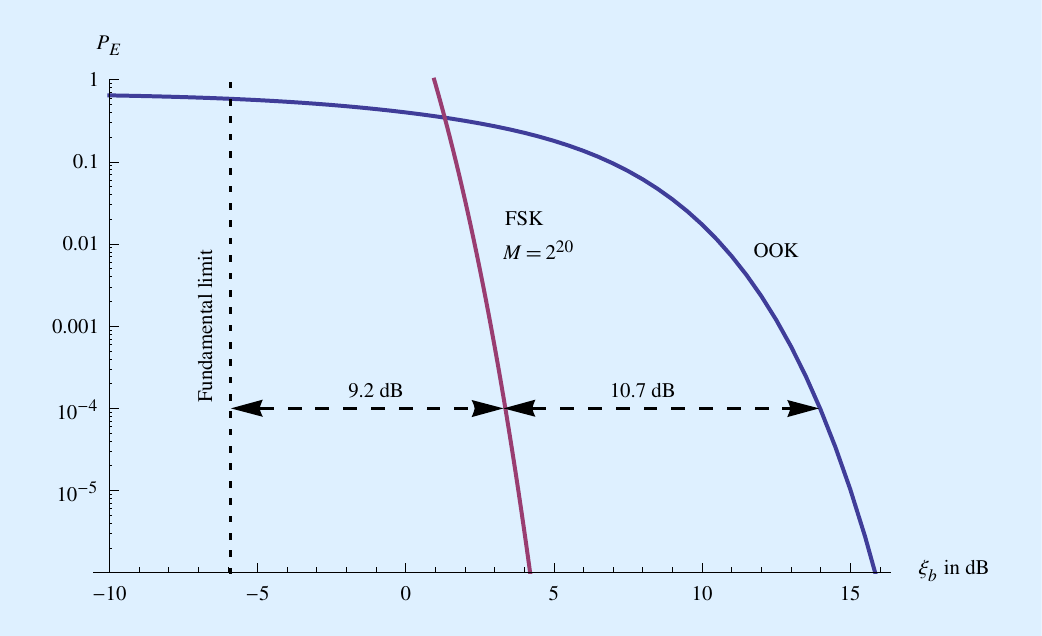}
	{1}
	{A log plot of a tight upper bound on the bit error probability $\pe$
	for both on-off keying (OOK) and M-ary frequency-shift keying (FSK)
	without time diversity on the AWGN channel without scintillation.
	This channel model applies to the outage strategy at the threshold of outage,
	with smaller $\pe$ achievable whenever the received signal flux $\flux$ is above average.
	The horizontal axis is the energy contrast ratio
	$\ecrb$ in dB.
	For M-ary FSK the value of $M$ shown is $M = 2^{20}$.
	The fundamental limit on power efficiency has been moved 4.3 dB to the left
	to account for the loss of information during outages.
	At $\pe = 10^{-4}$,
	M-ary FSK offers a 10.7 dB advantage over OOK (a linear factor of 11.6 in average power)
	and FSK suffers a penalty relative to the fundamental limit of 9.2 dB (linear factor of 8.3).
	The OOK penalty relative to the fundamental limit is 19.8 dB (a linear factor of 96.6).}

Recall from Figure \ref{fig:ecrbVsMdifferentRate}
that M-ary FSK plus time diversity has a penalty relative to the fundamental limit 
on power efficiency of a linear factor of $10.9$ for the same value of $M = 2^{20}$.
Thus, M-ary FSK without time diversity but employing an outage strategy is the slightly more
attractive option.
In view of its relative simplicity by eliminating the need for time diversity, this makes the outage stategy
an attractive solution on the interstellar channel.
However, it should be kept in mind that this is an ''apples and oranges'' comparison
between two very distinctive definitions of power efficiency.
The outage strategy achieves its small penalty at the expense of a deliberate 63\% average loss of
embedded information, and that loss can only be restored if the same message is transmitted
multiple times and observed over a long period of time.

A striking property of \eqref{eq:peFSKbound} is that when
\begin{equation}
\label{eq:ecrbbound}
\ecrb > 2 \, \log 2 \,,
\end{equation}
then $\pe \to 0 \ \ \ \text{as} \ \ \  M \to \infty$.
This simple coding can achieve arbitrarily high reliability as $M \to \infty$
as long as \eqref{eq:ecrbbound} is satisfied.
This lower bound on $\ecrb$ is a factor of two larger than the fundamental limit on the AWGN channel,
reflecting the loss in sensitivity due to using phase-incoherent detection.
Except for this 3 dB penalty, M-ary FSK can approach the fundamental limit on power
efficiency as $M \to \infty$ at the threshold of -outage.
Of course we need to
keep in mind that 63\% of the information is totally lost to outages, which adds an additional
4.3 dB penalty in effective power efficiency.
Thus, the total penalty relative to the fundamental limit on
effective power efficiency approaches 7.4 dB (a linear factor of 5.4) as $M  \to \infty$.

Also \eqref{eq:peFSKbound} can be solved for $\ecrb$ to
quantify the tradeoff between $\ecrb$ and $M$ at fixed $\pe$,
\begin{equation}
\label{eq:ecrbVsPe}
\ecrb >
2 \, \log 2 \cdot \left( 1 - \frac{\log (4 \, \pe)}{\log M} \right)
 \,.
\end{equation}
Note that the first term equals \eqref{eq:ecrbbound},
and the second term is the penalty relative to that limit due to using
a finite $M$, as a function of $\{ \pe, \, M \}$.
This is plotted in Figure \ref{fig:ecrbVsM} and compared to the asymptote of \eqref{eq:ecrbbound}.
This illustrates that there is a slowing rate of improvement as $M$ increases.
For example, increasing $M$ from $M = 2^{20}$ to $M = 2^{30}$ improves the
effective power efficiency by only 14\% (0.56 dB).

\incfig
	{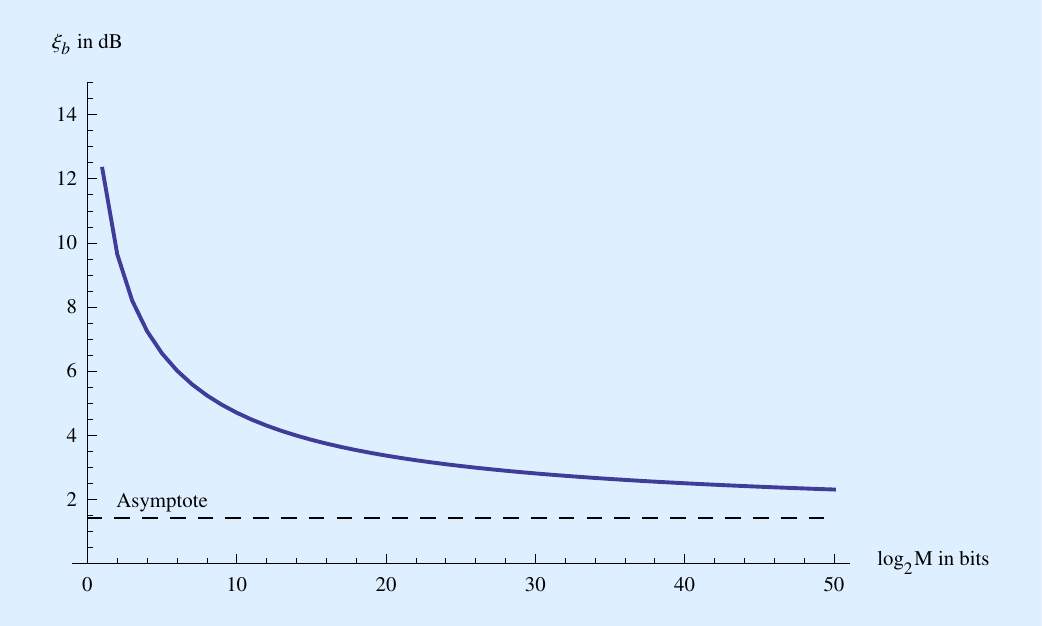}
	{1}
	{A plot of the energy contrast ratio per bit $\ecro$ in dB
	required to achieve a reliability $\pe = 10^{-4}$
	vs the number of bits
	per codeword $\log_{2} M$ for M-ary FSK.
	As $M$ increases the required $\ecrb$ decreases,
	approaching the asymptote $2 \, \log 2$, shown as a dashed line.}

\subsection{Adjusting information rate and average power}
\label{sec:adjrateandpower}

The previous development calculated the power efficiency for
a single codeword, but the information rate $\rate$ is determined by
how fast codewords are transmitted, and the average power $\power$
is proportional to $\rate$.
As multiple codewords are transmitted on a single carrier at a rate $F$,
the information rate is $\rate = F \cdot \log_{2} M$ and the
average power is $\power = F \cdot \energy_{h}$.
Thus, the power efficiency remains constant, independent of $F$.

If the transmitter wants to reduce $\power$ (at the expense of $\rate$),
it is advantageous to maintain $F = 1/T$, and increase $T$ as
$F$ is decreased.
This has the desirable effect of reducing the peak power as well as the
average power, by spreading the codewords out over a greater time.
However, there is a limit to how large $T$ can be increased while
maintaining the ICH time coherence constraint $T \le T_{c}$.
If further reductions in $\power$ are desired, then without affecting
$T$ a duty factor $\delta < 1$ can be introduced, so that the
peak power becomes higher than the average power $\power$.
This is illustrated in Figure \ref{fig:FSKvsDelta}.

Similarly, if an increase in $\rate$ is desired (at the expense of $\power$),
$T$ can be reduced while increasing $F$.
However, a different limit is reached when it is necessary to increase
bandwidth $B$, because eventually the ICH constraint $B \le B_{c}$ will be violated.
In that case,
multicarrier modulation (MCM) can be used as
described in Section \ref{sec:mcm} and illustrated in Figure \ref{fig:FSKvsCarriers}.
The desirable feature of equal peak and average powers is preserved by MCM.

 \incfig
 	{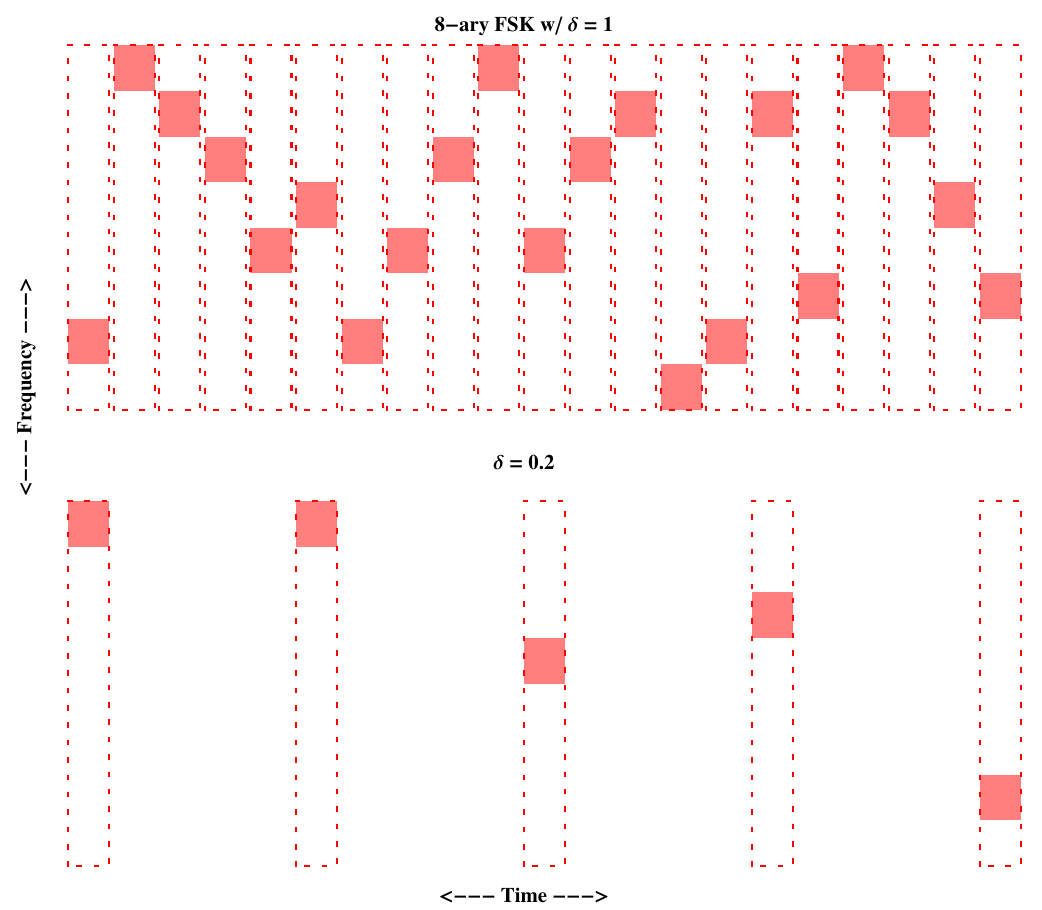}
	{.7}
	{
	Illustration of the location of energy bundles for M-ary FSK
	(with $M=8$), and how the information $\rate$ can be
	reduced by reducing the duty factor to $\delta = 0.2$ without changing $T$.
	The result is a reduction in both information rate $\rate$ and average power $\power$
	by a factor of five, while maintaining the same power efficiency.
	}

\incfig
	{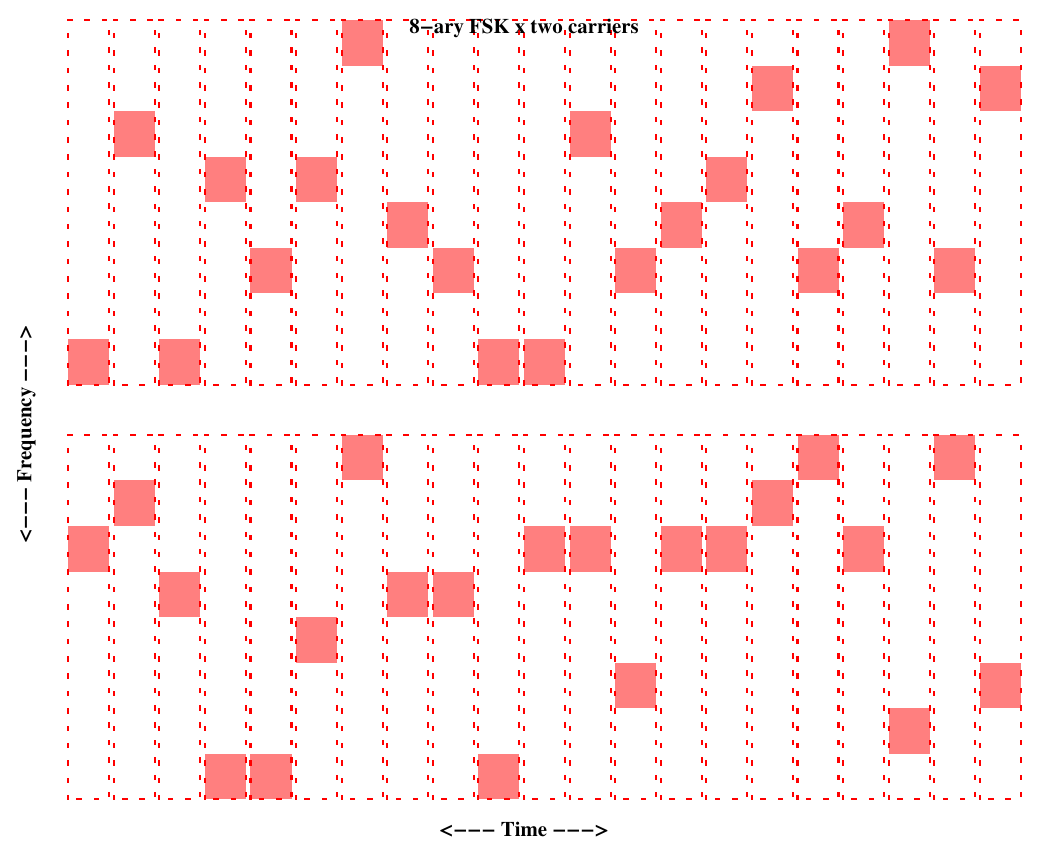}
	{.7}
	{
	Illustration of the location of energy bundles for M-ary FSK
	(with $M=8$), and how the information $\rate$ can be
	doubled by adding a second carrier.
	The average power $\power$ is also doubled, so the
	power efficiency is the same.
	Note that this is different than doubling $M$ to $M = 16$
	(even though the diagram would look similar)
	which would increase $\rate$ by $4/3$ and $\power$ not at all.
	The difference in the two cases is two vs. one energy bundles per codeword time duration.
	}

\subsection{Spectral efficiency}
\label{sec:FSKspectralefficiency}

As $M$ increases in M-ary FSK and $\powere$ increases as a result, 
a decrease in spectral efficiency $\spectrale$ is the price paid for higher power efficiency $\powere$.
To characterize $\spectrale$,
assume that a single carrier is used with a rate of transmission of energy bundles equal to $F$.
Each codeword time (time duration of $h(t)$) is $T$, and the duty factor is
\begin{equation}
\delta = T \cdot F \,.
\end{equation}
The bandwidth required for $h(t)$ is $B = K /T$, where $K$ is the degrees of freedom of the signal
defined in Section \ref{sec:channelmodeldof}.
To avoid overlap in the frequency domain, the minimum total bandwidth must be
\begin{equation}
B_{\text{FSK}} = M \cdot B \,.
\end{equation}
The information rate is
\begin{equation}
\label{eq:inforateR}
\rate = F \cdot \log_{2} M = \frac{\delta}{T} \cdot \log_{2} M
\end{equation}
and thus the spectral efficiency is
\begin{equation}
\spectrale = \frac{\rate}{B_{\text{FSK}}} = \frac{\delta}{K} \cdot \frac{\log_{2} M}{M} \,.
\end{equation}
Thus $\spectrale \to 0$ as $M \to \infty$ as expected.

Another useful insight is that the spectral efficiency is proportional to $\delta$.
Thus, from a spectral efficiency standpoint it is desirable to run each carrier at the
highest $\rate$ possible; that is, $\delta = 1$.
If this $\rate$ is not large enough, multiple carriers (MCM) can be used.
On the other hand, if the transmit power available to support $\delta = 1$ is not available,
$\power$ can be reduced by choosing a smaller $\delta$ at the expense of a smaller $\rate$.
Because of the fixed $\powere$, $\power \propto \delta$.
Finally, $\spectrale \propto 1/K$, 
so increasing the degrees of freedom $K$ reduces $\spectrale$ .

\section{Choice of $h(t)$}

In the development of this chapter thus far, nothing has been said about $h(t)$
other than the assumption that it has finite bandwidth $B$ and time duration $T$ and unit energy.
It has not been necessary to say more, because from the perspective of
reliability (reflected in the bit error probability $\pe$) and power efficiency $\powere$
the choice of the waveform $h(t)$ does not matter.
Another useful (and perhaps surprising) insight is that the power efficiency $\powere$
is not dependent on the parameters $\{ B_{c} , \, T_{c} , \, K_{c} = B_{c} \, T_{c} \}$.
Whatever these parameters may be, the transmitter can work around them while
maintaining fixed power efficiency through a combination of duty factor $\delta$
and number of carriers, and achieve any desired information rate $\rate$.

There are however a few practical considerations that weigh in on the
choice of $h(t)$, all of them related to the detection of energy bundles
during both discovery and communication.
This challenge is considered in Chapter \ref{sec:detection}, 
where considerations in the choice of $h(t)$ will be discussed.
These include the receiver designer's ability to guess $h(t)$ in order to
implement a matched filter,
the design of multi-channel spectral analyzers including the matched filter,
and most notably immunity to radio-frequency interference (RFI).

\section{History and further reading}

Multicarrier modulation (MCM) is widely used in terrestrial wireless communications.
The first commercial application (The Collins Radio Kineplex) in 1957
 exploited multiple carriers to counter selective fading in military communications \citeref{623}.
Because terrestrial bandwidth is a scarce resource, terrestrial applications use a clever form of MCM in which the subcarrier signals have a 50\% overlap in frequency without interference (this idea was patented in 1970 \citeref{611} and gained traction in the 1990's \citeref{612}). A specific realization called orthogonal frequency-division modulation (OFDM) \citeref{613} appears in wireless communication standards such as ''wireless fidelity'' (WiFi), ''worldwide interoperability for microwave access'' (WiMAX), and the 4G digital cellular standard ''long term evolution'' (LTE). 
These standards typically combine MCM with spread spectrum \citereftwo{615}{621} for RFI mitigation.
These numerous applications give us some additional confidence in this as an approach for
interstellar communication, as well as provides relevant design experience.

\section{Summary}

The concrete design of a communication system using the five principles of
power efficient design listed in Chapter \ref{sec:powerefficient} is undertaken,
based on a finite-complexity version of the channel code of
Chapter \ref{sec:fundamental} as well as a simplified outage strategy.
Additive white Gaussian noise (AWGN) is countered by using matched filtering in
the receiver,
and two methods for dealing with scintillation are compared.
The simpler method declares outages and does not attempt to recover
information during periods of low received signal flux due to scintillation.
The more sophisticated method uses time diversity combining rather than outages to counter scintillation,
and both employ M-ary FSK to counter the AWGN with high power efficiency.
For interstellar communication, the outage approach appears to be the
more attractive approach because there will be an opportunity to recover
the lost information from multiple replicas of a message, and because the achievable effective
power efficiency is actually a bit higher.

At the fundamental limit the information rate is proportional to the average received power,
and this concrete design attains the same tradeoff albeit with a penalty in the achieved
power efficiency relative to the fundamental limit.
The size of that penalty is determined.
A significant feature of the design is the way in which different rates can be achieved
by a combination of the rate $F$ at which codewords are transmitted and
the choice of parameter $M$.
After a 100\% duty factor is reached ($\delta = 1$), further increases in
rate can be achieved by increasing $M$.
An alternative (which achieves lower power efficiency) is to transmit two or more
carriers, which is called multicarrier modulation (MCM).


\chapter{Discovery}
\label{sec:discovery}

Interstellar communication cannot proceed unless and until
a receiver discovers the presence of an information-bearing signal.
Discovery is the grand challenge of interstellar communication because
of the lack of explicit transmitter/receiver coordination and the ''needle
in a haystack'' problem of searching for a viable signal over a large
range of spacial, temporal, and frequency parameters.
Even in the absence of reliable recovery of embedded information,
discovery of a signal
established as having a technological (as opposed to natural) origin
is valuable to the receiver,
because it reveals the existence and location of extraterrestrial life and a present or past
intelligent civilization possessing technological capability.
In this chapter the challenge of discovery of information-bearing signals designed for
power efficiency is analyzed, along with beacons designed strictly for discovery
without any information-bearing capability.
We start with some background discussion and a basic set of assumptions
that will form the basis of this analysis.

\section{Background}

Before attempting a serious attempt at discovery,
some basic questions must be cleared up.
What is the appropriate channel model for discovery?
What kind of signals are we trying to discover?
What distinguishes a ''better'' and ''worse'' design of a signal
from the perspective of discovery?
What are the tradeoffs that must be resolved in discovery?

\subsection{Channel model for discovery}

An interstellar channel model was developed in Chapter \ref{sec:channelmodel}.
This is the appropriate channel model for discovery as well as communication.

\subsubsection{Coherence hole}

In Chapter \ref{sec:fundamental} 
it was established that structuring the transmit signal around the
interstellar coherence hole (ICH) is necessary to 
approach fundamental limits on power efficiency.
The ICH is even more important to discovery.
There are several reasons for this.
Eliminating most ISM and motion impairments as
a consideration removes whole dimensions of search during discovery.
For example, the computationally intensive de-dispersion algorithms
used in pulsar astronomy \citeref{212} and Astropulse \citeref{631} are avoided, because unlike pulsars or transient phenomena of natural origin,
interstellar communication signals should be rendered
impervious to dispersion by design.
Eliminating whole dimensions of search in turn reduces the false alarm
rate, and thus allows discovery to operate at higher sensitivity.
Finally, the ICH size can be estimated based on the observation of physical quantities,
and thus it serves as a source of implicit coordination between transmitter
and receiver that constrains parameters of search, 
principally signal bandwidth and time duration.
Thus, the detection algorithms considered in this chapter assume
that the transmit waveform $h(t)$ has been chosen to render
impairments from the ISM and motion insignificant, save noise and scintillation.

\subsubsection{Scintillation}

During communication, the outage strategy of Chapter \ref{sec:infosignals}
assumed that the variation in received signal flux due to scintillation could be tracked,
in which case outage periods could be declared.
This is not a viable strategy for discovery, since there is no prior knowledge of a signal to
observe and to track.
The implication of scintillation for discovery is that there are periods during which the
received signal flux is below the average, and other periods when it is
above the average.
The latter periods are actually quite helpful for discovery,
since a receiver willing to make multiple observations will discover signals that would not be
visible in the absence of scintillation.

For purposes of discovery, neglecting the unknown carrier phase a model for a single energy bundle
after matched filtering and sampling is
from \eqref{eq:ICHmodel},
\begin{equation}
Z =
\begin{cases}
 \sqrt{\energy_{h}} \, S + N \,, & \text{signal energy present} \\
 N  \,, & \text{noise only} \,.
 \end{cases}
\end{equation} 
The statistics of $Z$ are not affected by an unknown phase.
The matched filter output is zero-mean Gaussian under both conditions, the only
difference being a larger variance an energy bundle is present,
\begin{equation}
\text{Var} [ Z ] = \expec{|Z|^{2}} =
\begin{cases}
 \energy_{h} \, \vs + N_0 \,, & \text{energy bundle present} \\
 N_0  \,, & \text{noise only}
 \end{cases}
 \,.
\end{equation}
The energy contrast ratio $\ecrs$ of \eqref{eq:ecrs}
is a measure of the fractional increase in variance due to
a scintillated energy bundle.

Even when an energy bundle is present,
in the presence of scintillation a single observation cannot achieve reliable detection,
since that observation stands a good change of occurring during
a period of below-average received flux
(when $\expec{|S|^2} = \flux$ is small).
Thus, to take advantage of the periods of higher-than-average flux,
it is necessary to make multiple observations in the same
location.
By ''location'', we mean the same line of sight and carrier frequency.
In the following, it is assumed that the receiver makes $L$ independent
observations.
By ''independent'', we mean that the observations are separated by a
time interval large relative to the scintillation coherence time,
so that the observations can be considered to be statistically independent.
This has been pointed out previously and it was found that
$L \approx 3$ suffices for high-power signals \citereftwo{122}{121}.
Here we show that a much larger $L$ is needed to
reliably discover power-efficient signals of either the information-bearing or
beacon's that conserve energy at the transmitter.
This is due to the sparsity of energy bundles at low signal powers,
as well as scintillation.

\subsection{Signal structure}

Past and present SETI observation programs have
emphasized discovery of a beacon, rather than recovery of information.
What is a beacon?
Here we define a beacon as a signal that is intended to be
discovered, but not empowered to convey information.
This chapter analyzes the discovery of both categories of signal, 
in particular looking for ways to design both an
information-bearing signal and a beacon that makes them ''easier''
to discover in terms of either complexity or reliability.

\subsubsection{Energy bundles}

This report in its totality addressed power-efficient and power-optimized design.
In this context, it can be assumed that beacons are structured around the
repeated transmission of energy bundles, as has been assumed for
information-bearing signals designed for power-efficiency.
 The energy bundle is a relatively short time-duration signal that bears some amount of energy.
 It is based on a waveform $h(t)$ that has a bandwidth $B$ and time duration $T$
 that falls in an interstellar coherence hole (ICH) as described further in Chapter \ref{sec:incoherence}.
 
 What is the justification for this assumption?
 In the case of information-bearing signals structured around energy bundles
 it is necessary for those energy bundles to fall in an ICH to approach the
 fundamental limit on power efficiency as described in Chapter \ref{sec:fundamental}).
In the case of a beacon, energy bundles
confined to an interstellar coherence hole (ICH) are desirable because they
can be detected without any processing related to impairments in the interstellar channel
other than noise.
This is significant since discovery is inherently processing-hungry due to the
''needle in a haystack'' challenge.
Utiliizing the ICH also improves the discovery reliability because the processing that is avoided
would in general enhance the noise and worsen the noise immunity and increase the false alarm probability.
There is no disadvantage to structuring a beacon around ICH energy bundles.
Finally,
basing both categories of signal on energy bundles allows the same front-end processing
(matched filtering combined with multi-channel spectral analysis)
to be employed in simultaneously seeking both categories of signal.

Here $h(t)$ is allowed to be any waveform that conforms to the
ICH constraints for a given line of sight and carrier frequency.
Generally this class of waveform includes sinusoidal signals.
In practical terms this implies an enhanced multi-channel spectral analysis
that also incorporates filtering matched to non-sinusoidal signals,
although certainly as a starting point it is reasonable to begin with sinusoidal signals.
A search for energy bundles based on sinusoids would employ the same
front end processing as is currently used following the
Cyclops prescription \citeref{142}.
Narrow pulses as are the target for the Astropulse search \citeref{631}
have a wide bandwidth that will be impaired by the frequency incoherence
of the channel, would require intensive de-dispersion processing,
and are not an appropriate signal design for energy efficiency.

\subsection{Stages of discovery}

Discovery of a signal structured around energy bundles becomes a two-stage process:
\begin{description}
\item[Energy detection.]
The first stage is the detection of a single energy bundle.
In making a large number of observations in different locations
(with spatial, temporal, and frequency variables), a single positive detection then triggers the pattern recognition stage.
With power-efficient and low-power signals a much larger number of observations with the same
spatial and frequency variables is necessary to have a reasonable chance to
successfully detect an energy bundle.
This is because (a) energy bundles are sparse and hence most observations
will not coincide with an energy bundle
and (b) energy bundles are less likely to be detected during periods of
low received signal flux due to scintillation.
\item[Pattern recognition.]
The receiver develops more convincing statistical evidence of a discovery
by detection of more than a single energy bundle 
along the same line of sight and at ''nearby'' carrier frequencies.
Contrary to a Cyclops beacon, short-term persistence of the signal is not expected.
That is,
an energy bundle is unlikely to be followed immediately by another energy bundle,
so the pattern recognition has to take account longer periods of time
and different (although relatively nearby) frequencies.
The pattern recognition algorithms are not looking for any prescribed deterministic pattern,
since in fact the location of energy bundles is likely to vary randomly in
accordance with embedded information.
\end{description}
In the energy-detection stage we are seeking a
\emph{signal of interest} (SoI).
This SoI is distinguished by its non-natural origin,
and could be either information-bearing or
a beacon.
At the energy-detection stage these two cases
will not be distinguishable, but during the pattern
recognition phase the relevant characteristics of the
SoI will be estimated.
At this stage an information-bearing signal can generally be statistically distinguished
from a beacon by its greater randomness in the location of energy bundles.

\subsection{Discovery reliability}
\label{sec:discoveryrel}

In evaluating any strategy for discovery, we
are primarily concerned with two probabilities,
\newcommand{\pa}{P_{\text{A}}}
\begin{align*}
&\pd = \prob{ \text{''SoI detected''} \left. \right| \text{''SoI is present''}} \\
&\pa= \prob{ \text{''SoI is present''} \left. \right| \text{''SoI detected''}} \,.
\end{align*}
The miss probability
$(1-\pd)$ tells us how frequently,
given that an SoI is present, the strategy fails to flag that signal location
as worthy of further examination.
Of course we desire to have $\pd \approx 1$.
The accuracy probability $\pa$ recognizes that sometimes an SoI detection
is in error, by which we mean the discovery strategy
flags that a SoI is present when in fact the observations consist of noise alone.
Of course we desire that $\pa \approx 1$ as well, because then we
can announce to the world with high confidence that an artificial signal
of technological origin has been established.

In order to determine $\pa$, it is necessary to know two other
probabilities,
\begin{align*}
&\pfa = \prob{ \text{''SoI detected''} \left. \right| \text{''noise alone''} }\\
&\pb = \prob{ \text{''SoI is present''} }
\end{align*}
The false alarm probability $\pfa$ tells us how often the discovery strategy
incorrectly flags an SoI when the observations consist of noise alone.
Of course we desire $\pfa \approx 0$.
The a priori probability that an SoI is present is $\pb$, and this
 is not a characteristic of the discovery strategy but rather an assumption
about the frequency with which an SoI can be expected will be present
in our observations.
There is no completely convincing way to justify
a specific value for $\pb$, but surely it must be very small.
Based on these probabilities, $\pa$ can be calculated as
\begin{equation}
\pa = 
\frac{\pd \cdot \pb}{\pd \cdot \pb + \pfa \cdot (1 - \pb )}
\approx \frac{1}{\displaystyle 1+\frac{\pfa}{\pb}}
\end{equation}
When $\pb \ll \pfa$,
most detections of an SoI are actually false alarms,
and thus most of the time a detected SoI is actually noise alone and $\pa \approx 0$.
The other case is $\pfa \ll \pb$, when
$\pa \approx 1$ and a detection is much more meaningful because
most of the time it flags that an SoI that is actually present.
The important conclusion is that
to avoid wasting a lot of effort further analyzing a signal location
(or to have the confidence to announce the discovery in a public press release)
$\pfa$ must be mighty small, and in particular small relative to $\pb$.

The two most direct metrics of the reliability of any particular discovery strategy and algorithm are
$\{ \pfa , \, \pd \}$.
The other probability of interest, $\pb$, is mostly a matter of judgement and not
dependent on our discovery design.
The analysis of $\{ \pfa , \, \pd \}$ can be applied at any stage of the discovery process. 
In the following analysis, this analysis will first be applied to an 
energy bundle detection stage consisting of $L$ independent observations.
At this preliminary stage,
$\pfa$ is defined as the probability that one or more energy bundles are
detected in the $L$ observations in the presence of noise alone.
$\pd$ is defined as the probability that one or more energy bundles are detected
 when in fact a SoI overlaps the $L$ observations.
In numerical examples the values $\pfa = 10^{-12}$ and $10^{-8}$ will be chosen
to get some idea of the sensitivity of the resulting $\{ \pd , \, L \}$ to $\pfa$.
These values of $\pfa$ are not nearly small enough to give us confidence in the discovery
of an SoI, but they may be reasonable objectives for the
energy bundle detection stage because they simply capture 
how often the pattern recognition stage is triggered.
Dramatically further reducing $\pfa$ is the function of the subsequent pattern recognition stage, which
observes the location and nearby locations for each discovered energy bundle over a longer period of time,
looking for additional energy bundles and eventually identifying a recognizable pattern.
Eventually, after a much longer period of observation
we can hopefully develop sufficient confidence that our SoI is actually an information
bearing signal or beacon.
This confidence requires a small enough $\pfa$ relative to $\pb$ to assert that $\pa \approx 1$.

\subsection{Power optimization for discovery}

In this report the focus is on design of a transmitted
signal with the goal of minimizing the average received power
that is necessary to discover that signal.
However, the ''minimize power'' criterion of design has
to be defined more carefully to distinguish discovery from communication.
During communication the power efficiency criterion was adopted in Chapter \ref{sec:fundamental}.
but this criterion has no relevance to discovery,
so a different metric called \emph{power optimization} is needed.

Unless the received power is very large and the duty factor $\delta = 1$, for the energy-detection stage
to achieve simultaneously $\pfa \approx 0$ and $\pd \approx 1$
requires that the number of observations be large, or $L \gg 1$.
An appropriate figure of merit for a signal design with respect to its
discoverability is the value of $L$ required to achieve a fixed $\pfa$ and $\pd$.
A signal design is said to be ''power-optmized for discovery''
when, for target values of $\pfa$ and $\pd$ and a given average power $\power$,
the required $L$ is minimized.
This criterion attempts to satisfy both transmitter and receiver
by jointly minimizing the discovery resources for each, the transmit power
and energy consumption required of the transmitter
and the processing resources and energy consumption required of the receiver,
subject to a constraint on the discovery reliability.

\subsection{Role of duty factor}

In designing an information-bearing signal that achieves a high power efficiency in Chapter \ref{sec:infosignals},
it was found that the reliability of information extraction was dependent on the energy per bundle $\energy_h$
and energy per bit $\energybit$.
Thus, in keeping reliability fixed, these quantities are held fixed.
The way to reduce the information rate $\rate$ and hence average received power $\power$
is to reduce the rate at which codewords are transmitted.
Within the constraint of energy bundle duration $T$ less than coherence time $T_c$,
$\power$ can be reduced by increasing $T$ while keeping $\energy_h$ fixed.
Larger reductions in $\power$ can be achieved by reducing the duty factor $\delta$,
so that the transmission is intermittent.
This raises two questions.
First, is this the best way to reduce the $\power$ while also doing a power optimization for discovery?
Second, a beacon can be thought of as an information-bearing signal in the limit of $\rate \to 0$.
Does this imply that the design of a beacon should follow the same prescription,
keeping the energy per bundle $\energy_h$ fixed while reducing the duty factor?
It will turn out that the answer to both questions is ''yes''.

This analogy to information-bearing signals suggests a \emph{periodic beacon}.
As illustrated in Figure \ref{fig:beaconDutyCycle}, this beacon design
consists of a periodic sequence of energy bundles.
Although it bears resemblance to an information-bearing signal,
it conveys no information since the location of the energy bundles is deterministic.
This lack of information content is manifested by all the energy bundles
being transmitted at the same carrier frequency.
There is not the same motivation for a large total bandwidth $\Btotal$
for a beacon, since the power efficiency objective is not relevant.

There are two possible methods of reducing the average power $\power$
for this beacon design.
One is to reduce the energy per bundle, and the other is to reduce the
rate at which energy bundles are transmitted.
The analogy to information-bearing signals suggests that the average power $\power$
should be reduced by increasing the period of the beacon, which can be
done by increasing $T$ (as long as the constraint $T \le T_c$ is met) 
or reducing the duty factor $\delta$.
The duty factor approach is illustrated in Figure \ref{fig:beaconDutyCycle}.

\incfig
	{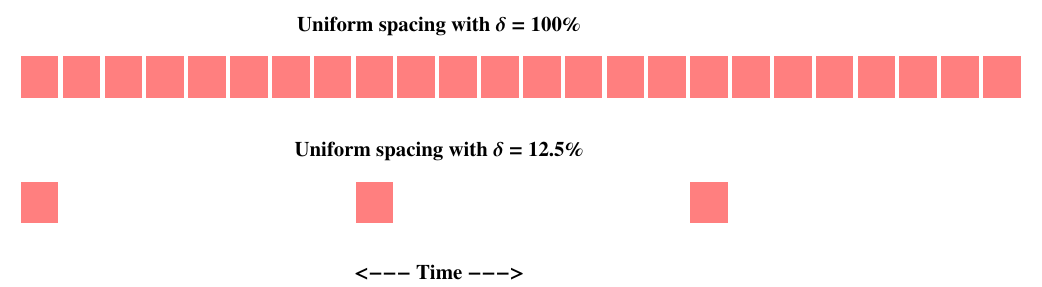}
	{.8}
	{
	An illustration of a periodic beacon with duty factor $\delta$ for $\delta = 1$ and $\delta = 1/8$.
	Assuming the energy of each bundle is the same in both cases, the average power $\power$
	of the lower duty factor case is reduced by a factor of $\delta$.
	The reliability of detection of each bundle that happens to overlap an observation
	is unchanged in the two cases, but a larger $L$ is required in the second case
	if the duration of one observation is less than one period of the beacon, since not all observations
	are guaranteed to completely overlap an energy bundle.
	It is expected that a larger $L$ will be required for a smaller $\power$, but the question is
	how much larger $L$ needs to be and whether the resulting tradeoff 
	between $\power$ and $L$ is the best that can be achieved.
	}

A continuously transmitted signal (equivalent to $\delta = 1$)
was prescribed in the Cyclops study \citeref{142} and
has been assumed in many current and past SETI searches.
This beacon design is a special case of the periodic beacon,
the special case with highest average power $\power$ and the minimum number
of observations $L$.
There is an opportunity to reduce the average received power at the expense of increasing $L$
by reducing the duty factor $\delta$.
This approach will reduce the average transmit power, and hence the transmitter's energy
consumption, by a factor of $\delta$.

Before concluding that reducing the duty factor is the best way to reduce power,
the second approach of reducing the energy of each bundle $\energy_{h}$ must be considered and rejected.
It will be established shortly that the energy per bundle does vary with $\power$
for an power-optimized periodic beacon, but the variation is small because of the high sensitivity
of detection probability for each energy bundle to the value of $\energy_{h}$.
Before proceeding to that argument, a subtle but important point should be highlighted.
The received value of $\energy_{h}$ depends on the
energy of the transmitted bundle,
the propagation loss, and the transmit and receive energy gains.
Thus, if a reduction in $\energy_{h}$ were feasible, that could be exploited in any of these
four places, and thus could directly benefit the receiver (through a smaller collection area in the
receive antenna) as well as the transmitter.
In contrast,
reducing $\power$ only through a reduction in duty factor $\delta$ can
only benefit the transmitter through lower average energy consumption.
This approach also does not reduce the peak power required at the transmitter, only the average power, and it cannot be exploited by using a lower-gain
transmit or receive antenna.

\section{Periodic beacons}

A periodic beacon consists of a periodic sequence of energy bundles,
with an average power $\power$ that is governed by the energy of each bundle $\energy_h$
and the rate at which bundles are transmitted $F$.
This simple structure is a good starting point for addressing 
the issues surrounding power-optimization for discovery.
These results are easily extended to power-efficient information-bearing signals,
which add randomness to the location of the energy bundles in either time or frequency.
Subsequently it will be argued that for a periodic beacon with any average power $\power$,
there is an information-bearing signal with the same $\power$
(and a commensurate information rate $\rate$) that can be essentially as easily discovered.

\subsection{Beacon structure and parameters}

The three parameters of a periodic beacon are illustrated in Figure \ref{fig:beaconParam}.
Each energy bundle is assumed to have duration $T$,
and bundles are spaced $T_{p}$ apart, implying a duty factor
\begin{equation}
\delta_{p} = \frac{T}{T_{p}} \,,
\end{equation}
where $0 < \delta_{p} < 1$.
If at the receiver each energy bundle harbors the same amount of energy $\energy_{h}$,
then the average received power is
\begin{equation}
\label{eq:avgbeaconpower}
\power = \frac{\energy_{h}}{T_{p}} \,.
\end{equation}
Note that if $\energy_{h}$ is held fixed, $\power$ decreases as the beacon period
$T_{p}$ is increased (the duty factor $\delta_{p}$ is decreased) or the energy per bundle $\energy_h$
is increased.

If the energy of each bundle is spread uniformly over the interval $T$,
then the peak power is
\begin{equation}
\label{eq:peakbeaconpower}
\powerpeak = \frac{\energy_{h}}{T} = \frac{\power}{\delta_{p}} \,.
\end{equation}
As expected the peak power $\powerpeak$ remains constant if average power $\power$
is reduced by reducing the duty factor $\delta_{p}$.
It is generally desirable in terms of the transmitter's capital costs to reduce $\powerpeak$,
and this can be accomplished by increasing $T$,
with the side effect of increasing $\delta_{p}$ without increasing $\power$.
However, this opportunity is limited by the coherence time of the channel to $T \le T_{c}$.

\incfig
	{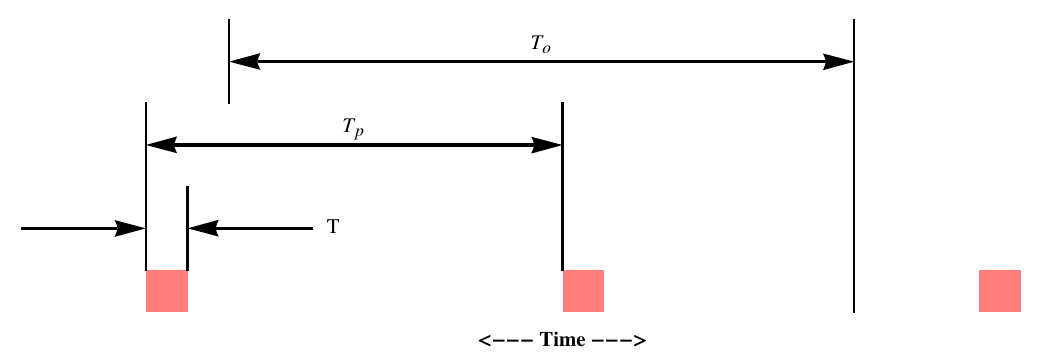}
	{1}
	{
	An illustration of the three parameters of a periodic beacon and its observation.
	The parameters determined at the transmitter are the duration of an energy bundle $T$ and
	the time interval between energy bundles $T_{p}$.
	The parameter determined at the receiver are duration of one observation $T_{o}$
	and the number of independent observations $L$ (not shown).
	}

\subsection{Characteristics of beacon observation}

Consider an observation of the received signal at a given carrier frequency
and starting time.
For purposes of studying the discovery probability, assume that
the carrier frequency is the correct one.
The parameters determined by the receiver designer are the
duration of a single observation $T_{o}$ and the number of independent
observations $L$,
with the total duration of the collective observations equal to $L \cdot T_{o}$.
Each observation is assumed to begin at a random point in the beacon period,
and the effect of scintillation on each observation is considered statistically independent of the others, 
meaning that the time interval between observations is large relative to the coherence time of the
channel as modeled in Chapter \ref{sec:incoherence}.

An individual energy bundle is said to overlap an observation if the
observation interval completely overlaps the time duration of the bundle,
so that an matched filter implemented to interpret the observation captures its full energy.
Then the average number of energy bundles overlapping a single observation is
\begin{equation}
\label{eq:avgbundles}
\frac{T_{o} - T}{T_{p}} = n_{o} + \delta_{o}
\end{equation}
where $0 \le n_{0}$ is an integer and $0 \le \delta_{o} <1$.
For example, there is always exactly one energy bundle in each 
observation ($n_{o} = 1$ and $\delta_{o} = 0$) when $T_{o} = T_{p} + T$.
When $T_{o} = 2 T$ the average number of energy bundles per observation is $\delta_{p}$
( $n_{o} = 0$ and $\delta_{o} = \delta_{p}$),
or equivalently the probability of an energy bundle overlapping one observation
is $\delta_{p}$.

It is generally useful to increase the observation time, because the
average number of bundles overlapping an observation also increases.
However, a point of diminishing returns occurs for $T_{o} > T_{p}+T$,
because for that case $1 \le n_{o}$ and there can be two or more energy bundles
overlapping a single observation.
Two energy bundles spaced so closely in time will experience
a correlated scintillation.
It is overall a more efficient use of cumulative observation time to keep 
the observations independent,
because that increases the likelihood of experiencing a large
received signal flux during periods overlapping one or more of the $L$ observations.
For our examples, therefore, we assume that $T_{o} \le T_{p}+T$,
in which case $n_{o} = 0$ and $0 < \delta_{o} \le 1$.

\subsection{Tradeoffs in detection reliability}

An analysis of the reliability of detection of a beacon is relegated to Appendix \ref{sec:Lobs}.
The result is expressed in a discovery relation that is fairly complicated and
can only be evaluated numerically.
Fortunately, some simple principles reveal much insight into to the tradeoffs in
power-optimized beacon design, and predict the results of the complete analysis
with a fair degree of accuracy.

\subsubsection{Opportunities for false alarms}

Consider the detection of a beacon using $L$ observations
with an overall probability of false alarm (one or more energy bundles
are falsely detected in noise alone) equal to $\pfa$ and overall
probability of detection (one or more energy bundles are detected in the
presence of a periodic beacon) equal to $\pd$.

In the absence of a beacon, the observations reflect noise alone.
Then $\pfa$ reflects the total number of degrees of
freedom $N$ in the observations, which equals the product of the
total time duration of the observations and the bandwidth of the observations,
\begin{equation}
\label{eq:discobsdof}
N = L \, T_{o} \cdot B = L \, K \, \left( \frac{\delta_{o}}{\delta_{p}} + 1 \right) \,.
\end{equation}
For example, the continuous-time noise signal at the output of an matched filter
has bandwidth no greater than $B$, and thus can be sampled at rate $B$
and the number of such samples $N$ is the total observation time $L \, T_{o}$ times $B$.
If the noise projected into the $N$ degrees of freedom are statistically
independent (the conditions for this are considered in Appendix \ref{sec:Lobs}),
then $\pfa$ is to good accuracy
equal to $N$ times the probability that a
false alarm occurs in a single degree of freedom.
Overall the detection is penalized for using a larger $L$ and $T_{0}$ by
requiring a smaller false alarm probability
$\pfa/N$ at the level of looking for an energy bundle in a single degree of freedom.

\subsubsection{Effect of duty factor alone}

Now consider the case where a beacon is actually present at the carrier frequency
and in the direction being observed.
There are two ways in which $L$ observations of duration $T_{0}$ can fail to detect
that signal.
Either no energy bundle overlaps \emph{any} of the $L$ observations,
or one or more bundles overlaps but they are all obscured by noise.
By ignoring the second possibility, a lower bound on the miss probability follows.
Ignoring the effects of noise, no bundles overlap one observation with probability $(1-\delta_{o})$
and thus no bundles overlap all $L$ observations with probability
\begin{equation}
\label{eq:Llowerbd}
1- \pd \ge \left(1-\delta_{o} \right)^{L} 
\ \ \ \ \text{or} \ \ \ \ 
L \ge \frac{
	\log \left( 
		\frac{1}{1 - \pd} 
		\right)
	}{
	\log \left( 
		\frac{1}{1-\delta_{o}} 
		\right)
	} \,.
\end{equation}
This relation reveals how large $L$ must be chosen to overcome
just the effect of $\delta_{o} < 1$.
Note that the larger $\pd$ we require, the larger $L$ must be.
Also, larger observation duty factor $\delta_{o}$ is helpful in reducing the
required number of observations $L$.

\subsubsection{Bundle energy hypothesis}

As noted earlier, the average beacon power $\power$ can be reduced by
a reduction in the energy $\energy_h$ of each bundle while
keeping the duty factor $\delta_p$ fixed, or by
decreasing $\delta_p$ while keeping $\energy_h$ fixed.
If the goal is to minimize the number of observations $L$, then the
 \emph{bundle energy hypothesis} asserts that reducing duty factor is preferable.
In other words, as the average power $\power$
is changed, the energy in each individual bundle $\energy_{h}$ should remain fixed.

\incfig
	{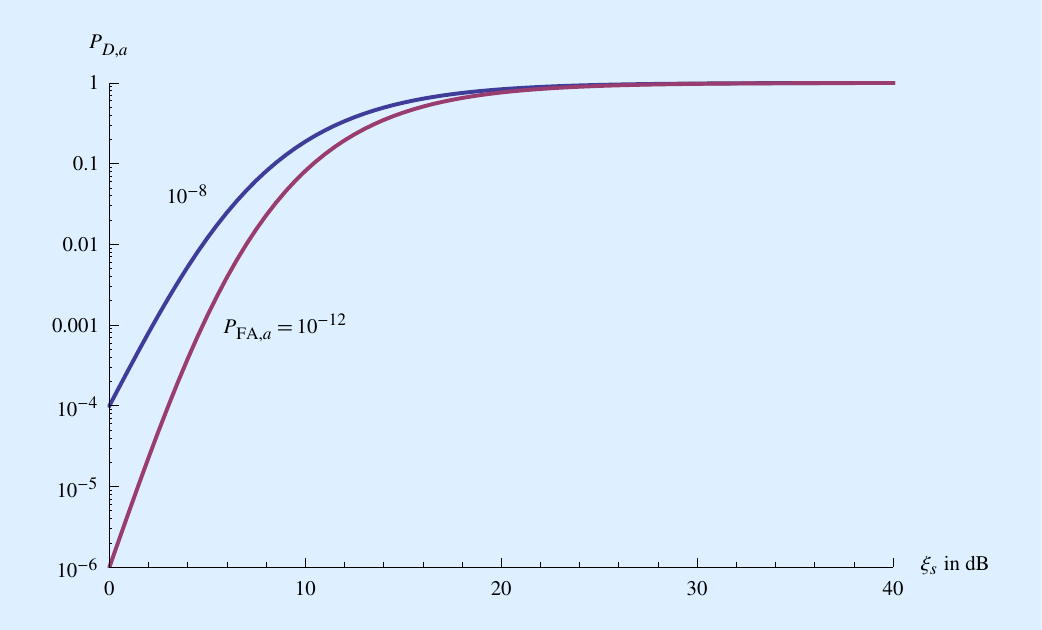}
	{.8}
	{An illustration of the reliability of detection of a single energy bundle
	in the presence of additive Gaussian noise and scintillation.
	The detection probability $\pda$ is plotted vs the
	energy contrast ratio $\ecrs$ in dB at the output of the matched filter.
	This is repeated for two different values of false alarm probability $\pfa$.}

The logic behind this hypothesis is that the probability of detection of an energy bundle that does
overlap an observation is a sensitive function of $\energy_{h}$ 
(or equivalently the energy contrast ratio $\ecrs$ of \eqref{eq:ecrs}).
In the detection of a single energy bundle, let $\pfaa$ and $\pda$ be respectively
be the false alarm and detection probabilities for the matched filter detection of a single
energy bundle in AWGN with power spectral density $N_{0}$ and with scintillation
with average scintillation flux $\vs$.
It is shown in Appendix \ref{sec:Lobs} that the relationship between these probabilities is
\begin{equation}
\pda = \pfaa^{\, \frac{ 1}{ 1+\ecrs}}
\end{equation}
and this relationship is plotted in Figure \ref{fig:detectOneBundle}.
For small $\ecrs$ the detection probability $\pd$ is small, so that
energy bundles are rarely detected.
There is a sharp threshold at about $\ecrs \approx 10$ to $15$ dB where
detection becomes fairly reliable.
For example, at $\ecrs = 15$ dB, an energy bundle is detected 57\% of the time
($\pd = 0.57$ at $\pfa = 10^{-12}$).
Thus, assuming that $\energy_h$ is already about as small as possible
to minimize beacon power,
allowing it to decrease further as $\power$ decreases would drastically reduce $\pda$, 
and thereby have an large impact in increasing the size of $L$.
The alternative of reducing $\delta_p$ also increases the required $L$
as in \eqref{eq:Llowerbd}, but in a less drastic way.
There is thus a tradeoff, but one that strongly favors reducing $\delta_{p}$ rather than $\energy_{h}$.

Maintaining fixed per-bundle energy can be accomplished by manipulating the
bundle duration $T$ and the duty factor $\delta_{p}$, since from \eqref{eq:avgbeaconpower}
\begin{equation}
\energy_{h} = \power \cdot T_{p} = \frac{\power \, T}{\delta_{p}} \,.
\end{equation}
Assuming $T$ is fixed, the bundle energy hypothesis predicts  that
$\delta_{p} \propto \power$, or the transmit duty factor should be reduced 
in proportion to the decrease in $\power$.
The bundle energy hypothesis also predicts that as $\power$ is decreased
reducing the duty factor $\delta_p$ can be avoided by increasing $T$.
However, the time incoherence of the channel limits this to $T \le T_c$, so once the
energy bundle duration is as large as possible
(from \eqref{eq:peakbeaconpower} this is also beneficial in reducing the peak power $\powerpeak$)
then further reduction in $\power$ requires reducing $\delta_{p}$ to maintain the same
detection reliability.

Note that reducing $\delta_{p}$ also increases the needed value of $L$ required to maintain
fixed reliability, either through increasing the length of an observation $T_{o}$
(increasing the false alarm probability)
or reducing the observation duty factor $\delta_{o}$
(reducing the probability that a bundle overlaps each observation).
The bundle energy hypothesis simply predicts that reducing the bundle energy $\energy_{h}$
would have a larger impact on $L$ and is to be avoided.

It is instructive to compare a periodic beacon and an M-ary FSK signal from the limited
perspective of the energy $\energy_{h}$ contained in each bundle, or equivalently the energy contrast ratio $\ecrs$.
For purposes of this comparison, assume that the outage strategy is employed
rather than time diversity.
The time diversity case will be considered separately in 
Section \ref{sec:discoveryWdiversity}.
For the beacon, $\ecrs$ is determined by the requirement on reliability of detection of the bundle
in the presence of noise and scintillation.
For the FSK signal, $\ecrs$ is determined by the requirement on probability of error
$\pe$ in the presence of noise and at a scintillation flux chosen to be at the threshold of outage. 
This comparison is germane to the question of how hard it is to discover an information-bearing signal
at an average power $\power$ sufficiently high to reliably recover information at the threshold of outage.
Rewriting the upper bound on $\pe$ of \eqref{eq:peFSKbound} in terms of $\ecrs$
rather than the energy contrast ratio per bit $\ecrb$,
\begin{equation}
\label{eq:peVsMandEcrs}
\pe \le \frac{M}{4} \, e^{- \ecrs /2} \,.
\end{equation}
\begin{changea}
This bound is plotted in Figure \ref{fig:peVsEcrso}, where we see that an FSK signal requires
 $\ecrs \approx 13$ to $18$ dB to achieve reliable information extraction, depending on the
value of $M$ and the desired $\pe$.
At smaller $M$
this is very similar to the $\ecrs \approx 13$ to $15$ that was required for reliable detection of a beacon.
At large $M$ the energy per bundle is overkill for reliable discovery, in order to overcome
the many more locations where energy bundles can be erroneously detected.
\end{changea}
The similarity of these contrast ratios is not surprising, since in both cases an energy
bundle is being detected.
In the case of the FSK signal at the threshold of outage, the scintillation is assumed to equal its average value,
whereas for the beacon the scintillation is random, sometimes attenuating and sometimes
amplifying the bundle energy.
This result anticipates the later argument
that a beacon offers no advantage over an information-bearing signal for ease of discovery.

\incfig
	{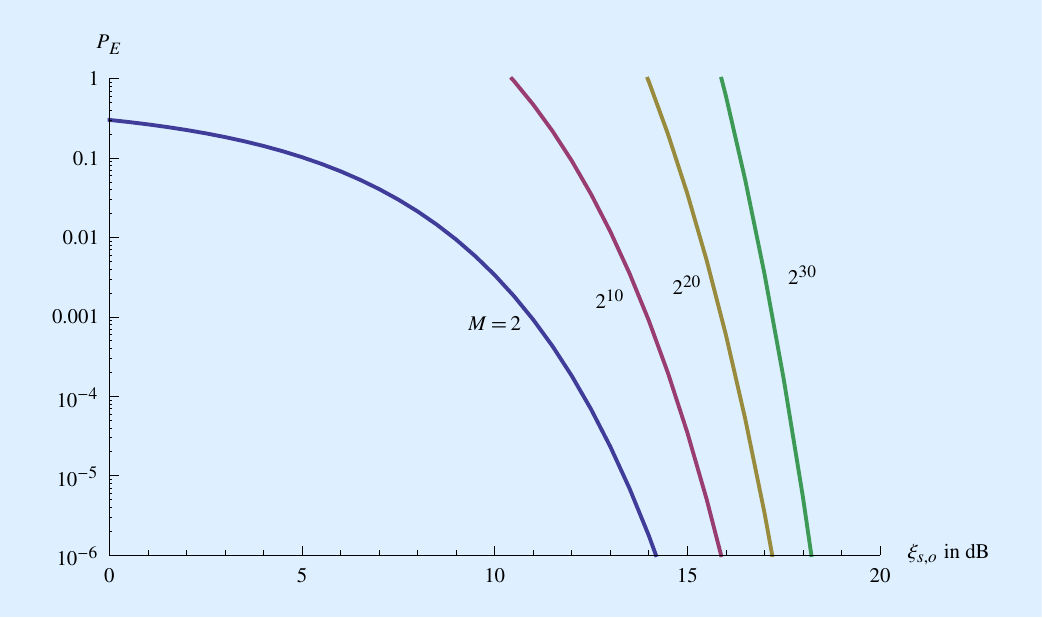}
	{1}
	{A log plot of an upper bound on the error probability for M-ary FSK 
	vs. the energy contrast ratio $\ecrs$ for an individual energy bundle making up that signal.
	At fixed $\ecrs$ the $\pe$ bound increases rapidly
	with $M$ because there are $(M-1)$ opportunities for a noise sample to be greater 
	than the energy bundle sample at the output of a bank of $M$ matched filters.
	$\pe$ is greater than unity for smaller $\ecrs$ because this is an upper bound that is
	accurate only for small $\pe$.}

\subsubsection{Total effective energy hypothesis}

The bundle energy hypothesis offers insight into the appropriate
choice of duty factor $\delta_p$ for a given beacon average power $\power$,
namely the $\delta_{p}$ required to achieve that power without reducing $\energy_{h}$.
On the other hand, it offers no insight into the values of $\{ L , \,T_{o} \}$ required to
achieve a certain overall level of reliability $\{ \pd , \, \pfa \}$ in detection of the beacon.
This is the role of the \emph{total effective energy hypothesis}.
The total effective energy $\energyeff$ is the total energy of all the bundles overlapping
the $L$ observations, each observation of duration $T_{o}$.
The total effective energy hypothesis states that
$\energyeff$ should be held approximately fixed as $\{ \power , \, L , \, T_{o} \}$ are varied.
The logic behind this hypothesis follows from an analogy to information-bearing signals, where
the reliability of information extraction
is determined by $\energybit$, the received energy per bit.
A beacon is designed to convey a single bit of information,
namely whether ''beacon is present'' or ''beacon is not present''.
Thus, it is a reasonable to assume that the total received energy
overlapping $L$ observations is germane to the reliability with which
that one bit of information embedded in the beacon can be extracted.

To apply this hypothesis,
the average number of bundles that are overlapping the collective observations is $L \cdot \delta_{o}$,
 the energy per bundle is $\energy_{h}$, and thus
\begin{equation}
\label{eq:totalenergy}
\energyeff = L \, \delta_{o} \, \energy_{h}
 \,.
\end{equation}
Note that $\energyeff$ is merely an average, since the actual
number of overlapping bundles is random.
It is also useful to define an energy contrast ratio $\ecreff$
expressed in terms of $\energyeff$,
\begin{equation}
\ecreff = \frac{\energyeff \, \vs}{N_{0}} = L \, \delta_{o} \, \ecrs \,.
\end{equation}
The analogy to information-bearing signals suggests that it is $\ecreff$
rather than $\energyeff$ alone that determines the required value of $L$,
or rather what matters is the ratio of total effective energy to noise power spectral density.

If $\energy_{h}$ (and hence $\ecrs$) is held fixed in accordance with the bundle energy hypothesis,
then a couple of notable observations can be made.
First,
$L \propto 1 / \delta_{o}$, or there is a direct tradeoff between
observation time $T_{o}$ and the number of observations $L$.
In particular the relation
\begin{equation}
\delta_{o} = \left( \frac{T_{o}}{T} - 1 \right) \cdot \delta_{p} \,.
\end{equation}
illustrates the possible tradeoffs as the average beacon power $\power$ is reduced,
and this is accompanied by a reduction in transmit duty factor $\delta_{p}$ in order to keep $\energy_{h}$ fixed.
One option is to keep the observation
time $T_{0}$ fixed, in which case $\delta_{o} \propto \delta_{p}$, and thus $L \propto 1/\delta_{p}$.
Another option is to keep $\delta_{o}$ (and hence $L$) fixed by some
combination of reducing $T$ and increasing $T_{0}$.
However the assumption that $T_{o} \le T_{p} + T$ puts a cap on this opportunity,
and ultimately it will be necessary to start reducing $\delta_{o}$ (and thus increasing $L$)
to achieve a further reduction in $\power$.

The second observation is that
$\ecreff$ can be made as large as desired simply by increasing $L$,
suggesting (correctly as it turns out) that there is no lower limit to the average power $\power$
of a beacon that can be reliably detected if we are willing to make
a sufficiently large number of observations.
This is due to the presumed long-term presence of the beacon, so that
a larger $L$ is always able to capture a larger total effective energy.
It is also due to the assumption that the energy per bundle should not be
reduced as the $\power$ is reduced, so that observations overlapping
an energy bundle are just as likely to detect it.

\subsubsection{How to reduce average beacon power}

The two hypotheses guide the reduction in beacon average
power without affecting detection reliability, namely
$\delta_p \propto \power$ and (oversimplifying a bit) $L \propto 1/\power$.
Achieving very low $\power$ requires action on the part of
both the transmitter and receiver designers.
The transmitter has to reduce the duty factor, and the
receiver has to increase the number of observations it makes
at each frequency and line of sight.
Fortunately these actions need not be coordinated.
The transmitter can reduce its average transmit power
as long as it also reduces $\delta_p$ in proportion,
keeping $\energy_h$ fixed.
This preserves the receivers chance at detecting the
beacon, although only those receivers that choose a
sufficiently large $L$ or sufficiently large observation time $T_o$
to compensate the smaller  $\delta_p$ will reliably detect the beacon.
From the receiver perspective, the antenna collection area
must be large enough to preserve an acceptable value for $\energy_{h}$,
and $T_o$ (up to a limit) or $L$ have to be sufficiently large
to allow beacons with a smaller $\power$ to be detected.

\subsection{Detection reliability for a periodic beacon}

The intuitive logic of the last section helps with understanding the tradeoffs,
but fails to take account of the quantitative tradeoff between the
number of energy bundles that overlap observations and
the relationship between $\energy_h$ and detection reliability.
Thus, these hypotheses give no hint as to the actual value of $L$ that may be necessary,
and also yields only qualitative and approximate results.
A complete analysis is given in Appendix \ref{sec:Lobs}.
The detection reliability is characterized by $\{ \pfa , \, \pd \}$,
where $\pfa$ is the probability of one or more detected energy bundles
out of $L$ observations in the presence of noise alone,
and $\pd$ is the same probability in the presence of a beacon.
It is then shown that the relationship between $\pfa$ and $\pd$ is
\begin{equation}
\label{eq:discoveryequation}
\pd [\, \pfa, \, \ecrs, \,  \delta_{o}, \, L \, ]
\approx 1 - \left[ 1-\delta_{o}  \, \left( \frac{\pfa}{N} \right)^{ \, \frac{1}{1+ \ecrs}} \right]^{\, L}
 \,,
\end{equation}
where $N$ is the degrees of freedom in all observations given by \eqref{eq:discobsdof}.
$\pd$ is monotonically increasing with $L$, and thus
the receiver always improves reliability by implementing a larger $L$.
For any $\pfa > 0$, $\ecrs > 0$, $0 < \delta_{p} \le 1$, and $K \ge 1$
\begin{equation}
\pd \underset{L \to \infty}{\longrightarrow} 1 \,.
\end{equation}
Thus, as suggested by the total effective energy hypothesis, $\pd \approx 1$ 
can always be achieved if $L$ is sufficiently large.
Another consequence of this principle is that for any reliability $\{ \pfa , \, \pd \}$
we can safely ask what $L$ is required to achieve that reliability without worrying
about the feasibility of achieving that reliability.

This property is a happy outcome of scintillation, since periods
of large flux reliably coincide with an energy bundle in at least one observation if $L$ is sufficiently large.
In this sense, scintillation is the savior of discovery, since
it permits discovery under the most adverse conditions.
However, in practical terms the size of $L$ could easily be too large.
For example, if $L$ is so large that it requires observations over a period of time large relative to the
existence (or for that matter patience) of a civilization, 
in practical terms the beacon is non-discoverable.
The transmitter designer must be aware of this and insure that the received signal flux
at the destination is sufficient for discovery with parameters
(antenna collection area, observation time, and number of observations) that it considers practical.

\subsection{Reducing average power by manipulating the energy per bundle}

As a reference point for considering variations in the parameters $\delta_{p}$ and $\delta_{o}$,
consider the \emph{Cyclops beacon} of the type prescribed by the Cyclops report \citeref{142}.
The primary characteristic that makes this beacon inefficient at low powers is its 100\% duty factor at all power levels.
The average power $\power$ of a Cyclops beacon can be reduced by increasing $T$ as the bundle energy $\energy_h$
is held fixed.
However, when $T$ reaches the limit of time coherence $T \le T_c$, the only way to reduce $\power$
further is to reduce $\energy_h$.
This runs counter to the bundle energy hypothesis, and as we will see results in a dramatic
inflation in the number of observations $L$.

The parameters for a Cyclops beacon with average power $\power$ are
\begin{align}
&\delta_{p} = 1 \\
&\power = \frac{\energy_{h}}{T} 
\ \ \ \ \text{or} \ \ \  
\ecrs =  \frac{\power \, T}{N_{0}}
\\
&T_{o} = 2 \, T 
\ \ \ \ \text{or} \ \ \
\delta_{o} = 1 \\
& N = 2 \, L \, K
\,.
\end{align}
The required $L$ as a function of $\ecrs \propto \power_{s}$ is illustrated in
Figure \ref{fig:noObsStandardBeacon},
where we see a very large $L$ is required to achieve low powers.
Recall from Figure \ref{fig:detectOneBundle} that the turning point where the detection probability for
a single energy bundle approaches unity is $\ecrs \approx 12$ to $15$, and the same
effect can be seen in Figure \ref{fig:noObsStandardBeacon}.
At $\ecrs = 15$ dB the required number of observations is $L = 19$ for $K = 1$
($L = 33$ for $K = 10^{6}$).
At lower powers than this $L$ increases very dramatically and the degrees of freedom $K$ starts
to have a larger effect.
At larger average powers, on the other hand, $L$ is reduced, reaching $L = 3$ at $\ecrs = 30$ dB.
Thus, even for the inefficient Cyclops beacon design, an increase in the number of observations
from $L = 3$ to $L = 19$ allows a reduction in the average beacon power by 15 dB, or a linear factor of 31.6.
Larger reductions in power this way become very costly in terms of $L$ because of the rapidly increasing
miss probability in the detection of individual energy bundles shown in Figure \ref{fig:detectOneBundle}.

\incfig
	{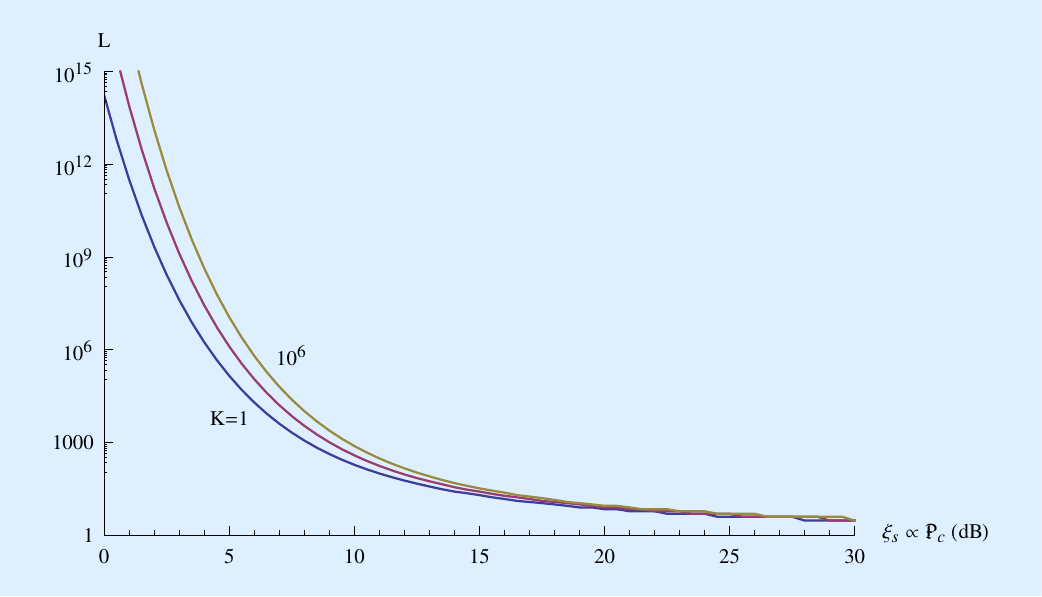}
	{.8}
	{For a Cyclops beacon with fixed 100\% duty factor ($\delta_{p} = 1$),
	a log plot of the required number of observations $L$ vs the energy contrast ratio
	$\ecrs \propto \power_{s}$ in dB for each energy bundle.
	The assumed detection reliability is $\pfa = 10^{-12}$ and $\pd = 10^{-4}$.
	Three values of degrees of freedom are plotted, $K=1$, $10^{3}$, and $10^{6}$.}

\subsection{Reducing average power by manipulating the duty factor}

The average power $\power$ can be reduced by reducing the energy contrast ratio $\ecrs$ while
keeping the transmit duty factor $\delta_{p}$ fixed (as just seen)
or by reducing $\delta_{p}$ while keeping $\ecrs$ fixed.
The latter approach follows from the bundle energy hypothesis, so our intuition suggests
that this is a more efficient way to reduce power.
There are two diametrically opposed strategies to detecting such a beacon at the receiver.
The first is to use the minimum observation interval $T_{o} = 2 \, T$,
which will maximize the number of observations $L$ required to achieve the desired detection reliability.
The parameters for a beacon with average power $\power$ are
\begin{align}
&\power = \frac{\energy_{h}}{T_{p}} 
\ \ \ \ \text{or} \ \ \  
\delta_{p} =  \frac{\power \, T}{\ecrs \, N_{0}} \\
&\delta_{o} = \delta_{p} \\
&N = 2 \, L \, K
\,.
\end{align}
The transmit duty factor $\delta_{p}$ is proportional to the average power $\power$.
The required $L$ for this approach as a function of $\delta_{p} \propto \power$ is illustrated in
Figure \ref{fig:noObsReduceDutyCycle}.
The reduction in $L$ is dramatic when compared with Figure \ref{fig:noObsStandardBeacon}.
Consistent with Figure \ref{fig:noObsStandardBeacon}, the reduction in $L$ is also dramatic
as $\ecrs$ decreases, at least up to $\ecrs \approx 15$ dB, after which comes a point of diminishing
returns where $L$ is close to its duty factor lower bound of \eqref{eq:Llowerbd}.
For example, at $\energy_{s}$

\incfig
	{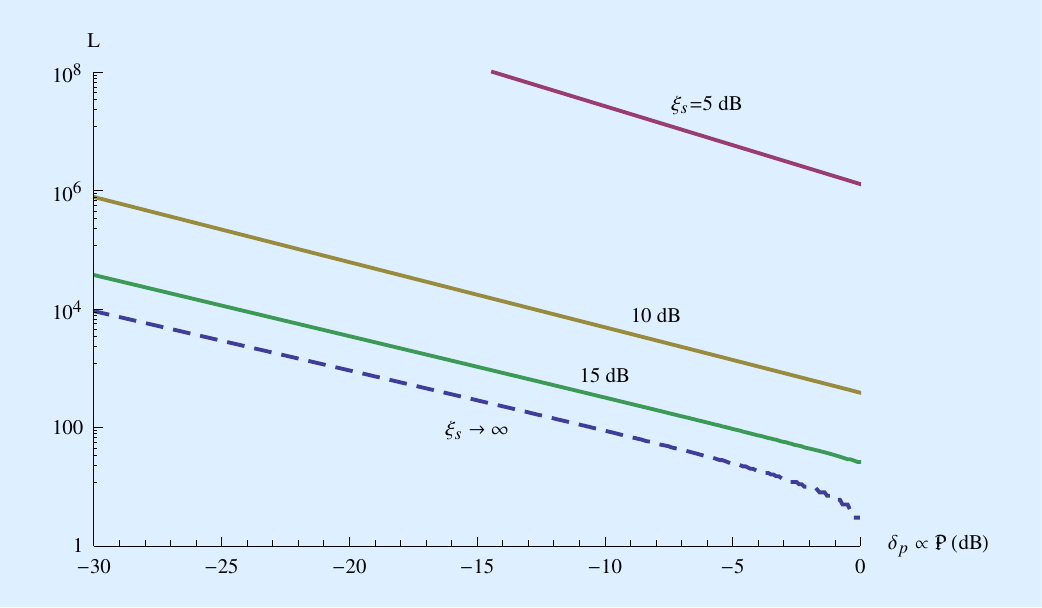}
	{.8}
	{To illustrate the reduction in average power $\power$ by reducing the
	transmit duty factor $\delta_{p}$,
	a log plot of the required number of observations $L$ vs the $\delta_{p}$ in dB
	is shown.
	The assumed detection reliability is $\pfa = 10^{-12}$ and $\pd = 10^{-4}$.
	A single values for the degrees of freedom $K=10^{3}$ is shown.
	The different curves show different values for the energy contrast ratio $\ecrs$,
	where the power $\power$ is also proportional to $\ecrs$.
	The dashed line is the lower bound on $L$ imposed by $\delta_{p}$ alone
	(ignoring the effects of noise) given by \eqref{eq:Llowerbd}.}

As an example of the dramatic reduction in $L$ from manipulating $\delta_{p}$ rather than $\energy_{s}$,
take as a baseline case a Cyclops beacon with $\ecrs = 30$ dB.
This beacon may actually be detected in current searches, since it requires only $L = 3$.
Now suppose the average power is reduced by 30 dB, or a linear factor of 1000.
Achieving that by reducing the bundle energy while retaining 100\% duty factor requires $L \approx 10^{16}$
from Figure \ref{fig:noObsStandardBeacon}.
This Cyclops design approach is clearly impractical.
However, achieving this by reducing $\delta_{p}$ alone
requires $\ecrs = 30$ dB and $\delta_{p} = - 30$ dB (a duty factor of 0.1\%).
This large an $\ecrs$ achieves close to the lower bound in Figure \ref{fig:noObsReduceDutyCycle}
(the dashed line) and thus requires $L \approx 10^{4}$.
This is still inefficient because the value $\ecrs = 30$ dB is considerably past the point of diminishing returns.
A more efficient approach would be to use a combination of a 15 dB reduction in $\ecrs$
(or equivalently a 15 dB reduction in energy per bundle $\energy_{h}$) and a 15 dB reduction in $\delta_{p}$.
That approach requires $L \approx 10^{3}$, or a further order of magnitude reduction relative to using
duty factor alone.
The total reduction in the number of observations required by the receiver is always a factor of about $10^{13}$.
The 15 dB reduction in duty factor (to 3.2\%) can only be implemented in the transmitter,
and the transmitter realizes as a result a 15 dB reduction in average transmit power 
(but no reduction in peak transmit power).
As far as the 15 dB reduction in energy per bundle, this could be implemented
at either the transmitter (by reducing the peak and average power by 15 dB)
or alternatively at the receiver (by reducing the receive antenna collection area by 15 dB or a linear factor of 31.6.

The diametrically opposite strategy at the receiver is to 
use the maximum observation interval $T_{o} = T_{p} + T$ ($\delta_{o} = 1$),
which will minimize the number of observations $L$ required to achieve the desired detection reliability
at the expense of greater observation time.
The parameters for a beacon with average power $\power$ are
\begin{align}
&\power = \frac{\energy_{h}}{T_{p}} 
\ \ \ \ \text{or} \ \ \  
\delta_{p} =  \frac{\power \, T}{\ecrs \, N_{0}} \\
&\delta_{o} = 1 \\
&N = L \, K \, \left( \frac{1}{\delta_{p}} + 1 \right)
\,.
\end{align}
Note that $\power$ is proportional to $\delta_{p}$.
The required $L$ for this observational strategy as a function of $\delta_{p} \propto \power$ is illustrated in
Figure \ref{fig:noObsLongObsTime}.
As expected, choosing a long duration for each observation eliminates the duty factor effect in $L$
and results in an $L$ that is both much smaller ($L = 28$ for $\ecrs=15$ dB and $\delta_{p} = -15$ dB)
and does not increase much as $\power$ is reduced as long as $\energy_{s}$ is large
enough to reliably detect individual energy bundles.
When $\ecrs$ is too small, the result is a much larger $L$ for all
duty factors and an $L$ that increases substantially as $\power$ is reduced reflecting the poor
reliability in detecting individual energy bundles.

\incfig
	{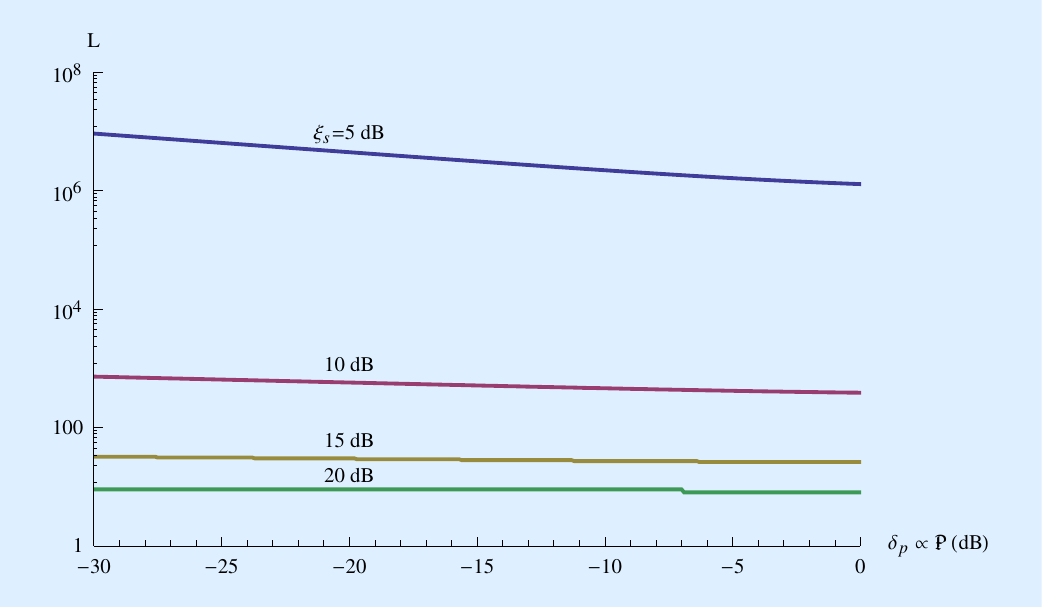}
	{.8}
	{Figure \ref{fig:noObsReduceDutyCycle} is repeated, except that each observation
	is over a full period of the periodic beacon ($T_{o}=T_{p}+T$, or $\delta_{o} = 1$).
	A longer observation interval has been exchanged for a smaller number of observations $L$.}

\subsection{Optimization of energy contrast ratio and duty factor}

Two diametrically opposed ways to reduce average power $\power$ have been considered.
The first is the Cyclops beacon,
where $\power$ is reduced by a reduction in the energy per bundle $\energy_{h}$ (and hence $\ecrs$) 
while keeping 100\% duty factor.
The second reduces the duty factor $\delta_{p}$ while keeping $\ecrs$ fixed.
In our earlier examples,
reducing $\ecrs$ proved effective at higher powers until $\ecrs \approx 15$ dB was reached, but after that further reductions
resulted in a dramatic increase in the number of observations $L$.
Reducing $\delta_{p}$ was much more effective at limiting the size of $L$ at low powers.
The size of $L$ can be kept very modest ($L \approx 30$ or so) at low powers by extending the
duration of each observation to equal the period of a incoming periodic beacon.

In summary, reducing $\ecrs$ was more effective at higher powers, and reducing $\delta_{p}$
was more effective at lower powers.
To truly capture the design tradeoffs, particularly in the transition region,
we must minimize $L$ by the \emph{joint} choice of $\ecrs$ and $\delta_{p}$.
This is easily accomplished by considering a beacon with fixed average power $\power$
and varying both the energy contrast ratio for each energy bundle $\ecrs$ and the duty factor $\delta_{p}$
to maintain that fixed $\power$.
The parameters of this fixed-average-power periodic beacon are
\begin{align}
&\ecrs = \frac{\xi_{0}}{\delta_{p}}
\ \ \ \ \text{where} \ \ \  
\xi_{0} = \frac{\power \, T}{N_{0}} \\
&N = L \, K \, \left( \frac{\delta_{o}}{\delta_{p}} + 1 \right)
\,.
\end{align}
The parameter $\xi_{0}$ is the per-bundle energy contrast ratio for the Cyclops beacon at 100\% duty factor.
Two choices for the receiver's observation strategy bracket the observation times of interest, 
the minimum per-observation duration ($\delta_{o} = \delta_{p}$)
and the maximum duration  ($\delta_{o} = 1$).

The resulting $L$ is plotted in Figure \ref{fig:Lvsp1},
where each curve corresponds to a fixed power $\power$.
For each curve,
as $\delta_{p}$ is decreased, the energy contrast ratio $\ecrs$
is increased to maintain a fixed $\power$.
For lower power beacons, there is an optimum $\delta_{p} < 1$
for which $L$ is minimum, and that $\delta_{p}$ decreases
as $\power$ decreases.
A beacon with this duty factor is power-optimum, in that it minimizes the number of observations $L$
required to achieve the target detection reliability.
For larger $\power$ the Cyclops beacon ($\delta_{p} = 1$) is power-optimum.
The lowest-power Cyclops beacon that is power-optimum has an energy
contrast ratio of approximately $\ecrs = 14$ dB for each energy bundle.

\incfig
	{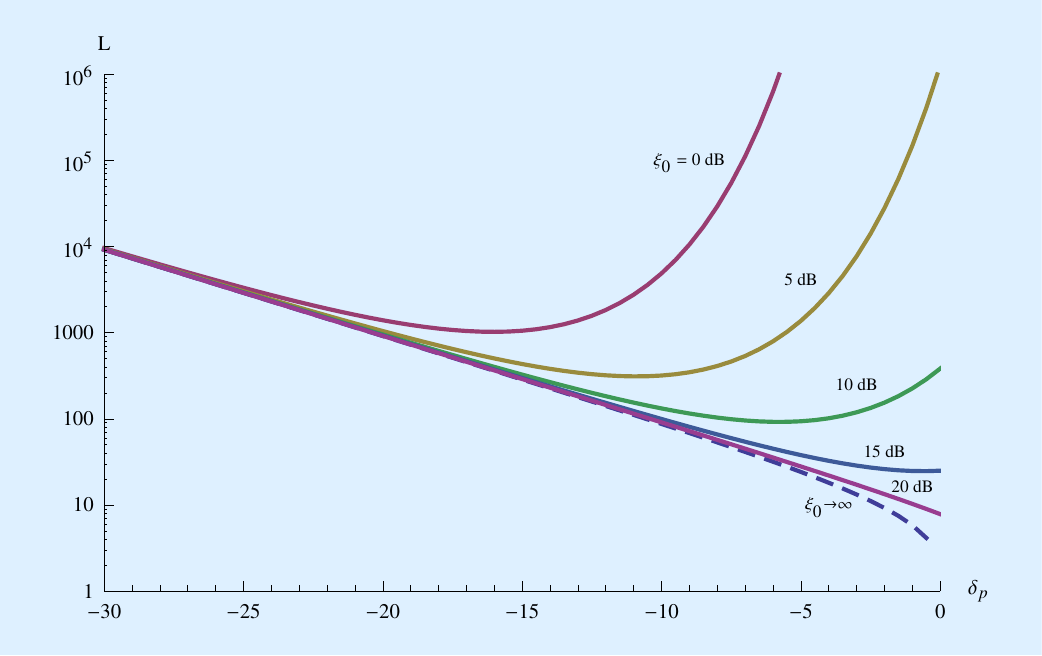}
	{1}
	{
	For the minimum per-observation window,
	a log plot of the number of observations $L$ vs the transmit duty factor $\delta_{p}$ in dB.
	The right endpoint is the Cyclops beacon with 100\% duty factor ($\delta_{p} = 0$ dB).
	For each curve, the average power $\power$ is kept fixed as $\delta_{p}$ is reduced.
	For all the plots, the reliability is fixed at $\pfa = 10^{-12}$ and $1-\pd = 10^{-4}$.
	The different curves correspond to different values of $\power$
	(represented by energy contrast ratio
	$\xi_0$ in dB at $\delta_{p} = 1$).
	The dashed curve is the lower bound on $L$ from \eqref{eq:Llowerbd}
	that captures only misses attributable to $\delta_{p} < 1$.
	The duration of each observation is $T_{o} = 2\,T$ ($\delta_{o} = \delta_{p}$).
	}

The bundle energy hypothesis predicts that $\delta_p \propto \xi_0$,
of course with the constraint that $\delta_p \le 1$.
The optimum value of $\delta_{p}$ 
(the minimum of the curves in Figure \ref{fig:Lvsp1})
as a function of $\xi_{0}$ is shown in Figure \ref{fig:optP1vsSNR},
where we see that it closely adheres to this hypothesis.
As one anchor,
the optimum duty factor is $\delta_p =$ 3.0\% at $\ecrs = 0$ dB.

\incfig
	{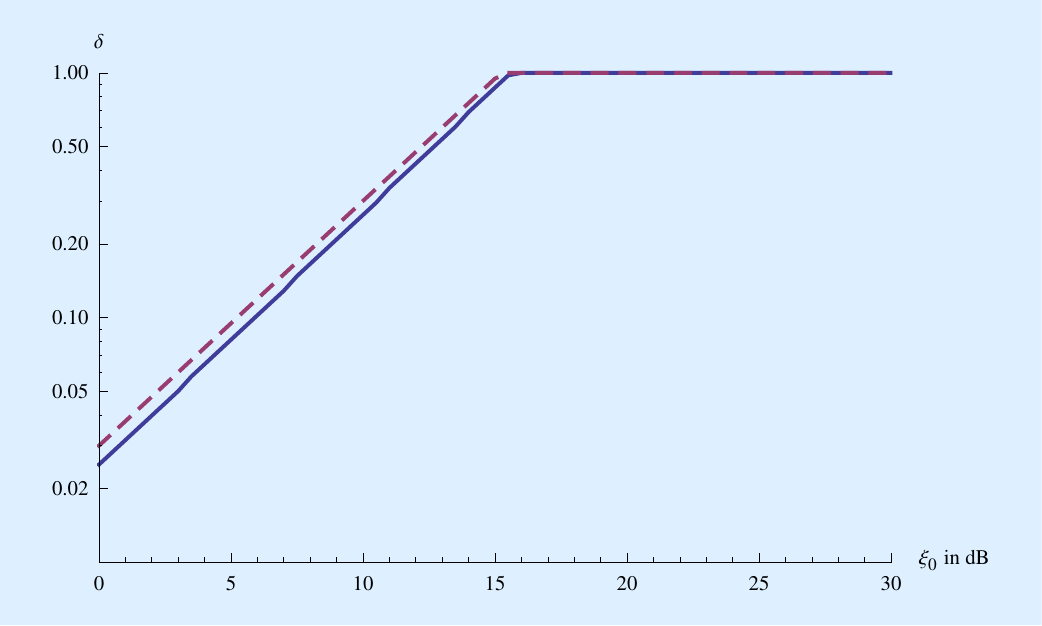}
	{.8}
	{A log plot of the optimum duty factor $\delta_{p}$ as a function of
	the beacon average power (represented by
	$\xi_{0}$ in dB) for the same conditions as in Figure \ref{fig:Lvsp1}.
	At $\xi_{0} = 0$ dB, the optimum value is $\delta_{p} = 0.030$ (3\%).
	The value of $\delta_p$ predicted by the
	bundle energy hypothesis ($\delta_p \propto \xi_0$)
	is shown as the dashed line.}

The bundle energy hypothesis is verified in a more direct way in Figure \ref{fig:beaconECRoneBundle}.
When energy contrast ratio $\ecrs$ is plotted, in the low-power regime it approximately constant
at $\ecrs \approx$ 15 dB.
In view of Figure \ref{fig:detectOneBundle} this is not surprising,
since an $\ecrs$ on this order is required to 
obtain even a 50\% detection probability $\pd$.
This is due to the scintillation, which adds another source of randomness
that adds to the randomness of the noise.
The small variation that does exist is in the opposite direction than we might expect, 
toward higher $\ecrs$ as $\delta_{p}$ (and hence $\power$) decreases.
Of course $\ecrs \propto \xi_{0}$ in the high-power region since $\delta_{p} = 1$ is fixed.

\incfig
	{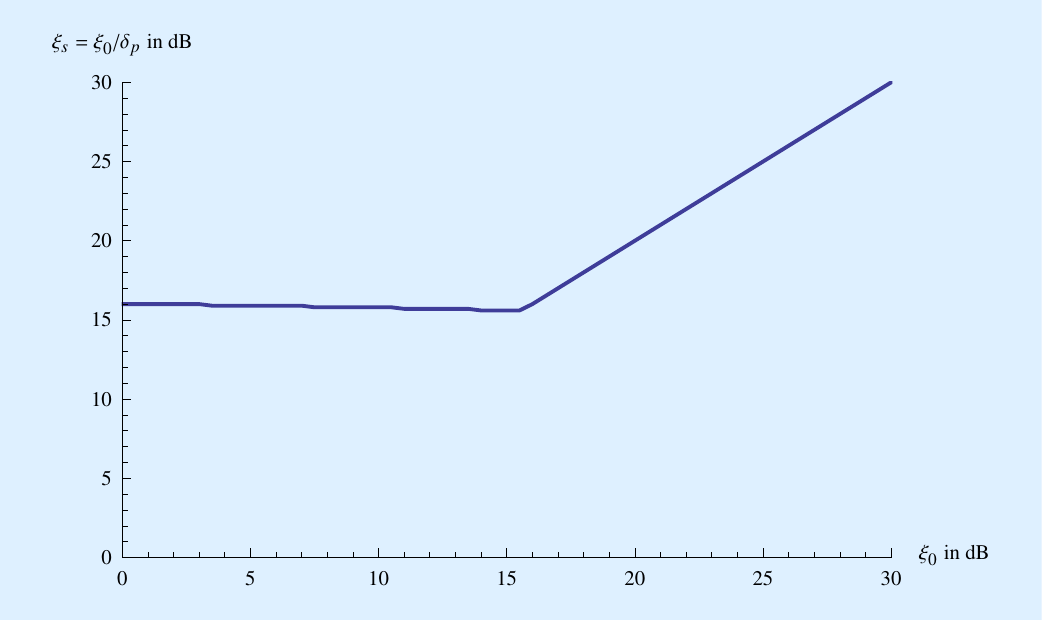}
	{.8}
	{For a beacon power-optimized for detection,
	the energy contrast ratio $\ecrs = \xi/\delta$ in dB 
	is plotted against $\xi_{0}$ in dB (which represents the beacon average power $\power$).
	This is a repeat of Figure \ref{fig:optP1vsSNR}, except it plots $\ecrs$ rather than $\delta_{p}$.
	At lower powers, where $\delta_{p} < 1$, $\ecrs$ is approximately constant as predicted by the
	bundle energy hypothesis.
	For larger powers, where the duty factor saturates at $\delta_p = 1$,
	the only way to increase the average power $\power$ is to increase $\energy_s$.
	}

The effect of reducing the duty factor on
the number of observations $L$ is illustrated in Figure \ref{fig:optLvsSNR}.
In the low-power regime, the effect of using a low transmit duty factor
is quite dramatic.
For example, at $\xi_{0} = 0$ dB, the value of $L$ is reduced by about
12 orders of magnitude (to $L=848$) by reducing the duty factor from $\delta_p = 1$
to $\delta_p = 0.03$ (3\%).
Clearly it is not practical to discover a Cyclops beacon at this low power,
but reducing the duty factor to 3\% saves the day as long as the receiver is
willing to implement hundreds of independent observations.
An alternative way to quantify the effect of duty factor is to measure horizontally
rather than vertically.
The reduction in signal
power that is possible for a fixed $L = 848$ observations attributable to reducing the
duty factor is $8.3$ dB, or an 85\% reduction in average beacon power.

\incfig
	{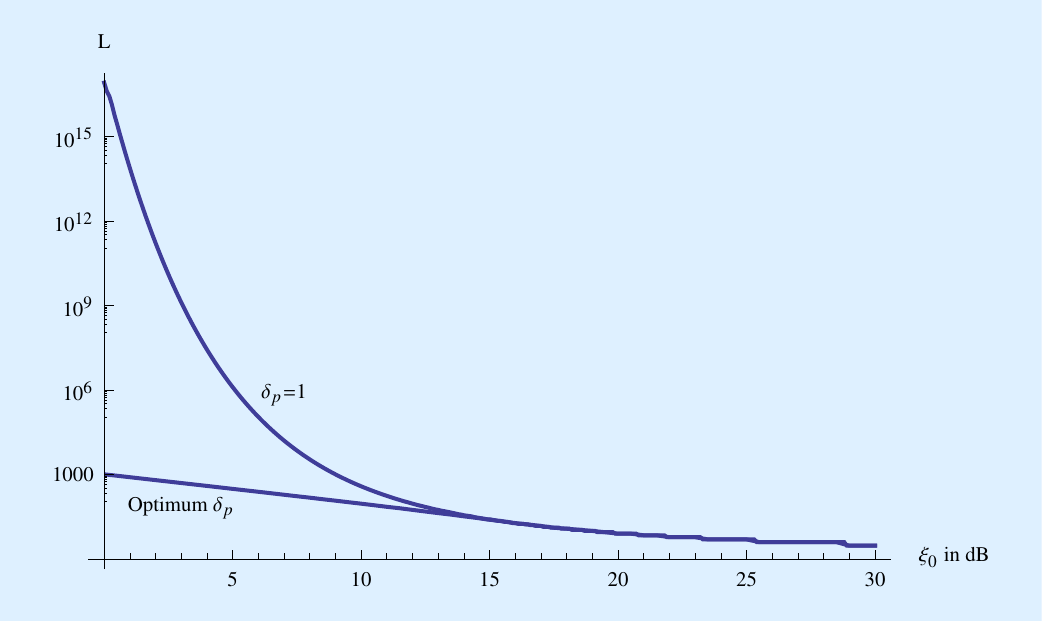}
	{.8}
	{A log plot of the number of observations $L$ required to achieve the same
	performance parameters as Figure \ref{fig:Lvsp1} vs
	the average beacon power (represented by $\xi_{0}$,
	which is the energy contrast ratio for a Cyclops beacon with the same average power).
	This is shown for both a Cyclops beacon with 100\% duty factor ($\delta_{p} = 1$)
	and for the optimum duty factor $\delta_{p}$.}

The level of reliability that is demanded in the detection of a beacon has an
impact on the required $L$ as illustrated in Figure \ref{fig:optLvsSNRlowDuty}.
If a lower reliability is demanded (larger $\pfa$ and smaller $\pd$), this naturally reduces the required $L$.
The prediction of the total effective energy hypothesis that $L \propto 1/\xi_0$
(modified slightly so that $L \to 1$ as $\xi_0 \to \infty$)
is shown by the dashed line.
Notably, however, the total effective energy hypothesis is only valid if $\delta_p$ is chosen optimally,
or in other words the bundle energy hypothesis is applied simultaneously.

\incfig
	{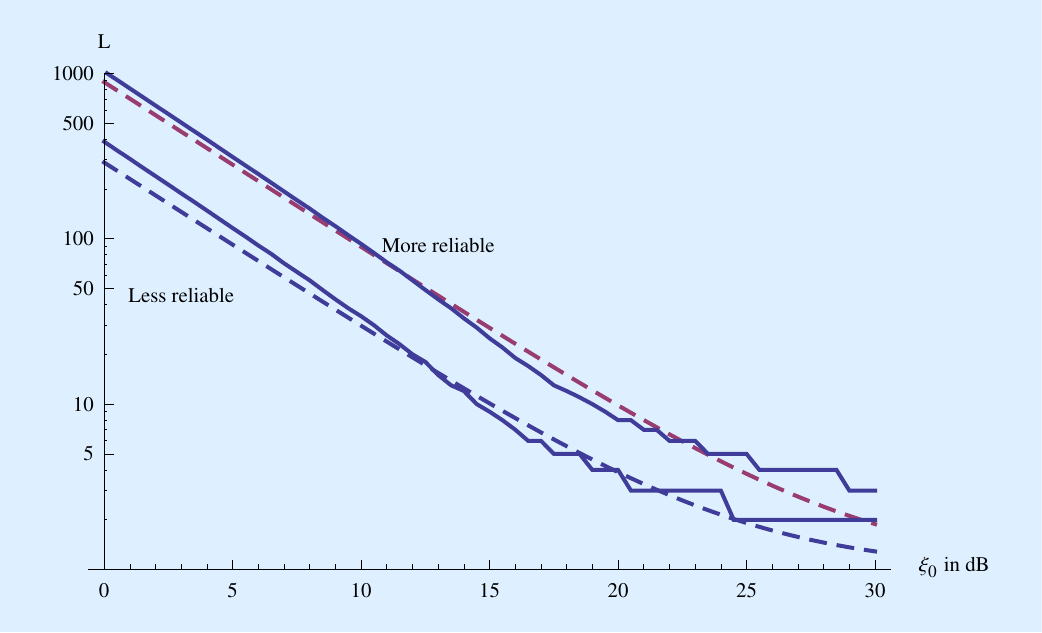}
	{.8}
	{Figure \ref{fig:optLvsSNR} is repeated showng only the
	number of observations required when the transmit duty
	factor $\delta_{p}$ is optimized to minimize $L$. Two levels of reliability are shown:
	''More reliable'' is  $\pfa = 10^{-12}$ and $\pd = 0.9999$
	(the same as in Figure \ref{fig:Lvsp1})
	and ''Less reliable'' is $\pfa = 10^{-8}$ and $\pd = .99$.
	Smaller $L$ results in reduced reliability.
	The dashed lines illustrate the prediction of the
	total effective energy hypothesis that $L \propto 1/\xi_0$.
	}

The overall conclusion is that imposing a multiple-observation requirement
on the receiver can have a dramatic impact in reducing the required transmit power.
For the detection reliability parameters $\{ \pfa, \, \pd \}$ chosen in Figure \ref{fig:Lvsp1},
increasing the number of observations from $L = 10$ to $L = 831$
reduced the required average transmit power by $17$ dB, or a linear
factor of about $50$.
The transmitter designer may view this imposition on the receiver
as relatively modest in light of its $50 \times$ reduction in average transmitted power,
especially since it is the receiver that derives all the value from successful 
beacon discovery.
A second conclusion is that a low duty factor $\delta_{p}$ is required in the transmitter
if $L$ is to be kept to a reasonable level in the low-power regime.
Of course small $\delta_p$ in itself implies a relatively large $L$, so there
is no avoiding a relatively large $L$ for low-power beacons.
Indeed, past and present SETI searches would have a negligible hope of success
in discovering a beacon power-optimized for discovery in the low-power regime.
	
\subsubsection{Long duration per observation}

The preceding results assumed the shortest possible duration of interest for a single observation,
$T_{o} = 2 \, T$.
The number of observations $L$ can be dramatically reduced by extending the duration of each
individual observation as shown in Figure \ref{fig:Lvsp1longObs}.
In this case, where $T_{o} = T_{p} + T$, $L$ decreases monotonically as the 
transmit duty factor $\delta_{p}$ decreases.
This suggests that if the receiver is going to make long individual observations,
the transmitter should choose the smallest duty factor possible.
However, this is misleading because in this case the total observation time
may be a better indicator of the processing resources required by the receiver than the
number of observations.
When the total observation time is plotted in Figure \ref{fig:Lvsp1totalObs}, the result is
very similar to the short individual observation in that the optimum duty factor $\delta_{p}$ is about the same.
The conclusion is that the transmitter designer can safely choose its duty factor according to the
average power and energy consumption objective without consideration of the receiver's strategy of using a small $T_{o}$
with large $L$ vs large $T_{o}$ with small $L$.

\incfig
	{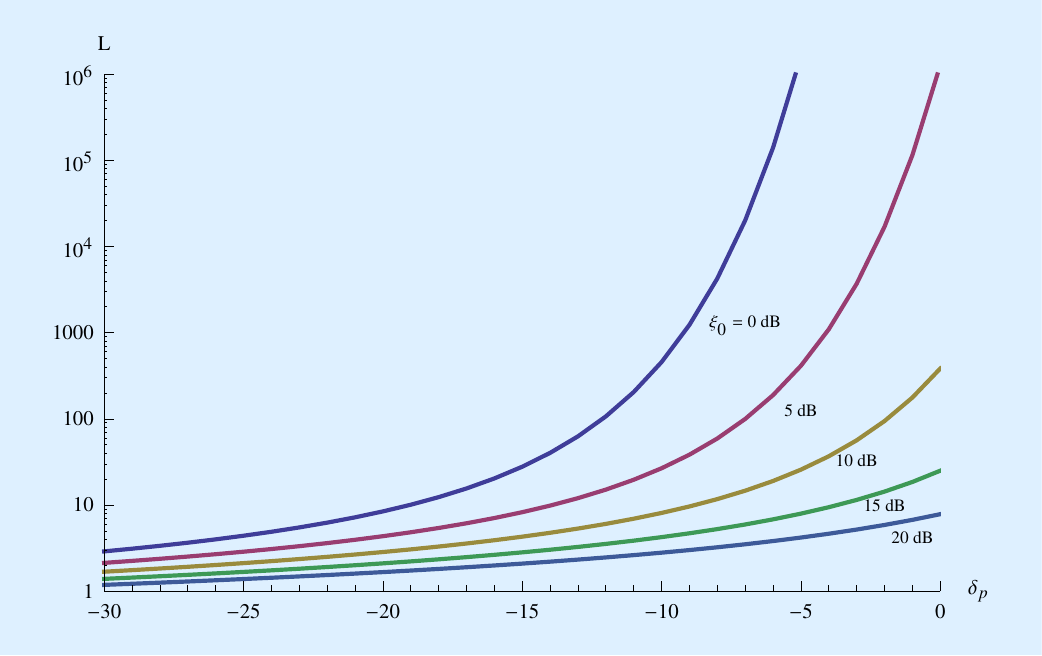}
	{.8}
	{Figure \ref{fig:Lvsp1} is repeated for the longest duration of each
	individual observation of interest, $T_{o} = T_{p} + T$ ($\delta_{o} = 1$).
	The number of observations $L$ is dramatically reduced for small $\delta_{p}$
	because the duty-factor effect on $L$ is eliminated.}

\incfig
	{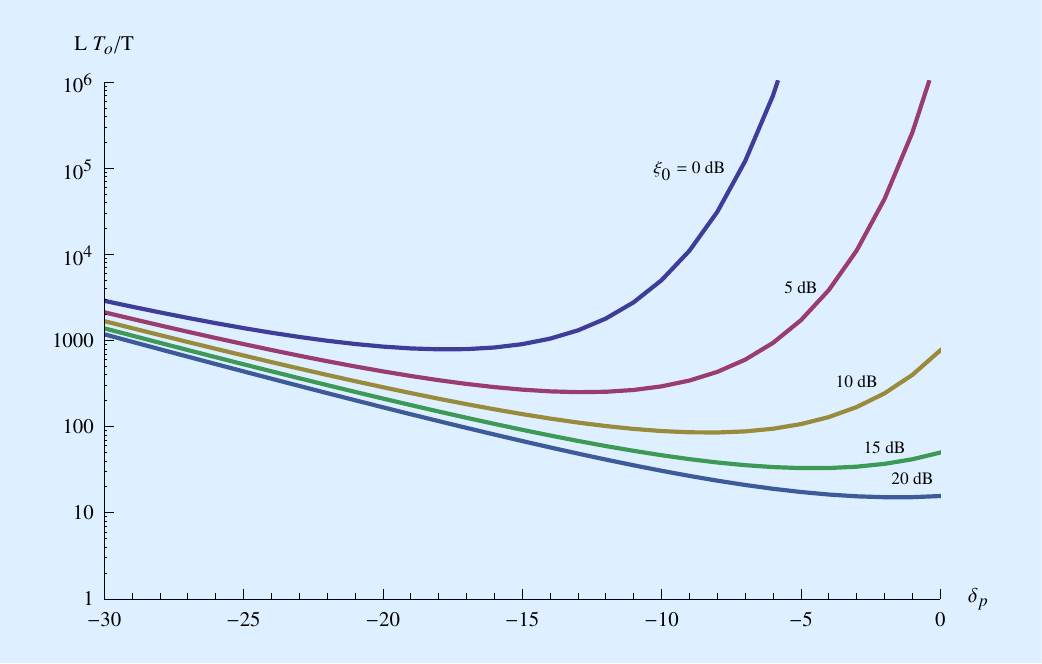}
	{.8}
	{Figure \ref{fig:Lvsp1longObs} is repeated, except that the vertical axes is not $L$, but rather
	the total observation time $L \, T_{o}$ normalized by the duration of one energy bundle $T$.
	With respect to minimization of the total observation time, the optimum value of $\delta_{p}$
	is about the same as in Figure \ref{fig:Lvsp1}.}

\subsection{Uncoordinated periodic beacon design}

The previous comparisons have assumed the joint optimization of the
beacon parameters.
In practice, the responsibility for choice of parameters is divided between transmitter
and receiver without coordination.
It is the transmitter that determines the parameters $\{ T, \, T_{p} \}$,
and it is the receiver that chooses the parameters $\{ T_{o},  \, L \}$.
Both the transmitter and receiver designers have influence over 
the energy contrast ratio $\ecrs$ of each energy bundle,
the transmitter through choice of the transmit energy per bundle
(which is related to the peak transmitted power $\powerpeak$) 
and transmit antenna gain,
and the receiver through choice of the receive antenna collection area.
Thus neither transmitter nor receiver has unilateral control over the reliability
with which individual energy bundles can be detected.

The partitioning of decision making can be characterized as follows:
\begin{description}
\item[Common knowledge.] 
The one parameter the transmitter and receiver can agree on (with some remaining degree of
uncertainty) in the duration $T$ of an energy bundle.
The receiver designer can reasonably assume that the transmitter wishes to minimize
the peak transmitted power, and will thus choose $T = T_{c}$, the coherence time of the
interstellar channel.
Since $T_{c}$ is based on observable physical parameters,
we can expect some consistency between the transmitter's and receiver's assumptions as to $T_c$.
\item[Transmitter designer's assumptions.]
The transmitter must make assumptions about the
\emph{minimum} resources devoted by the receiver.
In particular, a combination of (a) the propagation loss (based on knowledge of the distance), 
(b) the minimum assumed receive antenna collection area, and 
(c) an assumption as to the maximum
noise power power spectral density $N_{0}$ at the receiver together determine the
required transmitted energy per bundle.
In particular, the bundle energy hypothesis determines that these parameters collectively
must yield an energy contrast ratio $\ecrs$ at the receiver of about $\ecrs \approx 14$ dB or higher.

The transmitter also controls the average transmitted power through a choice of beacon period $T_{p}$.
The transmitter can make this power as small as it wishes, but must recognize that while
a larger $T_{p}$ beneficially decreases the transmit power, it also imposes a requirement
that the receiver devote a longer total observation time $L \cdot T_{o}$ commensurate with that lower power.
Thus, in making a choice as to transmitted power, an assumption is being made about the minimum
observation time devoted by the receiver.
\item[Consequences of receiver designer's choices.]
In general if the receiver makes resource decisions that exceed the minimum assumptions of the
transmitter, the beacon will be detected with high reliability.
Otherwise, the beacon is unlikely to be detected.
First, the receiver must choose a large enough receive antenna collection area to
achieve a sufficiently large energy contrast ratio $\ecrs$ for each energy bundle.
Second, the receiver must choose the duration $T_{o}$ that is compatible with the transmitter's
choice of $T$.
Generally this will be the easiest choice the receiver has to make, since $T_{o} = 2 \, T_{c}$
will work well for all transmitter choices of $T_{p}$ without affecting the total observation time.
Third, the receiver must also choose a sufficiently large
number of observations $L$ to accommodate the observation duty factor $\delta_{o}$,
which is determined by $\{ T , T_{p}, \, T_{o} \}$.
Otherwise, again, the beacon is unlikely to be detected.
\end{description}

This is a worst-case design methodology used to overcome the lack of explicit coordination.
The receiver makes an assumption about the minimum receive signal flux of the beacon,
and how that flux is partitioned between energy per bundle $\energy_{h}$ and
bundle repetition period $T_{p}$.
The receiver will be successful at detecting the beacon if these assumptions are
met or exceeded, and is unlikely to be successful otherwise.
Figure \ref{fig:beaconCompare} ($K = 1000$ was assumed) illustrated the tradeoff
between transmit resources (average power) and receive resources (processing, but not antenna capture area),
making the unrealistic assumption that the transmitter and receiver designs are coordinated.
A more realistic scenario is presented in Figure \ref{fig:beaconVsL}
with the assumption that the beacon and observation parameters are given by
\begin{align}
&\power = \frac{\energy_{h}}{T_{p}} 
\ \ \ \ \text{or} \ \ \  
\ecrs =  \frac{\xi_{0}}{\delta_{p}}
\ \ \ \ \text{for} \ \ \ 
\xi_{0} =  \frac{\power \, T}{N_{0}}
\\
&T_{o} = 2 \, T 
\ \ \ \ \text{or} \ \ \
\delta_{o} = \delta_{p} \\
&N = 2 \, L \, K
\,.
\end{align}
This assumes the minimum duration of each observation, thus relying on a larger number of observations $L$
rather than a longer observation time $T_{o}$ to compensate for a lower beacon power.
The energy contrast ratio $\xi_{0}$ corresponds to a Cyclops beacon with 100\% duty factor ($\delta_{p} = 1$)
with the same\footnote{
The horizontal axis of Figure \ref{fig:beaconCompare} was also $\xi_{0}$, although it wasn't
necessary to mention that level of detail in Chapter \ref{sec:powerefficient}
where only a relative comparison was being made.}
 $\power$.

In Figure \ref{fig:beaconVsL} it is assumed that the 
transmitter has reduced the transmit signal power by 15.2 dB (97\%)
relative to the Cyclops beacon design by choosing a 3\% duty factor
($\delta_p = 0.03$).
The receiver is not cognizant of this choice, but is using an antenna
with sufficiently large capture area to achieve the optimum received
energy contrast ratio $\ecrs$.
We then examine the consequences of using different values of $L$.
This illustrates that the receiver gets what it pays for, with increased
detection reliability when it devotes more processing resources to a larger number of observations $L$.
The transmitter design envisioned that $L = 1000$ could achieve a
miss probability of $10^{-4}$ at $\xi_{0} = 0$ dB.
In fact, as long as $L \ge 305$, the receiver can achieve that same miss probability, 
but it does require a larger $\xi_{0}$ and hence a larger receive antenna collection area.
In this case, the receiver can trade off less processing overhead for a larger antenna, or vice versa.
On the other hand, if the receiver implements $L = 3$ (as if this was
a high-power 100\% duty factor Cyclops beacon) it has virtually no chance
of detection even at considerably higher power levels than required for a 100\% duty factor Cyclops beacon
(which corresponds to $\xi_{0} \approx 15$ dB).
This is because so few observations fail to compensate for the low 3\% duty factor,
no matter how large the receive antenna collection area.

\incfig
	{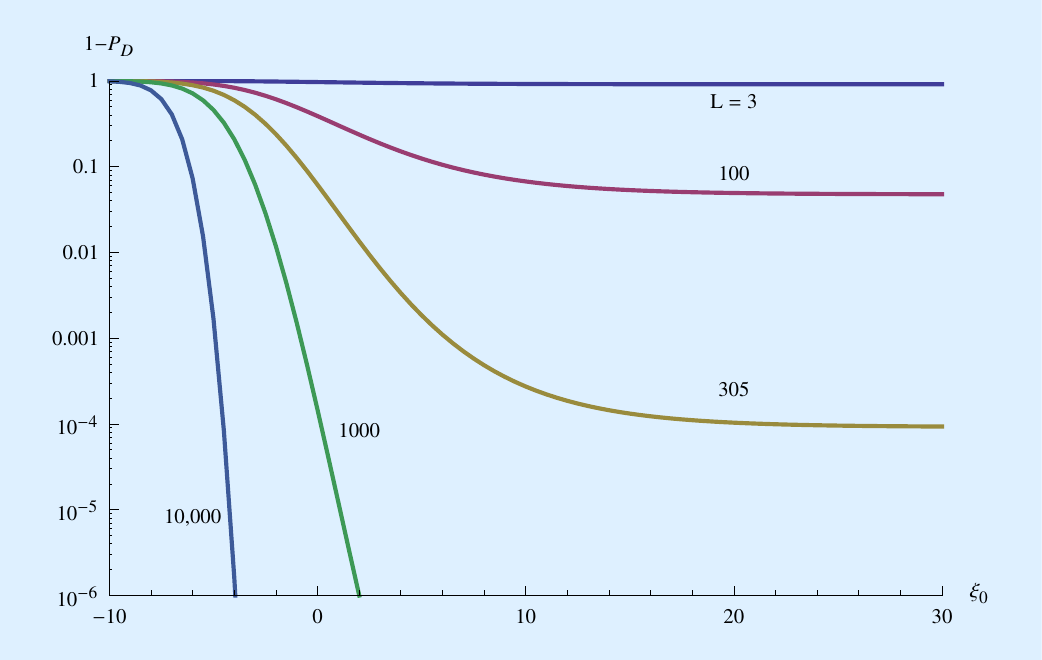}
	{.8}
	{
	A log plot of the miss probability $(1-\pd)$ vs the energy contrast ratio
	$\xi_{0}$ of a Cyclops beacon with 100\% duty factor at the same average power. 
	Each curve thus represents how the reliability in detection of the beacon
	varies with the average received power $\power$.
	It is assumed that the transmitter has chosen a 3\% duty factor of ($\delta_{p} = 0.03$)
	and the receiver has chosen the minimum duration for each observation 
	($T_{o} = 2 \, T$, which results in $\delta_{o} = 0.03$).
	The different curves represent different numbers of observations $L$.
	$L = 1000$ is more than sufficient to achieve a $1-\pd = 10^{-4}$ at $\xi_{0} = 0$ dB.
	The lower bound of \eqref{eq:Llowerbd} predicts that the same reliability can be achieved
	for large $\xi_{0}$ with $L = 305$.
	The remaining choices of $L$ illustrate that detection reliability is improved
	with larger $L$ and deteriorates if $L$ is chosen to be too small.
	}

Overall the conclusion is that the transmitter has two ways to reduce its average transmit power,
each imposing an additional resource burden on the receiver.
First, average power can be reduced significantly (with no reduction in peak power)
by reducing the duty factor $\delta_p$, if the receiver compensates for
this by a commensurate increase in its total observation time (as represented by $L$
in case of the minimum duration of each observation).
Second, both average and peak transmit power can be reduced by decreasing the
energy per bundle $\energy_h$,
if the receiver compensates by increasing its antenna collection area.

\begin{changea}
\section{Adding information to the periodic beacon}
\label{sec:beaconWinfo}

The challenges of discovery, and its importance as a necessary
enabler for subsequent communication, 
suggest that discovery should be in the driver's seat.
This strategy would focus on power-optimization of discovery, 
and then ask how information can be embedded in the signal
without affecting discovery, even if that means sacrificing some power efficiency.

This can be illustrated by an example.
Based on our earlier results
we could start with a periodic power-optimized beacon
with duty cycle $\delta$ and energy per bundle $\energy_h$.
Together these two parameters determine the average power $\power$,
and minimize the number of observations $L$ necessary in the receiver
to achieve a given probability of detection for that $\power$.
Having established these parameters, information can be embedded
by transmitting each beacon energy bundle at  one of $M$ different frequencies,
rather than relying on a single frequency.
If the receiver is performing a wideband spectral analysis
(as is the case today) then those energy bundles can be detected
as easily as if they were at a single frequency.
This has two implications in terms of discovery reliability:
\begin{itemize}
\item
There is an increase in the incidence of false detections
because of the increase in the number of frequency locations, 
but that increase can be attributed to
the wideband spectral analysis and rather than the M-ary FSK signal structure.
\item
If $M$ is sufficiently large so that the total bandwidth $\Btotal$
exceeds the scattering coherence bandwidth and therefore the scintillation
becomes frequency selective, then when
an observation overlaps the transmitted energy bundle
it may or may not experience a smaller received signal flux
depending on its frequency.
This phenomenon does not change the impact of scintillation
on discovery, since the probability of scintillation affecting
the result is the same.\footnote{
In so-called "polychromatic SETI" the signal would be transmitted
a multiple frequencies simultaneously \citeref{140}.
This desirably increases the probability of detection in a single
observation by taking advantage of the frequency selectivity of diversity,
but also increases the average power by a factor of $M$ as well,
so this is not a power-optimized solution.}
\end{itemize}

What power efficiency in information transfer can be accomplished this way?
This question can be answered with a simple reformulation of previous results.
Solving \eqref{eq:peVsMandEcrs} for $M$ in terms of $\{ \ecrs , \, \pe \}$,
at the threshold of outage or erasure $M$ is limited to
\begin{equation}
M \le 4 \, e^{\, \ecrs /2} \, \pe \,.
\end{equation}
Substituting the value $\ecrs = 15$ dB, which has been determined
to be the approximate value for discovery with the minimum number of receiver observations,
we get
\begin{equation}
M \le 3 \times 10^7 \, \pe
\end{equation}
For example, at $\pe = 10^{-4}$, it follows that $M \le 3000$.
On the other hand, using a larger value
like $M = 2^{20} \approx 10^3$ 
(which is desirable for power efficiency)
can only be achieved if $\pe \approx$ 3\%.

To summarize, if $M$ is large enough to achieve a high power efficiency
with the outage strategy, the energy per bundle 
$\energy_h$ is larger (by a few dB) than would be
ideal for discovery.
The good news is this makes discovery of each energy bundle
more reliable rather than less, but the bad news
is that discovery becomes 
moderately inefficient in its consumption of receiver observations.
If discovery were the only driver, it would be better
to reduce $\energy_h$ a bit and reduce the duty factor $\delta$,
achieving the same average power $\power$ with
fewer observations in the receiver.

\end{changea}

\section{Discovery of information-bearing signals}
\label{sec:discoveryoptimized}

\begin{changea}
If an end-to-end communication system is designed unconditionally
to maximize power efficiency, based on the last section we
can anticipate some compromise in the discovery power vs. observations
character of that signal.
\end{changea}

Fortunately information-bearing signals of the type considered in
Chapter \ref{sec:infosignals} do have a similar structure to periodic beacons.
They consist of a sequence of codewords, each of which
conveys $\log_2 M$ bits of information by transmitting an energy bundle in
one of $M$ locations in time or frequency.
By adjusting the rate at which these codewords are transmitted,
the total information rate $\rate$ and average power $\power$
can be adjusted in direct proportion to one another.
Note the parallel to adjusting the duty factor
of a beacon.
Each codeword consumes the same energy
(adhering to a bundle energy hypothesis), but the power is
adjusted by choosing a rate at which to transmit a codeword
(equivalent to an adjustment in duty factor $\delta$)
as described in Section \ref{sec:adjrateandpower}.

In view of this,
it is no surprise that discovery of power-efficient information-bearing signals
of the type considered in Chapter \ref{sec:infosignals}
is essentially the same challenge as the discovery of beacons that are
power-optimized for discovery.
The same tradeoff exists between transmitter and receiver resource allocation.
The average transmit power $\power$ can be reduced by a commensurate
reduction in information rate $\rate$, and the most efficient approach
is to reduce the rate at which codewords are transmitted,
while not changing the character of the codewords themselves, 
which is equivalent to reducing the duty factor.
The average and peak transmitted power $\powerpeak$ can both be reduced
by reducing the transmitted energy per bundle, but a certain level of energy
per bit $\energybit$ and hence energy per bundle must be maintained in the
receiver to achieve reliable extraction of information from the signal, 
and the receive antenna collection area has to be large enough to achieve this.

In contrast to a beacon, the objective is not only to discover the signal but also
to extract information from it during the communication phase.
These two challenges are considerably different, but normally the average received power $\power$
will be the same, unless the receive antenna is changed out in the transition from discovery to communication.
There are two very different interpretations of this common value of $\power$.
The difference can be summarized as follows:
\begin{description}
\item[Discovery.] 
$\power$ relates to the number of independent observations $L$ required
to detect a signal of interest.
The smaller $\power$, the larger the required $L$, and at all but very high average powers
$L$ must be relatively large (tens to hundreds).
This large $L$ is required for two reasons.
First, the receiver has no idea where an energy bundle may occur, so much of the
time it is observing during blank periods in the duty factor when there is no energy bundle present.
Second, the receiver has no idea as to the current state of scintillation, so much of the
time it is observing during periods of low received flux due to scintillation.
\item[Communication.]
During communication the receiver has two major advantages.
First, it is aware of the structure of the signal, and will observe only at times
when codewords are present.
The presence or absence of energy bundles at these times is determined
by the random information bits being communicated.
Second, it can form an estimate of the current level of received flux, and observe
only during periods of non-outage (if an outage strategy is in use).
Thus, it is not disadvantaged by the periods of low received flux in the same way as discovery.
The value of $\power$ is thus determined by the reliability with which the presence
or absence of an energy bundle can be detected by a \emph{single} observation
at the threshold of outage.
That outage threshold corresponds to a receive flux 
equaling its average value, which is the same value as it would have in the absence of scattering
and scintillation.
\end{description}
From this comparison, it is not clear whether a larger $\power$ is required during
discovery or during communication.
There are arguments in favor of both outcomes.
During discovery, detection is based on multiple observations, and during communication
it is restricted to a single observation.
This favors discovery, as the receiver can always increase the observation time $L\cdot T_{o}$ to reduce the
required $\power$.
During communication, the receiver may discard periods when the received flux is below
average as outages, and thus scintillation is aided by consistent amplification of the received flux during periods of non-outage. 

These competing factors can only be resolved by a quantitative comparison of the two cases.
An M-ary FSK signal is essentially identical to a periodic beacon
with the same duty factor, in that energy bundles occur in a regular
periodic factor with a duty factor $\delta$ determined by the information rate $\rate$,
which is also proportional to the average power $\power$.
The one difference between a periodic beacon and an FSK signal is that
these periodic energy bundles will be positioned at one of $M$ frequencies,
and that frequency will vary randomly (in accordance with the
embedded information) from one codeword to the next.
A discovery search algorithm that examines only one carrier frequency
at a time will thus see energy bundles that are not uniformly spaced,
but rather occur randomly at those uniformly spaced positions
but with a density that is a factor of $M$ lower.
This would have a substantial effect on detection, according to the
total effective energy hypothesis increasing the number of observations $L$
by a factor of $M$ to maintain the same detection reliability
(since the energy per bundle does not change with $M$).
In practice, however, a discovery search will typically be conducted by a multi-channel
spectral analysis combined with matched filtering to $h(t)$.
Thus discovery can proceed in essentially the same manner as
for a periodic beacon, with one modification.
Rather than defining a detection at one frequency, it can be defined
as a detection at any frequency within a range of $M$ frequencies.
This will affect the reliability parameters $\{ \pfa , \, \pd \}$, because the
receiver is observing in more places.
The effect is fairly minor, however.

For purposes of discovery, a power-efficient information-bearing M-ary FSK or PPM signal possesses
three parameters $\{ \ecrs, \, M, \, \delta_{p} \}$ that together determine the tradeoff between detection
reliability $\pd$ and the observation time $L \cdot T_{o}$ for a given false alarm probability $\pfa$.
On the other hand, the reliability of information extraction at the threshold of outage is measured by the
error probability $\pe$, and in Chapter \ref{sec:infosignals} this is shown to
be determined by $\{ \ecrb , \, M \}$ in \eqref{eq:ecrbVsPe},
where $\ecrb$ is the energy contrast ratio per bit $\ecrb$ measured at the threshold of outage.
The duty factor $\delta$ has no relation to the reliability of information recovery, but it
does determine the information rate $\rate$ as described in Section \ref{sec:FSKspectralefficiency}.
As shown in Chapter \ref{sec:infosignals}, the flux at the outage threshold is equal to the average flux,
and thus $\ecrb = \ecrb = \ecrs / \log_{2} M$.
Thus, from \eqref{eq:ecrbVsPe} the parameters of the FSK signal for purposes of discovery are
\begin{align}
&\ecrs = 2 \, \log_{e} \left( \frac{M}{4 \, \pe} \right) \\
&N = 2 \, L \, K \, M
 \,.
\end{align}

\incfig
	{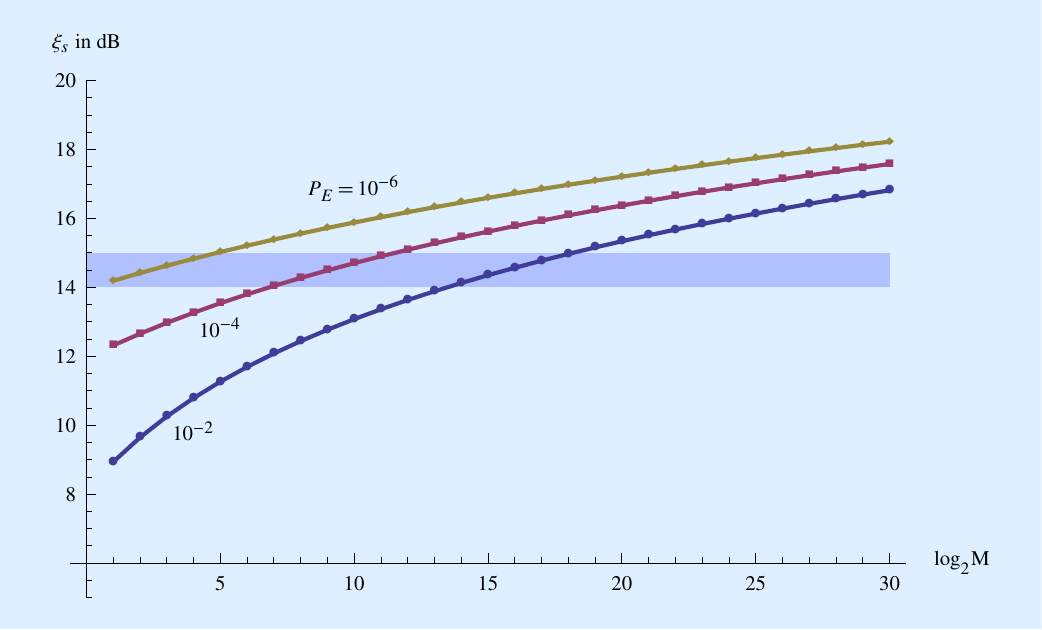}
	{1}
	{For an M-ary FSK signal, a plot of the energy contrast ratio $\ecrs$ for each
	energy bundle vs. $M$.
	Several values of probability of error $\pe$ are plotted.
	The shaded rectangle corresponds to $\ecrs = 14$ to $15$ dB, which is
	an ideal value to minimize the signal average power $\power$.
	Any higher value of $\ecrs$ makes inefficient use of bundle energy from a discovery standpoint,
	and any lower $\ecrs$ requires a substantial increase in the number of observations $L$.}

The most important characteristic of an information-bearing signal for
discovery is $\ecrs$, because it determines the probability of detection of
an individual energy bundle, as was illustrated in Figure \ref{fig:detectOneBundle}.
The bundle energy hypothesis predicts that $\ecrs$ should remain fixed
as duty factor $\delta_{p}$ is varied, and later numerical results
(for a particular target $\pfa$ and $\pd$) confirm this
hypothesis and indicates that $\ecrs \approx 14$ to $15$ dB offers the best
tradeoff between misses due energy bundles that go undetected and
observations that do not overlap an energy bundle.
Any smaller value causes an undue increase in $L$ due to misses, and any larger value
wastes energy on each bundle.
As an alternative way to view this tradeoff from Figure \ref{fig:detectOneBundle},
$\ecrs$ is plotted for different values of $M$ and $\pe$ in  Figure \ref{fig:ecrsVsM}.
As expected, $\ecrs$ increases with $M$ and as higher reliability
in information extraction (measured by $\pe$) is required.
Figure \ref{fig:ecrsVsM} reveals that for most cases, there is a pretty close
alignment between the energy per bundle (as reflected in the energy contrast ratio $\ecrs$)
desired for discovery and for information recovery.
This is not a surprising result, since both functions require the detection of an
individual energy bundle, and (for the chosen reliability parameters) a detection reliability that
is comparable.

Another difference between a periodic beacon and an FSK signal is the number
of locations where energy bundles may occur.
In the case of an M-ary FSK signal, that number of locations is $M$ times larger,
and this will increase the false alarm probability $\pfa$ relative to a beacon.
In particular, this increases the value of $N$, and
a larger $N$ will increase somewhat the number of observations $L$ required to achieve
a given $\{ \pfa , \, \pd \}$.
Thus discovery of an M-ary FSK signal needs to be examined more fully than
simply comparing the detection reliability for a single energy bundle.
The first question is how many observations are required to reliably detect the
FSK signal with an average power $\power$ that is sufficiently high to reliably extract
information from that signal after it is discovered.
Since $\power \propto \rate$,
signals with a larger information rate $\rate$ will be easier to discover (in the sense of requiring a smaller $L$).
Figure \ref{fig:discoveryDataVsL} shows how 
$L$ is related to the miss probability $(1-\pd)$
for a single-carrier signal with the highest possible $\rate$ ($\delta = 1$) for a single carrier.
For example, on the order of $L \approx 30$ suffices for  $M = 2^{20}$.
This is a much smaller $L$ than often required to discover a power-optimized beacon
in general (see Figure \ref{fig:optLvsSNRlowDuty}),
because this large-$\rate$ signal has relatively high average power $\power$ and, commensurate
with that, a $\delta = $ 100\% duty factor.

\incfig
	{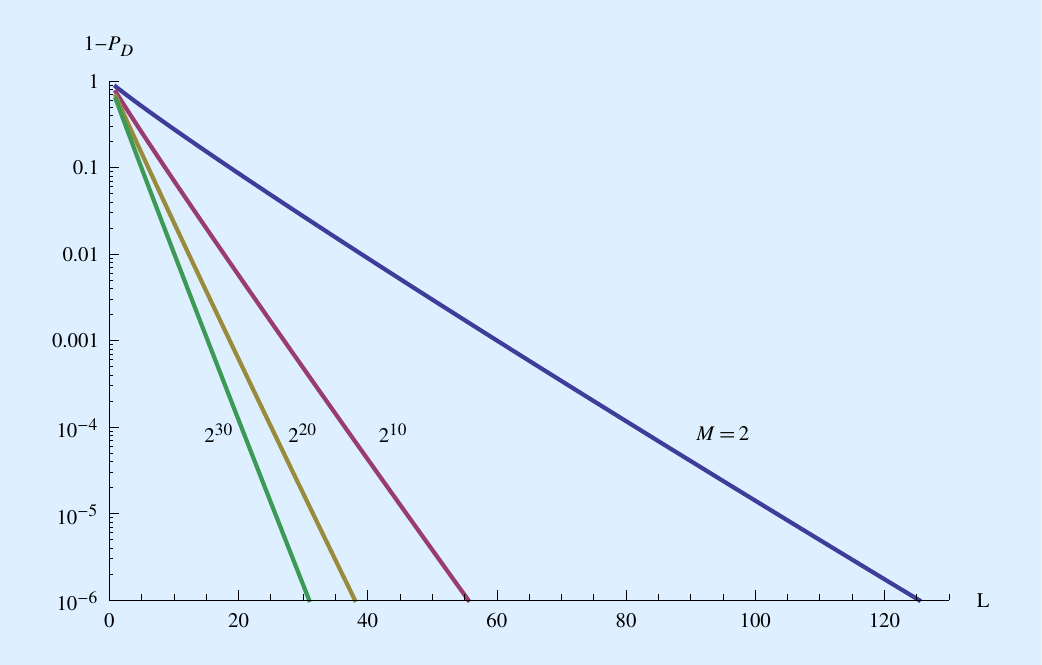}
	{1}
	{
For an information-bearing M-ary FSK or PPM signal, 
a log plot of the detection miss probability $(1 - \pd)$ vs the
number of observations $L$ for $M = 2$, $2^{10}$, $2^{20}$, and $2^{30}$.
The highest power $\power$ and rate $\rate$ for a single carrier is assumed
($\delta_{p} = 1$), an observation duration of $T_{o} = 2 \, T$ ($\delta_{o} = 1$),
a false alarm probability of $\pfa = 10^{-12}$, and a degrees of freedom $K = 1000$.
The average power $\power$ is chosen to be sufficiently high to extract information 
from the signal at the threshold of outage
with error probability $\pe = 10^{-4}$.
	}

The dependence on $M$ in Figure \ref{fig:discoveryDataVsL} predicts that
 signals become easier to discover as $M$ increases.
This reflects a tradeoff between larger energy per bundle $\energy_{h}$
(to maintain a fixed energy per bit) as $M$ increases,
against a larger $N \propto M$ to account for the larger number of locations to look for energy bundles.
This tradeoff is explored in more detail in Figure \ref{fig:discoveryDataVsM}.
Increasing $M$ can decrease the miss probability
over a range of one to two orders of magnitude.
The net effect of increasing $M$ is favorable for discovery as well as for power efficiency,
but of course also increases and average power $\power \propto \log_{2} M$.

\incfig
	{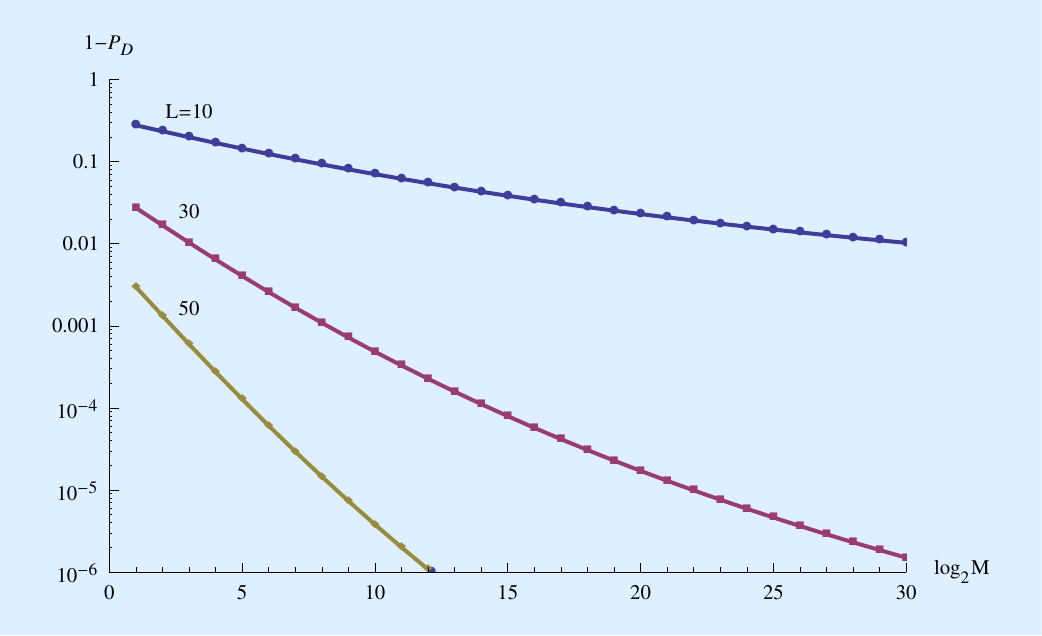}
	{1}
	{
A log plot of the miss probability $(1 - \pd)$ vs the
number of bits per codeword $\log_{2} M$ for several values of $L$.
The other conditions are the same as in Figure \ref{fig:discoveryDataVsL}.
	}

The preceding applies to an information rate $\rate$ (and hence power $\power$) that is the
highest possible for a single carrier.
The impact of reducing $\rate$
is shown in Figure \ref{fig:discoveryDataVsDelta}.
Reducing the rate by reducing $\delta_{p}$ renders discovery less reliable, which is not surprising
since the average signal power $\power$ is being reduced in proportion to 
$\rate$ and $\delta_{p}$.
This illustrates how the transmitter can reduce its average transmitted power $\power$,
paying two prices for this reduction.
First, there is a proportional reduction in information rate $\rate$, 
since at constant power efficiency $\power \propto \rate$.
Second, there
is an increase in the number of observations $L$ necessary in the receiver
that is inversely proportional to the power,
since by the total effective energy hypothesis $L \propto 1/\power$.
Ordinarily one would think that reducing $\rate$ would reduce the required
resources.
That is a valid assertion for the transmitter, but with regard to the discovery
phase and resources required in the receiver the effect is just the opposite.

\incfig
	{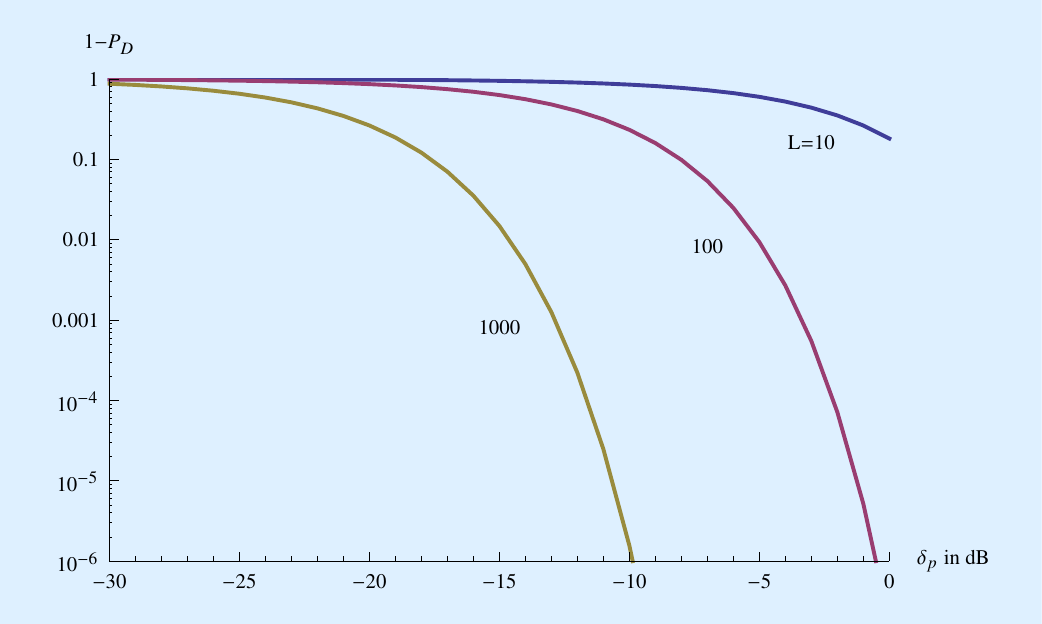}
	{1}
	{
A log plot of the miss probability $(1 - \pd)$ vs the duty factor $\delta_{p}$ in dB
for several values of $L = 80$.
As $\delta_{p}$ is reduced, the information rate also decreases as $\rate \propto \delta_{p}$.
Thus, a 10 dB reduction in $\delta_{p}$ reduces $\rate$ by one order of magnitude.
It is assumed that $M = 2^{20}$, and the remaining conditions are the same as in Figure \ref{fig:discoveryDataVsL}.
Smaller $\delta$ corresponds to smaller average signal power $\power \propto \delta_{p}$.
	}

Another tradeoff that should be kept in mind is between the transmitted energy per bundle
and the receive antenna collection area in order to maintain a large enough $\energy_{h}$
in the receiver for reliable discovery and communication.
What we have found is that the required $\energy_{h}$ is comparable for
reliable discovery and for reliable communication, and thus it is appropriate to use approximately
the same receive antenna collection area at both stages.
Relatively modest adjustments may be appropriate, depending on the specific
objectives for reliability.
In particular, increasing the antenna size in moving from discovery to communication
can dramatically reduce the probability of error $\pe$, which in turn makes the underlying message
semantics easier to interpret.
It may be relatively easy for the receiver designer to acquire additional resources at this stage,
since the existence of a information-bearing signal of interstellar origin has been established.

\section{Power-optimizing information-bearing signals for discovery}
\label{sec:discoveryoptimized}

The discovery of periodic beacons and information-bearing signals have been considered
separately, but with different emphasis.
Beacons were examined from the perspective of achieving the lowest receive power $\power$
consistent with a fixed number of observations $L$.
Since a beacon has no purpose other than to be discovered, 
the sole design criterion is to
minimize its power $\power$ for a fixed $L$ or 
equivalently minimize $L$ for a fixed $\power$.
Information-bearing signals have the additional objective of reliably conveying
information, and a higher information rate $\rate$ is generally desirable if the resulting
$\power$ is affordable.
Thus, aside from discovery there are two other criteria to consider,
the achievable information rate
$\rate$ and an energy per bit $\energybit$ that is sufficient for reliable
recovery of the information bits according to some error probability objective $\pe$.
For any  achievable power efficiency $\powere$, 
we know that $\power \propto \rate$ and there
is no lower limit on either $\power$ or $\rate$.
Thus, at any target $\power$, there is both a beacon and an
information-bearing signal with that average receive power.
At a small $\power$, the information-bearing signal will have
a commensurately low $\rate$.
The only remaining question is whether the information-bearing signal is as
discoverable as the beacon, and in particular whether it requires the same or
nearly the same $L$.

The manner in which we achieve a low $\power \propto \rate$
as described in Section \ref{sec:adjrateandpower} is compatible with efficient discovery.
Just as with the periodic beacon, $\power \propto \rate$ is
reduced by reducing the rate at which codewords are transmitted
or equivalently reducing the duty factor $\delta$.
Consistent with the bundle energy hypothesis for a beacon,
in order to preserve low $\pe$ the energy per bit and hence the
energy per bundle $\energy_h$ is kept fixed as the duty factor is reduced.
As shown in Figure \ref{fig:ecrsVsM}, that $\energy_h$ is comparable to a beacon,
depending on $M$ and the objective for $\pe$.

Starting with a periodic beacon,
turning it into an information-bearing signal requires the fairly minor step
of randomizing somewhat the location of energy bundles in either time or frequency.
That randomization is keyed to the information being conveyed.
This randomization need not affect the discovery of the signal much, since
that discovery focuses on detecting energy bundles at random observation times,
so some randomness in the time or frequency of the bundles
added to the randomness of the observation times has little added impact.
The randomness will inevitably increase the number of potential locations
in time and/or frequency where an energy bundle can occur, but in the
context of a multichannel spectral analysis this should not affect the
false alarm probability.

\subsection{M-ary FSK}

The results of Chapter \ref{sec:infosignals} can be reexamined from
a discovery perspective.
Suppose that an M-ary FSK signal is constrained to have an
energy contrast ratio of $\ecrs = 15$ dB, which is a good choice for
discovery,
and suppose that the outage strategy is used to deal with scintillation.
What is the reliability of information extraction from such a signal,
as measured by probability of error $\pe$?
For M-ary FSK the energy contrast ratio $\ecrb$ required
to achieve reliable information extraction is given by \eqref{eq:ecrbVsPe}
and plotted in Figure \ref{fig:ecrbVsM}.
However, $\ecrb = \ecrs / \log_{2} M$ is expressed in terms of the bundle energy per bit
in an $M$-bit codeword.
Expressing it instead in terms of the energy contrast ratio $\ecrs$ for the constituent energy bundles,
\begin{equation}
\log_{2} \pe \le \log_{2} M - \frac{\ecrs}{2 \, \log_{e} 2 } - 2 \,.
\end{equation}
Thus $\log \pe$ is a linear function of $\log_{2} M$, the number of bits per codeword.\footnote{
Note that $\pe > 1$ for sufficiently large $M$.
This is possible because this is merely an upper bound on $\pe$,
and the bound is tight only when $\pe \ll 1$.}
The error probability is plotted in Figure \ref{fig:peVsMpowopt}.
The error probability increases as $M$ increases because the energy contrast ratio
per bit $\ecrb$ grows smaller.
A physical explanation for this behavior is that as $M$ increases there are more opportunities for noise in
one or more of the $(M-1)$ locations that lack energy bundles to mimic an energy bundle.
The primary conclusion from Figure \ref{fig:peVsMpowopt} is that information can be extracted with high reliability
from an M-ary FSK signal that has been power-optimized for discovery.
This is especially true in the regime $\log_2 M \le 10$.
For larger $M$, the required value of $\ecrs$ grows larger, and this
makes discovery more reliable, but it also becomes somewhat inefficient for discovery
since smaller values of $\ecrs$ would suffice.

\incfig
	{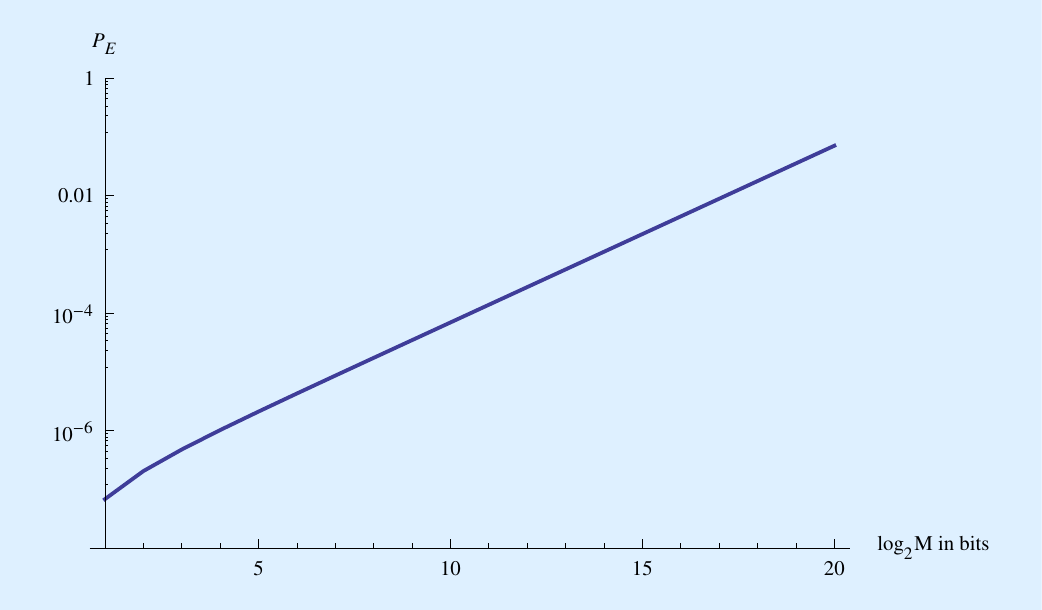}
	{.8}
	{A log plot of the probability of bit error $\pe$ vs the number of bits per codeword $\log_{2} M$
	between one and 20 for an energy contrast ratio $\ecrs = 15$ dB.
	This applies to
	an M-ary FSK information-bearing signal that has been power-optimized for discovery
	for low-power case where the duty factor $\delta <1$.
	The assumption $\ecrs = 15$ dB applies
	approximately independently of the average power $\power$.
	At higher powers, $\pe$ will decrease rapidly with average power $\power$ for any value of $M$.}

In conclusion, in the context of power-efficient design, there is little reason to consider
transmitting a beacon, since an information-bearing signal at the same power can be discovered as effectively at the receiver.
However, if we want to match the low power that we might seek in a beacon, that information-bearing
signal will have a low information rate commensurate with that low power.
Further, a low information-rate $\rate$ signal will require a commensurately larger
number of observations.
Low $\rate$ information-bearing signals are more difficult to discover
than high $\rate$, because they have lower average power $\power$.

\subsection{M-ary FSK with time diversity}
\label{sec:discoveryWdiversity}

The bundle energy hypothesis tells us that efficient discovery occurs when
the energy contrast ratio $\ecrs$ of individual bundles is sufficiently large
to enable their individual detection with a reasonably high probability $\pd$.
In the absence of time diversity, using an outage strategy
an M-ary FSK information-bearing signal 
with sufficient average power to reliably extract its information consists of
individual energy bundles that meet this criterion.
There is an issue, however, when time diversity is added to the mix.
A variation of the total energy hypothesis confirmed in Section \ref{sec:reliabilityFSKdiversity}
predicts that the required energy of each bundle decreases as the number of
time diversity copies of that bundle is increased, or $\energy_h \propto 1/J$.
This is because the diversity combining in the receiver ''gathers up'' all this
energy in $J$ copies, maintaining detection reliability.

This allowed reduction in $\energy_h$ creates a challenge for discovery.
To fully benefit from this energy in $J$ copies, the discovery strategy would
have to include diversity combining, which would introduce a whole
new dimension to the discovery search.
This added dimension would substantially increase the processing
requirements, and more importantly increases the false alarm probability.
Is there an alternative?
Yes, there is. Although it entails a penalty in average power,
the transmitter can deliberately use a larger energy
than necessary (from a information extraction perspective) for one or more of the
energy bundle copies, choosing that energy to be sufficiently large
for those individual high-energy bundles to be efficiently detected.
While this higher energy actually aids information extraction as well as discovery,
this is inefficient because the higher-energy bundles overlap periods of
lower received signal flux due to scintillation.
The result is that this approach will increase the penalty in average power
relative to the fundamental limit.

To quantify this effect,
assume that the total degree of time diversity is $J = J_0 + J_1$,
where $J_0$ is the number of replicas of the FSK energy bundle
with energy contrast ratio $\ecrszero$, and $J_1$ is the number with
ratio $\ecrsone$.
Assume that $\ecrszero > \ecrsone$ because $\ecrszero$ is
chosen to enable efficient discovery of those $J_0$ energy bundles,
while $\ecrsone$ is chosen so that the total energy
to sufficient to insure subsequent reliable extraction of information from the signal
using time diversity combining.
It follows that the energy contrast ratio per bit $\ecrb$ is
\begin{equation}
\ecrb = \frac{J_0 \, \ecrszero + J_1 \, \ecrsone}{\log_2 M} \,.
\end{equation}
An upper bound on the bit error probability $\pe$ is determined
in Appendix \ref{sec:pefsktd}, and plotted in Figure \ref{fig:diversityForDiscovery}
for one typical value of $M$ and $J$.
Choosing one energy bundle (out of $J$) to have sufficiently large energy
to permit reliable discovery creates a fraction of a dB penalty in
average power.
This penalty is multiplied as a larger number of energy bundles are
chosen to have the larger power, thereby reducing the number
of observations necessary in the receiver for reliable detection.

The penalty relative to the fundamental limit is fairly modest.
If time diversity is employed to counter scintillation while
minimizing average power, there is little alternative to this approach.
It can be thought of as embedding a beacon within an information-bearing signal,
although this perspective is not strictly valid since the large-energy bundles
participate fully in conveying information as well as attracting the
attention of the receiver.
These high-energy bundles also occur at data-dependent random frequencies,
rather than a fixed frequency as in the periodic beacon.

\incfig
	{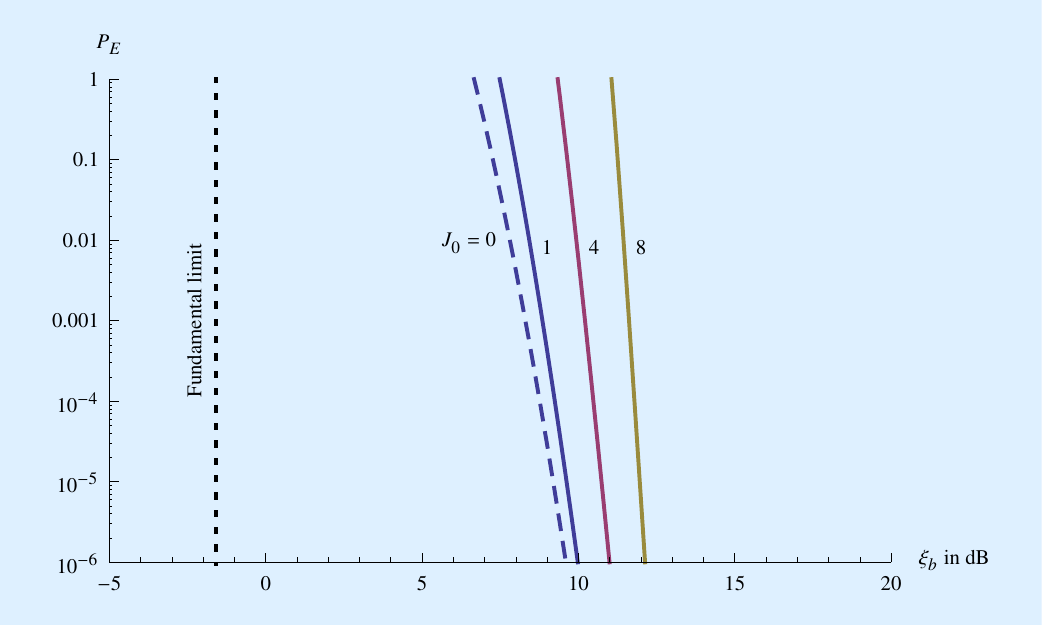}
	{1}
	{
	A comparison of an upper bound on the
	bit error probability $\pe$ for M-ary FSK with time diversity,
	not optimized for discovery (dashed curve) and
	optimized for discovery (solid curves) with some penalty
	in average power relative to the fundamental limit.
	In all cases $M = 2^{20}$ and $J = 50$, and $\pe$ is shown
	as a function of the energy contrast ratio per bit $\ecrb$.
	The dashed curve assumes all energy bundles have identical
	energy contrast ratio $\ecrs$, or $J_0 = 0$ and $J_1 = J$.
	The solid curves
	show the penally in $\ecrb$
	for making $J_0 = 1$, $4$ and $8$, where $J_0$ is the number of energy
	packets out of $J = 2^7$ that have sufficiently large 
	energy ($\ecrszero = 15$ db) to be more easily discovered.
	The value of $\ecrsone$ is inferred from the 
	values of $\{ \energybit , \, J_0, \, J_1, \, \ecrszero \}$.
	}

\textbox
	{Some suggestions for SETI programs}
	{
A question faced by SETI observation programs is whether to search for
beacons, which are information-free but serve to attract attention to the transmitting civilization,
or an information bearing signal.
Existing SETI programs have been influenced by terrestrial communications,
where power efficiency is sacrificed in favor of spectral efficiency.
In that domain, the classic argument that a beacon is worthwhile because it is
''obvious'' or ''easier to discover'' has some merit, because generally achieving
spectral efficiency requires complicated techniques that may generally interfere
with discovery and are likely to be very difficult to reverse engineer after discovery.
\boxbreak
The power-efficient regime has several conditions
favorable to interstellar communication.
One is that power-efficient signals are structured around on-off
signals, where detecting the presence of absence of an energy bundle is the
underlying receiver functionality.
This functionality is extremely simple, and the same signal processing
techniques apply to beacons and information-bearing signals.
As a result, it is simple to seek both types of signals simultaneously.
There is no ''simplicity'' argument in favor of a beacon.
A beacon has no advantage
over a low information rate signal in terms of the receiver resources
required to discover a signal at a given average receive power.
It is desirable to always embed information 
 in a signal at no extra cost to transmitter or receiver.
Our main issue would be whether to choose a very low information rate necessary to
achieve the lowest receive power, which also requires 
a larger number of observations to discover,
or a higher information rate associated with a higher power that requires less resources to discover.
\boxbreak
Current and past SETI searches for the Cyclops beacon will only be successful
if the transmitter designer has freely expended energy consumption.
There is an argument to be made for this, for example if that designer has available
inexpensive or free energy resources.
However, it seems unlikely that such a designer would fail to embed information,
since similar energy expenditure can achieve high information rates at no cost to discovery.
The sparsity in frequency required for
embedded information also compromises current and past discovery strategies,
irrespective of average power.
\boxbreak
In searching for more complex and wider-bandwidth signals, a big concern
expressed by researchers has been the interstellar impairments and the receiver processing
required to deal with those impairments.
In power-efficient design this is not a concern, because
a basic principle is to choose a transmit signal
that avoids these impairments altogether.
There is a resulting increase in bandwidth, so using \emph{more} bandwidth results in
\emph{less} impairment due to the interstellar medium, the opposite of a common assumption among
SETI researchers.
During discovery, no processing associated with interstellar impairments other than noise is required.
On the other hand, in the power-efficient regime and at lower average power
levels the receiver does have to
deploy the processing necessary to perform multiple independent observations at the
same location and at or near the same carrier frequency,
as well as pattern recognition algorithms to interpret the results.
}

\section{Discovery search strategies}

General strategies for discovery of power-efficient
information-bearing signals are now discussed.
These strategies apply with little modification to beacons
that have been power-optimized for discovery.

\subsection{Energy detection in discovery}

During the energy detection phase, energy bundles
that may be present are detected using a
multichannel spectral estimation combined with
filtering matched to $h(t)$.
In the degenerate case of single-frequency energy bundles,
where $h(t)$ is a constant,
the matched filtering comes for free as an outgrowth of
spectral estimation.
In all cases, the phase of the matched filter output is ignored,
and the magnitude-squared of the output is used as an estimate
of the energy of the bundle.
If the noise energy is known this value can be applied to a
threshold, or if there is some uncertainty in the noise level
the output can be normalized by the total energy within the
bandwidth and time duration of $h(t)$,
yielding an estimate of the energy contrast ratio rather than energy.
This detection algorithm is described and analyzed in Chapter \ref{sec:detection}.

In contrast to current SETI searches, to have hope of detecting
power-efficient signals it is necessary to revisit the same location
(line of sight and frequency) numerous times.
If after $L$ such observations no energy bundles have been detected,
this location can be permanently abandoned.
The larger $L$, the lower the average signal power $\power$ that will be reliably detected.
Thus, as a given location is revisited, in effect the sensitivity of the
detection (as measured by $\power$) improves.
If current technology does not permit a very large $L$, that is perfectly
acceptable because further observations in that location can be
performed at some future time.
The key requirement is to maintain a database of locations, and
for each location the cumulative number of observations that have been performed,
as well as relevant details of those observations such as observation times
and candidate choices of $h(t)$ including $T$ and $B$.
That database will serve as guidance to future searches as to
the least explored regions.
The transmitter should maintain the same carrier frequency, so that
the receiver maintains the greatest flexibility in conducting multiple observations
over the long term.

If it is assumed that the transmitter designer is motivated to minimize
its energy consumption through the use of low duty factor,
the resulting signals are unlikely to be detected in
existing SETI searches.
These existing searches do not realize the tens, hundreds, or thousands
of independent observations necessary to achieve a reasonable detection
probability for a power-efficient information-bearing signal or beacon
that has been power-optimized for discovery.
In addition, these existing searches typically assume short-term persistence of
the signal in an attempt to identify ''waterfall'' patterns (a carrier frequency
changing linearly with time) and to help rule out false alarms.
This existing approach is ineffective
for intermittent power-efficient signals of the type we have considered here.
It is possible that past SETI searches have in fact detected legitimate signals of interest
but incorrectly categorized them as false alarms because
they did not take account of the relatively low probability of detection
of a single energy bundle nor the possible intermittent nature of the signal.
These existing searches can, however, benefit
future searches for power-efficient signals by maintaining a location database
which captures for future analysis all detected energy bundles that have been classified
(perhaps incorrectly) as false alarm events.
These locations are promising starting locations for any future pattern-recognition analysis
while searching for power-efficient signals.

\subsection{Pattern recognition in discovery}

The previous analysis has presumed that detection of a beacon
or an information-bearing signal requires only the presence of a
single energy bundle in $L$ observations of duration $T_{o}$.
In the numerical examples the false alarm probability was set
fairly low, at $\pfa = 10^{-12}$ or $10^{-8}$.
Nevertheless, this form of detection will never suffice to declare
the discovery of a signal.
As discussed in Section \ref{sec:discoveryrel}, these values of
$\pfa$ are presumably still much larger than the a priori probability
of a signal being present, and thus false alarms will dominate
these detections leading to a low confidence in an actual discovery.

Once an energy bundle is discovered, it is appropriate to
continue observing in that same location (line of sight and frequency)
to see if other energy bundles are detected.
In essence, conditional on a successful detection
 the number of observations $L$ can be increased.
In view of the possibility of a time-varying frequency shift due to
uncompensated acceleration and due to information-bearing signals
that utilize more than one frequency (such as the M-ary FSK used
in our examples), these observations should be conducted
over a range of frequencies in the vicinity of detected energy bundles.

Should these continued observations detect more energy bundles,
the receiver can extend the observations at that location,
and switch to a pattern recognition mode in which discernible
patterns of energy are sought.
A recognizable and repeatable pattern presents even more significant
evidence of the presence of a signal.
The confidence in this evidence can be quantified by calculating the
probability that a given pattern represents a false alarm.
The location database should capture and document all observations that have been
made in the promising location and nearby frequencies,
as well as all detected energy bundles.

A multichannel spectral analysis may cover a much wider range of
frequencies than what would be expected for a single information-bearing signal,
or especially for a beacon which may well be narrowband.
Thus, the most efficient way to follow up on detected energy bundles may be to
pass observations to a separate processing that analyzes promising locations
over a narrower range of frequencies.

The location database can be analyzed separately, and repeatedly.
Statistical (rather than deterministic) pattern recognition is required since at low average power
levels the probability of detection of an individual energy bundle
is expected to be small (on the order of $\pd \approx 0.5$ at the limits of average power), and also because the
location of bundles is likely to be associated with a random stream of information bits.
Ideally a single database will be shared by all SETI observers worldwide,
so that the database represents the worldwide cumulative observations.
The database offers an attractive target for
future research and statistical analysis.
The literature and algorithms for statistical pattern recognition
appropriate for discovery of power-efficient signals is also an
interesting topic for future research,
and this research can benefit future analysis of the database.

\section{Summary}

The general issue of the discovery of artificial signals of interstellar origin
has been addressed from the perspective of minimizing energy consumption
in transmitter and receiver, and tradeoffs between resource allocation in the
transmitter and receiver.
It is argued that a signal structured around the repeated transmission of 
energy bundles, as has been proven to achieve the fundamental limit on
power efficiency in information transfer, is also meritorious for discovery.
Beacons, which transmit deterministic patterns of energy bundles, and
information-bearing signals, which transmit random patterns, have been studied.
Some general hypotheses about the amount of energy in the bundles
and the amount of energy overlapping the receiver's observation time
have been proposed and verified through more detailed statistical modeling. 

Our conclusion is that for any average received power, there is an information-bearing
signal that is as easy to discover as a beacon at that power.
A beacon is neither
easier to discover nor does it offer any other advantage such as lower average power
for a given reliability of discovery.
Beacons with a small average power are particularly compelling because they demand
lower energy resources at the transmitter.
Since average power is proportional to information rate for a power-efficient information-bearing signal,
an information-bearing signal with the same average power
as such a beacon has a low information rate.
An information-bearing signal with low information rate is more difficult
to discovery than one with a higher information rate, because its
average power is lower.
Any civilization seeking low energy consumption is likely to transmit an
information-bearing signal with a low information rate rather than a beacon
with the same average power.

In contrast to present and past SETI observation programs,
it has been concluded that multiple observations are needed to
reliably discover signals designed according to power-efficient principles.
Information-bearing signals at the highest rate for single carrier
and with sufficiently high power to extract information reliably require
on the order of tens of observations in the initial energy bundle detection phase.
As the power (and hence rate) of the information-bearing signal is decreased,
the number of required observations increases.
The regime of up to 1000 observations has been explored, but there is
no lower limit on the rate and power that can be discovered, and thus
no limit on the number of observations that is required.
The observation program at each carrier frequency should be viewed as
a long-term endeavor, with additional observations as permitted by
advancing technology and budgetary resources.

While existing SETI searches will fail to discover signals designed according to
energy-conserving principles,
these searches can be extended to this class of signals with fairly straightforward modifications.
Aside from multiple observations, strategies for separating true energy bundles
from false alarms in the pattern recognition phase have been discussed.
This is an area deserving of additional research.

An important conclusion is that a single worldwide database capturing
the location of all observations and all detection events as well as their
characteristics should be maintained.
Detection events in current searches that are characterized as false
alarms may in rare cases in fact represent actual low-average-power
beacons or information=bearing signals, but this can only be ascertained
by analyzing a longer-term set of observations in the same and
nearby locations.


\chapter{Detection}
\label{sec:detection}

Detection of the location of energy bundles is the first step in the
recovery of information from a power-efficient information-bearing signal
and detection of the presence of absence of an energy bundle
is the first stage of discovery.
There are clear distinctions between these two types of detection.
During communication, for assumed channel coding
(based on M-ary FSK), it is known that an energy bundle is present,
but the only question is its location from among $M$ possibilities.
During discovery, the role of detection is to determine whether an
energy bundle is present or not, absent any prior knowledge.
In both cases, the energy in a bundle is affected in an unknown
way by scintillation, although in the case of communication it will
be possible to track the state of scintillation with some degree of accuracy.

For both discovery and communication the best pre-processing for the
detection algorithm is a matched filter.
This is the best way to counter the AWGN, regardless of whether
scintillation is a factor or not.
In this chapter, the argument that the matched filter is the best pre-processing
will be made for the discovery case.
In that case, the appropriate metric of reliability of detection is
the false alarm probability $\pfa$ and the detection probability $\pd$.
In this chapter those two probabilities are calculated for a single
matched filter output applied to a threshold in the presence of scintillation.
Only the strong scattering case will be considered, for which the
scintillation follows Rayleigh (rather than Rician) statistics.
The argument that the matched filter is also the best front end for communication
follows similar lines and is omitted.

Another possible source of uncertainty is the noise power spectral density $N_0$.
For our assumed channel coding (based on M-ary FSK),
this is not an issue since the maximum algorithm of \eqref{eq:maxAlgorithm}
automatically adjusts to changing conditions on both $\energy_h$ and $N_0$,
taking advantage of the prior knowledge that exactly one energy bundle is present.
For discovery, on the other hand, knowledge of $N_0$ is important
in the interpretation of the energy estimate at the matched filter output.
In response to this challenge,
we will show how the detection can be modified (from the simple
matched filter followed by threshold)
to be sensitive to only the relative size of the signal level and the
noise level.
We will also show how choosing a larger degrees of freedom $K = B \, T$ for $h(t)$ is
helpful in this, giving an opportunity to implicitly and more accurately estimate the signal
and noise levels in the course of deciding whether signal is present or not.

\section{Matched filter}
\label{sec:matchedfilter}

The use of a matched filter as the basis for detection of a single energy bundle will now be justified.

\subsection{Continuous-time channel model}

To justify the use of a matched filter, a continuous-time model for the interstellar channel
at the baseband input to the receiver is needed.
Assuming that the carrier waveform $h(t)$ falls in an interstellar coherence hole (ICH),
the only impairments are AWGN and scintillation.
Neglecting any impairments other than these, the reception is
\begin{equation}
\label{eq:ctreception}
R(t) = 
\begin{cases}
\sqrt{\energy_{h}} \, S \, h(t) + N(t) \, & \ \text{energy bundle present} \\
N(t) \, & \ \text{noise only}
\end{cases}
\end{equation}
The random process $N(t)$ is AWGN with power spectral density $N_{o}$
and $S$ is a zero-mean complex-valued Gaussian scintillation
random variable with variance $\expec{| S |^{2}} = \vs = 1$.
The real- and imaginary parts of $S$ are identically distributed and statistically
independent, or equivalently $\expec{S^{2}} = 0$.
The magnitude-squared of the scintillation is the received flux $\flux = |S|^2$.

The matched filter followed by a sampler reduces
the continuous-time signal in \eqref{eq:ctreception} to a single complex-valued random variable $Z$,
given in the cross-correlation formulation by
\begin{equation}
\label{eq:mfcorr}
Z = \int_{0}^{T} R(t) \, \cc h (t) \, dt \,.
\end{equation}
Notice that the correlation interval is the same as the assumed duration $t \in [0,T]$ of $h(t)$.
We also assume that $h(t)$ is bandlimited to the frequency interval $f \in [0,B]$,
in which case \eqref{eq:mfcorr} implicitly incorporates an idea lowpass filter
that rejects all ''out of band'' noise.

The optimality of the matched filter as a pre-processing to reduce the continuous-time
reception of \eqref{eq:ctreception} to a single random variable for decision making
will now be made.
This optimality can be approached with varying degrees of sophistication,
but all roads lead to the same conclusion.
We follow two of these, motivated by the significant insight they
provide into the detection problem, and especially the way to deal
with not only an unknown signal magnitude (due to scintillation)
but also an unknown noise power $N_{0}$.
The following development is more intuitive than mathematical, 
so as to place understanding ahead of rigor.

\subsection{Sufficient statistic formulation}

The most powerful argument for using an matched filter as the pre-decision processing
is that the matched filter output of \eqref{eq:mfcorr} is a \emph{sufficient statistic} for reception \eqref{eq:ctreception}.
Roughly this argument says that the matched filter throws away much information, 
but none of that discarded information is statistically relevant to
the subsequent decision as to the presence or absence of an energy bundle.
It is simple to argue that the out of band noise is irrelevant, because it is
(a) not changed in any way if an energy bundle is present or absent and
(b) it is statistically independent of the in-band noise (because the noise is assumed to be white)
and thus offers no information about the in-band noise that could aid the detection.

Most of the remaining in-band noise is also irrelevant, namely that which is
not correlated with $h(t)$, although it is more work to formulate this argument.
As discussed in Section \ref{sec:channelmodeldof}, $h(t)$ has degrees of freedom $K = B \, T$.
Similarly, both the noise $N(t)$ and reception $R(t)$, following ideal bandlimiting to $f \in [0,T]$
and over time interval $t \in [0,T]$, have $K$ degrees of freedom.
The mathematical interpretation of this is that there exists a set of $K$ orthonormal
basis functions  $\{ g_{k} (t) , 1 \le k \le K \}$, the weighted summation of which
can represent either $h(t)$ or, after bandlimiting and time-limiting, $N(t)$ and $R(t)$.
By orthonormal, we mean
\begin{equation}
\int_{0}^{T} g_{m} (t) \, \cc g_{n} (t) \, dt = \delta_{m-n} \,.
\end{equation}
It doesn't matter what $\{ g_{k} (t) , 1 \le k \le K \}$ are chosen, so for
convenience choose as the first basis function $g_{1} (t) = h(t)$,
and the remaining basis functions arbitrarily to fill out a complete orthonormal basis.
Then the reception $R(t)$ (restricted to bandwidth $B$ and time duration $T$)
can be represented by a set of $K$ coefficients w.r.t. this basis,
\begin{equation}
\label{eq:receptionexpansion}
R_{k} = \int_{0}^{T} R (t) \, \cc g_{m} (t) \, dt
= \begin{cases}
\sqrt{\recenergy} \, S  + M_{1} \,, & k=1 \\
M_{k} \,, & 2 \le k \le K \,.
\end{cases}
\end{equation}
Note that $R_{1} = Z$, the output of a filter matched to $h(t)$
sampled at time $t = 0$, and the remaining coordinates are noise-only.
The noise components $\{M_{k} , 1 \le k \le K \}$ are readily shown to
be statistically independent, and thus the
$\{R_{k} , \, 2 \le k \le K \}$ offers no information as to either $\recenergy$
or $M_{1}$ and can be ignored just like the out of band noise.

What is notable about \eqref{eq:receptionexpansion} is the 
 concentration of signal energy in only one out of the total of $K$ coordinates.
This concentration is the signature of the presence of signal $h(t)$ within the
bandwidth $B$ and time duration $T$, and will be evident as long as
$\recenergy \, | S |^2 \gg N_{0}$ no matter what the specific values of
$\recenergy$ and $N_{0}$.
This concentration can be observed without the need to explicitly compute the $\{R_{k} , 2 \le k \le K \}$.
Define the total in-band energy of the reception as $\energy_{r}$, then
\begin{equation}
\energy_{r} = \sum_{k=1}^{K} |R_{k} |^{2} = \int_{0}^{T} R(t) \, g(-t) \, dt
\end{equation}
where $g(t)$ is the impulse response of an ideal lowpass filter that rejects all out of band noise energy.
Then a decision variable $Q$ that measures the concentration of energy
would normalize the matched filter output by the total energy,
\begin{equation}
\label{eq:QR1K}
Q = \frac{|R_{1} |^{2}}{\energy_{r}} \,.
\end{equation}
The value of $Q$ is constrained to fall in $0 \le Q \le 1$,
with larger values of $Q$ signaling a concentration of
energy in the first coordinate,
and a value on the order of $Q \approx 1/K$ signaling
a uniform distribution of energy across all the coordinates.
$Q$ is a desirable decision variable
when the size of $N_{0}$ is unknown or uncertain.
$Q$ can serve as the basis of a decision as to whether $h(t)$ is present or
absent by comparing it to a threshold $0 < \lambda < 1$,
where $Q > \lambda$ results in the decision that a signal is present.

\subsection{Least-squares formulation}
\label{sec:least squaressolution}

A least-squares formulation gives another perspective on the $Q$ of \eqref{eq:QR1K}
as a decision variable, and also shows that $Q > \lambda$ is a reasonable criterion
to decide the presence of an energy bundle even when
the AWGN assumption is not valid.
An example of when it is clearly not valid is the presence of radio-frequency interference (RFI).

Suppose the received signal with a known unit-energy waveform $\{ h(t ) , \, 0 \le t \le T \}$ is expected 
but in the reception $r(t)$ it is delayed by some unknown $\tau$,
multiplied by some unknown complex-valued factor $z$
representing an unknown magnitude and phase, and
corrupted by noise, interference, etc,
\begin{equation}
\label{eq:receptionGeneral}
r(t) = z \cdot h(t - \tau) + \text{noise} + \text{interference} + \dots \,.
\end{equation}
The least squares solution seeks to produce an estimate $\hat z$ of
the complex-vaued amplitude $z$ and an estimate $\hat \tau$ of time offset  $\tau$
by minimizing a distance metric between $r(t)$ and $\hat z \cdot h(t - \hat \tau)$.
The metric we chose in the least squares criterion is the energy in the error,
\begin{equation}
\label{eq:leastsquares}
\varepsilon (z,\tau ) =
 \int_{\tau} ^{\tau  + T} {\left| \, r(t) - z \cdot h(t - \tau) \, \right|^2 } \, dt \,.
\end{equation}
Our strategy is to determine, for each $\tau$, the value $z = \hat z$ that minimizes
$\varepsilon (z,\tau )$.
This eliminates one of the unknowns, leaving a resulting least squares error 
$\varepsilon_{\min} (\tau ) = \varepsilon (\hat z,\tau )$
that depends only on $\tau$.

It is shown in Appendix \ref{sec:lscriterion} that the resulting estimate and residual error are
\begin{align}
\label{eq:leastsquaressolnZ}
&\hat z (\tau) =  \int_{\tau} ^{\tau  + T} \, r(t) \, \cc h (t - \tau) \, dt = r(\tau) \convolve \cc h (- \tau)
 \\
\label{eq:leastsquaressolnE}
&\varepsilon_{\min} (\tau ) =   \varepsilon (0,\tau ) - | \hat z (\tau) |^{2} 
\ge 0
\end{align}
where $\convolve$ is the convolution operator.
The estimate $\hat z (\tau)$ as a function of time can be generated by
a filter with impulse response $\cc h(-t)$, which is by definition a filter
 matched to $h(t)$.
The size of $0 \le \varepsilon_{\min} (\tau ) \le  \varepsilon (0,\tau )$
is a direct indication of the goodness of the fit between $h(t)$ and $r(t)$.
For example, if $\varepsilon_{\min} (\tau ) \approx 0$, or equivalently
$| \hat z (\tau) |^{2} \approx \varepsilon (0,\tau )$, it follows that $r(t) \approx \hat z \, h(t - \tau)$.

A convenient way to measure the goodness of the fit is to calculate the decision variable
\begin{equation}
\label{eq:normdv}
Q =  \frac{\ | \hat z (\tau) |^{2}}{ \varepsilon (0,\tau )} \,.
\end{equation}
Then $Q$ is normalized to range $0 \le Q \le 1$, where the
value $Q=1$ indicates that $r(t)$ is totally aligned with $h(t)$
and $Q = 0$ indicates that $r(t)$ is orthogonal to $h(t)$.
Although the notation is different, \eqref{eq:normdv} and \eqref{eq:QR1K}
are mathematically identical definitions of $Q$ at $\tau = 0$.

\subsection{Detection performance}

A detector based on applying $Q$ to a threshold $0< \lambda <1$
does not need knowledge of either $\energy_{h}$ or $N_{0}$ or the
energy contrast ratio $\ecrs$,
which was defined in \eqref{eq:ecrs}.
We will now show that it does the reasonable thing, which is
respond to the size of $\ecrs$ (rather than $\energy_{h}$ or $N_{0}$ individually),
or in other words it is sensitive to the energy contrast ratio but not the
absolute level of the energy of a bundle or the energy of the noise.
Depending on threshold $\lambda$,
it will declare signal when $\ecrs$ is larger (bundle energy is larger than noise energy)
and will declare noise only when $\ecrs$ is smaller (bundle energy is smaller than noise energy)
or zero (there is no energy bundle).

The analysis proceeds by determining
 $\pfa$ and $\pd$ parameterized by $\lambda$, and
then eliminating $\lambda$ to find the direct relationship between $\pfa$ and $\pd$.
This analysis is performed in Appendix \ref{sec:discprob} for
strong scattering,\footnote{
For weak scattering such simple results as \eqref{eq:pdvspfa} are unattainable,
but it can be confirmed that discovery becomes easier (see Appendix \ref{sec:discoveryrician})
due to the presence of a specular component of flux.}
 finding that
\begin{equation}
\label{eq:pdvspfa}
\displaystyle
\pd = \left( 
\frac{\ecrs + 1}{\ecrs+ \pfa^{\, \frac{1}{ 1-K}}} \,
\right)^{K-1}
\,.
\end{equation}
Note that $\recenergy$ and $N_{0}$ enter only through their
ratio in $\ecrs$, so the detection algorithm automatically adjusts to
unknown values for $\recenergy$ and $N_{0}$, responding only to their ratio.
The detection algorithm implicitly estimates $N_{0}$ by examining the noise
within the ICH that falls in the $K-1$ coordinates that are absent signal,
and comparing the matched filter output (which is the first coordinate) to that noise estimate.
The detection probability is plotted for a fixed $\pfa$ in Figure \ref{fig:PdvsK}
as a function of $K$ and $\ecrs$.
The increasing $\pd$ with $K$ is because the estimate of $N_{0}$ built into \eqref{eq:normdv}
has increasing accuracy in estimating $N_{0}$ as there are more coordinates available to form the estimate.
The higher the actual energy contrast ratio $\ecrs$, the more likely the energy bundle is to be detected.
In Chapter \ref{sec:discovery} it was found that $\ecrs \approx 15$ dB is the goal for the
power optimization of discovery, and at that level $K=20$ is about the point of diminishing returns,
in the sense that further increases in $\pd$ are small. 

\incfig
	{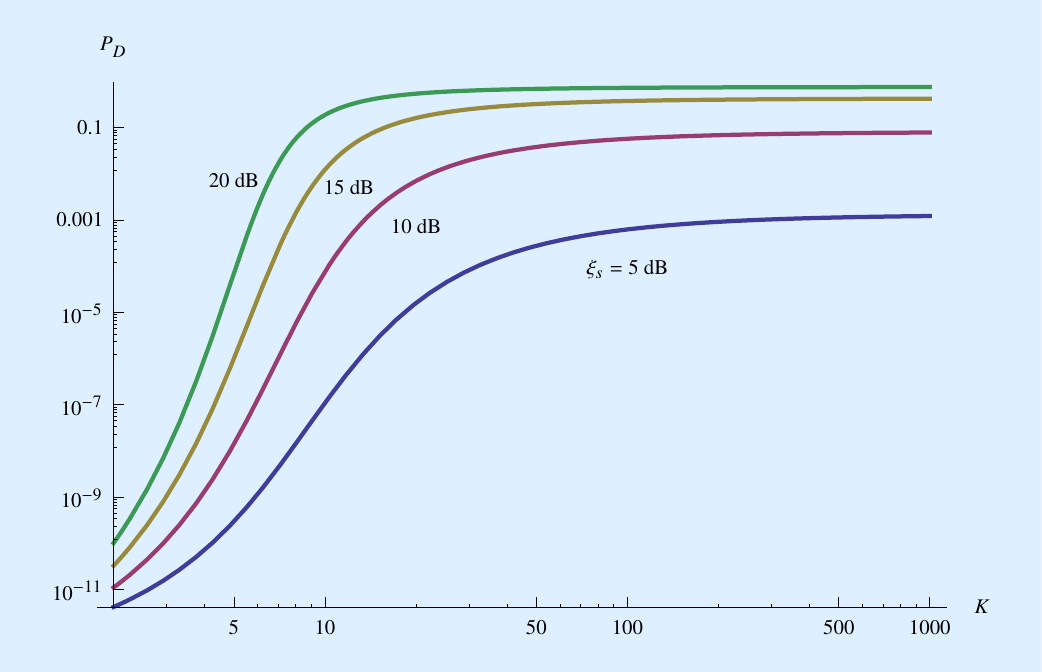}
	{.8}
	{A log-log plot of detection probability $\pd$ for an actual energy contrast ratio of $\ecrs =$ 5, 10, 15, and 20 dB.
	The horizontal axis is the degrees of freedom $K$ for a fixed
	$\pfa = 10^{-12}$.}

For any $\ecrs$, $\pd$ increases monotonically with $K$, approaching an asymptote of
\begin{equation}
\label{eq:asympPfaVsPd}
\pd \underset{K \to \infty}{\longrightarrow} \pfa^{\frac{1}{ 1+\ecrs}} <1 \,.
\end{equation}
Even $K \to \infty$ cannot achieve $\pd \approx 1$, except in the case where $\ecrs \to \infty$ or $\pfa \approx 1$.
This is the reason that multiple observations were found necessary in Chapter \ref{sec:discovery}
 to approach reliable detection.
 This can be blamed on scintillation, which sometimes results in periods with a low signal strength,
and during such ''outage'' periods the detector invariably declares ''noise'',
thereby failing to declare ''signal''.
Figure \ref{fig:PdvsPfa} illustrates how $\pd$ is monotonically increasing in $\pfa$ for this $K \to \infty$ asymptote.

\incfig
	{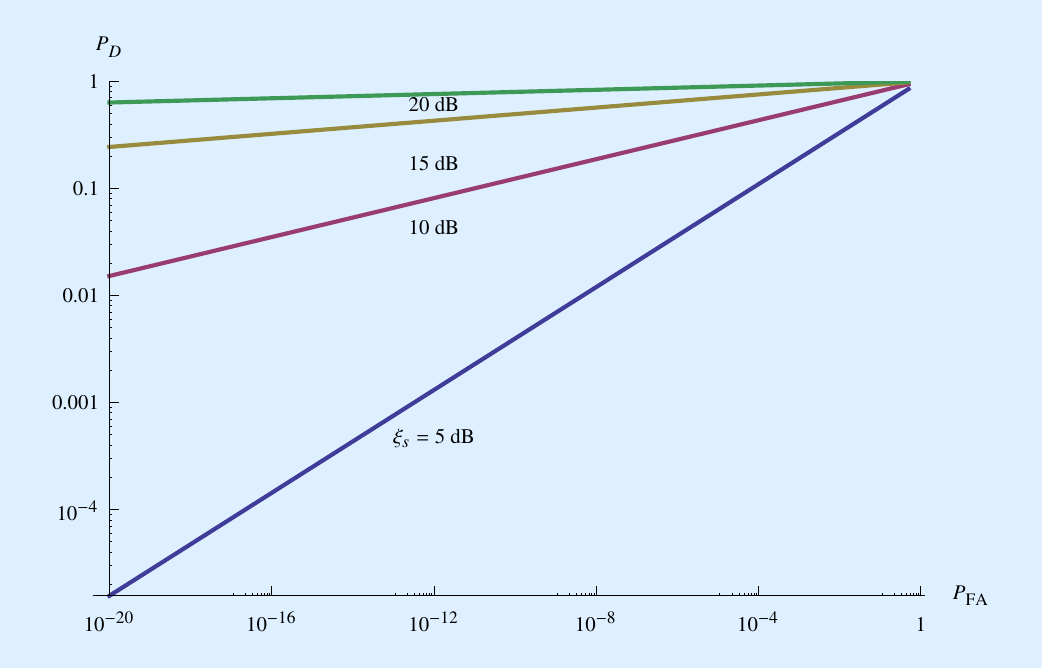}
	{.8}
	{A log-log plot of detection probability $\pd$ at the asymptote $K \to \infty$
	 for an actual energy contrast ratio of $\ecrs =$ 5, 10, 15, and 20 dB.
	The horizontal axis is the false alarm probability $\pfa$.}

\subsection{Known noise power spectral density}

If the noise power spectral density $N_{0}$ is known, then decision variable
\eqref{eq:QR1K} can be replaced by $Q = |R_{1}|^{2}$.
Because $N_{0}$ is known, the threshold $\lambda$ can be chosen
to control $\pfa$ based on that knowledge, and the resulting $\pd$ is determined unambiguously.
Not surprisingly, the relationship for $\pd$ vs $\pfa$ becomes  (Appendix \ref{sec:discprob})
the same as the $K \to \infty$ asymptotic result of \eqref{eq:asympPfaVsPd}.
This is because the implicit
estimate of $N_{0}$ built into \eqref{eq:QR1K} becomes increasingly accurate
(by the law of large numbers)
as $K \to \infty$.

The lesson to take away from Figure \ref{fig:PdvsK} is not that small $K$
precludes reliable detection.
Rather, the conclusion is that if $K$ is small then the receiver needs to
know $N_{0}$ in order to property set the threshold $\lambda$,
 or if prior knowledge is absent the receiver must estimate $N_{0}$ by some other means.

\textbox
	{Are sinusoidal signals special?}
	{
\begin{changea}
In the presence of AWGN, it has been shown that the
tradeoff between $\pfa$ and $\pd$ does not depend on the
waveform of $h(t)$, nor on any of its parameters like $B$ and $T$.
This includes a continuous-wave sinusoidal signal, which corresponds to
a d.c. baseband waveform $h(t) = \sqrt{1/T}$,
which has a small time-bandwidth product $B T \approx 1$.
It also includes a spread-spectrum signal with a large $B T \gg 1$,
which will desirably have a noise-like character \citeref{211}.
Any deterministic waveform $h(t)$ with unit energy is distinguishable
from AWGN (using matched filtering in the receiver) to the same degree.
As an aid to intuition on this significant point,
Appendix \ref{sec:cointossing} describes an analogy to noise generated by random coin tosses,
which is easier to understand but displays the same characteristics.
\boxbreak
The sinusoidal choice, which is advocated by the Cyclops report \citeref{142},
has special properties that suggest it should be included
as one of the options in any discovery search.
In order to implement a matched filter, $h(t)$ must be known or guessed.
Some would say that a constant $h(t)$ is obvious, and in support of this
it is the only choice that has no parameters left unspecified.
It is completely specified by its one degree of freedom $B T \approx 1$,
and this is an amplitude and phase parameter that does not need to be known or guessed.
\boxbreak
While there is a compelling argument for using a noise-like waveform
from the perspective of RFI immunity, such a waveform does have many more
degrees of freedom and thus many more parameters to guess.
As a practical manner, it is necessary to guess the algorithm by which
those parameters are generated rather than the parameters themselves.
In addition, for
$B T \gg 1$ a discovery search is forced to search over different values of $B$
or different values of $T$ (fortunately not both) with a fine granularity.
Fortunately the range of values can be narrowed substantially based
on physical parameters observable by both transmitter and receiver
and the resulting coherence time and coherence bandwidth.
A natural choice is the coherence time $T_{c}$, since regardless of the choice of
$B$ a transmitter designer wishing to minimize peak transmitted power will
choose the largest $T$ possible consistent with $T \le T_{c}$.
\end{changea}
	}

\section{Search for an energy bundle}
\label{sec:bundlesearch}

The basic building block for discovery is the filter matched (matched filter) to
waveform $h(t)$ defined in \eqref{eq:simplemf}.
To achieve high discovery sensitivity,
the receiver must guess the waveform $h(t)$ in use, 
or it may be able to search for a set of possible $h(t)$ waveforms
as long as the number of distinct possibilities is relatively small.
However, even making this guess successfully,
there are several degrees of uncertainty.
The actual waveform as observed at the receiver will be of the form
$h( \beta ( t - \tau) ) e^{\,i\,2 \pi f_1 t}$,
where the three uncertain parameters are $\{ \beta , \tau , f_1 \}$.
The parameter $\beta$ is the \emph{time dilation}, $\tau$ is the \emph{starting time},
and $f_{1}$ is the \emph{carrier frequency offset}.
The receiver must search over some range of these parameters,
because there is no possible way to infer them precisely.

The time dilation parameter $\beta$ reflects uncertainty in the
time scaling of the signal.
Aligning the signal with an ICH can
reduce (although not eliminate) this uncertainty.
For a reception $r(t)$, during discovery \eqref{eq:mfcorr} is replaced by
\begin{equation}
\label{eq:searchmatched filter}
Z(\tau, \beta , f_{1}) = \int R(t) \, \cc h ( \beta ( t - \tau ) )  e^{-\,i\,2 \pi f_1 t} \,,
\end{equation}
which calculates the matched filter for different values of $\tau$.
During the search the integral representing the matched filter must be repeated for different values of 
$\beta$ and $f_{1}$.
The baseband processing will typically operate with a signal of bandwidth
$J B$ for some large value of $J$, and use the fast Fourier transform
to simultaneously calculate a bank of matched filter outputs (representing different $f_{1}$) over time (representing different $\tau$).

One question is what granularity is necessary in searching these
three parameters, this relating directly to the needed processing resources.
The granularity of $\tau$ is, by the sampling theorem, $1/B$.
As discussed in Appendix \ref{sec:searchgranulatiry},
the required granularity of $\beta$ is on the order of $1/K$, and the
granularity of $f_{1}$ is on the order of $1/T$.
Thus, increasing either $K$ (to counter RFI, for example) or $T$ (to reduce peak transmitted power)
results in a finer granularity of search and increases the required processing resources.

\section{Choice of $h(t)$}

Any discussion of how to choose $h(t)$ has been deliberately deferred to last
to emphasize that with respect to all the concerns addressed thus far,
including discovery and communication in the presence of interstellar impairments,
the only properties of $h(t)$ that matter are its energy (assumed to be unity),
bandwidth $B$, and time duration $T$.
In particular, the waveform representing $h(t)$ does not matter.
Our design is tailored to a specific line of sight and $f_{c}$, which in turn influences the size of the ICH.
This places a restriction on 
$B \le B_{c}$ and $T \le T_{c}$, and $K = B T$
for the chosen waveform $h(t)$,
but within these constraints any $B$ and $T$ can be chosen.
All of our results depend on using the matched filter (matched filter) for detection,
and thus it is important to
choose an $h(t)$ that is relatively easy for the receiver to guess.

\subsection{Radio-frequency interference}
\label{sec:RFI}

A consideration that has not been discussed thus far is radio-frequency interference (RFI),
which is a serious and growing problem for terrestrially based radio telescopes.
The choice of $h(t)$ does have a substantial effect on RFI immunity, in the sense that
the response of the matched filter to the RFI affects the detection reliability.
If after bandlimiting to $B$ the RFI waveform is $r(t)$, then the response of the matched filter is
\begin{equation}
\int_{0}^{T} r(t) \, \cc h (t) \, dt \,.
\end{equation}
This value, which represents an unwanted bias in signal before application to
the detection threshold,
depends on both $r(t)$ and $h(t)$.
Analysis of this undesired interaction was covered in a previous paper \citeref{211}.
As is well known in terrestrial communications, the choice of an $h(t)$ with
some properties known as \emph{spread spectrum} result in a much more robust matched filter
response to the RFI, in the sense that the probability of a large response is greatly
diminished and the response is much more uniform over a wide range of interferers.
These properties are (a) a larger value for the degrees of freedom $K$, and
(b) a choice of $h(t)$ that spreads the energy uniformly over the
full time duration $T$ and bandwidth $B$.
Such a waveform has characteristics that mimic bandlimited and time-limited
white noise.
 As an aid to intuition on this point,
 in Appendix \ref{sec:cointossing} the analogy to a discrete-time 
 version of $h(t)$ generated by a pseudo-random coin toss generator is discussed.

For an $h(t)$ that falls in the ICH for a given line of sight and carrier frequency,
there is a limit to how large $K$ can be.
The maximum $K$ was plotted in Figure \ref{fig:dofICH} for a typical set of physical parameters of the ICH,
and over a range of carrier frequencies.
Fortunately the available $K$ is typically large, on the order of $10^{\,3}$ to $10^{\,5}$,
which offers considerable opportunity for RFI immunity.
If the strategy is to maximize $K$ in the interest of achieving greater immunity to RFI in the receiver.
then it is favorable to choose $B = B_{c}$ and $T = T_{c}$.
Given the inevitable uncertainty in estimating $B_{c}$ and $T_{c}$, 
the smallest credible values of $B_{c}$ and $T_{c}$ should be used.

\subsection{Implicit coordination}

In the context of interstellar communication, there is another argument for using a spread-spectrum
signal of this type, one that may be stronger than its advantages for RFI immunity \citeref{211}.
Due to the lack of coordination between the transmitter and receiver designs it is
important to have fundamental principles to guide the design, principles that
address concerns likely to be shared by both designers and concerns that easily yield
an optimal mathematical solution that can be arrived at by both designers.
In Chapter \ref{sec:powerefficient} the five principles of power-efficient design were cited
for exactly this reason.
They all address aspects of a set of shared concerns, namely a desire for
a low received average power $\power$ in relation to the information rate $\rate$ and a relatively few
independent observations required for reliable discovery of the signal.
These principles led directly to conclusions as to the likely choice of channel coding
(M-ary FSK or M-ary PSK) and power reduction through reduced duty cycle, but unfortunately only offer insight into the
bandwidth and time duration of $h(t)$ but not any other aspect of its waveform.
Since RFI immunity speaks directly to the choice of waveform, it can act as an
implicit form of coordination between transmitter and receiver that complements the
power-efficiency principles.

Simple optimization principles for RFI immunity lead to the conclusion that $h(t)$ should
statistically resemble white Gaussian noise, appropriately bandlimited and time-limited.
Further information about appropriate choices of $h(t)$ follow from
 pragmatic considerations  \citeref{211}.
These include the feasibility of guessing $h(t)$ at the receiver
 (dictating algorithmically-generated pseudo-random waveforms),
 and Occam's razor (a four-level QPSK signal is the minimum-complexity signal that
 meets the randomness requirements).
 In the interest of minimizing the search space during the discovery process, 
 a critical requirement is embedding, in which a shorter-length
 pseudo-random sequence must be the precursor to a longer sequence,
 eliminating one entire dimension of discovery search for the receiver.
 Either a search over $B$ is required (using a fixed $K$ and hence a variable $T = K/B$),
 or a search over $T$ is required (using a fixed $K$ and hence a variable $B = K/T$).
 In both cases, detection sensitivity is improved if some different values of $K$ are
 tried, but a relatively coarse granularity is sufficient.
 The choice between these two options might be governed by the estimated
 uncertainty in $B_{c}$ vs that in $T_{c}$.

\subsection{Example}

The desired statistical properties of $h(t)$ for RFI immunity can be achieved using
a pseudorandom sequence generator, so that the receiver need only guess the
\emph{algorithm} employed by the transmitter. 
The statistical characteristics dictated by RFI immunity can be met, in the simplest way,
by a binary sequence that replicates the statistics of a sequence of independent coin tosses,
but which also has a well-defined starting state.
Fortunately such signals are mathematically easy to generate, for example by the binary expansion of
an irrational number.
In particular, a readily guessable choice is based on a pseudorandom
sequence generated by the binary expansion of $e$, $\pi$, or $\sqrt{2}$ \citeref{211}.

Consider, for example, an $h(t)$ design based on the binary expansion of $\pi$.
As described in \citeref{211}, $h(t)$ is generated as a quadrature phase-shift keying (QPSK) signal 
(the lowest complexity meeting the requirement for statistical independence of the
real- and imaginary parts), two successive bits from a binary expansion $\pi$ can be mapped
onto a complex amplitude chosen from four QPSK points at rate $B_{c}$.
The resulting waveform $h(t)$ takes the concrete form
\begin{equation}
h(t) = \frac{1}{\sqrt{K}} \, \sum_{k=0}^{K_{c}-1} c_{k} \, g(t - k / B_{c}) \,,
\end{equation}
where each $c_{k} = (2 b_{k} -1) + i \, (2 b_{k+1} - 1)$ is chosen using two successive bits $\{ b_{k}, b_{k+1} \}$
from the binary expansion of $\pi$.
For example, the first 30 digits of $\pi$ equal
\begin{equation}
\{ b_{k} , \, 1 \le k \le 30 \} =
\{1,1,0,0,1,0,0,1,0,0,0,0,1,1,1,1,1,1,0,1,1,0,1,0,1,0,1,0,0,0\}\,,
\end{equation}
resulting in
\begin{equation}
\{ c_{k} , \, 1 \le k \le 15 \} =
\{1+i,0,1,i,0,0,1+i,1+i,1+i,i,1,1,1,1,0\} \,.
\end{equation}
The \emph{chip waveform} $g(t)$ is a sampling interpolation function
(and thus $g(0)=1$) for sampling rate $B$,
and the match between transmitter and receiver in the choice of $g(t)$ is non-critical.
Due to the embedded properties of the $\{ b_{k} \}$ sequence, the receiver is permitted to use
a different $K= K_{r}$ from the transmitter with some loss of signal energy if $K_{r} < K_{c}$,
or some unnecessary noise if $K_{r} > K_{c}$.

As an alternative, the simple sine wave or pulse could be used.
A sinusoid is especially attractive if the receiver designers are expected to
be astronomers or experimental physicists rather than communication engineers,
as spectral analysis is a familiar form of scientific inquiry
in many sub-disciplines of astronomy, physics, and chemistry.
The use of a short pulse as described in \citeref{578} is not attractive in our
power-efficient design because it is extremely broadband and will thus violate
the bandwidth coherency requirement for power-efficient design.
It also requires very high peak power transmission.

\section{Summary}

The detection of a single energy bundle
in the presence of additive white Gaussian noise (AWGN) has been considered.
Using two different criteria of optimality, the matched filter is shown to be
the best receiver processing.
When there is uncertainty about the power spectral density of the
AWGN, it is shown that a simple normalization of the matched filter
output results in a detector that automatically responds to the
ratio of signal energy to noise energy, rather than either signal energy or noise energy alone.
The detection performance does not depend on any property of the
waveform carrying the energy bundle except its total energy.
However, that waveform does strongly influence the immunity to
interference.
The design of a waveform chosen to maximize that immunity is illustrated.


\chapter{Sources of incoherence}
\label{sec:incoherence}

In Chapter \ref{sec:channelmodel} the impairments introduced in the ISM
and due to source/observer motion were divided into those that can
be circumvented by reducing the time duration and bandwidth of
the waveform $h(t)$ serving as the carrier for an energy bundle,
and those that can't.
The impairments that remain are a fundamental impediment to
interstellar communication, and those were modeled in more detail
in Chapter \ref{sec:channelmodel}.
The present chapter will model the remaining impairments,
those that are circumvented by restricting duration and bandwidth.
The goal is twofold.
First, these impairments are modeled with a sufficient detail
to establish that they can be circumvented.
This does not require a detail or sophistication of modeling
typical of the astrophysics literature, since the goal is much more modest.
Second, the typical size of the coherence time $T_c$ and coherence bandwidth $B_c$
of the interstellar coherence hole (ICH) is estimated.
Of particular interest is establishing that the resulting degrees of freedom
satisfies $K_{c} = B_{c} T_{c} \gg 1$, which is a necessary condition for
stable interstellar communication utilizing the power-efficient design methodologies
covered earlier in this report.

\section{Types of incoherence}

When the impairments are considered individually, two distinctive forms of incoherence are encountered:
\begin{description}
\item[Frequency incoherence.]
In the presence of the impairment, if the overall effect is linear and time-invariant,
it can be equivalently modeled in the frequency domain.
In this case, if there is a frequency-dependent change in amplitude or phase
that causes a distortion of the signal, this is called frequency incoherence.
This effect can usually be mitigated by reducing the bandwidth of the signal
until the frequency response is approximately constant over the signal bandwidth
and this source of signal distortion becomes insignificant.
\item[Time incoherence.]
Ifn the presence of the impairment, the effect varies with time, then
frequency-domain modeling is not useful.
This time dependence is called time incoherence.
This effect can usually be mitigated by reducing the time duration of the signal
until the this source of distortion is rendered insignificant.
\end{description}
Most of the impairments encountered in interstellar communication cause either a frequency-dependent
or time-dependent phase shift.
They are lossless, and thus do not cause a magnitude shift in either time or frequency.
Our general criterion for coherence is when the magnitude of the phase change is
less than $\alpha \pi$ for a parameter $0 \le \alpha \le 1$.
Choosing $\alpha$ smaller makes the criterion more strict.
At the limits of incoherence, there is generally both time and
frequency incoherence with additive phases, so the maximum
value $\alpha = 0.5$ insures that the total phase shift is
at most $\pi$.
The appropriate value of $\alpha$ depends on the
use to which the ICH is put, and the impact of incoherence on that use.
In all numerical examples, the conservative choice of $\alpha = 0.1$ is assumed,
except in a couple cases where the uncertainty is greater and we assume a smaller value
$\alpha = 0.01$.
The coherence bandwidth $B_c$ and coherence time $T_c$
are determined by the most stringent of the conditions that arise across all the impairments.

\section{Plasma dispersion}
\label{sec:dispersion}

Although it approaches an ideal vacuum, the ISM contains an extremely low-density 
plasma consisting of hydrogen ions and free electrons 
arising from the presence of hydrogen gas and ionizing radiation from the stars \citeref{213}.
The charged particles create a conductive medium that influences radio waves 
propagating over long distances.
This results in plasma dispersion, a frequency-dependent
group delay due to the free electons' lossless absorption and wavelength-dependent delayed emission of photons.

Plasma dispersion causes a frequency-dependent group delay $\tau (f)$.
Let the transfer function at passband be
\begin{equation}
\label{eq:passbanddispersiontf}
G(f) = e^{\,i\, 2 \pi \, \phi (f)}
\end{equation}
so that there is a phase shift $\phi (f)$ with no effect on signal magnitude.
The group delay $\tau(f)$ and $\phi(f)$ are related by
\begin{equation}
\label{eq:delayFromPhase}
\tau (f) = - \frac{1}{2 \pi} \cdot  \frac{\partial \phi (f)}{\partial f} \,.
\end{equation}
The standard model for this group delay used in pulsar astronomy is \citeref{212}
\begin{equation}
\tau(f) = \frac{\dc \, \dm}{f^{2}}
\end{equation}
As illustrated in Figure \ref{fig:delayAndPhase},
$\tau(f)$ decreases monotonically (and approximately linearly) with $f$.
While \emph{dispersion constant}
$\dc \approx 4.15 \times 10^{15}$ 
depends only on known physical constants
(and has units of $\text{Hz}^2 \ \text{pc}^{-1} \ \text{cm}^{3} \ \text{s}$),
the \emph{dispersion measure} $\dm$
summarizes all that needs to be known about
the current interstellar ''weather'' in order to predict the group delay profile for a given
direction of arrival (and has units of $\text{cm}^{-3} \ \text{pc}$).
Typically $\dm$ ranges over about two orders of magnitude, from about 10 to about 1000,
depending on the columnar electron density along the line of sight.

\incfig
	{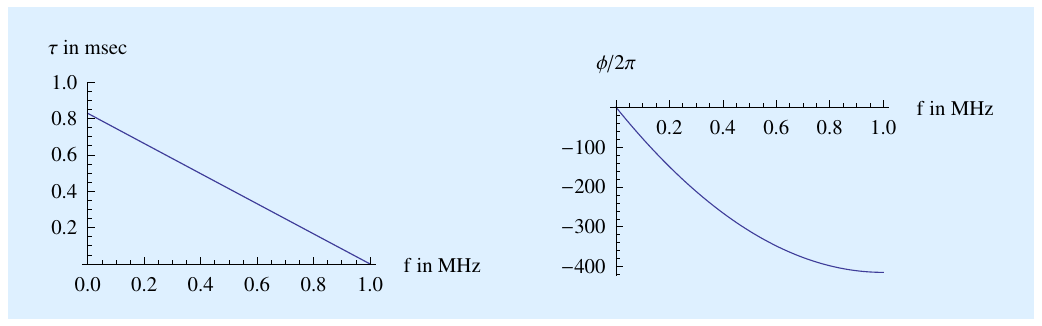}
	{1}
	{
	The dispersive group delay $ \tau(f)$ in msec
	(with $\tau_0$ chosen so that $\tau(B)=0$)  and resulting unwrapped phase
	$\phi(f)/2 \pi$ (with $\phi(f)$ in radians and $\phi_0=0$) 
	plotted against baseband frequency $f$ for
	$f_{c} = 1$ GHz, $B = 1$ MHz,
	and $\dm = 100$ $\text{cm}^{-3} \ \text{pc}$.
	The phase experiences more than 400 rotations across the bandwidth, and this
	phase variation will increase with 
	$\dm\uparrow$, $B\uparrow$, or $f_{c}\downarrow$.
	}

The ICH test is shown in Figure \ref{fig:dispersion}.
Over the frequency range $f \in [f_{c}, f_{c} + B]$, if $B$ is kept small
enough $\phi (f)$ can be modeled by a linear dependence on $f$,
\begin{equation}
\label{eq:appxphase}
\phi (f) \approx \phi (f_{c}) - \tau (f_{c}) \cdot f
\end{equation}
which corresponds to an unknown phase $\phi (f_{c})$ plus a
fixed (not frequency-dependent) group delay $\tau (f_{c})$.
If the deviation of the actual $\phi (f)$ from \eqref{eq:appxphase}
is less than $\alpha \, \pi$ radians over the entire range $f \in [f_{c}, f_{c} + B]$
then the distortion of $h(t)$ due to nonlinear phase can be neglected.

It is shown in Appendix \ref{sec:plasmaappx} that 
approximation \eqref{eq:appxphase} is accurate within $\alpha \, \pi$ radians when
\begin{equation}
\label{eq:dispersioncoherencebandwidth}
B <  B_{0} = \rho \, f_{c}^{\, 3/2}
\ \ \ \text{for} \ \ \ \ 
\rho
= \sqrt{\frac{\alpha}{2 \, \dc \, \dm}} \,.
\end{equation}
As illustrated in Figure \ref{fig:noDispersion}, $B_{0}$
decreases (dispersion becomes more severe) as $\dm$ increases
or $f_c$ decreases.
The value of $B_{0}$ varies over about four orders of magnitude for the parameters shown.
The line of sight ''interstellar weather conditions'' also influence $B_{0}$ through the $\sqrt{1/\dm}$ term in $\rho$.

\incfig
	{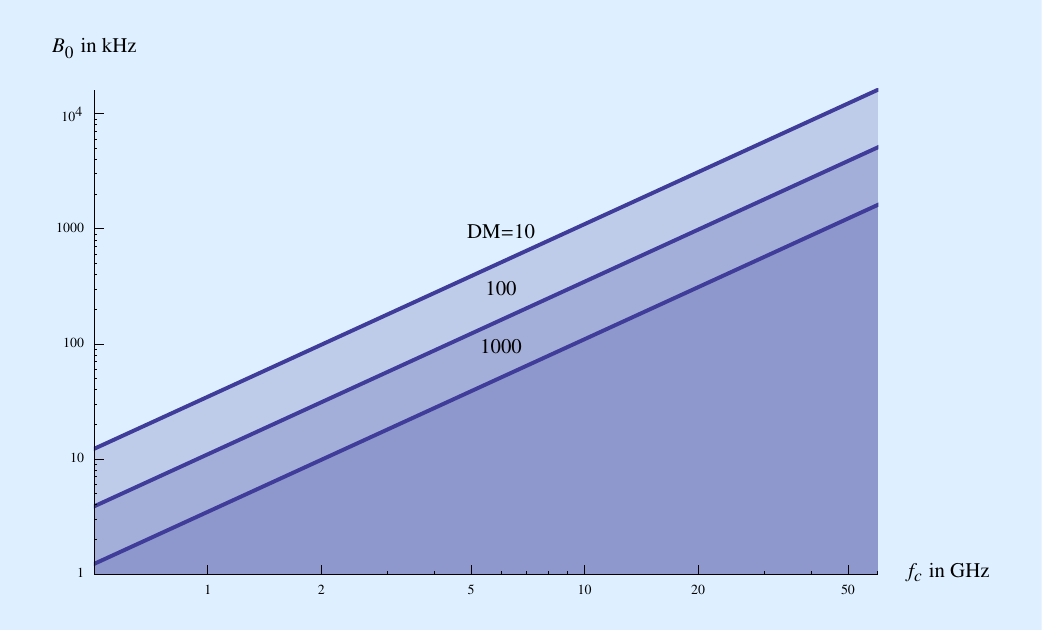}
	{1}
	{
	A log-log plot of the coherence bandwidth
	$B_{0}$ of \eqref{eq:dispersioncoherencebandwidth} in kHz 
	vs carrier frequency $f_c$ in GHz for low,
	moderate, and severe dispersion.
	A $\alpha = 0.1$ criterion for coherence is used.
	}

The baseband model for dispersion when \eqref{eq:dispersioncoherencebandwidth} is satisfied
is illustrated in Figure \ref{fig:dispersion}.
The result is the phase shift $\phi (f_{0})$ which is unknown and
added to unknown carrier phase $\theta$ together with a
baseband delay equal to the plasma dispersion delay at the carrier frequency
$\tau (f_{c})$.
In the time domain, this model becomes
\begin{equation}
|z(t)| \approx  \left| R_{h} ( t - \tau (f_{c}) ) \right| \,.
\end{equation}
The phase shift does not affect $|z(t)|$, and 
flat delay $\tau (f_{c})$ can be neglected since it is added to an already-large
and unknown propagation delay.
When $B > B_{0}$, however,
 the MF output becomes smeared over time and, as a result reduced in amplitude.
This is illustrated in Figure \ref{fig:IRvsB}, where the effect of dispersion on
$R_{h} (t)$ as in Figure \ref{fig:dispersion} is illustrated for different values of $B$.
Violating  \eqref{eq:dispersioncoherencebandwidth}
reduces the sensitivity of a detector that depends on the statistic $|z(0)|$.

\incxy
	[@C=10pt] 
	{dispersion}
	{
	A model for plasma dispersion within an ICH.
	At passband (top diagram) the effect is a frequency-dependent group delay
	$\tau (f)$.
	The bottom baseband model results when approximation \eqref{eq:appxphase} is applied,
	 the resulting passband filter is referred to baseband,
	 and then the order of the dispersion and matched filters is reversed.
	}
	{
 	h(t) \ar@{=>}[r] &
	*{\bigotimes} \ar@{=>}[r]^<<<<<{\displaystyle x(t)} &
	*+<1pc>[F-,]{
	\begin{matrix}
	\text{Group delay} \\
	\displaystyle \tau(f)
	\end{matrix}
	} \ar@{=>}[r]^>>>>>{\displaystyle y(t)} &
	*{\bigotimes} \ar@{=>}[r] &
	*+<1pc>[F-,]{\cc h(-  t)} \ar@{=>}[r] & 
	z(t)
	\\
	& \displaystyle e^{\, i \, ( 2 \pi \, f_{c} \, t + \theta)} \ar@{=>}[u] &&
	\displaystyle e^{ - \, i \, 2 \pi \, f_{c} \, t } \ar@{=>}[u] \\
	R_{h} (t) \ar@{=>}[rr] &&
	*+<1pc>[F-,]{
	\begin{matrix}
	\text{Phase} \\
	\displaystyle \theta + \phi(f_{c})
	\end{matrix}
	} \ar@{=>}[r]&
	*+<1pc>[F-,]{
	\begin{matrix}
	\text{Delay} \\
	\displaystyle \tau(f_{c})
	\end{matrix}
	} \ar@{=>}[rr]&&
	z(t)
	}

\incfig
 	{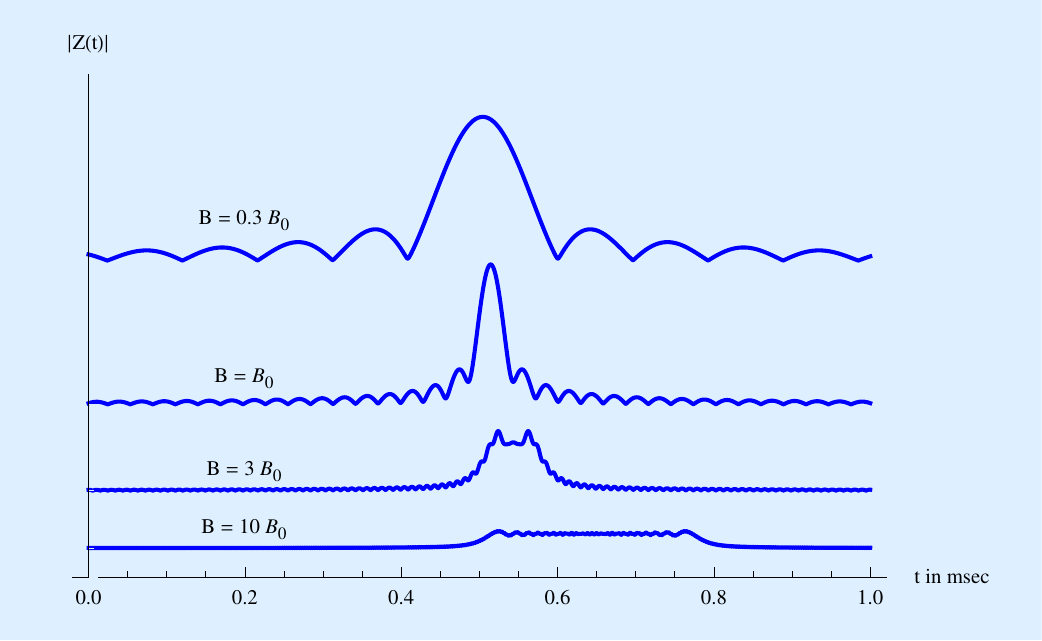}
	{.8}
	{The response $z(t)$ to input $R_{h} (t)$ in Figure \ref{fig:dispersion}
	shown for difference values of $B$.
	The parameters are the same as Figure \ref{fig:delayAndPhase}
	except the bandwidth is much smaller ($B$
	satisfying  \eqref{eq:dispersioncoherencebandwidth} is 34.7 kHz).
	Plotted is magnitude $| f(t) \convolve R_h (t) |$ (but not phase) for different
	values of $B$ ranging from $B = 0.3 \, B_0$ to $B = 10 \, B_0$.
	For $B < B_0$ the pulse broadens due to reduced bandwidth
	rather than dispersion.
	For $B > B_0$ the pulse is smeared out in time
	and reduced in magnitude.
	The effect of dispersion at $B = B_0$ in terms of smearing
	and magnitude reduction is visible but 
	has no material effect on $|z(0)|$.}

There is an alternative interpretation of \eqref{eq:dispersioncoherencebandwidth}
that  lends additional insight,
and follows the convention in wireless communications \citeref{574}
of focusing on group delay rather than phase.
The effect of
$B \gg B_0$ can be observed in the impulse response of the dispersion,
which is illustrated in Figure \ref{fig:impulseResponse} for the same
parameters as Figure \ref{fig:delayAndPhase}.
After removing the smallest fixed band-edge group delay $\tau (f_{c}+B)$,
the remaining group delay is approximately distributed
uniformly through $[ 0, \tau_s]$.
Thus the magnitude of the impulse response is approximately spread uniformly
over time $[ 0, \tau_s]$, where $\tau_{s}$ is called the \emph{delay spread},
defined as the range of group delays across the bandwidth
\begin{equation}
\label{eq:delayspread}
\tau_{s} = \tau (f_{c}) - \tau(f_{c} + B)
 \,.
\end{equation}
The phase, however, is chaotic and very sensitive to $\dm$.
Delay spread $\tau_{s}$ is a measure of how much the dispersion increases the time duration of a signal.
It is easily shown that
\begin{equation}
B \, \tau_s = \left( \frac{B}{B_0} \right)^2
\ \ \ \text{for} \ \ \ \
\alpha = 1
 \,.
\end{equation}
This is strikingly similar to \eqref{eq:delaycbw} because
$B_{0} \tau_{s} = 1$.
A coherence bandwidth defined as the inverse of the delay spread
is a fundamental characteristic of wireless propagation \citeref{574}.
	
\incfig
	{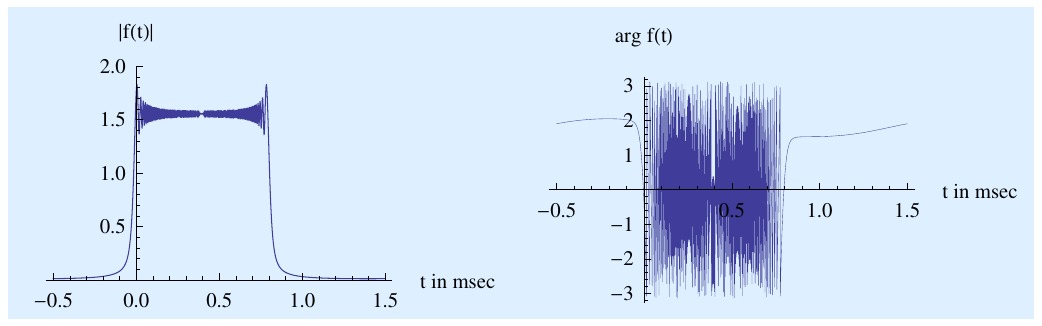}
	{1}
	{
	The magnitude and phase of the dispersive impulse response $f(t)$ for the
	 $F(f)$ illustrated in Figure \ref{fig:delayAndPhase}
	 after bandlimiting to $f \in [0,B]$.
	The magnitude is roughly uniformly spread between 0 and $0.83$ msec,
	corresponding to a delay spread $\tau_s =0.83$ from \eqref{eq:delayspread}.
	The oscillations and non-causality in $| f(t) |$ are caused by 
	ideal band limiting of the impulse response rather than dispersion,
	whereas the rapidly varying phase 
	(reflecting the 1 MHz bandwidth) is due primarily to dispersion.
	}

There are two related issues related to transmitter/receiver coordination
that are addressed in Appendix \ref{sec:equalization}.
First, based on pulsar observations both the transmitter and receiver can estimate $\dm$ along the line-of-sight.
However, both estimates have some uncertainty, both individually and jointly.
Second, it is possible for either the transmitter or the receiver to equalize
the dispersion, presumably for the lowest $\dm$ estimate.
Either may do this because they are motivated to increase the worst-case
coherence bandwidth $B_{0}$, which is dominated by the highest $\dm$ estimate
(possibly reduced by the equalization of the lowest $\dm$ estimate).
This has implications for the receiver strategy in estimating a value for $B_{0}$ to inform
its search strategy.

\section{Source/observer motion}
\label{sec:doppler}

Source-observer motion is potentially an important source
of impairment in the form of Doppler shifts in frequency,
time dilation, and fading.
Although elementary physics texts describe motion effects in terms of
a shift in frequency, this is a monochromatic approximation and the
effect on nonzero-bandwidth signals is a little more complicated.
This effect is now modeled, making the approximation that the
velocities are small relative to the speed of light so that
relativistic effects can be neglected.
All of the results of this section can be made more accurate by
including relativistic effects and more accurate modeling of
orbital dynamics, but no significant new insights would result.

\subsection{Time-varying distance and passband signals}

Assume that the component of distance $d_{\parallel} (t)$ between source and observer
in the direction of the line of sight is a function of time.
Generally we will consider only the differential variation in $d_{\parallel} (t)$, 
due for example to the orbital dynamics of a planet, and not include the
very large constant distance between two stelar systems.
This time-varying distance results in a time-varying group
delay $d_{\parallel} (t) / c$, where $c$ is the free-space speed of light.
The effect of this delay on a passband signal of the form of \eqref{sec:passbandsignal} is
\begin{equation}
y(t) = 
\Re \left\{
\exp \left\{ \,i\, 2 \pi \, f_{c} \left( t - \frac{d_{\parallel} (t)}{c} \right)  \right\}
\cdot h \left( t - \frac{d_{\parallel} (t)}{c} \right)
\right\}
\end{equation}
Thus, the equivalent complex-valued baseband signal is
\begin{equation}
\label{eq:dopplerbaseband}
\exp \left( - \,i\, 2 \pi \, \frac{d_{\parallel} (t)}{\lambda_{c}} \right)
\cdot h \left( t - \frac{d_{\parallel} (t)}{c} \right) \,.
\end{equation} 
The impact is a time variation in the energy bundle waveform,
and this can affect the time coherence.

\begin{changea}
There are two distinct effects to consider.
The first term in \eqref{eq:dopplerbaseband} is
a time-varying phase shift affecting $h(t)$ due to the effect of velocity on the carrier.
This phase shift can be neglected if $T$ is sufficiently small.
The total change in phase will be less than $\alpha \pi$ if
\begin{equation}
\label{eq:tccarrier}
\frac{\left|  d_{\parallel} (t+t_1) - d_{\parallel} (t) \right|}{\lambda_{c}} \ll \frac{\alpha}{2} 
\ \ \ \text{for} \ \ \ t_1 \le T
\,.
\end{equation}
Depending on the motion effects that affect $d_{\parallel} (t)$, this
can be turned into a criterion on $T$, and the result is a value for the time coherence due to motion.
Specifically time coherence is achieved when the distance change 
over $t \in [0,T]$ is small relative to one wavelength $\lambda_c$.

The second term in \eqref{eq:dopplerbaseband}
is a time-variation in the group delay experienced by the baseband waveform $h(t)$.
This is a kind of non-linear warping of the time axis as observed in the receiver.
Fortunately this effect can always be neglected because the bandwidth of $h(t)$
is never large enough to be able to resolve this non-linear warping effect.
To see this, assume that  \eqref{eq:tccarrier} is satisfied, in which case
the corresponding delay spread affecting $h(t)$ is
\begin{equation}
\tau_{s} = \max_{0 \le t_1 \le T} \ 
\frac{\left| d_{\parallel} (t+t_1) - d_{\parallel} (tT) \right|}{c} \ll \frac{\alpha}{2 \, f_{c}} 
\,.
\end{equation}
Since it will always be true that $B \ll f_{c}$,
it follows that $2 B \tau_{s} \ll \alpha \, B / f_{c} \ll 1$ and thus by the criterion of \eqref{eq:delaycbw}
this delay spread can be neglected.
In this case the baseband signal of \eqref{eq:dopplerbaseband} becomes
\begin{equation}
\exp \left( - \,i\, 2 \pi \, \frac{d_{\parallel} (t)}{\lambda_{c}} \right)
\cdot h (t) \,.
\end{equation} 

The effect of motion on the coherence time is thus always dominated by the criterion \eqref{eq:tccarrier},
and time-warping effects can safely be neglected.
Condition  \eqref{eq:tccarrier} will now be investigated for typical motion scenarios.
\end{changea}

\subsection{Constant velocity}

The simplest case is a constant velocity.
Assume that $d_{\parallel} (t) = v_{\parallel} t$, over the time interval $[0,T]$.
Then the baseband signal over that interval is
\begin{equation}
\exp \left( - \,i\, 2 \pi \, \frac{v_{\parallel}}{\lambda_{c}} \, t \right)
\cdot h(t)
\end{equation}
The effect is a shift in the carrier frequency by $v_{\parallel} / \lambda_{c}$  Hz
which is of no significance
since the receiver has no absolute reference
for the carrier frequency, which can be freely chosen by the transmitter.

On the other hand, suppose that two energy bundles experience different
velocities, $v_1$ and $v_2$. Then the frequency offsets are different for the two energy bundles.
Clearly this cannot be ignored when a receiver is, for example,
doing time diversity combining.
In this case it would be necessary to estimate the change in velocity with time.
These different velocities signal the presence of acceleration, which
will be caused by the solar system dynamics.

The foregoing applies to the component of velocity parallel to the line of sight $v_{\parallel}$.
In contrast,
the velocity transverse to the line of sight $v_{\bot}$ is the direct cause
of scintillation as considered in Section \ref{sec:fading}.

\subsection{Diurnal rotation and orbital motion}

While two solar systems may have a substantial relative velocity,
we have seen that this results in a carrier frequency offset and 
time dilation of $h(t)$ that cannot be distinguished from parameter choices
made by the transmitter.
Generally the only germane origin of \emph{variation}
in velocity is the diurnal rotation of a planet about its axis
and its orbital motion about its host star.
Stars are accelerating as well, for example due to the increasing rate of expansion of the
universe, but this effect can be neglected.
Of course both the transmitter and receiver will typically contribute to
these effects, unless either or both locates its equipment on a spacecraft
that is not subject to rotation and orbital motion.

To qualitatively understand the effect this has on the baseband signal,
the effects due to rotation and orbital motion at transmitter and at receiver
will be considered separately.
Also make the simplification that the resulting motion is circular and
the worst-case assumption that it falls in a plane containing the line of sight.
The time-variation of distance for a simple circular motion is
\begin{equation}
\label{eq:circulardist}
d_{\parallel} (t) = R \cdot \sin \left( 2 \pi \, \frac{t}{P} \right)
\end{equation}
where $R$ is the radius of the circular motion and $P$ is the period.
Approximate values for these parameters for Earth are give in Table \ref{tbl:ctaccel}.

Consider one energy bundle extending over time interval $t \in [ t_1, t_1 + T ]$.
Assuming the distance change is small over this time interval,
a Taylor series expansion with only lower-order terms is useful.
This expansion about $t = t_1$ yields
\begin{align}
d_{\parallel} (t) =&
R \,  \sin \left(\frac{2 \pi  t_1}{P}\right) +
\frac{2 \pi \, R \, (t-t_1) \, \cos \left(\frac{2 \pi t_1}{P}\right)}{P} -
\frac{2 \pi ^2 \, R \,  (t-t_1)^2 \, \sin \left(\frac{2 \pi t_1}{P}\right)}{P^{\,2}} \\
&-
\frac{4 \pi ^3 \, R (t-t_1)^3 \, \cos \left(\frac{2 \pi t_1}{P}\right)}{3 \, P^{\,3}} +
\, \dots \, + \,.
\end{align}
The zero-order fixed-distance term can be neglected.
The first-order term is a constant velocity, which can be neglected within
a single ICH.
Within an ICH, the most significant source of impairment is the acceleration term $(t-t_0)^{2}$.
This is potentially a significant source of time incoherence, but one that
can be rendered negligible by choosing
$T$ sufficiently small.
A maximum phase shift less than $\alpha \pi$ determines the maximum $T$ as
\begin{equation}
\label{eq:dopplercoherence}
\frac{2 \pi^{2} \, R \, T^{2}}{\lambda_{c} \, P^{2}} \ll \frac{\alpha}{2}
\ \ \ \text{or} \ \ \ \ 
T \ll \frac{P}{2 \pi} \cdot \sqrt{\, \frac{\alpha \, \lambda_{c}}{R} \, }
\end{equation}
From Table \ref{tbl:ctaccel},
the coherence time $T$ decreases as $f_{c}^{\,-1/2}$, and acceleration becomes
an increasingly significant phenomenon for larger $f_{c}$.
The resulting coherence times are plotted as a function of $f_{c}$ in Figure \ref{fig:coherenceTimeAccel}.

\doctable
	{\small}
	{ctaccel}
	{The approximate radius $R$ and period $P$ for Earth
	and the resulting coherence time $T_{c}$ and frequency shift $\Delta f_{c}$, where
	$f_{\text{GHz}}$ is the carrier frequency in GHz.
	Since many planets or moons in the Milky Way may have a smaller radius or shorter period,
	a safety factor of $\alpha = 0.01$ on the maximum phase shift is assumed
	in calculating $T$.
	}
	{l c c c}
	{
	\textbf{Case} & \textbf{Parameter} & \textbf{Coherence time} & \textbf{Frequency shift} \\
	\toprule
	\addlinespace[3pt]
	Diurnal rotation & $R = 6.39 \times 10^{6}$ m \\
	& $P = 8.6 \times 10^{4}$ sec & $T \ \text{sec} = 3 \, f_{\text{GHz}}^{\,- 1/2}$
	& $| \Delta f | \ \text{kHz} = 3 \, f_{\text{GHz}} $
	\\
	\cmidrule{1-4}
	Orbital motion & $R = 1.5 \times 10^{11}$ m \\
	& $P = 3.15 \times 10^{7}$ sec & $T \ \text{sec} = 7 \, f_{\text{GHz}}^{\,- 1/2}$
	& $| \Delta f |  \ \text{kHz}  = 200 \,f_{\text{GHz}}$
	\\
	\addlinespace[3pt]
	\bottomrule
	}
	
\incfig
	{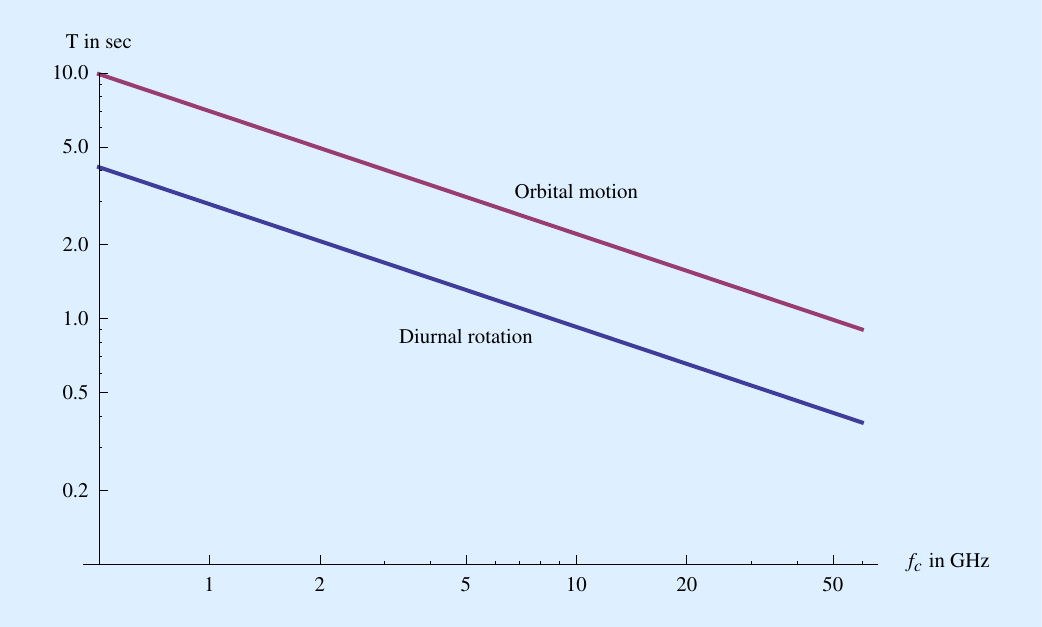}
	{1}
	{A log plot of the coherence time $T$ vs. the carrier frequency $f_c$.
	The dynamics of Earth are assumed with an additional safety factor of $\alpha = 0.01$
	using the formulas from Table \ref{tbl:ctaccel}.}

\subsection{Motion compensation}
\label{sec:motioncomp}

\begin{changea}
Assume that two stars are in motion with constant velocity
relative to one another, where that velocity may not be accurately known.
Under these conditions the only source of acceleration between
transmitter and receiver is the solar-system
dynamics of the transmitter and receiver relative to their local stars.
Measuring or calculating that acceleration does not require knowledge
of the relative velocity in the stars.
Assume that the transmitter and receiver are located at a fixed position
on a planet within the respective solar systems.
\end{changea}
The dynamics of each solar system is well known to the local inhabitants.
For example, on Earth the HORIZONS System from the
Jet Propulsion Laboratory provides real-time data on
the current status of the solar system \citeref{636}.
There is an opportunity for the transmitter and the
receiver to compensate for this known dynamics.
The full correction within an ICH and among ICH's can be
performed at baseband by multiplying by the conjugate of the time-varying
phase factor of \eqref{eq:dopplerbaseband},
\begin{equation}
\exp \left( \,i\, 2 \pi \, \frac{d_{\parallel} (t)}{\lambda_{c}} \right) \,.
\end{equation}
If the transmitter and receiver both do this independently, it
will eliminate the component of motion 
aligned with the line of sight (at least that source due to solar system dynamics)
as a source of incoherence
and relax the constraint on $T$ due to this physical phenomenon.
A simpler compensation would be to adjust for the 
frequency offset among different ICH's without compensating for the
time-varying phase shift within an ICH.
The remaining time incoherence within an ICH can be rendered negligible
by choosing $T$ in accordance with \eqref{eq:dopplercoherence}. 

\section{Scattering}
\label{sec:scattering}

Interstellar scattering has been investigated by a variety of techniques,
including relatively simple geometric optics \citerefthree{628}{629}{630}
and numerical evaluation of turbulent gas models \citeref{627}, always evaluating the models
by comparison to empirical observations of pulsar signals.
Here we employ the simplest possible methodology, geometrical optics 
and ray tracing assuming a one-dimensional
scattering screen.
A recommended tutorial on scattering that assumes a two-dimensional screen is \citeref{598}.
Of course reality is much more complicated, as the scattering medium is dispersed
through three-dimensional space, but three-dimensional modeling is necessarily 
based on computational approaches, which are less useful as a source of intuition,
and while this can improve the accuracy of numerical results it does not introduce any major new
issues or insights.

Our motivation in using a deliberately simple model is threefold.
First, the goal is to convey the mechanisms of scattering in the simplest fashion,
to emphasize intuition over accuracy.
Second, the limited objective is to convince the reader of the existence of the ICH
and convey an appreciation of the physical mechanisms that determine its size.
Third, the needs of interstellar communication are much different from astrophysics.
Astrophysics must work backwards to estimate the characteristics of
a natural source after propagation through the interstellar medium,
whereas communication seeks to design an artificial signal that circumvents
the impairments introduced in that propagation.
If this design is successful, the detailed modeling of those impairments is unnecessary
because they can be neglected.
Having said this, a serious effort at quantifying the size of the ICH
in the context of an interstellar communication system design
will want to draw upon the full suite of modeling and observations represented in the astrophysics literature,
rather than relying on the simple model illustrated here, in order to obtain the
most numerically accurate results.

\subsection{Geometry of scattering}

For the interstellar medium, where scales are vastly larger than one wavelength,
scattering can be qualitatively explained in terms of the simple ray model of
Figure \ref{fig:scattering-geometry}, where a thin one-dimensional scattering screen
is assumed \citeref{598}.
The actual ISM has variations in electron density, which manifest themselves as spatial
variations in the plasma dispersion effects.
In the thin screen model, these variations are represented by a variation in
group delay and phase shift with lateral distance $x$.
The variations in electron density are typically quite small, and thus have little effect on the
group delay of propagation through the screen, but do cause a substantial variation in phase shift.
Phase is very sensitive to small perturbations in electron density.

\incfig
	{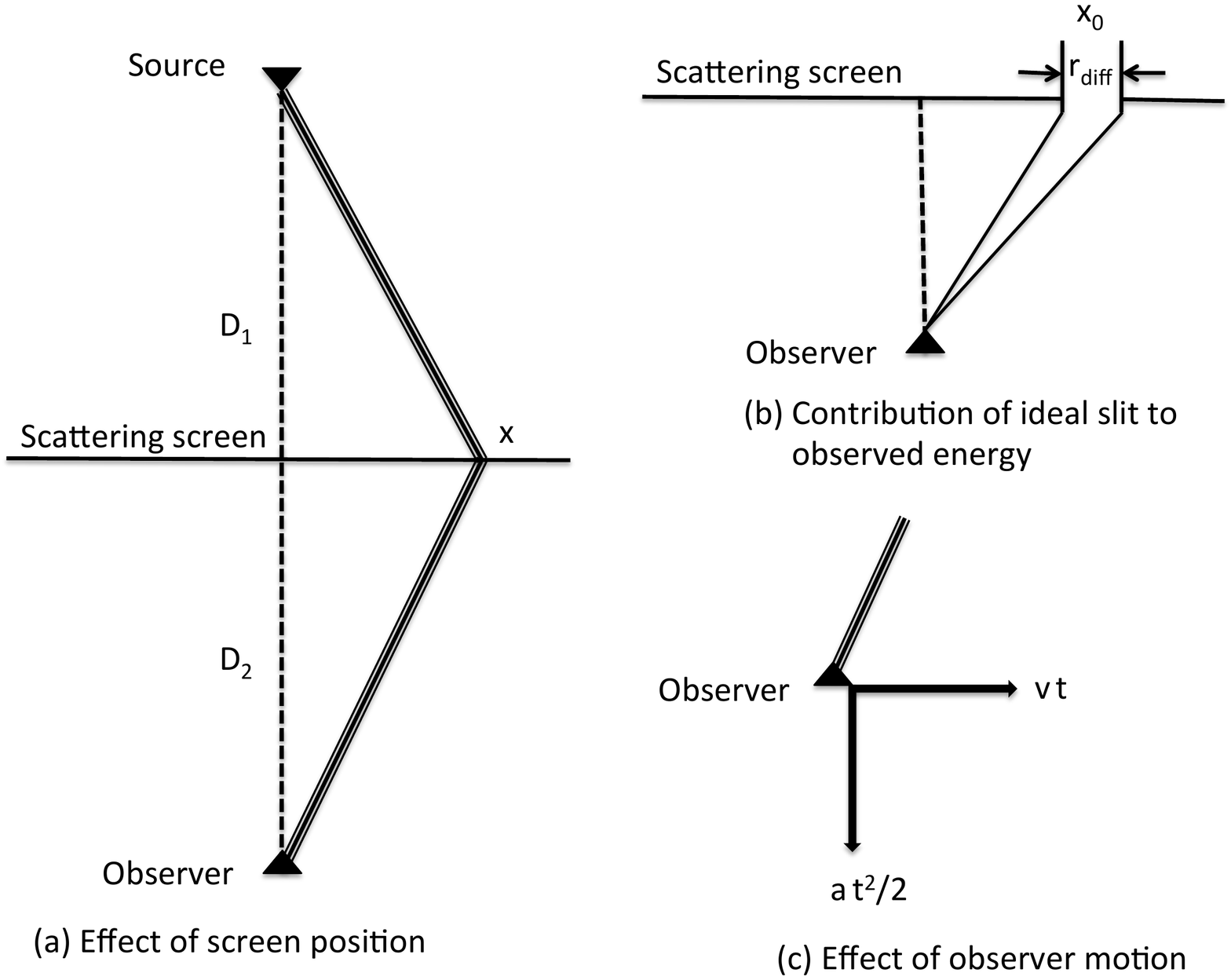}
	{.8}
	{Illustration of the geometry of scattering using a simplified one-dimensional
	scattering screen. (a) Rays passing through the screen at different lateral distances $x$
	encounter different group delays and phase shifts due to small variations in electron density.
	The total distance from source to observer is $(D_{1}+D_{2})$, divided between
	$D_{1}$ from source to screen and $D_{2}$ from screen to observer.
	(b) An ideal slit is opaque to all rays except over the range $[x_{0}-\rdiff/2, x_{0}+\rdiff/2]$. 
	(c) Motion of the observer in the direction of the line of sight with acceleration $a$ 
	and transverse to the line of sight with velocity $v$.}

A slightly refined model to Figure \ref{fig:scattering-iD-illustration} is shown in
Figure \ref{fig:scattering-geometry}a.
It is still one dimensional and assumes a thin screen, but it captures
the distance from source to screen and
screen to observer.
Consider a single ray passing through point $x$ on the screen as shown in
Figure \ref{fig:scattering-geometry}a.
Define the excess distance travelled along the ray (relative to $x = 0$) as $d(x)$,
and that $|x| \ll D_{1}$ and $|x| \ll D_{2}$.
The Fresnel approximation assumes that third and higher powers of $x$ can be neglected,
\begin{equation}
d(x) = \sqrt{D_{1}^{\,2} + x^{2}} + \sqrt{D_{2}^{\,2} + x^{2}} - (D_{1} + D_{2})
\approx \frac{x^{2}}{2 D}
\ \ \ \text{for} \ \ \ \ 
|x| \ll D
\end{equation}
where $D$ is defined by
\begin{equation}
\label{eq:d12}
\frac{1}{D} = \frac{1}{D_{1}} + \frac{1}{D_2}
 \,.
\end{equation}
Thus, following the Fresnel approximation the simpler model of Figure \ref{fig:scattering-iD-illustration}
still applies, with $D$ defined by \eqref{eq:d12}.

The excess propagation delay associated with a ray passing through
point $x$ on the screen is thus approximately $d(x)/c$.
This delay has frequency transfer function
\begin{equation}
\label{eq:delaytf}
\exp \left\{ - \,i\, 2 \pi \, f \, \frac{d(x)}{c} \right\} 
\approx
\exp \left\{ - \,i\, \pi \, \frac{f x^2}{c \, D} \,\right\}
=
\exp \left\{ - \,i\, \pi \, \frac{x^2}{\rf^{\,2}} \right\}
 \,.
\end{equation}
The Fresnel scale $\rf$ (defined in \eqref{eq:rf})
is the scale $0 \le x \le \rf$ over which there
is a $\pi$ radian phase shift in the transfer function due to geometric effects only.
Notably $\rf (f)$ is a function of frequency $f$, but this dependency has
been suppressed in \eqref{eq:delaytf} because the variation is negligibly small
over the band $f \in [f_{c}, f_{c} + B ]$.
To see this, expand in a Taylor series,
\begin{equation}
\frac{\rf ( f_{c} + \Delta f)}{\rf ( f_{c})} = 1 - \frac{\Delta f}{2 \, f_{c}} - \dots \,,
\end{equation}
so that $\rf (  f_{c} + \Delta f ) \approx \rf (  f_{c} )$ for $\Delta f \le B \ll f_{c}$.

When the screen itself introduces an extra phase shift $\phi(f,x)$,
the transfer function of \eqref{eq:delaytf} should be multiplied by
$\exp \left\{ \,i\, 2 \pi \,\phi (f,x) \right\}$.
This phase can be a function of both $x$
(because of inhomogeneities in the turbulent clouds of ionized gasses)
and $f$ (because of plasma dispersion).

The idea behind ray tracing in geometric optics is that the total response at the observer is
the superposition of the responses due to individual rays 
corresponding to different values of $x$.
The total response is given by the Fresnel-Kirchoff integral,
which integrates the transfer function of screen plus delay over 
all values of $x$,
\begin{equation}
\label{eq:fresnelkirchoff}
\psi (f) = \frac{e^{- \, i \, \pi/4}}{\rf} \int_{-\infty}^{\infty} \exp \left\{ i \, \phi(x,f) - i \, \pi \, \frac{x^{2}}{\rf^{\,2}} \right\} \, dx \,.
\end{equation}
Due to the identity
\begin{equation}
 \int_{-\infty}^{\infty} \exp \left\{ - i \, \pi \, \frac{x^{2}}{\rf^{\,2}} \right\} \, dx = \rf \, e^{ \, i \, \pi/4} \,,
\end{equation}
the normalization constant in \eqref{eq:fresnelkirchoff}
results in $\psi (f) \equiv 1$ (no attenuation or phase shift)
for the special case $\phi (x,f) \equiv 0$
(no scattering).
In particular this normalization renders the attenuation to be frequency-independent,
 in agreement with the plane wave
solution of Maxwell's equations for freespace propagation.
It does, however,
ignore the propagation loss (which increases as the square of distance) .
Thus, \eqref{eq:fresnelkirchoff} isolates the effects of the scattering screen
while omitting propagation loss.

\subsection{Scattering within a coherence hole}

The scattering model of \eqref{eq:fresnelkirchoff} can be simplified
considerably within an ICH.
Although scattering may place even more restrictive conditions,
the bandwidth $B$ of the signal is restricted by
\eqref{eq:dispersioncoherencebandwidth} in order to limit the
frequency incoherence due to plasma dispersion.
Thus, within the range of frequencies of the ICH,
the resulting model for $\phi(f,x)$ simplifies to that
of Figure \ref{fig:dispersion}.
Namely, the phase through the scattering screen consists of a
fixed phase $\theta + \phi (f_c,x)$ together with a group delay $\tau(f_c,x)$
that depends on carrier frequency but is otherwise frequency-independent.
In addition, within the turbulent ionized gasses, the inhomogeneity
in electron density over the scales of interest is quite small, so that the group delay
$\tau (f_c,x)$ can be considered to be independent\footnote{
Even if there were a variation of $\tau (f_c,x)$ with $x$, the resulting
effect would be negligible within an ICH if $2 B \tau_{s} \ll 1$ for
a total delay spread $\tau_{s}$.}
of $x$,
and further can be neglected because it is added to the very large
propagation delay.
The net effect of these assumptions is that
the phase can be considered frequency-independent, or $\phi (f,x) \approx \phi (x)$.
The fixed unknown carrier phase $\theta$ can also be neglected
since it does not affect the observed flux.

In view of these approximations, any waveform distortion of signal
$h(t) \, e^{ \, i \, 2 \pi \, f_c t }$ passing through the screen
on a ray through point $x$ can be
neglected, but rather $h(t)$ is merely subject to an excess group delay $d(x)/c$
introduced by the geometry and shifted in phase by $\phi(x)$ due to the screen.
The resulting observed passband signal is
\begin{equation}
e^{\,i\,\phi (x)} \, 
h \left( t - \frac{d (x)}{c} \right) \,
\exp \left\{ \, i \, 2 \pi \, f_c \left( t - \frac{d (x)}{c} \right) \right\} \,,
\end{equation}
with corresponding baseband signal
(after substituting for $d(x)$)
\begin{equation}
\label{eq:htafterscreen}
h \left( t - \frac{d (x)}{c} \right) \cdot \exp \left\{  i \, \phi(x)  - \, i \, 2 \pi \, \frac{x^2}{\rf^{\,2}} \right\} \,.
\end{equation}

Within an ICH, the bandwidth of $h(t)$ is small enough that the variation in delay
of $h(t)$ as a function of $x$ can be neglected.
In that event, the total observation is the superposition of this signal over all $x$, or
$\sqrt{\flux} \cdot h(t)$ for scintillation flux
\begin{equation}
\label{eq:fresnelkirchoffs}
\flux =
\left| \,
\frac{1}{\rf} \int_{-\infty}^{\infty} \exp \left\{ i \, \phi(x) - i \, \pi \, \frac{x^{2}}{\rf^{\,2}} \right\} \, dx
\, \right|^{2}
 \,.
\end{equation}
Again, \eqref{eq:fresnelkirchoffs} has been normalized so that
$\flux = 1$ when the phase shift of the scattering screen is a constant, or $\phi(x) \equiv \phi(0)$.
Also note the parallel between this model and the simplified modeling of Section \ref{sec:scintillation},
with the replacement of a sum by the Fresnel-Kirchoff integral.

\subsection{Coherent patches on the scattering screen}
\label{sec:coherentpatch}

The size of integral \eqref{eq:fresnelkirchoffs} depends on the spatial coherency of the
phase.
In regions of $x$ for which that phase is changing rapidly, the contribution to the
integral will be small, while the most significant contributions to $\flux$ come from
areas of slow phase variation.
There are two contributors to phase variation, the random phase due to inhomogeneities
in electron density $\phi(x)$, and the geometric variation due to path length changes
that is proportional to $x^{2}$.
The latter increases in importance with large $x$, and its greater rate of variation with increasing $x$
effectively limits the scale over which a significant contribution to $\flux$ can occur.

Statistically $\phi (x)$ displays ''coherence patches''
over distances in $x$ over which the variation in $\phi(x)$ is small
roughly equal to its diffraction scale $\rdiff$ \citeref{598}.
It is these coherent patches that contribute most of the energy in $\flux$.
As illustrated in Figure \ref{fig:scattering-geometry}b,
the effect can be captured qualitatively by modeling a single coherent patch located
at $x = x_{0}$ as an ideal slit $[x_{0}-\rdiff/2,x+\rdiff/2]$, 
within which the phase shift introduced by the screen is a constant $\phi(x) = \phi(0)$.
The contribution of such a slit to $\flux$ is
\begin{equation}
\label{eq:fresnelkirchoffslit}
\flux =
\left| \,
\frac{1}{\rf} \int_{-\rdiff/2}^{\rdiff/2} 
\exp \left\{ - i \, 2 \pi \, \frac{x_{0} \, u}{\rf^{\,2}} \right\} 
\, \exp \left\{ - i \, \pi \, \frac{u^2}{\rf^{\,2}} \right\}
\, du \,
\right|^{2}
\,.
\end{equation}
When modeling a small scale patch on the screen, when $\rdiff \ll \rf$,
the Fraunhofer approximation neglects the $u^{2}$ term.
In this case the integration yields
\begin{equation}
\label{eq:fresnelkirchofffraun}
\flux \approx
 \left| \,  \frac{\rdiff}{\rf} \cdot \text{sinc} \left( \frac{\pi \, \rdiff \, x_{0}}{\rf^{\,2}} \right) \, \right|^{2}
\,.
\end{equation}
This spells out how, as the location $x = x_{0}$  of an ideal slit (with width $\rdiff$) is changed,
the magnitude of the contribution of that slit to the flux observed at $x = 0$ changes.
This is a ''reverse'' diffraction pattern, capturing the effect of a variation in slit location
on the flux arriving at a single point of observation.
This pattern defines the new refraction scale at the screen
\begin{equation}
\label{eq:rreffdef}
\rref = \frac{\rf^{\,2}}{\rdiff} \,.
\end{equation}
Written in terms of $\rref$, \eqref{eq:fresnelkirchofffraun} becomes
\begin{equation}
\label{eq:fresnelkirchofffraunslit}
\flux \approx
\left| \frac{\rdiff}{\rf} \cdot  \text{sinc} \left( \frac{\pi \, x_{0}}{\rref} \right) \right|^{2}
\,.
\end{equation}
The scale $\rref$ equals
half the width of the main lobe in the pattern, and thus
the region of the screen $|x| \le \rref$ is the
primary origin of flux observed at $x = 0$.
Of course, some flux originates from outside the screen's refraction scale, but it is smaller.
The energy $\flux$ in \eqref{eq:fresnelkirchofffraunslit} is 
plotted in Figure \ref{fig:observationDiffraction} for two
 strong scattering scenarios.
From \eqref{eq:fresnelkirchofffraunslit} the energy arriving at the observer
for $x_{0} = 0$ is greater for a wider slit (larger $\rdiff$) because a wider slit allows more
total energy to pass the screen.
The nulls occur at multiples of $\rref$, so these nulls are closer together
for a wider slit (larger $\rdiff$).

\incfig
	{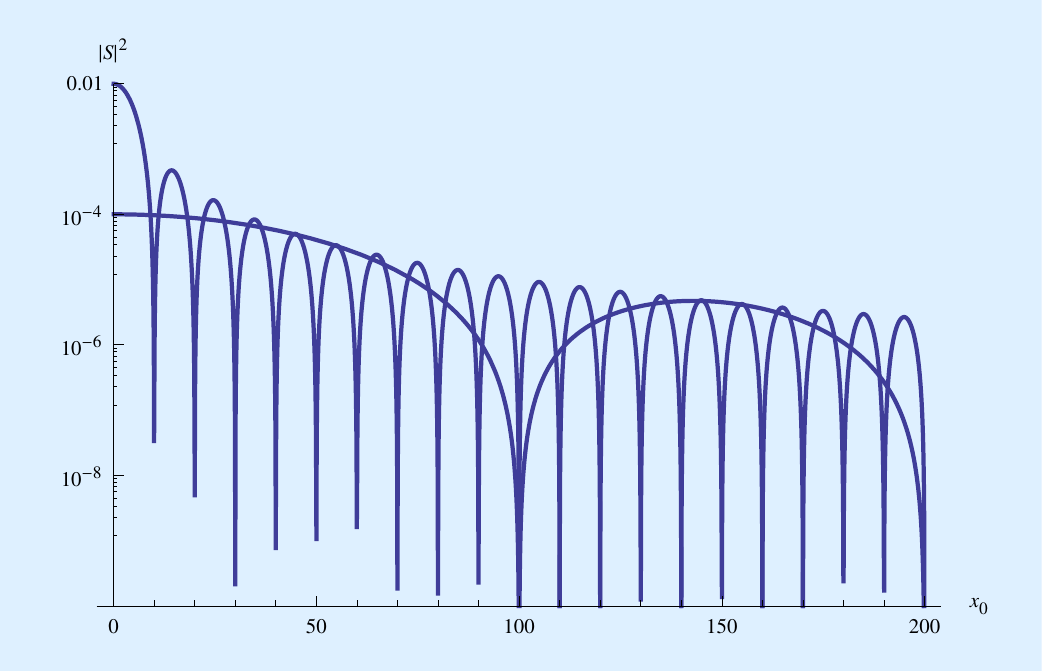}
	{.8}
	{A log plot of the energy $|S|^{2}$ arriving at the observer and emanating 
	from an ideal slit of width $\rdiff$ located at point $x_{0}$
	on the screen. 
	It is assumed that $\rf=1$, so in effect the abscissa has
	units of one Fresnel scale.
	Two values of $\rref$ are shown: $\rdiff=0.1$ (or $\rref=10$) has its first
	null at $x_{0}=10$, and $\rdiff=0.01$ (or $\rref=100$)  has its first
	null at $x_{0}=100$. The amplitude at $x_{0} = 0$ is higher for the
	wider slit ($\rdiff=0.1$) because more incident energy is able to pass through.}

\subsection{Focusing by the scattering screen}
\label{sec:lensing}

Coherent patches cannot explain the behavior $\flux > 1$, since in the limit
of dense coherent patches with the same constant phase,
$\flux \approx 1$ (this becomes equivalent to freespace propagation).
Focusing by the screen offers a physical explanation for $\flux >1$.
In \eqref{eq:fresnelkirchoffs} the contribution to $\flux$ from a given region can
 be large, even for very large $x_{0}$, if the screen-induced phase approximately cancels out the
pathlength-induced phase, or\footnote{
If \eqref{eq:lensphase} were valid for all $x$ then $\flux \to \infty$,
which indicates a problem with the model.
This non-physical behavior is because the Fresnel approximation has been used,  $|x| \to \infty$ violates
this approximation.
This is not of practical concern because \eqref{eq:lensphase} is unlikely to be satisfied for any
range of $x$ much larger than $\rdiff$.}
\begin{equation}
\label{eq:lensphase}
\phi(f_c,x) \approx \phi_0 + \pi \, \frac{x^{2}}{\rf^{\,2}} 
\ \ \ \text{or} \ \ \ \
\frac{\partial \phi(f_c,x) }{\partial x} \approx 2 \pi \frac{x}{\rf^{\,2}}
\,.
\end{equation}
This is, of course, the definition of a lens that focuses the energy on the
observer's reference point.
While \eqref{eq:lensphase} is unlikely to be satisfied over significantly large
scales, it is certainly possible within the diffraction scale $\rdiff$.
Note that over the range $[x_{0}-\rdiff/2,x+\rdiff/2]$,
$\phi(f_{c}, x)$ must vary over the range of $2 \pi x_{0} / \rref$
to satisfy \eqref{eq:lensphase}.
Outside the refraction scale $|x_{0}| > \rref$ this variation must be larger than $2 \pi$,
which makes it statistically unlikely.
When lensing does occur, it's contribution to $|S|$ is the same
as the ideal slit with the same aperture located at $x_{0} = 0$,
regardless of whether the lensing occurs inside or outside the refraction scale.

\subsection{Frequency coherence of scattering}
\label{sec:delayspread}

\begin{changea}
Scattering potentially introduces frequency incoherence to an energy bundle
due to the delay spread $\tau_s$ resulting from multiple propagation paths from
transmitter to receiver through different points $x$ on the scattering screen.
The geometry of scattering coupled to the statistics of the phase shift $\phi (x)$
determine $\tau_s$, which in turn determines the scattering coherence bandwidth.
This is the maximum bandwidth $B$ above which the individual paths become
resolvable, and as a result the frequency response is no longer fixed
over all frequencies.
The astrophysics literature has a different name for this, the ''de-correlation bandwidth" \citeref{598}.
The scattering coherence bandwidth has a second interpretation
described in Section \ref{sec:relationship}.
\end{changea}

Approximation \eqref{eq:fresnelkirchoffs} assumes that $h(t)$ has bandwidth $B$ less
than the scattering coherence bandwidth.
The starting point for determining the scattering coherence bandwidth is
the relationship between this bandwidth and the delay spread
given in \eqref{eq:delaycbw}, which says that $2 B \tau_s < \alpha$, where $\alpha$
 is a chosen parameter that controls the maximum allowable phase shift $\alpha \pi$ across bandwidth $B$.
We merely have to determine the delay spread $\tau_s$, which depends on the strength of the scattering.

\subsubsection{Strong scattering}

Consider first the strong scattering case.
The main lobe of the interference pattern at the observer corresponds to $|x| \le \rref$.
If we include the energy from a scale $n$ times larger than this to be conservative,
the delay spread becomes
\begin{equation}
\tau_{s} = \frac{d( n \, \rref )}{c} \approx \frac{n^{2}}{2 \, f_{c}} \cdot
\left( \frac{\rf}{\rdiff} \right)^{\,2} \,.
\end{equation}
Thus frequency coherence is achieved when
\begin{equation}
\label{eq:scatterngcb}
B < \frac{ \alpha }{n^{2}} \cdot f_{c} \left( \frac{\rdiff}{\rf} \right)^{2} \,.
\end{equation}
Taking into account the frequency dependence of $\rdiff$ and $\rf$,
the $B$ in \eqref{eq:scatterngcb} increases as $f_{c}^{4.4}$, which is
much more rapidly
than the coherence bandwidth for plasma dispersion.
Thus,
the relative importance of plasma dispersion increases with $f_{c}$.

\subsubsection{Weak scattering}

For weak scattering, most of the scattered energy arrives from the first Fresnel zone $|x| \le \rf$,
and again to be conservative by including $n$ times this scale,
\begin{equation}
\tau_{s}  = \frac{d( n \, \rf )}{c} \approx \frac{n^{2}}{2 \, f_{c}} \,.
\end{equation}
The criterion for frequency coherence is then
\begin{equation}
B < \frac{ \alpha }{n^{2}} \cdot f_{c} \,.
\end{equation}
Thus, once we move into the weak scattering regime, the coherence
bandwidth increases more slowly with $f_{c}$.
However, the coherence bandwidth is so large in this regime that 
frequency incoherence is unlikely to be an issue.

Arriving at a more accurate estimate for the $B$ that guarantees
frequency coherence depends on the detailed statistics of $\phi(x)$
and the resulting distribution of the energy arriving as a function of delay.
For energy arriving within a scale $|x| \le r$,
the scattering amplitude seen at the MF output is given by
\eqref{eq:delayspread},
\begin{equation}
\label{eq:fresnelkirchoffsx}
\flux =
\left| \, \frac{1}{\rf}  \cdot \int_{-r}^{r} \exp \left\{ i \, \phi(x) - i \, \pi \, \frac{x^{2}}{\rf^{\,2}} \right\} \, dx \, \right|^{\,2} \,.
\end{equation}
This model is a considerable improvement to the ideal slit in that it 
captures the effects of coherent patches and lensing
as well as other detailed effects.
A methodology used in the astrophysics literature is to generate a large set
of samples functions of phase $\phi(x)$ based on models of wavenumber spectrum,
and perform a Monte Carlo simulation of the resulting flux $\flux$.

\section{Fading and scintillation}
\label{sec:fading}

In a static model of scattering of Section \ref{sec:scattering}, time coherence is not an issue because
the model for $S$ is not time-dependent.
In reality, there are two sources of nonstationarity.
One is changes in $\phi(x)$ with time due to the turbulence of the interstellar clouds.
This is a phenomenon that plays out over days or weeks, and thus is not a factor
in defining the size of an ICH.
Over shorter time scales, the dominant source of non-stationarity is
the interaction of source/observer motion with scattering.
As $T$ is decreased
the total opportunity for variation shrinks and eventually becomes insignificant.

Within an ICH, fading is a velocity rather than acceleration phenomenon.
In Figure \ref{fig:scattering-geometry}c,
the velocity of the receiver causes a carrier frequency offset that was
neglected in Section \ref{sec:doppler}.
However in the presence of scattering, velocity cannot be neglected because
the component of receiver motion in the direction of each incoming ray
from the scattering screen varies slightly with $x$.
This results in a small relative frequency offset among the rays, and that in turn
causes the constructive and destructive interference mechanism of refraction
to vary with time and turn cause a variation in scintillation amplitude.
This effect cannot be compensated
at the receiver as argued in Section \ref{sec:doppler} by a simple carrier-frequency offset.
The superposition of the Doppler-shifted rays occurs
\emph{before} the receiver's observation, and thus there is no single correction that can
simultaneously take care of all the Doppler frequency shifts for all the rays.

The effect of observer motion
can be modeled by assuming two components of velocity,
 $v_{\parallel}$ parallel to the line of sight and $v_{\bot}$ transverse to the line of sight.
 As pictured in Figure \ref{fig:scattering-geometry}c, only $v_{\bot}$
 is a significant source of fading.
 The variation in velocity
 for $t \in [0,T]$ due to acceleration is very small and can be neglected.
The total excess length of a ray through point $x$ on the screen then becomes, as a function of time,
\begin{equation}
\label{eq:dwv}
\sqrt{D_{1}^{\,2} + x^{2}}
+ \sqrt{(D_{2}+ v_{\parallel} t )^{2} + (x - v_{\bot} t)^{2}} 
- ( D_{1} + D_{2} )
\approx
d(x) + v(x) \cdot  t \,.
\end{equation}
In Appendix \ref{sec:fadinganalysis} a Fresnel approximation is applied to \eqref{eq:dwv},
and it is shown that the resulting effect at baseband is a time varying phase
that depends on $x$.
If $T$ is kept sufficiently small, the total phase change can be kept small.
This phase variation is less than $\alpha \pi$ radians when
\begin{equation}
\label{eq:tcf}
\left| \,
\frac{f_{c} \, v_{\bot} \, x \, T} {c \, D_{2}} 
\, \right| < \alpha \,
\end{equation}
 for all values of $x$.
 In particular \eqref{eq:tcf} must be satisfied for the largest $x$ that is a significant contributor
to energy arriving at the observer.

Consider first the strong scattering regime.
Following Section \ref{sec:delayspread},
suppose that only $|x| \le n \, \rref$ is a significant source of energy,
in which case the most stringent condition on $T$ for \eqref{eq:tcf} becomes
\begin{equation}
\label{eq:tcfr}
T \ll \frac{\alpha \, c \,D_{2}}{f_{c} \, | v_{\bot} | \, n \, \rref} 
\,.
\end{equation}
Since $D \le D_{2}$, a slightly more stringent version of \eqref{eq:tcfr} is
\begin{equation}
\label{eq:sctsc}
T < \frac{\alpha \, c \, D}{f_{c} \, | v_{\bot} | \, n \, \rref} 
= \frac{\alpha \, \rdiff}{n \, | v_{\bot} |} \,.
\end{equation}
Thus, significant time variations in scintillation amplitude occur on spatial
scales of observer motion that are greater than a multiple of the diffraction scale $n \, \rdiff$.
Since whatever happens at the plane of motion of the observer is a mirror of the
scattering screen (with appropriate geometric corrections), this significant change
over a scale that mirrors significant phase change at the screen is not surprising.
Since $\rdiff \propto f_{c}^{\,6/5}$, this coherence time increases (albeit relatively slowly)
with $f_{c}$.

For the weak scattering regime, the significant range is  $|x| \le n \, \rf$.
The coherence time becomes \eqref{eq:sctsc} with $\rdiff$ replaced by $\rf$, or
\begin{equation}
T < \frac{\alpha \, \rf}{n \, | v_{\bot} |} \,.
\end{equation}
This case (which for a given $D$ corresponds to larger $f_{c}$) sees
$T \propto f_c^{- 1/2}$, which decreases with increasing $f_{c}$.

\section{Numerical example}

While some impairments have been deemed to be insignificant,
the operative sources of incoherence are summarized in Table \ref{tbl:ICHsources}.
A safety factor of $\alpha \le 1$ permits an adjustment of the stringency of the criterion.
There are two primary contributions to frequency incoherence,
both resulting from a spreading of the arriving energy over time.
One is the frequency-dependence of group delay that is characteristic of
plasma dispersion, and the other is the spatial-dependence of group delay due to scattering.
Two primary contributors to time incoherence have been identified,
both due to time-varying phase shift caused by relative source/observer motion.
One is the component of acceleration of
source and observer along the line of sight, and the other is the interaction between the
velocity of the observer transverse to the line of sight and the spatial distribution of angle of arrival
due to scattering.
Rendering all these effects insignificant within an ICH determines the
coherence time $T_{c}$ and coherence bandwidth $B_{c}$
of the ICH, and thus the degrees of freedom $K_{c}$.
The most important physical determinants are the 
distribution of electron density in the ISM
along the line of sight and the relative motion of source and observer.
There is also a strong dependence on the carrier frequency $f_{c}$.

\doctable
	{\small}
	{ICHsources}
	{The significant sources of time and frequency incoherence in interstellar communication,
	distinguishing between the strong and weak scattering regimes.}
	{l l c c}
	{
	\textbf{Type} & \textbf{Source} & \textbf{Strong} & \textbf{Weak}\\
	\toprule
	\addlinespace[3pt]
	Frequency & Plasma dispersion 
	& $B <  \rho \, f_{c}^{\, 3/2}$ 
	& \text{Same}
	\\
	\addlinespace[3pt]
	\cmidrule{2-4}
	& Scattering delay spread 
	& $\displaystyle B < \frac{\alpha}{n^{2}} \cdot f_{c} \, \left( \frac{\rdiff}{\rf} \right)^{2}$
	& $\displaystyle B < \frac{\alpha}{n^{2}} \cdot f_{c} $
	\\
	\addlinespace[3pt]
	\cmidrule{1-4}
	Time & Acceleration 
	& $\displaystyle T < \frac{P}{2 \pi} \cdot \sqrt{\, \frac{\alpha \, \lambda_{c}}{R}}$
	& \text{Same}
	\\
	\addlinespace[3pt]
	\cmidrule{2-4}
	& Scintillation 
	& $\displaystyle T < \frac{\alpha \, \rdiff}{n \, | v_{\bot} |}$
	& $\displaystyle T < \frac{\alpha \, \rf}{n \, | v_{\bot} |}$
	\\
	\addlinespace[3pt]
	\bottomrule
	}

The coherence bandwidth and coherence time are plotted in
Figure \ref{fig:btCoherence} for the numerical values given
in Table \ref{tbl:assumptions}.
A value $\alpha = 0.1$ is used, except for acceleration
where a more conservative factor $\alpha = 0.01$ is used
to account for the strong possibility that another planet or moon
may experience a stronger acceleration than Earth, and also to
account for the possible accumulation of accelerations due to transmitter
and receiver.
Over the entire range of carrier frequencies, one impairment
is more constraining than the other.
In the case of frequency coherence, plasma dispersion
determines the ICH coherence bandwidth $B_c$.
For time coherence, acceleration dominates the coherence
time $T_c$.
The resulting $B_c$ increases with $f_c$, and $T_c$
decreases with $f_c$.
As shown in Figure \ref{fig:dofICH}, the net effect
is a $K_c = B_c T_c$ that starts at $K_{c} \gg 1$
and increases monotonically with $f_c$. 

\incfig
	{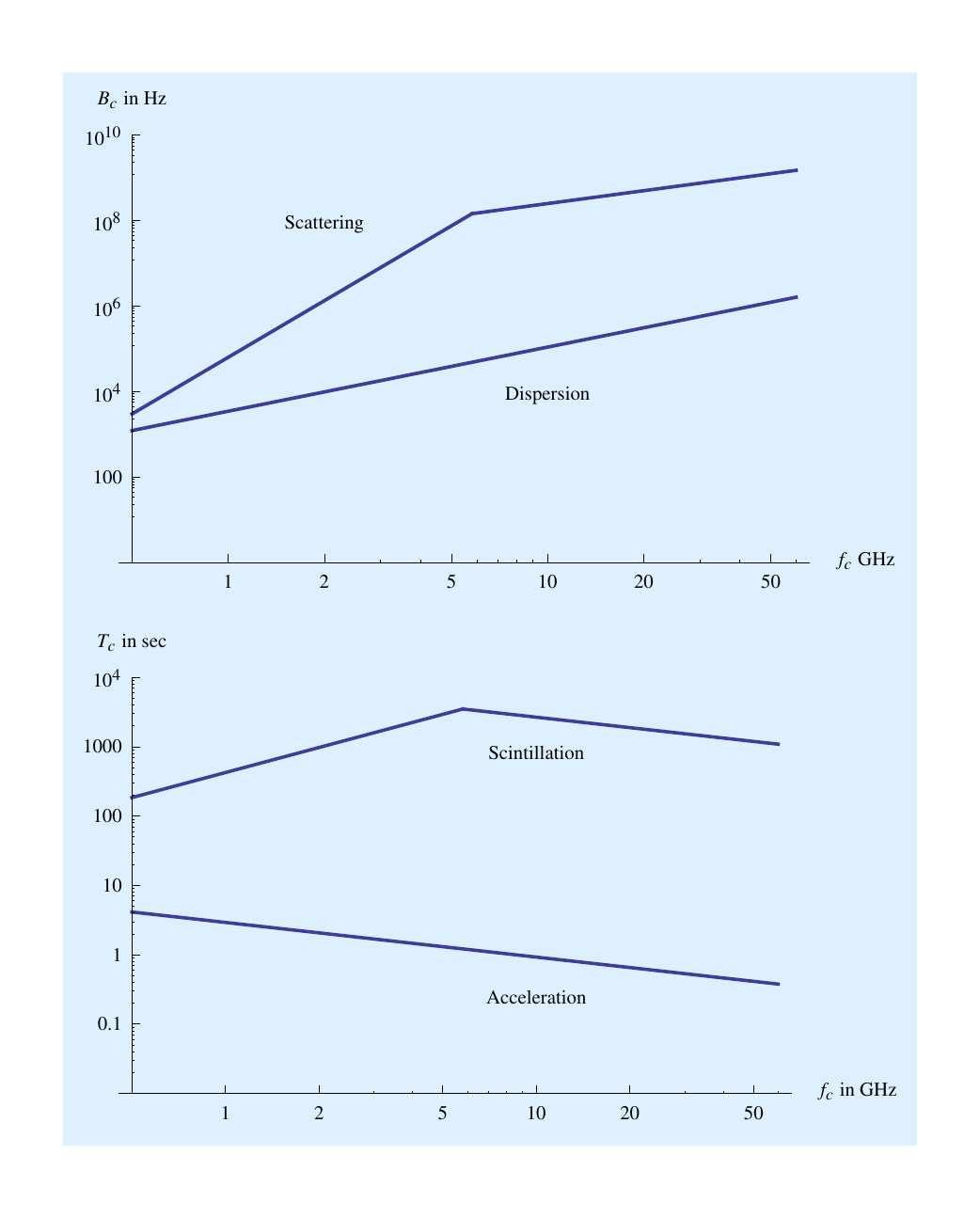}
	{.8}
	{Log-log plots of coherence bandwidth $B_c$ above and coherence
	time $T_c$ below, plotted against carrier frequency $f_{c}$
	for the assumed physical parameters of Table \ref{tbl:assumptions}.
	In all cases, it is assumed that $\alpha = 0.1$
	and $n = 2$.
	For scattering phenomena, the break in the slope corresponds to the
	transition from strong scattering (at lower $f_{c}$) to weak scattering (at higher $f_{c}$).
	In each case there are two impairments
	that limit the size of an ICH, but one always is dominant}
		
Any receiver designer should be cognizant of two possible corrections that may be performed in the transmitter.
In the interest of increasing the coherence bandwidth $B_{c}$, the transmitter may perform
partial equalization for the minimum dispersion.
In the interest of increasing the coherence time $T_{c}$, the transmitter may
correct for the component of acceleration along the line of sight.
Interestingly, these are the very same impairments that
dominate $B_c$ and $T_c$, and thus there is motivation to
make these corrections if there is a desire to increase $K_c$,
for example to obtain greater immunity to RFI.

\begin{changea}

\section{Relationship of non-overlapping energy bundles}
\label{sec:relationship}

All the channel codes considered in Chapters \ref{sec:fundamental} and \ref{sec:infosignals}
consist of patterns of energy bundles that do not overlap one another in time, 
in frequency, or both.
In order to analyze the channel code, it is necessary to model the joint statistics of
the treatment afforded these distinct energy bundles by the interstellar channel.
The conclusions of this section were summarized in Section \ref{sec:jointtreatment}.

\subsection{Energy bundles at different frequencies}

Consider an energy bundle that is transmitted during the time interval $t \in [0, T]$
but may be modulated to one of two different frequencies $f_1 < f_2$.
The two candidate transmitted baseband waveforms are
$h(t) \, e^{\, i \, 2 \pi \, f_1 \, t}$ and $h(t) \, e^{\, i \, 2 \pi \, f_2 \, t}$ where
$h(t)$ falls in an ICH (or in other words its bandwidth $B$ is less than the
coherence bandwidth $B_c$, or $B \le B_c$) and where $f_2 - f_1 > B$
to avoid frequency overlap between the two bundles.
The question is under what conditions the interstellar propagation treats these two bundles differently.
The answer to this question depends on the numerical value of  $(f_2 - f_1)$
in relation to the scattering coherence bandwidth.

Figure \ref{fig:ICHmodel} the transfer function experienced by the two energy bundles must take the form
\begin{align*}
& G(f_c + f_1 + \Delta f ) = S_1 \cdot \exp \{ \, i \, \phi_1 - i \, 2 \pi \, \Delta f \, \tau_1 \} \\
& G(f_c + f_2 + \Delta f ) = S_2 \cdot \exp \{ \, i \, \phi_2 - i \, 2 \pi \,  \Delta f \, \tau_2 \} \,,
\end{align*}
where  $\{ S_1, \, \tau_1 \}$ and $\{ S_1, \, \tau_1 \}$ are the propagation parameters
experienced by the two energy bundles.
The question is the relationship between $\{ S_1, \, \tau_1 \}$ and $\{ S_1, \, \tau_1 \}$.
We will always have that $\tau_1 > \tau_2$ due to the differential group delays of plasma dispersion.
Thus, we expect there to be a small time offset between the two energy bundles as observed by the receiver,
unless it was compensated by the transmitter.
Equipped with partial or complete knowledge of the dispersion measure $\dm$,
the transmitter could easily compensate for this group delay by
introducing a frequency-dependent time offset.

The more interesting and important question is the relationship of $S_1$ and $S_2$.
For the case where $( f_2 + B_c  - f_1)$ is \emph{less} than the scattering coherence bandwidth
determined in Section \ref{sec:delayspread},
the frequency transfer function is essentially constant over the
full frequency range $f \in [f_c + f+1, f_c + f_2 + B_c ]$.
For this case, therefore, we must have that $S_1 \approx S_2$.
On the other hand, if $( f_2 + B_c  - f_1)$ is becomes \emph{larger} than the
scattering coherence bandwidth, then the constant transfer function approximation
for the scattering coherence bandwidth
begins to break down, and we expect that the difference between $S_1$
and $S_2$ becomes material.
If $( f_2 + B_c  - f_1)$ is \emph{much} larger than the scattering coherence bandwidth,
then $S_1$ and $S_2$ will have no relationship
(they are statistically independent).

Recall from Figure \ref{fig:scatVsdispCB} that the scattering coherence bandwidth
is typically much larger than the dispersion coherence bandwidth, where the latter
dominates the ICH coherence bandwidth $B_c$.
Further, the bandwidth $B$ of an energy bundle can be less (even significantly less)
than $B_c$.
Thus, there is quite a bit of headroom to ''pack in'' energy bundles at different frequencies
(as in M-ary FSK) and expect that they will experience essentially the same scintillation.
This cannot be taken too far, however, as illustrated by the ratio of scattering to
dispersion coherence bandwidths that was illustrated in Figure \ref{fig:ratioCB}.
As $M$ is increased in M-ary FSK eventually the different frequencies start to
experience significantly different scintillation, and the receiver must employ a
specific strategy to deal with this.

\subsection{Energy bundles at different times}

Consider now two energy bundles $h(t - t_1 )$ and $h(t - t_2)$ at the same carrier frequency
but at different times $t_2 < t_1$.
As $(t_2 - t_1)$ becomes larger than the scintillation coherence time the scintillation
experienced by these two energy bundles will start to diverge, becoming statistically independent.
In terms of the channel codes considered in Chapter \ref{sec:infosignals} this is never a problem,
and in fact time diversity works optimally when the redundant
energy bundles are sufficiently separated in 
time so that the scintillation parameters are independent.

The two energy bundles will experience different effects due to acceleration, and the
magnitude of this effect depends on whether the transmitter or the receiver (or both)
compensate for their local contribution to acceleration as described in Section \ref{sec:motioncomp}.
For example, in the presence of ideal circular motion
at one end of the line of sight
the uncompensated frequency offset as a function of time is
from \eqref{eq:circulardist}
\begin{equation}
\Delta f (t) = \frac{v_\parallel (t)}{\lambda_c} = 
\frac{\displaystyle 2 \pi \, R \, \sin \left( 2 \pi \, \frac{t}{P} \right)}{\lambda_c \, P} \,.
\end{equation}
This will result in different frequency offsets at different times, with a maximum differential
shift in carrier frequency between two non-overlapping energy bundles equal to
\begin{equation}
\label{eq:freqoffetworse}
 | \Delta f | \le \frac{4 \pi \, R}{\lambda_c \, P} \,.
\end{equation}
From Table \ref{tbl:ctaccel}, $ | \Delta f | $ is larger for orbital motion,
and increases in proportion to $f_{c}$.
However, the frequency shift will be much smaller 
than predicted by \eqref{eq:freqoffetworse} over time scales short
relative to $P$ (one day or one year).
\end{changea}

\section{Summary}

The sources of time and frequency incoherence have been modeled in sufficient detail
to be convincing that they can be circumvented in the interstellar coherence hole (ICH)
by restricting the time duration and bandwidth of a waveform used for an individual energy bundle.
Included in the analysis have been plasma dispersion and scattering due to
clouds of interstellar gasses within the interstellar medium, as well as
relative motion of transmitter and receiver due to solar system dynamics.
The resulting coherence time and bandwidth have been estimated for typical values of
governing physical parameters.
The results show that the available degrees of freedom are much larger than unity,
providing assurance that the coherence of the interstellar channel is sufficient to support
power-efficient communication.
This also opens up the possibility of choosing noise-like waveforms in an energy bundle,
for example as a means to gaining immunity to interference.
If the transmitter and receiver both compensate for their respective contribution to
acceleration, the coherence time and degrees of freedom can be substantially increased.

\begin{changea}
There are two district physical phenomena that determine the coherence time
(acceleration and scintillation, both due to motion) and likewise the coherence bandwidth
(dispersion and scattering, both due to propagation effects).
In each case one of these dominates and determines the coherence bandwidth of the
interstellar coherence hole.
In order to model the effect of the interstellar channel on two energy bundles
at non-overlapping times or at non-overlapping frequencies,
it is necessary to consider these phenomena separately.
\end{changea}


\chapter{Conclusions}

This report has addressed the end-to-end design of an interstellar communication
system taking account of all known impairments from the interstellar medium
and induced by the motion of source and observer.
Both discovery of the signal and post-discovery communication have been addressed.
It has been assumed that the transmitter and receiver designers
wish to jointly minimize the cost, and in that context we have considered
tradeoffs between resources devoted in the transmitter vs. the receiver.
The resources considered in minimizing cost are principally energy consumption
for transmitting the radio signal and energy consumption for signal processing
in a discovery search.

Thus study is applicable to communication with
interstellar spacecraft (sometimes called ''starships'') in the future, although it will probably be quite some time
before these spacecraft wander far enough to invoke many of the impairments considered here.
In the present and nearer term this study if relevant to communication with other
civilizations.
In the case of communication with other civilizations, we of course
will be designing either a transmitter, or a receiver, which in either
case has to be interoperable with a similar design by the other civilization.
In either case, it is very helpful to consider the end-to-end design
and resource tradeoffs, as we have attempted here.
In our view, a major shortcoming of existing SETI observation programs
(which requires the design of a receiver for discovery)
is the lack of sufficient attention to the issues faced by the transmitter
in overcoming the large distances and the impairments introduced in interstellar space.

One profound conclusion of this study is that if we assume that the transmitter
seeks to minimize its energy consumption, which is related to average transmit power, 
then the communication design becomes relatively simple.
The fundamental limit on power efficiency in interstellar communication has been
determined, that limit applying not only to us but to any civilization no matter how advanced.
Drawing upon the early literature in power-efficient communication design,
five simple but compelling design principles have been identified.
Following these principles has been shown to permit designs that approach the
fundamental limit.
Although the same fundamental limit does not apply to the discovery process,
we have defined an alternative resource-constrained design approach that minimizes
processing resources for a given power level, or power level for a given processing resource.
Again, application of a subset of the five principles leads to a design
that can achieve dramatic reductions in the transmitter's energy consumption
 relative to the
type of Cyclops beacons that have been a common target of SETI observation programs,
at the expense of a more modest increase in the receiver's
energy consumption through an increased observation time in each location.

If we assume that a transmitter designer considers energy to be a scarce
resource, and bandwidth to be a relatively ample resource, 
then the type of signals the transmitter will be deemed to be false alarm events
in current and past SETI observation programs, and will likely be dismissed.
Fortunately, however, fairly straightforward modifications to these search strategies
can open up the possibility of detection of such signals.
These modifications include a substantial increase in the number of observations
conducted on each line of sight and carrier frequency, and a more sophisticated
statistical pattern recognition algorithm.

The results here has direct bearing on a topic of relatively recent interest,
the so-called broadband SETI.
As we have shown by invoking the fundamental limits, information-bearing signals designed for
power efficiency are intrinsically and inevitably inefficient in their use of bandwidth.
A concern in broadband SETI has been the dispersive impairments of the
interstellar medium, which tend to be invoked with increasing bandwidth.
We have shown that in spite of their higher-bandwidth requirement,
information-bearing signals designed according to power-efficiency principles need \emph{not}
invoke greater impairments.
In fact, they \emph{must} not invoke greater impairments.
To that end have identified and quantified the interstellar coherence hole (ICH), which
allows the building blocks of a transmit signal (the energy bundles) to be designed to have
wider bandwidth and still not be subject
to any interstellar impairments other than noise and scintillation.
By transmitting patterns of these energy bundles in time and frequency,
the broadband requirement for power-efficient designs to approach the fundamental
limit can be satisfied without invoking interstellar impairments other than noise and scintillation.
No processing in either transmitter nor receiver related to other impairments is necessary,
save the relatively minor burden of compensating for acceleration.
Adhering to the constraints of the ICH in the transmit design is one of the
five principles of power-efficient design.
A commonly stated disadvantage of broadband SETI is the receiver processing requirements
associated with overcoming interstellar impairments.
We have shown that these impairments can instead be circumvented at the transmitter
through the judicious choice of signal waveform and without
processing overhead.
On the other hand, the discovery of power-efficient signals at the receiver
does require processing resources devoted to multiple observations, and the
 sensitivity of the discovery to low-power signals is directly related to the processing
resources devoted to these observations.

Finally, it has been shown that in the power-efficient regime information-free beacons offer essentially no
advantage over information-bearing signals with respect to the average power
and energy consumption or the difficulty of discovery.
Searches targeted at information-bearing signals can readily discover beacons, 
although the reverse is not true.
Thus it makes sense to design our discovery searches that seek out power-efficient and
power-optimized signals to simultaneously seek both types of signals.

\chapter*{Acknowledgements}
\addcontentsline{toc}{chapter}{Acknowledgements}

This research was supported in part by a grant from the
National Aeronautics and Space Administration to the SETI Institute.
The author is indebted to the following individuals for
generously sharing their time for discussion and comments:
James Benford,
Samantha Blair, 
Gerry Harp,
Andrew Siemion,
Jill Tarter,
David Tse,
and
Dan Werthimer.
Ian Morrison has been particularly helpful in maintaining an ongoing
dialog on these issues and reading and commenting on drafts of this work.


\appendix
\addappheadtotoc


\chapter{Interstellar link budget}
\label{sec:linkbudget}

In an interstellar communication link,
the required average transmit power can be
determined by first establishing the receive power $\power_R$
necessary for a desired reliability,
and then referring that power to the transmitter and determining
the corresponding average transmit power $\power_T$.
In both cases the power is proportional to the information rate $\rate$,
with a constant of proportionality equal to the energy per bit $\energybit$.

\subsection{Receive energy per bit}

The fundamental limit for the interstellar channel,
taking into account noise and scintillation, is given by
\eqref{eq:fundlimit}.
This places a lower
bound on the energy per bit measured at baseband in the receiver,
\begin{equation*}
\energybit > N_0 \, \log 2 
\ \ \ \text{joules}
\,.
\end{equation*}
The value of the noise power spectral density
introduced in the receiver circuitry is $N_0 = k \, T_n$,
where $k$ is Boltzmann's constant
(numerically equal to $k = 1.38 \times 10^{\, - 23}$ joules per degree kelvin)
and $T_n$ degrees kelvin is the  equivalent system noise temperature.
White noise is a good approximation at radio frequencies, where
sources of thermal noise can be considered white and quantum effects
can be ignored.

The state of the art in receiver design on Earth can achieve a
receiver internal noise temperature of 5 degrees kelvin.
This would be a representative value that a transmitter would
have to assume in order to communicate reliably with us.
If we were to construct our own transmitter, it might be reasonable
to assume a lower value.
Added to this is cosmic background radiation at 3 degrees kelvin,
applicable to more advanced civilizations as well as ourselves,
giving a total noise temperature of $T = 8$ degrees kelvin.
This is the most optimistic case, since it does not take into
account star noise.
Based on an assumption  of 8 degrees kelvin,
the fundamental limit for received energy per bit is
\begin{equation*}
\energybit > k \, T_n \, \log 2
= 7.66 \times 10^{\,-23} \ \ \text{joules} \,.
\end{equation*}
For example, at a bit rate equal to $R = 1$ bit per second,
the corresponding average power is
\begin{equation*}
\power_R = 7.66 \times 10^{\,-23} \ \ \text{watts}
\ \ (-221.2 \ \text{dBW})
\,.
\end{equation*}

\subsection{Transmit energy per bit}

A power link budget for a typical interstellar communication system
can be developed.
Assume that the transmitter and receiver are separated by a
distance $D$, with average power $\power_T$ transmitted
and $\power_R$ received.
Then for an ideal receive aperture antenna with area $A_R$,
 the power loss is given by
\begin{equation*}
\frac{\power_R}{\power_T} = G_T \cdot \frac{A_R}{4 \pi \, D^{\,2}}
\,,
\end{equation*}
where $G_T$ is the gain of the transmit antenna.
For the isotropic case (where $G_T = 1$) the fraction
of the transmit power that is captured is the ratio of
$A_R$ to the area of a sphere with radius $D$. 
If the transmit antenna is also an aperture antenna with area $A_T$,
the maximum gain is
\begin{equation*}
G_T = \frac{4 \pi \, A_T}{\lambda_c^{\,2}}
\end{equation*}
where $\lambda_c$ is the wavelength at carrier frequency $f_c$.
Defining a similar gain $G_R$ for the receive antenna
(with area $A_R$ substituted for $A_T$),
the Friis transmission equation is
\begin{equation*}
\frac{\power_R}{\power_T} 
= G_T \, G_R \, \left( \frac{\lambda_c}{4 \pi \, D} \right)^2 \,.
\end{equation*}
While $G_T$ and $G_R$ are geometrical parameters of the antenna designs,
the term in parenthesis captures the effect of propagation.
For a given average received power $\power_R$,
the average transmit power $\power_T$ scales as $\power_T \propto D^{\,2}$,
and (taking into account the wavelength dependence of the antenna gains) $\power_R \propto f_c^{\,-2}$.

For a numerical example, assume that $D = 1000$ light years
and $f_c = 5$ GHz ($\lambda_c = 6$ centimeters).
For both transmit and receive antennas, use our own
Arecibo antenna as a prototype, with
a 305 meter diameter and 
assume that inefficiency reduces the effective antenna area by 50\%
at both transmitter and receiver.
For these parameters, the antenna gains are
$G_T = G_R = 81.1$ dB,
and the path loss is 425.9 dB.
At the fundamental limit, the required transmit energy per bit is
\begin{equation*}
\energybit= 1.85 \times 10^4 \ \ \text{joules} \,.
\end{equation*}
For an energy contrast ratio $\ecrs = 15$ dB, the value of $\energy_h$
is a factor of $10^{15/10} / \log 2$ larger, or for the same assumptions
\begin{equation*}
\energy_h= 8.43 \times 10^5 \ \ \text{joules} \,.
\end{equation*}

\chapter{Matched filtering}

This appendix provides an analysis and justification of the matched filter
as the optimum receiver processing in the presence of AWGN alone.

\section{Modulation and demodulation}
\label{sec:moddemod}

To prepare this signal for radio transmission, it is multiplied by $\sqrt{\recenergy}$
to give it total energy $\recenergy$, and then modulated by
carrier frequency $f_{c}$ before taking twice the real part $\Re \{ \cdot \}$.
The modulating carrier has a phase $\theta$ that is unknown to the receiver.
The resulting real-valued signal
\begin{equation}
\label{sec:passbandsignal}
2 \, \Re \left\{ \sqrt{\recenergy} \, h(t) \, e^{\,i\,\theta} \, e^{\,i\, 2 \pi \, f_{c} t } \right\}
\end{equation}
is suitable for transmission by electromagnetic means through  the
interstellar medium (ISM), and is confined to frequency range $|f| \in [f_{c}, f_{c} + B ]$.

For purposes of defining the ICH,
the front-end of a receiver is optimized for AWGN
while pretending that there are no other ISM impairments.
The first step is to demodulate the received signal to baseband,
and in the absence of any impairments from the ISM this yields
\begin{equation*}
r(t) =
\text{LPF} \left\{
2 \Re \left[ \sqrt{\recenergy} \, h(t) \, e^{\,i\,\theta} \, e^{\,i\, 2 \pi \, f_{c} t }  \right\}
\, e^{- \,i\, 2 \pi \, f_{c} t }
\right]
= \sqrt{\recenergy} \, h(t) \, e^{\,i\,\theta} \,,
\end{equation*}
where the low pass filter (LPF) eliminates
frequency components in the vicinity of $2 f_{c}$ and
eliminates all noise outside the bandwidth of $h(t)$.

The optimum baseband receiver processing to counter AWGN is
the matched filter (see Section \ref{sec:matchedfilter} for a derivation of this).
As illustrated in Figure \ref{fig:ICHidea} a filter matched to $h(t)$ has
impulse response $\cc h( - t )$, yielding output signal $z(t)$, where
\begin{equation}
\label{eq:simplemf}
z(t) = \int_{- \infty}^{\infty} r(\tau) \, \cc h (\tau - t) \, d \tau \,.
\end{equation}
In particular 
\begin{equation*}
z(0) =  \sqrt{\recenergy} \, e^{\, i \, \theta} + \text{noise} \,.
\end{equation*}
Although $\theta$ is unknown to the receiver,
its effect can be eliminated by the energy estimate
\begin{equation*}
\left| z(0) \right|^{2} = \recenergy + \text{noise} \,,
\end{equation*}
yielding an estimate of the received energy $\recenergy$.

\section{The LS solution}
\label{sec:lscriterion}

Minimizing \eqref{eq:leastsquares} can be accomplished by differentiation with respect to $z$.
However, \eqref{eq:leastsquares} is not an analytic function of $z$ because it includes both $z$ and $z^*$. 
Written in the following form,
$\varepsilon$ is a real-valued analytic function of $z$  and $\cc z$  considered as independent variables.
\begin{equation}
\label{eq:LSerror}
\varepsilon (z, \cc z,\tau ) = \int_{\tau} ^{\tau  + T} 
\left( \, r(t) - z \cdot h(t - \tau) \right)
\left( \, \cc r (t) - \cc z \cdot \cc h (t - \tau) \right)
 \, dt \,.
\end{equation}
As shown in \citeref{176},
a stationary point can be determined by setting
\begin{equation*}
0 = \frac{\partial }{ {\partial \cc z } }\varepsilon (z, \cc z,\tau )
= -  \int_{\tau} ^{\tau  + T} 
\left( \, r(t) - z \cdot h(t - \tau) \right)
\cc h (t - \tau)
 \, dt \,.
\end{equation*}
Solving for $z$ yields \eqref{eq:leastsquaressolnZ},
and substituting $\hat z$ into \eqref{eq:LSerror} yields \eqref{eq:leastsquaressolnZ}.

\chapter{Statistical model for a coherence hole}
\label{sec:modelICH}

\section{Scintillation without modulation}
\label{sec:withoutmod}

The statistical properties of the model for an ICH given in \eqref{eq:ICHmodel}
will now be established.
Define the random variable $S$ as
\begin{equation*}
S = \sum_{n=1}^{N} r_n \, e^{\,i\, (\Phi_{n}}
\end{equation*}
where the $\{ r_n \}$ are fixed but unknown quantities
and the $\{ \Phi_n \}$ are independent and uniformly distributed.
By the Central Limit Theorem we can approximate $S$ as Gaussian
for reasonably large $N$.
It follows from independence and $\expec{e^{\,i\, \Phi_{m}}} = 0$ that $\expec{S}= 0$ and
\begin{equation*}
\expec{S^{2}}  = 
\expec{ \sum_{n=1}^{N} \sum_{m=1}^{N} r_n \, r_m \, e^{\,i\, ( \Phi_{m} + \Phi_{n}} ) }
= \sum_{n=1}^{N} r_n^{\,2} \, \expec{e^{\,i\, 2 \Phi_{m}}} = 0 \,.
\end{equation*}
This establishes that the $\Re \{ S \}$ and $\Im \{ S \}$ have equal variance and are uncorrelated and hence independent.
The variance of $S$ is
\begin{equation*}
\expec{ | S |^{2}} = 
\expec{ \sum_{n=1}^{N} \sum_{m=1}^{N} r_n \, r_m \, e^{\,i\, ( \Phi_{m} - \Phi_{n}} ) }
= \sum_{n=1}^{N} r_n^{\,2} \,.
\end{equation*}
For the $n \ne m$ terms, the expected values are zero due to independence, leaving
only the $n = m$ terms.

From \eqref{eq:noiseICH}, that $E |M|^2 = N_0$ is easily established.
More interesting is the conclusion that $\Re \{ M \}$ and $\Im \{ M \}$ have equal variance and are uncorrelated and hence independent.
This follows from
\begin{equation*}
\expec{ M^2 } = 
N_0 \int_0^T e^{-\,i\,4 \pi \, f_c t} \, \left( \cc h (t) \right)^2 \, dt = 0 \,,
\end{equation*}
where $\left( \cc h (t) \right)^2$ has bandwidth $2 B$, and after
modulation with carrier frequency $2 f_c$ its spectrum is guaranteed not to overlap
$f = 0$ and thus the integrand averages to zero.

\section{Scintillation and amplitude modulation}
\label{sec:withmod}

Suppose $h(t)$ is amplitude modulated by $A$,
where the flux for weak scattering given by \eqref{eq:ricianflux} is assumed.
Then the MF output is
\begin{equation*}
Z = \sqrt{\recenergy} \cdot \left( \sqrt{\gamma} \, 
e^{\,i\,\Psi} \, \sigma_{s} + \sqrt{1 - \gamma} \, S \right) \, A + M \,,
\end{equation*}
where $0 \le \gamma \le 1$ is the fraction of signal that arrives in the specular path.
Conditioned on a particular specular phase $\Psi = \psi$,
and writing the data symbol in polar coordinates $A = a \, e^{\,i\,\theta}$,
$Z$ becomes Gaussian with independent real- and imaginary parts,
and with first and second order statistics
\begin{align}
&\expec{\Re \{ Z \}} = \sqrt{\, \recenergy \, \gamma \,}  \, a \, \sigma_{s} \, \cos ( \psi + \theta )
\ \ \ \text{and} \ \ \ \  \expec{\Im \{ Z \}} = \sqrt{\, \recenergy \, \gamma \, } \, a \, \sigma_{s} \, \sin ( \psi + \theta )
\\
& \var{\Re \{ Z \}} = \var{\Im \{ Z \}} = \frac{\recenergy \, a^{2} \, (1 - \gamma ) \, \vs + N_{0}}{2} \,.
\end{align}
Observe that only $\expec{Z}$ (and not $\var{Z}$) is dependent on $\theta$.
When $\psi$ is unknown and uniformly distributed, $\expec{Z}$
yields no information on the value of $\theta$.
The conclusion is that any phase component in $A$ is rendered unobservable by scintillation
for any $\kappa$ and $\gamma$.
On the other hand $Q = |Z|^{2}$ makes a good real-valued decision variable
since its mean value includes a component of $a^{2}$ for any $0 \le \gamma \le 1$
(and in particular for strong scattering where $\gamma = 0$)
and any $\psi$, since
\begin{equation*}
\expec{|Z|^{2}} = \recenergy \, a^{2} \, \vs + N_{0} \,.
\end{equation*}
Particularly notable is the lack of dependence on $\gamma$,
or in other words the strength of the scattering.

\chapter{Derivation of fundamental limits} 
\label{sec:capacity}

We begin by reviewing a couple of bounding techniques that will prove indispensable,
and then proceed to proving some results on the fundamental limits on power efficiency.

\section{Probabilistic bounds}
\label{sec:bounding}

\subsection{Union bound}
\label{sec:unionbound}

A simple bound that often proves useful in bounding an event of the form
of ''A or B or C or $\dots$'' is the \emph{union bound}.
Mathematically, the term ''or'' turns into ''union''.
For any set of events $\{ A_i , \, 1 \le i \le N \}$,
\begin{equation*}
\prob{ \bigcup_{i=1}^N A _i } \le \sum_{i=1}^N \prob{A_i}
\end{equation*}
with equality if and only if the events are disjoint ($A_i \bigcap A_j = \varnothing$ for all $i \ne j$).

\subsection{Weak law of large numbers}
\label{sec:weaklaw}

Let $X$ be a random variable with average value $\expec{X} = \mu$.
Suppose that $\{ X_1, \, X_2 , \, \dots X_J \}$ is a set of $J$ identically distributed independent random variables
with the same distribution as $X$.
The weak law of large numbers asserts that for any $\epsilon > 0$
\begin{equation*}
\prob{ \left| \frac{1}{J} \sum_{k=1}^{J} X_k  - \mu \right| > \epsilon } \to 0 \ \ \ \ \text{as} \ \ \ \  J \to \infty \,.
\end{equation*}
Thus, as $J$ increases the probability distribution of the sum becomes increasingly concentrated about $\mu$.

\subsection{Chernoff bound}
\label{sec:chernoff}

The \emph{Chernoff bound} is one of the favorite tools of communications engineering, because it yields exponentially tight bounds on the tail probability for a sum of independent random variables.
Let $u(x)$ be the unit step function, and note that $u(x) \le e^{\, \rho \, x}$ for all $x$ and any $\rho \ge 0$.
Then for any random variable $X$, threshold $\lambda$, and $\rho \ge 0$,
\begin{align}
&\prob{X \ge \lambda} = \expec{u(X-\lambda)} \le \expec{e^{\, \rho ( X- \lambda )}}
= e^{\, - \rho \, \lambda} \, \expec{e^{\, \rho \, X}}
= e^{\, - \rho \, \lambda} \, \mgf{X}{\rho} \\
&\prob{X < \lambda} \le e^{\, \rho \, \lambda} \, \mgf{X}{- \rho}
\,,
\end{align}
where $\mgf{X}{t}$ is the moment generating function of $X$.
The bound can be minimized by the choice of $\rho \ge 0$.

Suppose that $\{ X_1, \, X_2 , \, \dots X_J \}$ is a set of $J$ identically distributed independent random variables
with the same distribution as $X$.
The Chernoff bound is particularly convenient for bounding the tail probability for the sum of independent
random variables, because the moment generating function is readily shown to be
\begin{equation*}
\expec{ \exp \left\{ \sum_{k=1}^{J} X_k \right\} } = \mgf{X}{\rho}^J  \,.
\end{equation*}

\section{Channel code based on M-ary FSK and time diversity}
\label{sec:capacitynoknowledge}

The  result stated in \eqref{eq:fundlimit2} will now be shown to be a consequence
of the strong law of large numbers applied to the decision variable $Q_m$ defined in \eqref{eq:diversitydecisionvar}.
This proof follows \citerefWloc{638}{Section II} closely, except for the added assumption that
energy bundles conform to the constraints of the ICH.
We also show that this result applies whether or not scintillation is present; that is,
for AWGN alone as well as the AWGN with scintillation.
The scintillation is assumed to be the result of strong scattering, so it is governed by Rayleigh statistics.

Without loss of generality, assume that the actual energy bundle falls in decision variable $Q_M$,
while $\{ Q_m , \, 1 \le m < M \}$ consist of noise only.
For a given threshold $\lambda$ there are two types of error that can occur
in the detection of the location of the energy bundle.
The first is a miss in the case of $m = M$, and the other is a false alarm in the case of $1 \le m < M$.
Assuming a threshold $\lambda$, the probabilities of these two types of error are
\begin{align}
&1- \pd = \prob{Q_M < \lambda } \\
& \pfa = \prob{ Q_m \ge \lambda} \,,
\end{align}
where $\pd$ is the probability of detection.
Since the $\{ Q_m , \, 1 \le m \le M \}$ are statistically independent, the probability of 
a symbol error is
\begin{equation}
\label{eq:peFSKpfapd}
\psymbol = 1 - \pd \cdot \left( 1 - \pfa \right)^{\,M-1} \,.
\end{equation}
A symbol error is defined as either a choosing the wrong location for the energy bundle
or an ambiguous outcome (an energy bundle is detected at two or more frequencies).
We will establish conditions under which $\psymbol \to 0$, which implies that the
probability of bit error $\pe \to 0$.
The union bound of Appendix \ref{sec:unionbound} yields a simpler expression,
\begin{equation*}
\psymbol \le 1 - \pd + (M-1) \, \pfa < 1 - \pd + M \, \pfa \,.
\end{equation*}
To show that $\psymbol \to 0$, it is only necessary to show that $\pd \to 1$ and $M \, \pfa \to 0$.

Address first $m = M$, where we are interested in showing that $\pd \to 1$.
The expected value of $Q_M$ is
\begin{equation}
\label{eq:diversitymeans}
\expec{Q_M} = \begin{cases}
\energy_h \, \vs + N_0  \,, & \text{AWGN + scintillation} \\
\energy_h + N_0  \,, & \text{AWGN alone} \,.
\end{cases}
\end{equation}
Assuming that $\vs = 1$, which is true for interstellar scattering,
the value of $\expec{Q_M}$ is identical with and without scintillation.
The weak law of large numbers of Appendix \ref{sec:weaklaw}
insures that $\pd \to 1$ as $J \to \infty$ as long as the threshold is chosen to
be less than the mean, or in particular the value
\begin{equation}
\label{eq:thresholdepsilon}
\lambda = N_0 + (1-\epsilon) \,  \energy_h
\end{equation}
for any value $0 < \epsilon < 1$ will suffice.
If we were to remove time diversity and let $J = 1$, then $\pd < 1$.
Even in the absence of scintillation this is true, and thus time diversity is necessary ($J \to \infty$) for the
results that follow, even in the absence of scintillation.\footnote{
An upper bound on the error probability $\pe$ for M-ary FSK on
an AWGN channel without scintillation is determined
in Appendix \ref{sec:peppm}.
The conclusion is that as long as the energy contrast ratio per bit $\ecrb > 2 \log 2$
then $\pe \to 0$ as $M \to \infty$.
Thus, in the absence of scintillation there is a 3 dB penalty in power efficiency
for setting $J = 1$, attributable to the phase-incoherent detection.
Time diversity in conjunction with low duty cycle are necessary to avoid this 3 dB penalty even in the absence of scintillation.
}

Address next the case $1 \le m < M$, where the $Q_m$ consist of noise
alone and our interest is in demonstrating conditions under which $M \, \pfa \to 0$.
For this purpose, since $Q_m$ is the sum of $J$ independent random variables
the Chernoff bound of Appendix \ref{sec:chernoff} is useful.
The moment generating function of $|Z|^2$ where $Z$ is a zero-mean complex-valued
Gaussian random variable with independent real and imaginary parts
and $\expec{|Z|^2} =  \sigma^2$,
\begin{equation}
\label{eq:mgfchisquare}
\mgf{|Z|^2}{\rho} = \frac{1}{1 - \sigma^2 \, \rho}
\ \ \ \text{for} \ \ \ \ 
\rho < \frac{1}{\sigma^2} \,.
\end{equation}
In the case of $\{ Q_m \,, \, 1 \le m < M \}$, $\sigma^2 = N_0$.
Thus the Chernoff bound (after choosing the optimum $\rho$) is
\begin{equation*}
\pfa \le \left(\frac{\lambda \, e^{1-\frac{\lambda
   }{N_0}}}{N_0}\right)^J \,.
\end{equation*}
Substituting for threshold $\lambda$,
\begin{align}
M \, \pfa < M \left(
e^{- \ecrs \, (1 - \epsilon)}
( 1+ \ecrs \, (1 - \epsilon))
\right)^J
\,,
\end{align}
where $\ecrs$ is the energy contrast ratio defined by \eqref{eq:ecrs}.
Substituting for $M$ from \eqref{eq:rate0} and $\energy_h$ from \eqref{eq:power0},
\begin{equation*}
M \, \pfa <
\left(
2^{\frac{\rate \, T}{\delta}} 
e^{- \frac{\power \, T \vs \, (1 - \epsilon)}{N_0 \, \delta }}
\left( 
1+\frac{\power \, T \, (1-\epsilon)}{\delta \, N_0}
\right)
\right)^{\,J} \,.
\end{equation*}
Thus, when the condition
\begin{equation}
\label{eq:conditRoverP}
\rate \, \log 2 <  \frac{\power \, (1 - \epsilon)}{N_0}
- \frac{\delta}{T} \, \log \left( 
1+\frac{\power \, (1-\epsilon)}{ N_0 \, \delta}
\right)
\end{equation}
is satisfied, $M \, \pfa \to 0$ as $J \to \infty$.
The term containing $\delta$ is positive and monotonically approaches $0$ as $\delta \to 0$.
Thus as long as
\begin{equation*}
\frac{\rate}{\power} <
\frac{\log_2 e}{N_0} \,,
\end{equation*}
then \eqref{eq:conditRoverP} is satisfied for sufficiently small values of $\epsilon$ and $\delta$.

That $\epsilon \to 0$ is necessary for $\psymbol \to 0$ shows that approaching the fundamental
limit requires precise knowledge of $\energy_h$ and $N_0$ in order to establish the threshold.
That $\delta \to 0$ implies that it is necessary for $\energy_h \to \infty$.
This latter fact explains why phase-incoherence detection is not an obstacle to approaching
the fundamental limit, even on the AWGN channel where phase-coherent detection would be possible.
As the signal energy grows without bound the noise is swamped out so the distinction 
between coherent and incoherent disappears.
	
\chapter{Coin tossing analogy}
\label{sec:cointossing}

Although it doesn't apply directly to interstellar communication, the
 analogy between the discovery of a continuous-time signal $h(t)$ in AWGN
 and the discovery of a specific binary sequence in the presence of noise created by
 random coin tosses may be helpful to intuition.
 This analogy explains in a simple and less-mathematical way why the shape of $h(t)$ \emph{doesn't} matter for
 detection in noise, and also why the shape of $h(t)$ \emph{does} matter with respect to interference immunity.
 
\section{Discrete matched filter}

Consider a reception which consists of a sequence of $K$ 
elements, where each element is either a HEAD or a TAIL,
\begin{equation*}
r_{k} =
\begin{cases}
\ \ h_{k} \,, & \text{signal} \\
\ \ n_{k} \,, & \text{noise} \,.
\end{cases}
\end{equation*}
We know that $n_{k}$ is generated randomly by nature, and can be modeled as
a sequence of statistically independent fair coin tosses.
Thus, each possible sequence $\{ n_{k} , \, 1 \le k \le K \}$ is equally likely
with probability $2^{\,- K}$.
In the other case, signal $\{ h_{k} , \, 1 \le k \le K \}$ is presumed to have been generated
by some technological means, rather than a natural process.
Our goal is to observe $\{ r_{k} , \, 1 \le k \le K \}$ and make a decision between
signal and noise.
One might surmise that one distinguishing characteristic of a technologically-generated
signal is that it was generated by an algorithm, or sequence of steps that could
be executed by a software program or generated by a piece of Boolean hardware.

Unfortunately there is no practical way to identify whether 
$\{ r_{k}  \}$ was generated by an algorithm, without making much
stronger assumptions about the characteristics of that algorithm.
For example, if the Boolean hardware consisted of a memory storing a sequence
$\{ h_{k} \}$, then that hardware could generate literally any
sequence $\{ r_{k} \}$.
If no assumptions are made, any $\{ r_{k} \}$ could have been generated
by noise with probability $2^{\,-K}$.

If on the other hand, if it is known that the signal is a single unique sequence
(or this is correctly guessed),
then the detection algorithm can simply compare $\{ r_{k} \}$ to that $\{ h_{k} \}$
and declare signal if there is agreement in all $K$ positions of the sequence.
In that case, the probability of detection is $\pd = 1$.
On the other hand, the detector will incorrectly declare signal if $\{ n_{k} \}$ happens to
replicate $\{ r_{k} \}$ precisely, which occurs with probability $2^{\,-K}$.
Thus, the false alarm probability is $\pfa = 2^{\,-K}$.
The important point is that this detector performance does not depend on
the signal sequence $\{ h_{k} \}$, because the noise $\{ n_{k} \}$ is completely random
and it can replicate any $\{ h_{k} \}$ with the same probability.
This is analogous to our result that the waveform $h(t)$ does not matter
in the presence of AWGN (which is also completely random).

Suppose due to our uncertainty about  $\{ h_{k} \}$, we allow the signal to be
any one of $N > 1$ specific sequences (or corresponding algorithms).
In this case it remains that $\pd = 1$ if one of those sequences is valid.
However, $\pfa$ increases with $N$, since  $\{ n_{k} \}$ can more easily replicate
one of $N$ sequences. In the limit as $N \to 2^{K}$, $\pfa \to 1$.
This illustrates why the detector must keep $N$ small to keep $\pfa$ under control.

\section{Immunity to discrete interference}

Consider now a very different challenge,
\begin{equation*}
r_{k} =
\begin{cases}
\ \ h_{k} \,, & \text{signal} \\
\ \ i_{k} \,, & \text{interferece} \,,
\end{cases}
\end{equation*}
where the noise sequence has been replaced by
$\{ i_{k} , \, 1 \le k \le K \}$, an unknown interference.
(Interference, when present,
is likely to be much larger than the noise, and thus
it is not unreasonable to assume that the receiver has to deal
with interference, or noise, but not both at the same time.)
The question is how we can choose signal $\{ h_{k} \}$
to make it unlikely to be replicated by $\{ i_{k} \}$,
but without knowledge of $\{ i_{k} \}$.
The transmitter designers have no knowledge of the origin of $\{ i_{k} \}$,
and thus are unwilling to even make any assumptions about its statistics.
We define the transmitter's challenge as somehow choosing $\{ h_{k} \}$
so that the false alarm probability $\pfa = 2^{\,-K}$ is a constant,
independent of $\{ i_{k} \}$ and with the same $\pfa$ as we saw with noise.
As just seen, the choice of $\{ h_{k} \}$ doesn't matter when it comes to noise,
so the transmitter is free to choose $\{ h_{k} \}$ any way we choose.

The following way of choosing $\{ h_{k} \}$ satisfies our goals.
Choose $\{ h_{k} \}$ randomly by performing $K$ independent fair coin tosses.
Having determined $\{ h_{k} \}$ in this manner, the choice of $\{ h_{k} \}$ can be shared with the receiver.
(In practice $\{ h_{k} \}$ would be generated by a pseudo-random coin toss generator,
and the receiver would have to guess or infer that algorithm and generate the same
sequence $\{ h_{k} \}$ itself.)
 where it is compared with $\{ r_{k} \}$, our discrete MF algorithm.
Now $\pfa$ is the probability that $K$ coin tosses exactly match
$\{ i_{k} \}$, which is $\pfa = 2^{\,-K}$, the same for any $\{ i_{k} \}$.
Thus, the detector performance is identically the same for noise and interference,
regardless of the interference.
If $\{ h_{k} \}$ where chosen in any other way, such as another random ensemble
with different statistics, then some interferers $\{ i_{k} \}$ would
be replicated by $\{ h_{k} \}$ with higher probability than others.

\chapter{Communication reliability}
\label{sec:performancecalc}

Throughout a random variable $V$ will be a normalized complex-valued Gaussian random variable with
\begin{equation}
\label{eq:normalcomplexgaussian}
\expec{V} = \expec{V^{2}} = 0 \ \ \ \text{and} \ \ \ \expec{|V|^{2}} =2 \,.
\end{equation}
Then the random variable
\begin{equation*}
X = \left| \sqrt{\xi} + V \right|^{2}
\end{equation*}
is a non-central $\chi^2$ random variable with non-centrality parameter $\xi$
and two degrees of freedom.
For $\xi = 0$ it is a central $\chi^2$ random variable with two degrees of freedom.)
Write this as $X \sim \chi^{2} (\xi)$, and let the
probability density and cumulative distribution functions be
$f (\xi,x)$ and $F (\xi,x)$.
The moment generating
function is
\begin{equation*}
\mathcal M (\xi, t) = \expec{e^{t X}} = \frac{\displaystyle e^{\frac{ \xi  \, t}{1-2 \, t}}}{1-2 \, t}
\ \ \ \text{for} \ \ \  2\,  t < 1 \,.
\end{equation*}

\section{Multilevel on-off keying (OOK)}
\label{sec:multillevelOOK}

It is easy to see that multilevel modulation is less power efficient
than two-level OOK.
Consider a more data symbol constellation
$A_{k} \in \{  n , \, 0 \le n \le N-1 \}$, where
$N = 2$ corresponds to OOK, and the spacing between levels is
kept constant to keep the noise immunity fixed.
If each point in the constellation is equally likely, then the average
energy in each symbol is
\begin{equation*}
\recenergy \cdot \expec{A^{2}} = \recenergy \cdot \frac{(2 N - 1)(N-1)}{6}
\end{equation*}
and the information embedded in each symbol is $\log_{\,2} N$.
The power efficiency is the
ratio the latter to the former, and it is maximized by $N = 2$.
For example, choosing $N=4$ (two bits per symbol) reduces the
power efficiency relative to $N=2$ by a factor of 3.5 (5.4 dB).

\section{Probability of error for OOK}
\label{sec:peook}

On-off keying (OOK) provides a baseline case similar to the
types of signals often assumed in the SETI literature.
They therefore provide a baseline against which to compare
channel codes that seek the greatest power efficiency.
The bit error probability will now be determined in the presence of
AWGN, with and without Rayleigh fading
(which models scintillation in the presence of strong scattering).

\subsection{Additive white Gaussian noise (AWGN)}
\label{sec:OOKawgnOnly}

In the absence of fading,
the decision variable is $Q = | \sqrt{\energy_h} + N  |^2$, where
$\energy_h = 0$ for a zero bit and $\energy_h > 0$ for a one bit.
In contrast to FSK, we do not know that there is an energy bundle
present, so there is no way to avoid the need for knowledge of $N_0$.
Assume that $Q$ is applied to a threshold $\lambda$, it is natural to choose
$\lambda = \energy_h + N_0$, which is halfway between the two mean values.
The probabilty of a false alarm
(energy bundle detected when there is none present) is
\begin{equation}
\label{eq:pfape}
\pfa =
\prob{ |N|^2 > \lambda } = e^{\, -1 - \ecrs} \,.
\end{equation}
The probability of a miss (energy bundle not detected when there is one present) is
\begin{equation*}
1 - \pd =
\prob{ | \sqrt{\energy_h} + N |^2 < \lambda } \,.
\end{equation*}
There is no closed-form expression for $\pd$, but the
Chernoff bound of Appendix \ref{sec:chernoff} gives us an exponentially tight bound,
\begin{align*}
1 - \pd &\le \expec{ \exp \left\{ - \rho \left(  | \sqrt{\energy_h} + N |^2 - \lambda \right)  \right\} }
= e^{\, \rho \, \lambda } \, \mgf{ | \sqrt{\energy_h} + N |^2}{- \rho} \\
&= \frac{ \exp \left\{ \rho \,  \left(-\frac{2 \, \energy_h}{N_0 \, \rho
   +1}+\energy_h+N_0\right) \right\} }{N_0 \, \rho +1}
\end{align*}
for any $\rho > 0$.
Choosing $\rho$ to minimize the right side,
\begin{equation*}
\label{eq:pmisspe}
1 - \pd \le \frac{2 \, (\ecrs+1) \,  \exp \left(\frac{\ecrs \, \left(-3 \,
   \sqrt{8 \, \ecrs^{\,2}+8 \, \ecrs+1}+8 \, \ecrs+5\right)}{\sqrt{8 \, \ecrs^{\,2}+8 \, \ecrs+1}+1}\right)}{\sqrt{8 \, \ecrs^{\,2}+8 \,  \ecrs+1}+1}
\end{equation*}
Assuming that the two cases are equally likely, $\pe$ is half the sum of \eqref{eq:pfape} and \eqref{eq:pmisspe}.
This result is plotted in Figure \ref{fig:gapFundamental}.

\subsection{AWGN and scintillation}

The decision variable is $Q = | \sqrt{\energy_h} \, S + N  |^2$, where
$\energy_h = 0$ for a zero bit.
This case is easier because the statistics are $\chi^2$ rather than
noncentral $\chi^2$, and closed form results are easy to obtain.
Apply $Q$ to a threshold $\lambda = N_0 + a$ for some value of $a > 0$.
Then assuming the two cases are equally likely, the error probability is
\begin{equation*}
\pe = \frac{1}{2} \left( 1- \pd + \pfa \right) =
\frac{1}{2}
   \left(1 -e^{-\frac{a+N_0}{\energy_h+N_0}}+e^{-\frac{a+N_0}{N_0}}\right) \,.
\end{equation*}
Choosing $a$ to minimize $\pe$, this becomes
\begin{equation*}
\pe =
\frac{1}{2}
   (\ecrs+1)^{-\frac{\ecrs+1}{\ecrs}}
   \left((\ecrs+1)^{\frac{1}{\ecrs}}+\ecrs
   \left((\ecrs+1)^{\frac{1}{\ecrs}}-1\right)\right) \,.
\end{equation*}
This relation is plotted in Figure \ref{fig:gapFundamental}.

\section{Error probability for M-ary FSK and time diversity}
\label{sec:pefsktd}

The probability of bit error $\pe$ will now be upper bounded
for the combination of M-ary FSK with time diversity using
the maximum algorithm of \eqref{eq:maxAlgorithm}.
This differs from a similar development in Appendix \ref{sec:capacitynoknowledge},
where a threshold algorithm is used instead.
While the threshold algorithm is necessary to 
asymptotically approach the fundamental limit,
the maximum algorithm has greater practical appeal
for finite $M$ and $J$ because the detection algorithm requires no knowledge of
the signal parameters $\{ \energy_h , \, N_0 , \, \ecrs \}$.
A bound is determined because calculation of the exact
error probability requires messy numerical integrations.
The Chernoff bounding technique of Appendix \ref{sec:chernoff}
is employed because it gives exponentially tight bounds for small $\pe$.

The more general formulation of
Section \ref{sec:discoveryWdiversity} is used, since the
simpler case of Section \ref{sec:diversitystrategy} is simply a special case for
$J_0 = 0$ and $J_1 = J$.
Let $\psymbol$ be the symbol error probability, or in other words the probability that the
maximum detection algorithm chooses the wrong location.
Since $\psymbol$ does not depend on the actual location, assume the actual energy bundle
is located at $m = M$.
Then
\begin{equation*}
\psymbol =
\prob{
\left( \max_{1 \le m \le M-1} Q_m \right) > Q_M
}
=  \prob{
\bigcup_{m = 1}^{M-1} \left( Q_m > Q_M \right)
}
\,.
\end{equation*}
The union bound of Appendix \ref{sec:unionbound}
replaces the union by a sum,
\begin{equation*}
\psymbol \le
\sum_{m = 1}^{M-1}  \prob{ Q_m > Q_M}
= (M-1) \cdot \prob{ Q_1 > Q_M}
 \,,
\end{equation*}
since all the $\{ Q_m , \, 1 \le m < M \}$ are identically distributed.

Each term can be upper bounded using the Chernoff bound.
For any $\rho > 0$,
\begin{equation*}
\prob{Q_1 > Q_M } = \prob{Q_1 - Q_M > 0} \le \expec{\exp \{ - \rho \, (Q_1 - Q_M)  \}}
= \mgf{Q_1}{- \rho} \cdot \mgf{Q_M}{\rho} \,.
\end{equation*}
The moment generating function $\mgf{|Z|^2}{\rho}$ where $Z$ is complex-Gaussian
with variance $\sigma^2$ is given by \eqref{eq:mgfchisquare}.
Each random variable $\{ Q_m , \, 1 \le m \le M \}$ is the sum of $(J_0 + J_1 )$ 
independent random variables, so the moment generation functions are
\begin{align*}
&\mgf{Q_1}{\rho} = \left( \frac{1}{1 - \rho \, N_0}  \right)^{J_0 + J_1}
\\
&\mgf{Q_M}{\rho} =  \left(  \frac{1}{1 - \rho \, (\energy_{h,0} \, \vs + N_0)}  \right)^{J_0}
 \cdot \left(  \frac{1}{1 - \rho \, (\energy_{h,1} \, \vs + N_0)}  \right)^{J_1} 
\end{align*}
for $\rho < 1/(\energy_{h,0} \, \vs + N_0)$.
The final bound on $\psymbol$ is determined by minimizing over $\rho < 1/(\energy_{h,0} \, \vs + N_0)$.
We omit the final result, which is complicated.

Now let's relate $\psymbol$ to $\pe$.
Consider the set of $M$ possible symbols.
To calculate the bit error probability, assume there has been a random symbol error.
For any particular bit position in the symbol, $M/2$ of the symbols disagree
and $M/2$ symbols (including the correct one) agree.
Assuming that all $M-1$ incorrect symbols are equally likely, the
conditional probability of bit error is $M/2$ divided by $M-1$, and
\begin{equation}
\label{eq:symboltobit}
\pe = \frac{M/2}{M-1} \cdot \psymbol \,.
\end{equation}
Specializing to $J_0 = 0$ results in the simple expression of \eqref{eq:peVecrsFSKtd}.

\section{Error probability for M-ary FSK}
\label{sec:peppm}

The analysis of Appendix \ref{sec:pefsktd} can be repeated for the outage strategy.
One difference is the absence of time diversity ($J_0 = 0$ and $J_1 = 1$),
which allows us to avoid bounding in favor of a closed form result.
The second difference is non-central $\chi^2$ statistics for $Q_M$.

For the normalized MF outputs the symbol error probability
$\psymbol$ equals
the probability that one or more of the first $(M-1)$ matched filter outputs is larger in
magnitude than the $M$-th output,
\begin{align}
\psymbol &=
\prob{
\left( \max_{1 \le m \le M-1} | V_m |^2 \right) > \left| \sqrt{2 \, \log_2 M \cdot \ecro} + V_{M} \right|^2
} \\
&= \prob{
\bigcup_{m = 1}^{M-1} \left( |V_m|^2 >  
\left| \sqrt{2 \, \log_2 M \, \ecro} + V_{M} \right|^2 \right)
} \\
&= \int_{0}^{\infty} \prob{ \bigcup_{m = 1}^{M-1} \left( |V_m|^2 >  x \right) }
f ( 2 \, \log_2 M \, \ecro , \, x ) \, dx
\,.
\end{align}
The union bound
replaces the union by a sum,
\begin{align}
\notag
\psymbol   &\le
\sum_{m = 1}^{M-1}  \int_{0}^{\infty} \prob{ |V_m|^2 > x } f ( 2 \, \log_2 M \, \ecro , x ) \, dx \\
\notag
 &= (M-1) \cdot  \int_{0}^{\infty} (1 - F(0,x) ) \cdot  f ( 2 \, \log_2 M \, \ecro  , x ) \, dx \\
 \notag
&= (M-1) \cdot \mathcal M ( 2 \, \log_2 M \, \ecro  \, , - 1/2 ) \\
\label{eq:bound1}
&= \frac{1}{2} \, (M-1) \, M^{-\frac{\ecro }{2 \, \log 2}}
 \,.
\end{align}
Combining \eqref{eq:bound1} and \eqref{eq:symboltobit} the result is \eqref{eq:peFSKbound}.

\subsection{Error Probability for (M,N)-ary FSK}
\label{sec:pemnaryfsk}

After normalization, each of the $M$ matched filter outputs takes the form
\begin{align*}
&\left| \sqrt{\frac{2 \, \energy_h}{N_0}} + V \right|^{\,2} =
\left| \sqrt{2 \, \xi} + V \right|^{\,2} \\
&\xi = \frac{\energybit \log_2 \begin{pmatrix} M\\L \end{pmatrix} }{L \, N_0}
\,.
\end{align*}
For $(M-L)$ of the $M$ outputs, $\energy_h = 0$.
The probability of a single noise-only estimate being larger than a single signal-plus-noise
estimate is $e^{\,- \xi /2}/2$.
There are $L (M-L)$ different ways this type of error can occur, so by the
union bound of Appendix \ref{sec:unionbound} the probability of symbol error is bounded by
\begin{equation*}
\psymbol \le L \, (M-L) \, e^{\,- \xi /2}/2
\end{equation*}
and the bit error probability $\pe \approx \psymbol / 2$.
Thus we get that for a given value of $\pe$
\begin{equation*}
\ecrb = \frac{\energybit}{N_0} \ge \frac{ 2 L \, \log \left( \frac{L \, (M-L) }{4 \pe} \right) }{ \log_2 \begin{pmatrix} M\\L \end{pmatrix} }
\end{equation*}

The maximum spectral efficiency $\spectrale$ occurs when the duty factor $\delta = 1$,
in which case
\begin{equation*}
\spectrale = \frac{ \log_2 \begin{pmatrix} M\\L \end{pmatrix} / T }{M \, B} 
= \frac{ \log_2 \begin{pmatrix} M\\L \end{pmatrix}}{M}
\,,
\end{equation*}
where the choice $B T = 1$ also maximizes $\spectrale$.

\chapter{Discovery reliability}
\label{sec:discprob}

\section{Detection probability for single observation}
\label{sec:discprobsingle}

Consider the decision variable $Q$ is given by \eqref{eq:QR1K}.
An equivalent condition to $Q > \lambda$ is
\begin{equation}
\label{eq:discoverytest1}
(1 - \lambda) \cdot 
\left| \, \sqrt{\recenergy} \, \left( \sqrt{\,\gamma \,} \, \sigma_{s} \, e^{\,i\,\psi} + \sqrt{\,1-\gamma\,} \, S \right) + M_{1} \, \right|^{2} > \lambda \cdot \sum_{k=2}^{K} \left| M_{k} \right|^{2} 
\,.
\end{equation}
Specular phase $\psi$ does not affect the statistics, so set $\psi = 0$.
Define a normalized noise $V_{k} = M_{k} / \sqrt{N_{0}/2}$ with
unit-variance real and imaginary parts.
Then \eqref{eq:discoverytest1} becomes
\begin{equation}
\label{eq:discoverytest2}
\frac{1 - \lambda}{\lambda} \cdot V > W
\end{equation}
where
\begin{align*}
V &= \left| \, \sqrt{\frac{2 \, \recenergy}{N_{0}}} \, \left( \sqrt{\,\gamma \,} \, \sigma_{s} + \sqrt{\,1-\gamma\,} \, S \right) + V_{1} \, \right|^{2} \\
W &=  \sum_{k=2}^{K} \left| V_{k} \right|^{2} 
\end{align*}

The distribution $W$ is $\chi^{2}$ with $(2 \, K - 1)$ degrees of freedom,
and only the statistics of $V$ depend on the parameters of the model.
Since the statistics of $W$ are divorced from all the parameters, this
relation demonstrates how $Q$ is sensitive to the signal-to-noise ratio
$\ecrs$ rather than signal and noise energies individually.

\subsection{Rayleigh fading}
\label{sec:discoveryrayleigh}

For $\gamma = 0$, closed-form results for
false alarm probability $\pfa$ and detection probability $\pd$ can be obtained.
The simplification for this case is that $( \sqrt{2 \, \recenergy/N_{0}} \, S + V_{1} )$ is a zero-mean
Gaussian random variable with variance $2 \, (1+\ecrs)$.
Hence \eqref{eq:discoverytest2} can be rewritten as
\begin{equation}
\label{eq:V1W}
\alpha \cdot \left| U_{1} \right|^{2} 
 > W
 \ \ \ \text{for} \ \ \ \
\alpha = \frac{1-\lambda}{\lambda} \cdot (1 + \ecrs)
\end{equation}
where $U_{1}$ is also a Gaussian random variable
with unit-variance real- and imaginary parts.
In \eqref{eq:V1W}
$\left| U_{1} \right|^{2}$ is $\chi^{2}$ with two degrees of freedom
and $W$ is $\chi^{2}$ with $(2 \, K - 2)$ degrees of freedom.
The probability of event \eqref{eq:V1W} is readily determined,
\begin{equation}
\label{eq:probQgl}
\pd
= \int F \left( \frac{w}{\alpha}, \, 2 \right) \, f (w, \, 2 K - 2) \, dw
= \left( \frac{\alpha}{1+\alpha} \right)^{K-1} 
 \,,
\end{equation}
where $F (x,n)$ and $f (x,n)$ are the cumulative distribution
and probability density functions for a $\chi^{2}$ distribution with $n$ degrees of freedom.
Then $\pfa$ equals \eqref{eq:probQgl} for $\ecrs = 0$, and
eliminating  $\lambda$ from these two relations yields \eqref{eq:pdvspfa}.

When $N_{0}$ is known, the probability of $Q = |R_{1}|^{2} > \lambda$ is
\begin{equation*}
\pd = F \left( \frac{2 \, \lambda}{\recenergy \, \vs + N_{0}} , \, 2 \right) =
\exp \left\{ - \frac{\lambda}{\recenergy \, \vs + N_{0}} \right\} \,.
\end{equation*}
$\pfa$ is obtained by setting $\recenergy = 0$.
Eliminating $\lambda$ to solve for $\pd$ in terms of $\pfa$ yields
\begin{equation}
\label{eq:PfaVsPd}
\pd = \pfa^{\,\frac{1}{1+\ecrs}} \,,
\end{equation}
which is repeated as \eqref{eq:asympPfaVsPd}.

\subsection{Rician fading}
\label{sec:discoveryrician}

When $\gamma > 0$, $V$ in \eqref{eq:discoverytest2} is non-central $\chi^{2}$
with two degrees of freedom.
There is no easy closed-form expression for $\pd$ as in the Rayleigh fading case.
Our intuition that Rayleigh fading is ''worst case'' in the sense of being more
difficult case to discover (because of the absence of a specular component)
can be confirmed by calculating a signal-to-noise ratio.
The first and second moments of $V$ are
\begin{align*}
&\expec{V} = 2 \, ( 1 + \ecrs ) \\
&\var{V} =  4 \, \left( 1 + 2 \, \ecrs + (1 - \gamma^{2} ) \, \ecrs^{\,2} \right) 
\,.
\end{align*}
A larger $\pd$ and smaller $\pfa$ are associated with
a bigger difference between mean values (signal plus noise vs noise alone) and
a smaller standard deviation.
These effects can be captured in a signal-to-noise ratio measure
\begin{equation*}
\snr_{\text{Rice}} =
\frac{\left( \expec{V} - \left. \expec{V} \right|_{\recenergy=0} \right)^{\,2}}{\var{V}} 
= \frac{\ecrs^{\,2}}{1 + 2 \, \ecrs + (1 - \gamma^{2}) \, \ecrs^{\,2}}
\,.
\end{equation*}
$\snr_{\text{Rice}}$ is plotted in Figure \ref{fig:snrRician} for different values of $\gamma$.
It is notable how small $\snr_{\text{Rice}}$ is for small values of $\gamma$.
This is the influence of the increasing prominence of outages when the
Rayleigh case is approached.
However, $\snr_{\text{Rice}}$ improves as $\gamma$ gets larger, reflecting the
increasing discovery reliability, particularly for larger $\ecrs$.

\incfig
	{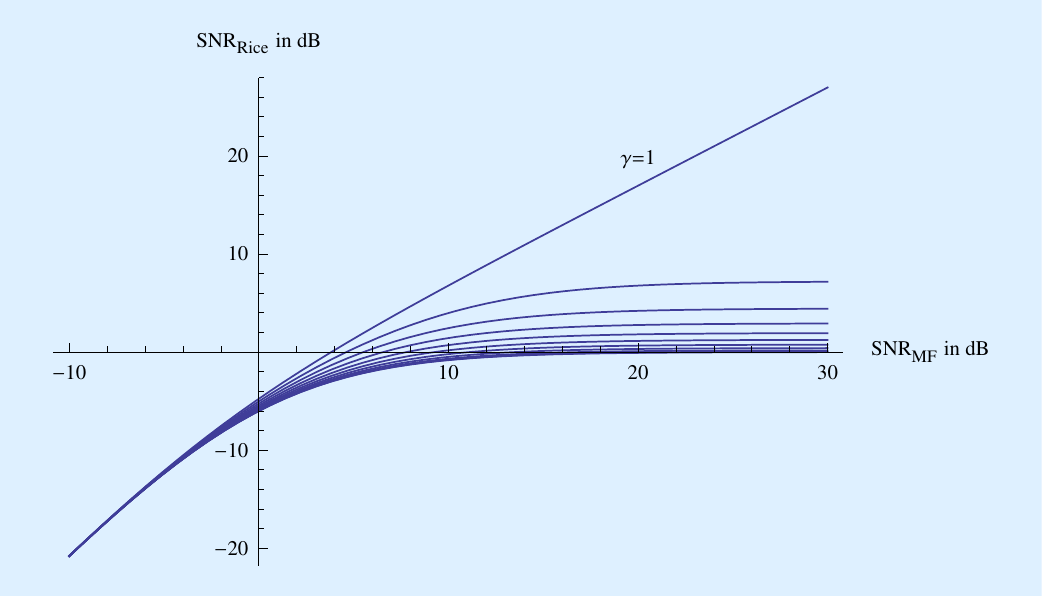}
	{.8}
	{
	A plot of $\snr_{\text{Rice}}$ in dB vs $\ecrs$ in dB for $\gamma=0$ to $1$
	in steps of $0.1$. For a larger specular component ($\gamma$ larger)
	$\snr_{\text{Rice}}$ gets larger, reflecting the increasing reliability in signal flux
	due to the increasing proportion of flux arriving in the specular component.
	}

\section{Detection probability for multiple independent observations}
\label{sec:Lobs}

The discovery relation of \eqref{eq:discoveryequation} will now be derived.
Consider three granularities of observation:
(a) individual energy bundle, (b) one contiguous observation window, and (c)
the collection of $L$ observations.
It is assumed that the received flux remains constant over a contiguous observation window,
but is statistically independent from one observation window to the next.
Let the probabilities at granularity (a) be  $\{ \pfaa, \, \pda \}$
and at granularity (c) $\{ \pfa, \, \pd \}$.
Thus,  $\{ \pfa, \, \pd \}$ reflect the overall sensitivity of the $L$-observation detector,
where a ''detection'' is defined as one or more detected energy bundles (at the level of (a))
in the $L$ observations.
We limit our analysis to the simpler Rayleigh fading case,
where closed-form results are obtainable.

\subsection{False alarm probability}

Suppose there is no beacon present
If $h(t)$ has bandwidth $B$, then the output of an MF also has bandwidth $B$
and can be sampled at a rate $B$ without losing information.
Then the total number of samples in $L$ observations, each of duration $T_{o}$, is
\begin{equation}
\label{eq:NvsDeltaK}
N = L \,T_{o} \, B = L \, K \left( \frac{\delta_{o}}{\delta_{p}} + 1 \right) \,.
\end{equation}
$N$ is the collective degrees of freedom in all the observations.
The false alarm probability in one of these samples is $\pfaa$,
and if they are statistically independent then
the false alarm probability at granularity (c) is
\begin{equation}
\label{eq:pfavspfaa}
1- \pfa = (1 - \pfaa)^{N}  \ \ \ \text{or}\ \ \ \ \pfa \approx N \cdot \pfaa \,.
\end{equation}
Typically $\pfaa$ is very small, so the approximation is very accurate.

The independence of these samples depends on $h(t)$.
Let $h(t)$ have Fourier transform $H(f)$, and let the autocorrelation
function of $h(t)$ be $R_{h} (t)$, which is the inverse Fourier transform of $| H(f) |^{2}$.
Then the noise at the output of the MF has a power spectral density $| H(f) |^{2}$
and autocorrelation function $R_{h} (t)$.
The samples of noise at the MF output are Gaussian and independent if
and only if $R_{h} (k/B) = \delta_{k}$, or equivalently
\begin{equation*}
\sum_{m} \, \left| H \left( f - \frac{m}{B} \right) \, \right|^{2} =
\frac{1}{B} \,,  \ \ \ 0 \le f < B \,.
\end{equation*}
For example, if $h(t)$ is ideally band limited to $f \in [0,B]$, then this condition becomes
\begin{equation*}
| H (f ) | =
\begin{cases}
\displaystyle
\frac{1}{\sqrt{B}}, & \ \ 0 \le f < B \\
0, & \ \ \text{otherwise} \,.
\end{cases}
\end{equation*}
Thus, we can say that the samples at the output of the MF
are statistically independent (\eqref{eq:NvsDeltaK} is valid)
if, roughly,
the energy of $h(t)$ is uniformly spread over the range of
frequencies $f \in [0,B]$.
Equivalently the magnitude of $H(f)$ is an ideal lowpass filter with an arbitrary
phase response.

\subsection{Detection probability}

Suppose now that a periodic beacon with period $T_{p}$ is present.
The relationship between $\pfaa$ and $\pda$ for a filter matched to an
individual energy bundle is given by \eqref{eq:PfaVsPd},
\begin{equation}
\label{eq:PfaaVsPda}
\pda = \pfaa^{\,\frac{1}{1+\ecrs}} \,.
\end{equation}
Now consider a single observation of duration $T_{o}$.
If the observation time is limited to $T_{o} \le T_{p} + T$, then
there is either zero (probability $1- \delta_{o}$) or one  (probability $\delta_{o}$)
energy bundle overlapping this observation.

There are two cases to consider.
If there is an energy bundle overlapping this observation,
out of the $B \, T_{o}$ samples at the output of a MF, we can assume
that all but one harbor noise only.
(Again this is for the case $R_{h} (k /B) = \delta_{k}$.)
Strictly speaking a detection can occur due to (a) a detection with
probability $\pda$ for the one sample containing signal or (b)
a false alarm with probability $\pfaa$ in any of the other samples.
In practice $\pfaa \approx 0$ is very small relative to $\pda \approx 1$, so we
can ignore those false alarms and accurately approximate the detection probability
for a single observation overlapping an energy bundle by $\pda$.
For the case where there is no energy bundle overlapping the observation,
we can accurately approximate the detection probability as $0$.
Thus, the probability of detection in one observation is
\begin{equation*}
\prob{ \text{energy packet detected} } \approx
\pda \cdot \delta_{o} + 0 \cdot (1-\delta_{o}) =
\delta_{o} \, \pda
\,.
\end{equation*}

Finally, consider the aggregate of $L$ observations.
Assuming those observation times are random relative to the period of the beacon,
 the $L$ binary variables representing ''detection'' or ''no detection''
in each of the $L$ observations are statistically independent.
With that assumption, the number of ''detections'' has a binomial distribution.
A ''miss'' occurs only in the case where all $L$ observations yield ''no detection'',
which occurs with probability
\begin{equation}
\label{eq:PdaAndDeltaVsPd}
1 - \pd = \left( 1 - \delta_{o} \, \pda \right)^{L}
\,.
\end{equation}
Combining \eqref{eq:pfavspfaa}, \eqref{eq:PfaaVsPda}, and \eqref{eq:PdaAndDeltaVsPd},  
\eqref{eq:discoveryequation} follows.

\chapter{Plasma dispersion}

\section{Coherence bandwidth $B_c$}
\label{sec:plasmaappx}

Integrating both sides of \eqref{eq:delayFromPhase},
the phase in \eqref{eq:passbanddispersiontf} is, for $f > f_c$,
\begin{equation*}
\phi(f) = \phi (f_{c}) - 2 \pi \, \int_{f_{c}}^{f} \tau(u) \, du
= \phi (f_{c}) -  2 \pi \, \dc \, \dm \, \left( \frac{1}{f_{c}} - \frac{1}{f} \right) \,.
\end{equation*}
Expanding $\phi (f)$ in a Taylor series about $f = f_{c}$,
\begin{equation}
\label{eq:taylor}
\phi(f) =
\phi (f_{c}) -  2 \pi \, \dc \, \dm
\left(
\frac{f-f_{c}}{f_{c}^{\,2}} - \frac{(f-f_{c})^{2}}{f_{c}^{\,3}}  + \frac{(f-f_{c})^{3}}{f_{c}^{\,4}}  - \dots
\right) \,.
\end{equation}
The linear term is benign because it corresponds to a fixed group delay
added to a very large (and not accurately known) propagation delay,
so the actual distortion is caused by the second-order and higher terms.
Neglecting the third-order and higher terms
(which are smaller than the second-order term by a power of $B / f_c \ll 1$),
the second-order term 
has variation less than $\alpha \pi$ for $f \in [f_{c},f_{c}+B]$ if
\begin{equation*}
2 \pi \, \dc \, \dm \,
\frac{ B^{\,2}}{f_{c}^{\,3}} \ll \alpha\, \pi
\end{equation*}
which is yields \eqref{eq:dispersioncoherencebandwidth}.
In that case, \eqref{eq:taylor} becomes for $f \in [f_{c},f_{c}+B]$
\begin{equation*}
\phi(f) \approx
\phi (f_{c}) -  2 \pi \, \tau (f_{c}) \cdot ( f - f_{c} ) \,.
\end{equation*}
Thus, \eqref{eq:passbanddispersiontf} becomes
\begin{equation*}
G(f)
\approx
e^{ \, i \, \phi (f_{c})}
\cdot
e^{ \, - i \, 2 \pi \, (f - f_{c})  \, \tau (f_{c})}
\,.
\end{equation*}
The equivalent baseband transfer function is $G(f+f_{c})$, and this
too is illustrated in Figure \ref{fig:dispersion}.

\section{Partial equalization}
\label{sec:equalization}

Even though $\dm$ refers to a measurable physical property
pertaining to the line of sight between transmitter and receiver,
the state of knowledge of $\dm$
may be different at the transmitter and the receiver.
Astronomical observations provide
a priori knowledge about $\dm$ \citeref{213} in the form
of upper and lower bounds $\dm_{L} \le \dm \le \dm_{U}$.
The resulting uncertainty is captured by $\deltadm = (\dm_{U} - \dm_{L})$.
A more advanced civilization may possess tighter bounds
and a smaller $\deltadm$, but can probably never reduce the uncertainty entirely.

There is the possibility of equalization for any specific
value of $\dm$ by introducing by adding an artificial group delay such that
the artificial group delay plus the group delay corresponding to $\dm$
is constant over $f \in [0,B]$.
This can be done in either transmitter or receiver, and
the combined effect is indistinguishable from a benign fixed delay
which adds to the very large propagation delay.
Specifically a causal allpass filter with response
$F_{\text{eq}} (f) = \exp \{ i \, 2 \pi \, \phi_{eq} (f) \}$, where
\begin{equation}
\label{eq:preeq}
\tau_{\text{eq}} (f) = \dc \cdot \dm \cdot \left( \frac{1}{f_{c}^{\,2}} - \frac{1}{(f+f_{c})^{2}} \right)
\ \ \ \text{and} \ \ \ \ 
\frac{\phi_{\text{eq}} (f)}{2 \pi} = - \dc \cdot \dm \cdot \frac{f^{\,2}}{f_{c}^{\,2} \, (f+f_{c})} 
\end{equation}
placed before or after the dispersion results 
overall in a constant group delay $\dc \cdot \dm / f_{c}^{\,2}$.
Done in the transmitter, this pre-equalization has no effect on
the noise.
Done in the receiver, this post-equalization 
does change the noise \emph{waveform} but has no 
effect on the noise \emph{statistics} since $F_{\text{eq}} (f)$ is
an all pass (phase-only) filter \citerefWloc{43}{Chap. 8}.

An issue for the receiver to consider is whether or not the transmitter
has performed pre-equalization
$\dm_L$, after which $0 \le \dm \le \deltadm$.
Since $\deltadm < \dm_U$, this would increase the dispersion coherence bandwidth $B_0$.
Thus, a receiver estimating a range of $B_{0}$ for a given line of sight
prior to discovery may wish to consider two cases: with and without pre-equalization.

\chapter{Coherence time with fading}
\label{sec:fadinganalysis}

The Fresnel approximation
expands the left side of \eqref{eq:dwv} in a Taylor series and retains terms 
up to $x^{2}$ and $t^{2}$.
The constant term is a flat delay and is discarded.
The result is the right side of \eqref{eq:dwv}
\begin{equation}
\label{eq:vappx}
v(x) = v_{\parallel} - \frac{v_{\bot} x}{D_{2}} - \frac{v_{\parallel} x^{2}}{2 \, D_{2}^{\,2}}
\approx 
v_{\parallel} - \frac{v_{\bot} x}{D_{2}} \,.
\end{equation}
The fact that the net velocity $v(x)$ depends on $x$ limits the time coherence,
because different rays have slightly different Doppler shifts which causes their
superposition to change with time.

Within an ICH, where the distortion of $h(t)$ by plasma dispersion can be neglected,
the observer sees, associated with a ray through point $x$, the passband response
\begin{equation*}
\Re \left\{ h \left( t - \frac{d(x) - v(x) \, t}{c} \right) 
\exp \left\{ \,i\,2 \pi \, f_{c} \, \left( t - \frac{d(x) - v(x) \, t}{c} \right) \right\} \right\} \,.
\end{equation*}
Within an ICH, the effect of delay on $h(t)$ can be neglected, and thus the 
equivalent complex baseband signal is
\begin{equation}
\label{eq:threeterms}
h \left( t \right) \, \exp \left\{ - \,i\,2 \pi \, \frac{ f_{c} \,d(x) }{c}) \right\} \,
\exp \left\{ \,i\,2 \pi \, \frac{f_{c} \,v(x) \, t}{c} \right\} \,.
\end{equation}
The first two terms in \eqref{eq:threeterms} capture the effect for a stationary observer,
but the third velocity-dependent term can be approximating using \eqref{eq:vappx},
\begin{equation*}
 \exp \left\{ \,i\,2 \pi \, \frac{f_{c} \,v(x) \, t}{c} \right\} \approx
 \exp \left\{ \,i\,2 \pi \, \frac{f_{c} \,v_{\parallel} \, t}{c} \right\} \exp \left\{ - \,i\,2 \pi \, 
 \frac{f_{c} \,v_{\bot} \, x \, t }{c \, D_{2}} \right\} \,.
\end{equation*}
The effect of $v_{\parallel}$ can be compensated by a simple carrier frequency offset
as described in Section \ref{sec:doppler}.
That $v_{\parallel}$ can be rendered impotent in this manner is not surprising,
since this component of velocity (aligned with the light of sight) affects all rays
in the same way.
The second term is a time-varying phase at baseband which affects the
time coherence
unless the total phase shift it causes over $t \in [0, T]$ is limited to $\alpha \pi$, which is condition \eqref{eq:tcf}.

\chapter{Search granularity in dilation and frequency}
\label{sec:searchgranulatiry}

Any mismatch between the time dilation parameter $\beta$ used in the
implementation of the MF and the actual value of $\beta$ results in incoherence.
The same is true for any mismatch between the carrier frequency $f_{1}$ and reality.
These sources of incoherence are totally under the control of the receiver in its
choice of a granularity in a search over these parameters.
Hence we don't put them in the same category as natural sources of incoherence.

\section{Time dilation error}

Since a search over $\beta$ will generally need a small range,
write it as $\beta = 1 + \delta$, and ask how $\delta \ne 0$ affects coherence.
This can be studied by assuming that $r(t) = h(t)$ but the MF used
is $\cc h (- (1+\Delta \delta) t)$.
 the uncertain range in order to find a close enough
The largest acceptable value of $| \Delta \delta |$ depends to some extent on $h(t)$.
We now show that for a signal $h(t)$ with a flat magnitude spectrum over bandwidth $0 \le f \le B$,
the MF output is affected very little when $| \Delta \delta | \ll 1/ \pi K$.
Thus the fractional dilation offset is on the order of $1/\pi K$, which implies that
an $h(t)$ with larger $K$ requires a finer-grain search over the dilation parameter.
Time dilation error is a factor in both time and frequency coherence, since a given error $\Delta \delta$
places an error on the product $K = B T$.

The Fourier transform of $h ((1+ \Delta \delta) t)$ is $H(f/ (1+ \Delta \delta))$.
Assume that $\Delta \delta > 0$ (the final result is the same for $\Delta \delta < 0$).
Then the MF output is
\begin{equation*}
Z = \int_{0}^{B/(1+\Delta \delta)} H(f) \, \cc H \left( \frac{f}{1+\Delta \delta}  \right) \, df
=  \frac{1}{B} \, \int_{0}^{B/(1+\Delta \delta)} \exp \{ i \left( \phi(f) - \phi \left( \frac{f}{1+\Delta \delta} \right) \right) \, df
\end{equation*}
The group delay $\tau (f)$ is defined by \eqref{eq:delayFromPhase}.
Since $h(t)$ is confined to $t \in [0,T]$, it follows that $0 \le \tau (f) \le T$.
The sensitivity of $Z$ to $\Delta \delta$ can be gauged by the derivative
\begin{equation*}
v = \left. \frac{\partial Z}{\partial \Delta \delta} \right|_{\Delta \delta = 0} 
= \frac{i}{B} \, \int_{0}^{B} f \frac{\partial \phi (f)}{\partial f}  \, df - 1
= \frac{- i \, 2 \pi}{B} \, \int_{0}^{B} f \, \tau (f) \, df -1
\end{equation*}
where
\begin{equation*}
0 \le  \int_{0}^{B} f \, \tau (f) \, df \le \frac{B^{2} T}{2} 
\ \ \ \ \text{or} \ \ \ \ \ 
1 \le | v |^{2} \le 1 + \pi^{2} \, K^{2} \,.
\end{equation*}
Thus, $|Z|$ is relatively little affected if  $\Delta \delta | v | \ll 1$ or
$\Delta \delta \ll 1/ \pi K$.

\section{Carrier frequency offset}

As with delay and dilation, the receiver will calculate the MF output for
discrete values of carrier frequency.
There may be a mismatch $\Delta \ecrb$ between reality and the receiver's assumption.
In this case, the MF output becomes 
\begin{equation*}
Z = \int_{0}^{T} | h(t) |^{2} e^{-\,i\,2 \pi \, \Delta \ecrb \, f_{c} \, t} \, dt \,.
\end{equation*}
The offset will have an insignificant effect of $Z$ if
the total phase shift is much less than $\pi$,
or $| \Delta \ecrb | \ll 1/2 f_{c} T$.
Time coherence is maintained if
the absolute carrier frequency spacing is small relative to $1/T$.

\end{makefigurelist}



\addcontentsline{toc}{chapter}{Bibliography}
\bibliographystyle{IEEEtran}
\small
\bibliography{SETI-references,patents}

\newpage
\normalsize
\chapter*{AUTHOR}
\addcontentsline{toc}{chapter}{Author}
  
David G. Messerschmitt is the Roger A. Strauch Professor Emeritus of Electrical Engineering and Computer Sciences (EECS) at the University of California at Berkeley. At Berkeley he has previously served as the Chair of EECS and the Interim Dean of the School of Information. He is the co-author of five books, including \emph{Digital Communication} (Kluwer Academic Publishers, Third Edition, 2004). He served on the NSF Blue Ribbon Panel on Cyberinfrastructure and co-chaired a National Research Council (NRC) study on the future of information technology research. His doctorate in Computer, Information, and Control Engineering is from the University of Michigan, and he is a Fellow of the IEEE, a Member of the National Academy of Engineering, and a recipient of the IEEE Alexander Graham Bell Medal recognizing ``exceptional contributions to the advancement of communication sciences and engineering''.


\end{document}